\documentstyle[12pt,uwmthesis,graphics,psfig]{report}
\newcommand{\half}{\frac{1}{2}}
\newcommand{\om}{\omega}
\newcommand{\bom}{{\bar\omega}}
\newcommand{\ep}{\epsilon}
\newcommand{\bi}{\begin{itemize}}
\newcommand{\ei}{\end{itemize}}
\newcommand{\be}{\begin{equation}}
\newcommand{\ee}{\end{equation}}
\newcommand{\ba}{\begin{array}}
\newcommand{\ea}{\end{array}}

\newcommand{\ds}{\displaystyle}

\newcommand{\bea}{\begin{eqnarray}}
\newcommand{\eea}{\end{eqnarray}}
\newcommand{\bean}{\begin{eqnarray*}}
\newcommand{\eean}{\end{eqnarray*}}
\newcommand{\sinY}{\sin\theta \partial_{\theta} Y_l^m}
\newcommand{\cosY}{\cos\theta Y_l^m}

\newcommand{\nn}{\nonumber}
\newcommand{\tmr}{\frac{2M_0}{R}}
\newcommand{\rx}{\frac{r}{R}}
\newcommand{\ry}{1-\frac{r}{R}}
\newtheorem{thm}{Theorem}
\def\lesssim{\mathrel{\mathchoice {\vcenter{\offinterlineskip\halign{\hfil
$\displaystyle##$\hfil\cr<\cr\sim\cr}}}
{\vcenter{\offinterlineskip\halign{\hfil$\textstyle##$\hfil\cr<\cr\sim\cr}}}
{\vcenter{\offinterlineskip\halign{\hfil$\scriptstyle##$\hfil\cr<\cr\sim\cr}}}
{\vcenter{\offinterlineskip\halign{\hfil$\scriptscriptstyle##$\hfil\cr<\cr\sim\cr}}}}}
\def\grsim{\mathrel{\mathchoice {\vcenter{\offinterlineskip\halign{\hfil
$\displaystyle##$\hfil\cr>\cr\sim\cr}}}
{\vcenter{\offinterlineskip\halign{\hfil$\textstyle##$\hfil\cr>\cr\sim\cr}}}
{\vcenter{\offinterlineskip\halign{\hfil$\scriptstyle##$\hfil\cr>\cr\sim\cr}}}
{\vcenter{\offinterlineskip\halign{\hfil$\scriptscriptstyle##$\hfil\cr>\cr\sim\cr}}}}}
\newcommand{\apj}{{\it Astrophysical J.}}
\newcommand{\aap}{{\it Astron. and Astrophys.}}
\newcommand{\cmp}{{\it Commun. Math. Phys.}}
\newcommand{\grg}{{\it Gen. Rel. Grav.}}
\newcommand{\lr}{{\it Living Reviews in Relativity}}
\newcommand{\mnras}{{\it Mon. Not. Roy. Astr. Soc.}}
\newcommand{\pr}{{\it Phys. Rev.}}
\newcommand{\prl}{{\it Phys. Rev. Lett.}}
\newcommand{\prd}{{\it Phys. Rev. D}}
\newcommand{\prsl}{{\it Proc. R. Soc. Lond. A}}
\newcommand{\ptrsl}{{\it Phil. Trans. Roy. Soc. London}}
\newcommand{\rmp}{{\it Rev. Mod. Phys.}}
\begin{document}
%
\title{Stability and Rotational Mixing of Modes in Newtonian
and Relativistic Stars}
\author{\bf Keith H. Lockitch}
\program{Physics}
\majorprof{John L. Friedman}
\degree{Doctor of Philosophy}
\submitdate{August, 1999}
\copyrightyear{1999}
\copyrighttrue
\doctoratetrue
\figurespagetrue
\tablespagetrue
\dblabstractfalse
\multivolumesfalse
\multiminorsfalse

\Abstract{Almost none of the r-modes ordinarily found in 
rotating stars exist, if the star and its perturbations 
obey the same one-parameter equation of state; and rotating 
relativistic stars with one-parameter equations of state 
have no pure r-modes at all, no modes whose limit, for a 
star with zero angular velocity, is an axial-parity
oscillation.  Rotating stars of this kind similarly have 
no pure g-modes, no modes whose spherical limit is a 
perturbation with polar parity and vanishing perturbed 
pressure and density. Where have these modes gone?

In spherical stars of this kind, r-modes and g-modes form a 
degenerate zero-frequency subspace.  We find that rotation 
splits the degeneracy to {\it zeroth} order in the star's 
angular velocity $\Omega$, and the resulting modes are 
generically hybrids, whose limit as $\Omega\rightarrow 0$ 
is a stationary current with axial and polar parts.  
Lindblom and Ipser have recently found these hybrid modes 
in an analytic study of the Maclaurin spheroids.  We present 
the first calculation of these modes in relativistic stars.  

Because each mode has definite parity, its axial and polar
parts have alternating values of $l$.   We show that each 
mode belongs to one of two classes, axial-led or polar-led, 
depending on whether the spherical harmonic with lowest 
value of $l$ that contributes to its velocity field is axial 
or polar.  We numerically compute these modes for slowly 
rotating newtonian polytropes and Maclaurin spheroids, and 
for slowly rotating relativistic stars with uniform density.
Timescales for the gravitational-wave driven instability and 
for viscous damping are computed for the hybrid modes of the 
newtonian models using assumptions appropriate to neutron stars.  
The instability to nonaxisymmetric modes is, as expected, 
dominated by the $l=m$ r-modes with simplest radial dependence, 
the only modes which retain their axial character in newtonian 
isentropic models.  For relativistic isentropic stars, these 
$l=m$ modes are replaced for $l\geq2$ by axial-led hybrids.
We find analytically the post-newtonian corrections to these 
modes for uniform density stars.}
\beforepreface
\prefacesection{Acknowledgements}

\vspace{0.5in}
\begin{center}
{\it To Robert and Gillian Lockitch, my first teachers.}
\end{center}

\vspace{0.5in}
\noindent
It is a pleasure to thank my advisor, John Friedman, for 
his guidance throughout my graduate studies and
for his constant encouragement during our collaboration.
For my graduate education in theoretical physics I am also 
indebted to Leonard Parker, Bruce Allen, Yutze Chow and the 
late Nick Papastamatiou. 

I am particularly grateful to Nils Andersson, Lee Lindblom 
and Sharon Morsink for numerous discussions and for helpful 
comments on various aspects of this research.  I also thank Jim 
Ipser, Yasufumi Kojima, Kostas Kokkotas, Ben Owen, Bernard Schutz 
and Nick Stergioulas for helpful discussions and for sharing 
related work in progress, and Warren Anderson for his assistance 
in producing Fig. 1.  

This work was partly conducted at the Albert Einstein 
Institute during the workshop ``Neutron star dynamics and
gravitational wave emission.''  I am grateful to the AEI for 
their generous hospitality.  This work has also been supported
in part by fellowships from the UWM Graduate School and by
NSF grant PHY-9507740.

My family has been a constant source of love and support
and I thank them with all my heart. In particular, I am
deeply grateful to my wife, Cornelia, whose patience and 
encouragement have sustained me throughout my graduate
studies.

\afterpreface


\chapter{Introduction}

\section{Background and Motivation}

This dissertation examines a new class of oscillation modes 
of rotating stars.  The work has been motivated by the recently 
discovered r-mode instability (see below), and answers a number 
of previously unresolved questions concerning the nature of the 
r-mode spectrum in newtonian and relativistic stellar models.

The structure and stability of rotating relativistic stars has 
recently been reviewed in detail (Stergioulas \cite{s98}, Friedman 
\cite{f96, f98}, Friedman and Ipser \cite{fi92}), and a general 
discussion of the small oscillations of relativistic stars may be 
found in a recent review article by Kokkotas \cite{k96} (see also 
Kokkotas and Schmidt \cite{ks99}).  In this work, we will focus our 
attention on non-radial oscillations, which were first studied 
in relativistic stars by Thorne and collaborators (Thorne and 
Campolattaro \cite{tc67}, Price and Thorne \cite{pt69}, Thorne 
\cite{t69a,t69b}, Campolattaro and Thorne \cite{ct70}, Ipser and 
Thorne \cite{it73}).  

The spherical symmetry of a non-rotating star implies that its
perturbations can be divided into two classes, polar or axial, 
according to their behaviour under parity.  Where polar tensor 
fields on a 2-sphere can be constructed from the scalars $Y_l^m$
and their gradients $\nabla Y_l^m$ (and the metric on a 2-sphere),
axial fields involve the pseudo-vector $\hat r\times \nabla Y_l^m$, 
and their behavior under parity is opposite to that of $Y_l^m$.  
That is, axial perturbations of odd $l$ are invariant under parity, 
and axial perturbations with even $l$ change sign. 

It is useful to further divide stellar perturbations into subclasses
according to the physics dominating their behaviour.  This
classification was first developed by Cowling \cite{c41} for the polar 
perturbations of newtonian polytropic models.  The f- and p-modes
are polar-parity modes having pressure as their dominant restoring 
force.  They typically have large pressure and density perturbations
and high frequencies (higher than a few kilohertz for neutron stars).  
The other class of polar-parity modes are the g-modes, which are 
chiefly restored by gravity.  They typically have very small pressure 
and density perturbations and low frequencies.  Indeed, for isentropic 
stars, which are marginally stable to convection, the g-modes are all
zero-frequency and have vanishing perturbed pressure and density 
(see Sect. 2.1).  Similarly, all axial-parity perturbations of 
newtonian perfect fluid models have zero frequency in a non-rotating 
star. The perturbed pressure and density as well as the radial component 
of the fluid velocity are all rotational scalars and must have polar 
parity.  Thus, the axial perturbations of a spherical star are simply 
stationary horizontal fluid currents (see Sect. 2.1).

The analogues of these modes in relativistic models of neutron stars
have been studied by many authors.  More recently, an additional class 
of outgoing modes has been identified that exist only in relativistic 
stars.  Like the modes of black holes, these are essentially associated 
with the dynamical spacetime geometry and have been termed w-modes, or 
gravitational wave modes.  Their existence was first argued by Kokkotas 
and Schutz \cite{ks86}.  The polar w-modes were first found by Kojima 
\cite{k88} as rapidly damped modes of weakly relativistic models, while 
the axial w-modes were first studied by Chandrasekhar and Ferrari 
\cite{cf91b} as scattering resonances of highly relativistic models.  
(See the reviews by Kokkotas \cite{k96} and Kokkotas and Schmidt 
\cite{ks99}.)

In general, this classification of modes also describes the 
oscillations of rotating stars, although the character of the modes
may be significantly affected by rotation.  Because a rotating star 
is also invariant under parity, its perturbations can be classified 
according to their behaviour under parity.  If a mode varies 
continuously along a sequence of equilibrium configurations that 
starts with a spherical star and continues along a path of increasing 
rotation, the mode will be called axial if it is axial for the 
spherical star.  Its parity cannot change along the sequence, but $l$ 
is well-defined only for modes of the spherical configuration.
 
Rotation imparts a finite frequency to the axial-parity perturbations 
of newtonian models.  Because these modes are restored by the Coriolis 
force, their frequencies are proportional to the star's angular velocity, 
$\Omega$.  These rotationally restored axial modes were first 
studied by Papaloizou and Pringle \cite{pp78}, who called them r-modes
because of their similarity to the Rossby waves of terrestrial 
meteorology.  For a normal mode of the form 
$e^{i(\sigma t+m\varphi)}$, Papaloizou and Pringle found the 
r-mode frequency to be,
\be
\sigma+m\Omega = \frac{2m\Omega}{l(l+1)}.
\label{p&p_freq}
\ee

It is only rather recently that the oscillation modes of rotating 
relativistic stars have begun to be accessible to numerical 
study (see below). Early work on the perturbations of such stars 
focused mainly on the criteria for their stability, and led to the 
surprising discovery that all rotating perfect fluid stars are 
subject to a non-axisymmetric instability driven by gravitational 
radiation.  The instability was discovered by Chandrasekhar 
\cite{ch70} in the $l=m=2$ polar mode of the uniform-density, 
uniformly rotating Maclaurin spheroids.  Although this mode is 
unstable only for rapidly rotating models, by looking at the 
canonical energy of arbitrary initial data sets, Friedman and 
Schutz \cite{fs78b} and Friedman \cite{f78} showed the instability 
to be a generic feature of rotating perfect fluid stars.  

In essence, the CFS (Chandrasekhar-Friedman-Schutz) instability 
operates by converting the rotational energy of the star partly
into the oscillation energy of the perturbation and partly into 
gravitational waves.  For a normal mode of the form 
$e^{i(\sigma t+m\varphi)}$ this nonaxisymmetric instability 
acts in the following manner.  

In a non-rotating star, gravitational radiation removes positive 
angular momentum from a forward moving mode and negative angular 
momentum from a backward moving mode, thereby damping all 
time-dependent, non-axisymmetric modes. In a star rotating 
sufficiently fast, however, a backward moving mode can be dragged 
forward as seen by an inertial observer; and it will then radiate
positive angular momentum.  The mode continues to carry negative 
angular momentum because the perturbed star has lower total angular 
momentum than the unperturbed star.  As positive angular momentum
is removed from a mode with negative angular momentum, the angular 
momentum of the mode becomes increasingly negative, implying that
the amplitude of the mode increases.  Thus, the mode is driven by 
gravitational radiation.

Since the instability acts on modes that are retrograde with
respect to the star, but prograde as seen by an inertial observer, 
a mode will be unstable if and only if its frequency satisfies the 
condition,
\be
\sigma(\sigma+m\Omega) < 0.
\label{criterion}
\ee
For the polar f- and p-modes, the frequency is large and 
approximately real.  Condition (\ref{criterion}) will be met 
only if $|m\Omega|$ is of order $|\sigma|$, so that for a given 
angular velocity the instability will set in first through modes 
with large $m$. 

The CFS instability spins a star down by allowing it to radiate
away its angular momentum in gravitational waves.  However, to
determine whether this mechanism may be responsible for limiting
the rotation rates of actual neutron stars, one must also 
consider the effects of viscous damping on the perturbations.  
Detweiler and Lindblom \cite{dl77} suggested that viscosity would
stabilize any mode whose growth time was longer than the viscous
damping time, and this was confirmed by Lindblom and Hiscock 
\cite{lh83}. Recent work has indicated that the 
gravitational-wave-driven instability can only limit the rotation 
rate of hot neutron stars, with temperatures above the superfluid 
transition point, $T\sim 10^9\mbox{K}$, but below the temperature 
at which bulk viscosity apparently damps all modes, 
$T\sim 10^{10}\mbox{K}$. (Ipser and Lindblom \cite{il91}; Lindblom 
\cite{l95} and Lindblom and Mendell \cite{lm95})  
Because of uncertainties in the temperature of the superfluid phase 
transition and in our understanding of the dominant mechanisms for 
effective viscosity, even this brief temperature window is not 
guaranteed.

To calculate the timescales associated with viscous and radiative
dissipation it is necessary to compute explicitly the normal modes 
of oscillation.  Until recently, the polar f- and p-modes were 
expected to dominate the CFS instability through their coupling 
to mass multipole radiation.  As we have already noted, all 
axial-parity fluid oscillations are time-independent in a 
spherical model, and therefore do not couple to gravitational 
radiation at all. (Thorne and Campolattaro \cite{tc67})
In a rotating star, the rotationally restored r-modes do couple
to current multipole radiation.  However, their low frequencies 
and negligible perturbed densities in newtonian stars made it 
seem implausible that their contribution to gravitational
radiation would compare to that of the polar-parity modes.

Indeed, apart from studies of the axial-parity oscillations of 
models of the neutron star crust (van Horn \cite{vh80}, Schumaker and 
Thorne \cite{st83}), the axial modes were almost universally ignored 
in the early research on perturbations of relativistic stars.  
It has only been recently that interest in these modes has been 
revived, following the work of Chandrasekhar and Ferrari on the 
resonant scattering of axial wave modes \cite{cf91b} and on the 
coupling between axial and polar modes induced by stellar rotation 
\cite{cf91a}.  (Further recent studies of axial modes are reviewed
by Kokkotas \cite{k96} and Kokkotas and Schmidt \cite{ks99}.)

Thus, the first explicit calculations of the dissipative 
timescales associated with neutron star oscillations focused
on the $l=m$ f-modes (Ipser and Lindblom \cite{il91}, Lindblom 
\cite{l95}, Lindblom and Mendell \cite{lm95}), and until very 
recently it was only these modes that had been studied in 
connection with the CFS instability.  It has long been hoped 
that some neutron stars rotate sufficiently fast to be subject 
to the CFS instability, and that the gravitational radiation produced 
might be detectable by gravitational wave observatories.  However, 
based on the initial studies of dissipation in f-mode oscillations, 
the prospects were rather unpromising.  Neutron stars formed from 
stellar collapse would certainly be hot enough to pass through
the temperature window at which viscous damping is apparently 
suppressed, but there was little evidence that neutron stars formed 
in supernovae (or by the accretion-induced collapse of white dwarves) 
rotate rapidly enough for the onset of instability. Following an 
early suggestion of Papaloizou and Pringle \cite{pp78b}, Wagoner 
\cite{w84} had proposed another scenario in which an old, accreting 
neutron star, spun up past the onset of nonaxisymmetric instability 
would achieve an equilibrium state with angular momentum acquired by 
accretion balanced by angular momentum radiated in gravitational waves. 
This scenario, too, appeared to have been ruled out by the strength of 
damping by mutual friction and viscosity at the temperatures expected 
for such stars, $T\sim 10^8\mbox{K}$.

Very recently, however, a series of surprising results have 
emerged that dramatically improve these prospects.  

The first surprise was the discovery that the r-modes are CFS unstable 
in perfect fluid models with arbitrarily slow rotation.  First indicated 
in numerical work by Andersson \cite{a97}, the instability is implied
in a nearly newtonian context by the newtonian expression for the r-mode 
frequency (\ref{p&p_freq}), which satisfies the CFS instability 
criterion, (\ref{criterion}), for arbitrarily small $\Omega$,
\be
\sigma(\sigma+m\Omega) = 
- \frac{2(l-1)(l+2)m^2\Omega^2}{l^2(l+1)^2} < 0.
\ee
A computation by Friedman and Morsink \cite{jfs97} of the canonical 
energy of initial data showed (independent of assumptions on the existence
of discrete modes) that the instability is a generic feature of 
axial-parity fluid perturbations of relativistic stars.

As we have just observed, the generic instability of perfect fluid 
models will be of no astrophysical importance if, in actual stars, the 
unstable modes are damped by viscous dissipation.  Studies of the viscous 
and radiative timescales associated with the r-modes (Lindblom et al. 
\cite{lom98}, Owen et al. \cite{o98}, Andersson et al. \cite{aks98}, 
Kokkotas and Stergioulas \cite{ks98}, Lindblom et al. \cite{lmo99}) have 
revealed a second surprising result:  The growth time of r-modes driven
by current-multipole gravitational radiation is significantly shorter 
than had been expected.  In fact, it has turned out be so short for some 
of the r-modes that their instability to gravitational radiation reaction 
easily dominates viscous damping in hot, newly formed neutron stars.  
A neutron star that is rapidly rotating at birth now appears likely to 
spin down by radiating most of its angular momentum in gravitational waves.  
(See, however, the caveats indicated below.)

Hot on the heels of these theoretical surprises was the discovery
by Marshall et.al. \cite{m98} of a fast (16ms) pulsar in a supernova 
remnant (N157B) in the Large Magellanic Cloud.  Estimates of the initial
period put it in the 6-9ms range, thus providing the long-sought
evidence of a class of neutron stars that are formed rotating 
rapidly.  Hence, the newly discovered instability appears to set
the upper limit on the spin of the newly discovered class of
neutron stars!


The current picture that has emerged of the spin-down of a hot, 
newly formed neutron star can be readily understood in terms of 
a model of the r-mode instability due to Owen, Lindblom, Cutler, 
Schutz, Vecchio and Andersson (hereafter OLCSVA) \cite{o98}.  
Since one particular mode (with spherical harmonic indices $l=m=2$ 
and frequency $\sigma = -4\Omega/3$) is expected to dominate the 
r-mode instability, the perturbed star is treated as a simple system 
with two degrees of freedom: the uniform angular velocity $\Omega$ of 
the equilibrium star, and the (dimensionless) amplitude $\alpha$ of 
the $l=m=2$ r-mode. 
Initially, the neutron star forms with a temperature large 
enough for bulk viscosity to damp any unstable modes, 
$T \grsim 10^{10}\mbox{K}$; the star is assumed to be rotating 
close to its maximum (Kepler) velocity, $\Omega_K\sim \sqrt{M/R^3}$, 
the angular velocity at which a particle orbits the star's equator.
The star then cools by neutrino emission at a rate given by a 
standard power law cooling formula (Shapiro and Teukolsky 
\cite{st83b}). Once it reaches the 
temperature window at which the $l=m=2$ r-mode can go unstable, 
the system is assumed to evolve in three stages.  

First, the amplitude of the r-mode undergoes rapid exponential 
growth from some arbitrary tiny magnitude. Using conservation of 
energy and angular momentum, OLCSVA derive the following equations 
for the evolution of the system in this stage.
\be
\frac{d\Omega}{dt} = -\frac{2\Omega}{\tau_V}
\frac{\alpha^2 Q}{1+\alpha^2 Q}
\label{owen1}
\ee
\be
\frac{d\alpha}{dt} = \frac{\alpha}{|\tau_{GR}|}
-\frac{\alpha}{\tau_V} \frac{1-\alpha^2 Q}{1+\alpha^2 Q}
\label{owen2}
\ee
Here, $\tau_{GR}$ and $\tau_V$ are, respectively, the 
timescales for the growth of the mode by gravitational radiation 
reaction and the damping of the mode by viscosity (see Sect. 2.6).  
(The parameter $Q$ is a constant of order 0.1 related to the 
initial angular
momentum and moment of inertia of the equilibrium star.)
Since the initial amplitude $\alpha$ of the mode is so small, 
the angular momentum changes very little at first 
(Eq. (\ref{owen1})).  That this stage is characterized by
the rapid exponential growth of $\alpha$ is the statement
that the first term in Eq. (\ref{owen2}) (the radiation 
reaction term) dominates over the second (viscous damping).

Eventually the mode will grow to a size at which linear
perturbation theory is insufficient to describe its behaviour.
It is expected that a non-linear saturation will occur, halting
the growth of the mode at some amplitude of order unity, 
although the details of these non-linear effects are poorly 
understood at present.  When this saturation occurs, the system 
enters a second evolutionary 
stage during which the mode amplitude remains essentially 
unchanged and the angular momentum of the star is radiated 
away.  During this stage OLCSVA evolve their model system 
according to the equations
\be
\alpha^2=\kappa
\label{owen3}
\ee
\be
\frac{d\Omega}{dt} = - \frac{2\Omega}{|\tau_{GR}|}
\frac{\kappa Q}{1-\kappa Q}
\label{owen4}
\ee
where $\kappa$ is constant of order unity parameterizing 
the uncertainty in the degree of non-linear saturation.
The star spins down by Eq. (\ref{owen4}), radiating away most 
of its angular momentum while continuing to cool gradually.  

When its temperature and angular velocity are low enough that 
viscosity again dominates the gravitational-wave-driven 
instability, the mode will be damped.  During this third
stage, OLCSVA return to Eqs. (\ref{owen1})-(\ref{owen2}) to 
continue the evolution of their system.  That the mode amplitude 
decays is the statement that the second term in Eq. (\ref{owen2})
(the viscous damping term) dominates the first (radiation
reaction), at this temperature and angular velocity.

The net effect of this three-stage evolutionary process is
that the newly formed neutron star is left with an angular
velocity small compared with $\Omega_K$.  This final 
angular velocity appears to be fairly insensitive to the 
initial amplitude of the mode and to its degree of non-linear 
saturation. 
A final period $P\grsim5-10\mbox{ms}$ apparently rules out
accretion-induced collapse of white dwarves as a mechanism 
for the formation of millisecond pulsars with 
$P\lesssim 3\mbox{ms}$.

The r-mode instability has also revived interest in the Wagoner 
\cite{w84} mechanism, involving old neutron stars spun up by 
accretion to the point at which the accretion torque is balanced 
by the angular momentum loss in gravitational radiation.
Bildsten \cite{b98} and Andersson, Kokkotas and Stergioulas 
\cite{akst98} have proposed that the r-mode instability might 
succeed in this regard where the instability to polar modes 
seems to fail.  However, the mechanism appears to be highly
sensitive to the temperature dependence of viscous damping.
Levin \cite{l98} has argued that if the r-mode damping is a 
decreasing function of temperature (at the temperatures expected
for accreting neutron stars, $T\sim 10^8\mbox{K}$) then viscous
reheating of the unstable neutron star could drive the system 
away from the Wagoner equilibrium state.  Instead, the star would 
follow a cyclic evolution pattern.  Initially, the runaway reheating 
would drive the star further into the r-mode instability regime and 
spin it down to a fraction of its angular velocity. Once it has 
slowed to the point at which the r-modes become damped, it would 
again slowly cool and begin to spin up by accretion.  Eventually, 
it would again reach the critical angular velocity for the onset of 
instability and repeat the cycle.  Since the radiation
spin-down time is of order 1 year, while the accretion spin-up time 
is of order $10^6$ years, the star spends only a small fraction of the
cycle emitting gravitational waves via the unstable r-modes.  
This would significantly reduce the likelihood that detectable 
gravitational radiation is produced by such sources. 
On the other hand, if the r-mode damping is independent of - or
increases with - temperature (at $T\sim 10^8\mbox{K}$) 
then the Wagoner equilibrium 
state may be allowed (Levin \cite{l98}).  Work is currently in 
progress (Lindblom and Mendell \cite{lm99}) to investigate the r-mode 
damping by mutual friction in superfluid neutron stars, which was the 
dominant viscous mechanism responsible for ruling out the Wagoner 
scenario in the first place (Lindblom and Mendell \cite{lm95}).

Other uncertainties in the scenarios described above are still 
to be investigated.  There is substantial uncertainty in the 
cooling rate of neutron stars, with rapid cooling expected if stars 
have a quark interior or core, or a kaon or pion condensate. Madsen 
\cite{m98} suggests that an observation of a young neutron star with 
a rotation period below $5-10\mbox{ms}$ would be evidence for a quark 
interior; but even without rapid cooling, the uncertainty in the 
superfluid transition temperature may allow a superfluid to form at 
about $10^{10} \mbox{K}$, possibly killing the instability.  
We noted above the expectation that the growth of the unstable 
r-modes will saturate at an amplitude of order unity due to 
non-linear effects (such as mode-mode couplings); however,
this limiting amplitude is not yet known with any certainty and
could be much smaller.  In particular, it has been suggested that 
the non-linear evolution of
the r-modes will wind up the magnetic field of a neutron star, 
draining energy away from the mode and eventually suppressing the
unstable modes entirely (Rezzolla, Lamb and Shapiro \cite{rls99};
see also Spruit \cite{s98}). 

The excitement over the r-mode instability has generated a large 
literature.
(Andersson \cite{a97},
Friedman and Morsink \cite{jfs97},
Kojima \cite{k98},
Lindblom et al. \cite{lom98},
Owen et al. \cite{o98},
Andersson, Kokkotas and Schutz \cite{aks98},
Kokkotas and Stergioulas \cite{ks98},
Andersson, Kokkotas and Stergioulas \cite{akst98},
Madsen \cite{mad98},
Hiscock \cite{h98},
Lindblom and Ipser \cite{li98},
Bildsten \cite{b98},
Levin \cite{l98},
Ferrari et al. \cite{fms98},
Spruit \cite{s99},
Brady and Creighton \cite{bc98},
Lockitch and Friedman \cite{lf98} (see Ch. 2),
Lindblom et al. \cite{lmo99}, 
Beyer and Kokkotas \cite{bk99}, 
Kojima and Hosonuma \cite{kh99}, 
Lindblom \cite{l99}, 
Schneider et al. \cite{sfm99},
Rezzolla et al. \cite{rea99}, 
Yoshida and Lee \cite{yl99})
It has also generated a number of questions which have not been 
properly answered, some of which are addressed in this dissertation.

Despite the sudden interest in the r-modes they are not yet 
well-understood for stellar models appropriate to neutron
stars.  A neutron star is accurately described by a perfect
fluid model in which both the equilibrium and perturbed 
configurations obey the same one-parameter equation 
of state.  Hereafter, I will call such models isentropic, 
because isentropic models and their adiabatic perturbations 
obey the same one-parameter equation of state.

For stars with more general equations of state, the r-modes 
appear to be complete for perturbations that have axial-parity.  
However, this is not the case for isentropic models. Early work 
on the r-modes focused on newtonian models with general equations 
of state (Papalouizou and Pringle \cite{pp78}, Provost et al. 
\cite{pea81}, Saio \cite{s82}, Smeyers and Martens \cite{sm83})
and mentioned only in passing
the isentropic case.  In isentropic newtonian stars, one finds 
that the only purely axial modes allowed are the r-modes with 
$l=m$ and simplest radial behavior.
(Provost et al. \cite{pea81}\footnote{An appendix in this paper 
incorrectly claims that no $l=m$ r-modes exist, based on an 
incorrect assumption about their radial behavior.};
see Sect. 2.3.1)  It is these r-modes only that have been 
studied (and found to be physically interesting) in connection 
with the gravitational-wave driven instability.  

The first part of this dissertation (Ch. 2) addresses the 
question of the missing modes in isentropic newtonian models.
(Lockitch and Friedman \cite{lf98})
The disappearance of the purely axial modes with $l>m$ occurs 
for the following reason.  We have already noted that all axial
perturbations of a spherical star are time-independent 
convective currents with vanishing perturbed pressure and density.  
We have also noted that in spherical isentropic stars the 
gravitational restoring forces that give rise to the g-modes 
vanish and they, too, become time-independent convective currents 
with vanishing perturbed pressure and density.  Thus, the space of 
zero frequency modes, which generally consists only of the axial
r-modes, expands for spherical isentropic stars to include 
the polar g-modes.  This large degenerate subspace of zero-frequency 
modes is split by rotation to zeroth order in the star's angular 
velocity, and the corresponding modes of rotating isentropic stars 
are generically hybrids whose spherical limits are mixtures of 
axial and polar perturbations.  These hybrid modes have already been 
found analytically for the uniform-density Maclaurin spheroids by 
Lindblom and Ipser \cite{li98} in a complementary presentation that 
makes certain features transparent but masks properties (such as their
hybrid character) that are our primary concern. Lindblom and Ipser 
point out that since these modes are also restored by the Coriolis 
force, it is natural to refer to them as rotation modes, or 
generalized r-modes.

Having found the missing modes in isentropic newtonian stars, I then 
turn to the corresponding problem in general relativity.  The r-modes
of rotating relativistic stars have been studied for the first time 
only recently (Andersson \cite{a97}; Kojima \cite{k98}; Beyer and 
Kokkotas \cite{bk99}; Kojima and Hosonuma \cite{kh99}), 
but none of these calculations have found the modes in the isentropic 
stellar models appropriate to neutron stars.  
As in the newtonian case, a spherical
isentropic relativistic star has a large degenerate subspace of
zero-frequency modes consisting of the axial-parity r-modes and 
the polar-parity g-modes.  Again, the degeneracy is split by 
rotation and the generic mode of a rotating isentropic star is a 
hybrid whose spherical limit is a mixture of axial and polar 
perturbations.  The second part of this dissertation (Chs. 3-4) 
presents the first calculation finding these 
modes in isentropic relativistic stars.  Although isentropic 
newtonian stars retain a vestigial set of purely axial modes 
(those having $l=m$), rotating relativistic stars of this type 
have {\it no} pure r-modes, no modes whose limit for a spherical 
star is purely axial.  Instead, the newtonian r-modes with 
$l=m\geq 2$ acquire relativistic corrections with both axial and 
polar parity to become discrete hybrid modes of the corresponding 
relativistic models (see Sects. 3.4.1-3.4.2).  

This dissertation examines the hybrid rotational modes of 
rotating isentropic stars, both newtonian and relativistic.  
Sect. 1.2 begins with a brief summary of the theory of 
self-gravitating perfect fluids and their linearized 
perturbations. Ch. 2 considers the hybrid modes of newtonian 
stars, first proving that the time-independent modes of 
spherical isentropic stars are the r- and g-modes (Sect. 2.1), 
and then moving to consider rotating stars (Sect. 2.2).
Sect 2.3 distinguishes two types of modes, axial-led and
polar-led, and shows that every mode belongs to one of the 
two classes.  Sects. 2.4-2.5 deal with the computation of 
eigenfunctions and eigenfrequencies for modes in each class,
adopting what appears to be a method that is both novel and 
robust.  For the uniform-density Maclaurin spheroids, these 
modes have been found analytically by Lindblom and Ipser. I
find machine precision agreement with their eigenfrequencies and 
corresponding eigenfunctions to lowest nontrivial order in the 
angular velocity $\Omega$.  I also examine the frequencies and 
modes of $n=1$ polytropes, finding that the structure of the modes
and their frequencies are very similar for the polytropes and the
uniform-density configurations.  The numerical analysis is 
complicated by a curious linear dependence in the Euler equations, 
detailed in Appendix B.  The linear dependence appears in a power 
series expansion of the equations about the origin.  It may be 
related to difficulty other groups have encountered in searching 
for these modes.  Finally, Sect. 2.6 examines unstable modes, 
computing their growth time and expected viscous damping time.  
The pure $l=m=2$ r-mode retains its dominant role, but the 
$3\leq l=m\lesssim10$ r-modes and some of the fastest growing 
hybrids remain unstable in the presence of viscosity.

Chs. 3 and 4 are concerned with the hybrid modes of isentropic
relativistic stars, which turn out be very similar in character 
to their newtonian counterparts.  In Sect. 3.1, the proof that the 
time-independent modes of spherical isentropic stars are the r- and 
g-modes is generalized to relativity. In Sect. 3.2 the perturbation 
equations governing the hybrid modes in slowly rotating stars are 
derived; their structure parallels the corresponding newtonian 
equations of Sect. 2.2.  This similarity between the newtonian and 
relativistic equations leads to an identical structure of the mode 
spectrum and to a parallel theorem in Sect 3.3 that every non-radial 
mode is either an axial-led or polar-led hybrid (the result
has so far been proven only for slowly rotating relativistic stars).
This chapter concludes with a discussion of the boundary 
conditions appropriate to the hybrid modes and the construction
of some explicit solutions (Sect 3.4).  I show that there are
no modes in isentropic relativistic stars whose limit as 
$\Omega\rightarrow 0$ is a pure axial perturbation with 
$l\neq 1$.  In particular, the newtonian r-modes having  
$l=m\geq 2$ do not exist in isentropic relativistic stars and
must be replaced by axial-led hybrid modes (Sect. 3.4.1).
I explicitly construct these particular modes to first 
post-newtonian order in slowly rotating, uniform density
stars (Sect. 3.4.2).  

Finally, Ch. 4 involves the computation of eigenfunctions 
and eigenfrequencies, applying essentially the same numerical 
method as was used in the newtonian calculation.  A set of
modes from each parity class is constructed for uniform density 
stars and compared with their newtonian counterparts.  The
relativistic corrections turn out to be small for the modes and
stellar models considered.  As in the newtonian calculation, 
the numerical analysis is complicated by a curious linear 
dependence in the perturbation equations.  The linear dependence,
again, appears in a power series expansion of the equations about 
the origin, and is discussed in Appendix D.  

\vspace{0.3in}

Throughout this dissertation I will work in geometrized units, 
($G=c=1)$, except in Sect. 2.6 where $G$ and $c$ are restored 
to their cgs values for the explicit computation of dissipative 
timescales.  I use the conventions of Misner, Thorne and 
Wheeler \cite{mtw} for the metric signature $(-,+,+,+)$ and the
sign of the curvature tensors.  I adopt the abstract index 
notation (see, e.g., Wald, Sect 2.4 \cite{wald}) with latin
spatial indices and greek spacetime indices; components of 
tensors will always be written with respect to a choice of
coordinates $(t,r,\theta,\varphi)$. 

\section{Self-Gravitating Perfect Fluids and Their 
Linearized Perturbations}

We will be considering stationary perfect fluid stellar
models in both newtonian gravity and in general relativity.
We construct both rotating and non-rotating equilibrium
stars and then study the equations of motion, linearized
about these equilibria, governing their small oscillations.
We will make use of both the eulerian and lagrangian
perturbation formalisms, which we now briefly review.

\subsection{Newtonian Gravity}

In the newtonian theory of gravity, a complete description
of an isentropic perfect fluid configuration is provided 
by the fluid density $\rho$, the pressure $p$, the fluid 
velocity $v^a$ and the newtonian gravitational potential 
$\Phi$.  These must satisfy a barotropic (one-parameter) 
equation of state,
\be
p=p(\rho),
\label{N:eos}
\ee
the equation of mass conservation,
\be
\partial_t \rho + \nabla_a(\rho v^a) = 0,
\label{N:cons}
\ee
Euler's equation,
\be
(\partial_t + v^b\nabla_b) v_a 
+ \frac{1}{\rho} \nabla_a p + \nabla_a\Phi = 0,
\label{N:euler}
\ee
and the newtonian gravitational equation,
\be
\nabla^2 \Phi = 4\pi\rho.
\label{N:grav}
\ee

An equilibrium stellar model is a time-independent solution
$(\rho, p, v^a, \Phi)$ to these equations.  The small perturbations 
of such a star may be studied using either the eulerian or the 
lagrangian perturbation formalism (see, e.g., Friedman and Schutz 
\cite{fs78a}).  If 
$[\bar\rho(\lambda), \bar p(\lambda), 
\bar v^a(\lambda), \bar\Phi(\lambda)]$ is a smooth family of 
solutions to the exact equations (\ref{N:eos})-(\ref{N:grav}) 
that coincides with the equilibrium solution at $\lambda=0$,
\[
[\bar\rho(0), \bar p(0), \bar v^a(0), \bar\Phi(0)]
= (\rho, p, v^a, \Phi),
\]
then the eulerian change $\delta Q$ in a quantity $Q$ may be 
defined (to linear order in $\lambda$) as,
\be
\delta Q \equiv \left. \frac{dQ}{d\lambda}\right|_{\lambda=0}.
\ee
Thus, an eulerian perturbation is simply a change 
$(\delta\rho, \delta p, \delta v^a, \delta\Phi)$ 
in the equilibrium configuration at a particular 
point in space.  

In the lagrangian perturbation formalism (Friedman and Schutz 
\cite{fs78a}), on the
other hand, perturbed quantities are described in terms of a 
lagrangian displacement vector $\xi^a$ that connects fluid 
elements in the equilibrium and perturbed star.  The lagrangian 
change $\Delta Q$ in a quantity $Q$ is related to its eulerian 
change $\delta Q$ by
\be
\Delta Q = \delta Q + \pounds_{\xi} Q,
\ee
with $\pounds_{\xi}$ the Lie derivative along $\xi^a$. 
The fluid perturbation is then entirely determined by the 
displacement $\xi^a$:
\be
\Delta v^a = \partial_t \xi^a
\ee
\be
\frac{\Delta p}{\gamma p} = \frac{\Delta \rho}{\rho} = - \nabla_a \xi^a,
\label{N:Del_etc}
\ee
(where $\gamma$ is the adiabatic index) and the corresponding Eulerian 
changes are
\bea
\delta v^a  &=& (\partial_t+\pounds_v) \xi^a \label{N:del_v} \\
\delta \rho &=& - \nabla_a(\rho \xi^a) \label{N:del_rho} \\
\delta p    &=& - \gamma p \nabla_a\xi^a - \xi^a\nabla_a p
\label{N:del_p}
\eea
with the change in the gravitational potential determined by
\be
\nabla^2 \delta \Phi = 4\pi \delta \rho.
\ee

\subsection{General Relativity}

In general relativity, a complete description of an 
isentropic perfect fluid configuration is provided by a 
spacetime with metric $g_{\alpha\beta}$, sourced by an 
energy-momentum tensor,
\be
T_{\alpha\beta} = (\epsilon+p)u_\alpha u_\beta + p g_{\alpha\beta},
\ee
where the fluid 4-velocity $u^\alpha$ is a unit timelike vector
field,
\be
u^\alpha u_\alpha = -1,
\label{unit_u}
\ee
and $\ep$ and $p$ are, respectively, the total energy density 
and pressure of the fluid as measured by an observer moving with 
4-velocity $u^\alpha$.  The metric and fluid variables must, again, 
satisfy a barotropic equation of state,
\be
p = p(\ep),
\ee
as well as the Einstein field equations,
\be
G_{\alpha\beta}=8\pi T_{\alpha\beta}.
\label{einstein}
\ee


An equilibrium stellar model is a stationary solution
$(g_{\alpha\beta}, u^\alpha, \ep, p)$ to these equations.  
The small perturbations of such star may be studied using 
either the eulerian or the lagrangian perturbation formalism 
(Friedman \cite{f78}; see also Friedman and Ipser \cite{fi92}).  
As in the newtonian case, an eulerian perturbation
may described in terms of a smooth family, 
$[\bar g_{\alpha\beta}(\lambda), \bar u^\alpha(\lambda), 
\bar \ep(\lambda), \bar p(\lambda)]$, of
solutions to the exact equations (\ref{unit_u})-(\ref{einstein}) 
that coincides with the equilibrium solution at $\lambda=0$,
\[
[\bar g_{\alpha\beta}(0), \bar u^\alpha(0), \bar \ep(0), \bar p(0)] 
= (g_{\alpha\beta}, u^\alpha, \ep, p).
\]
Then the eulerian change $\delta Q$ in a quantity $Q$ 
may be defined (to linear order in $\lambda$) as,
\be
\delta Q \equiv \left. \frac{dQ}{d\lambda}\right|_{\lambda=0}.
\ee
Thus, an eulerian perturbation is simply a change 
$(\delta g_{\alpha\beta}, \delta u^\alpha, \delta \ep, \delta p)$ 
in the equilibrium configuration at a particular point in 
spacetime.  

In the lagrangian perturbation formalism (Friedman \cite{f78};
see also Friedman and Ipser \cite{fi92}), on the
other hand, perturbed quantities are expressed in terms of 
the eulerian change in the metric 
$h_{\alpha\beta} \equiv \delta g_{\alpha\beta}$, and a lagrangian 
displacement vector $\xi^\alpha$, which connects fluid elements 
in the equilibrium star to the corresponding elements in the 
perturbed star.  The lagrangian change $\Delta Q$ in a quantity 
$Q$ is related to its eulerian change $\delta Q$ by
\be
\Delta Q = \delta Q + \pounds_{\xi} Q,
\label{DelQ}
\ee
with $\pounds_{\xi}$ the Lie derivative along $\xi^\alpha$. 

The identities,
\begin{eqnarray}
\Delta g_{\alpha\beta} &=& h_{\alpha\beta} 
+ 2\nabla_{\left(\alpha\right.}\xi_{\left.\beta\right)} 
\label{Del_metric} \\
\Delta\varepsilon_{\alpha\beta\gamma\delta} &=&
\half\varepsilon_{\alpha\beta\gamma\delta} \
g^{\mu\nu} \Delta g_{\mu\nu}
\end{eqnarray}
then allow one to express the fluid perturbation in terms of
$h_{\alpha\beta}$ and $\xi^\alpha$,
\be
\Delta u^\alpha = 
\half u^\alpha u^\beta u^\gamma\Delta g_{\beta\gamma}  
\ee
\be
\frac{\Delta p}{\gamma p} = \frac{\Delta \epsilon}{\epsilon + p}
= \frac{\Delta n}{n} = -\half q^{\alpha\beta}\Delta g_{\alpha\beta}
\label{Del_etc} \\
\ee
where $n$ is the baryon density and 
$q^{\alpha\beta} \equiv g^{\alpha\beta} + u^\alpha u^\beta$.
Using Eqs. (\ref{DelQ})-(\ref{Del_etc}), it is 
straightforward to express the corresponding eulerian changes
also in terms of $h_{\alpha\beta}$ and $\xi^\alpha$.


\chapter{Newtonian Stars}

The oscillation modes considered in this dissertation are
dominantly restored by the Coriolis force and have frequencies
that scale with the star's angular velocity, $\Omega$.  Thus, 
all of these modes will be degenerate at zero frequency in a 
non-rotating star.  To study properly the spectrum of these 
rotationally restored modes, we must first examine the 
perturbations of a non-rotating star, and find all of the 
modes belonging to its degenerate zero-frequency subspace.

\section{Stationary Perturbations of Spherical Stars}

We consider a static spherically symmetric, self-gravitating perfect 
fluid described by a gravitational potential $\Phi$, density $\rho$ 
and pressure $p$. These satisfy an equation of state of the form
\be
p = p( \rho), \label{eos}
\ee 
as well as the newtonian equilibrium equations
\be
\frac{1}{\rho}\nabla_a p + \nabla_a\Phi = 0  \label{equil}
\ee
\be
\nabla^2 \Phi = 4\pi \rho. \label{newton} \label{equil_end}
\ee

We are interested in the space of zero-frequency modes, the linearized 
time-independent perturbations of this static equilibrium.  This 
zero-frequency subspace is spanned by two types of perturbations: (i)
perturbations with $\delta v^a\neq 0$ and $\delta \rho = \delta p = 
\delta \Phi = 0$, and (ii) perturbations with $\delta \rho$, $\delta p$ 
and $\delta \Phi$ nonzero and $\delta v^a=0$.  If one assumes that no 
solution to the linearized equations governing a static equilibrium is 
spurious, that each corresponds to a family of exact solutions, then 
the only solutions (ii) are spherically symmetric, joining neighboring 
equilibria.

The decomposition into classes (i) and (ii) can be seen as follows. 
The set of equations satisfied by 
$(\delta\rho, \delta p, \delta \Phi, \delta v^a)$ are the perturbed mass 
conservation equation,
\be
\delta \left[ \partial_t \rho + \nabla_a(\rho v^a) \right] = 0,
\ee
the perturbed Euler equation,
\be
\delta \left[ (\partial_t + \pounds_v) v_a 
	+ \frac{\mbox{$1$}}{\rho}\nabla_a p 
	+ \nabla_a(\mbox{$\Phi - \half v^2$} ) \right] = 0,
\ee
the perturbed Poisson equation, $\delta\mbox{[Eq. (\ref{newton})]}$,
and an equation of state for the perturbed configuration (which may, 
in general, differ from that of the equilibrium configuration). 

For a time-independent perturbation these equations take the form
\be 
\nabla_a(\rho \delta v^a) = 0, 
\label{continuity}
\ee
\be
\frac{1}{\rho}\nabla_a \delta p 
- \frac{\delta\rho}{\rho^2}\nabla_a p 
+ \nabla_a\delta\Phi = 0,  
\label{pert_hydro}
\ee
and
\be
\nabla^2 \delta\Phi = 4\pi \delta\rho.
\label{pert_newton}
\ee

Because Eq. (\ref{continuity}) for $\delta v^a$ decouples from 
Eqs. (\ref{pert_hydro}) and (\ref{pert_newton}) for 
$(\delta \rho, \delta p, \delta \Phi)$, any solution to Eqs. 
(\ref{continuity})-(\ref{pert_newton}) is a superposition of a solution
$(0, 0, 0, \delta v^a)$ and a solution 
$(\delta \rho, \delta p, \delta \Phi, 0)$. 
This is the claimed decomposition.

The theorem that any static self-gravitating perfect fluid is spherical 
implies that the solution $(\delta \rho, \delta p, \delta \Phi, 0)$ is 
spherically symmetric, to within the assumption that the static 
perturbation equations have no spurious solutions 
(``linearization stability'')\footnote{We are aware of a proof of this 
linearization stability for relativistic stars under assumptions on the 
equation of state that would not allow polytropes (K\"{u}nzle and Savage 
\cite{ks80}).}.

Thus, under the assumption of linearization stability we have shown that 
all stationary non-radial ($l>0$) perturbations of a spherical star have 
$\delta \rho = \delta p = \delta \Phi = 0$ and a velocity field 
$\delta v^a$ that satisfies Eq. (\ref{continuity}).

A perturbation with axial parity has the form (see, e.g., 
Friedman and Morsink \cite{jfs97}),
\be
\delta v^a = U(r) \epsilon^{abc}  \nabla_b 
			Y_l^m \nabla_c r, 
\label{form_ax}
\ee
and automatically satisfies Eq. (\ref{continuity}).

A perturbation with polar parity perturbation has the form,
\be
\delta v^a = \frac{W(r)}{r} Y_l^m \nabla^a r
			+ V(r) \nabla^a Y_l^m; 
\label{form_po}
\ee
and Eq. (\ref{continuity}) gives a relation between W and V,
\be
\frac{d}{d r} (r \rho W) - l(l+1) \rho V = 0.
\label{N:sph_cont}
\ee
These perturbations must satisfy the boundary conditions of 
regularity at the center, $r=0$ and surface, $r=R$, of the star.  
Also, the lagrangian change in the pressure (defined in the next 
section) must vanish at the surface of the star.  These boundary 
conditions result in the requirement that
\be
W(0) = W(R) = 0;
\ee 
however, apart from this restriction, the radial functions $U(r)$ 
and $W(r)$ are undetermined.

Finally, we consider the equation of state of the perturbed star.
For an adiabatic oscillation of a barotropic star (i.e., a star
that satisfies a one-parameter equation of state, $p=p(\rho)$) 
Eq. (\ref{N:Del_etc}) implies that the
perturbed pressure and energy density are related by
\be
\frac{\delta p}{\gamma p} = \frac{\delta \rho}{\rho} 
+ \xi^r \left[\frac{\rho'}{\rho} - \frac{p'}{\gamma p}
\right]
\label{N:ad_osc}
\ee
for some adiabatic index $\gamma(r)$ which need not be the function
\be
\Gamma(r) \equiv \frac{\rho}{p}\frac{dp}{d\rho}
\ee
associated with the equilibrium equation of state.  Here, 
$\xi^a$ is the lagrangian displacement vector and is related 
to our perturbation variables by Eq. (\ref{N:del_v}), which becomes
\be
\delta v^a = \partial_t \ \xi^a,
\ee
or (taking the initial displacement (at $t=0$) to be zero)
\be
\xi^r = t \ \delta v^r.
\label{N:form_xi_r}
\ee

For the class of perturbations under consideration, we have
seen that $\delta p = \delta\rho = 0$, thus Eqs. (\ref{N:ad_osc}) and
(\ref{N:form_xi_r}) require that
\be
\delta v^r \left[\frac{\rho'}{\rho} 
- \frac{p'}{\gamma p}\right] = 0.
\label{N:eos_req}
\ee
For axial-parity perturbations this equation is automatically
satisfied, since $\delta v^a$ has no $r$-component (Eq. (\ref{form_ax})).  
Thus, a spherical barotropic star always admits a class of zero-frequency 
r-modes.

For polar-parity perturbations, $\delta v^r \propto W(r) \neq 0$,
and Eq. (\ref{N:eos_req}) will be satisfied if and only if 
\be
\gamma(r)\equiv\Gamma(r)=\frac{\rho}{p}\frac{dp}{d\rho}.
\ee
Thus, a spherical barotropic star admits a class of zero-frequency
g-modes if and only if the perturbed star obeys the same 
one-parameter equation of state as the equilibrium star.  We call 
such a star isentropic, because isentropic models and their 
adiabatic perturbations obey the same one-parameter equation
of state.

Summarizing our results, we have shown the following.  A spherical
barotropic star always admits a class of zero-frequency r-modes 
(stationary fluid currents with axial parity); but admits
zero-frequency g-modes (stationary fluid currents with polar parity)
if and only if the star is isentropic.  Conversely, the zero-frequency
subspace of non-radial perturbations of a spherical isentropic 
star is spanned by the r- and g-modes - that is, by convective fluid 
motions having both axial and polar parity and with vanishing perturbed 
pressure and 
density.\footnote{Note that for spherical stars, nonlinear
couplings invalidate the linear approximation after a time 
$t\sim R/\delta v$, comparable to the time for a fluid element to 
move once around the star. For nonzero angular velocity, the linear 
approximation is expected to be valid for all times, if the amplitude 
is sufficiently small, roughly, if  $|\delta v|< R\Omega$.}
Being stationary, these modes do not couple to gravitational radiation. 
One would expect this large subspace of modes, which is degenerate at 
zero-frequency, to be split by rotation, so let us now consider the 
perturbations of rotating stars.

\section{Perturbations of Rotating Stars}

We consider perturbations of an isentropic newtonian star, rotating
with uniform angular velocity $\Omega$.  No assumption of slow rotation
will be made until we turn to numerical computations in Sect. 2.4.  The
equilibrium of an axisymmetric, self-gravitating perfect fluid is
described by the gravitational potential $\Phi$, density $\rho$,
pressure $p$ and a 3-velocity
\be
v^a = \Omega \varphi^a,
\ee
where $\varphi^a$ is the rotational Killing vector field.

We will use the lagrangian perturbation formalism reviewed in 
Sect. 1.2.1.  Since the equilibrium spacetime is stationary 
and axisymmetric, we may decompose our perturbations into modes 
of the 
form\footnote{We will always
choose $m \geq 0$ since the complex conjugate of an $m<0$ mode with 
frequency $\sigma$ is an $m>0$ mode with frequency $-\sigma$.  
Note that $\sigma$ is the frequency in an inertial frame.}
$e^{i(\sigma t + m \varphi)}$.  In this case, the Eulerian change 
in the 3-velocity (\ref{N:del_v}) is related to the lagrangian 
displacement $\xi^a$ by,
\be
\delta v^a  = i(\sigma + m \Omega) \xi^a.
\ee


We can expand this perturbed fluid velocity in vector 
spherical harmonics
\be 
\delta v^a = \sum_{l=m}^{\infty} \left\{ \frac{1}{r}
	W_l Y_l^m \nabla^a r + V_l \nabla^a Y_l^m 
	- i U_l \epsilon^{abc} \nabla_b Y_l^m \nabla_c r 
	\right\} e^{i \sigma t},        \label{v_exp}
\ee
and examine the perturbed Euler equation.    

The lagrangian perturbation of Euler's equation is
\bea
0 &=& \Delta [ (\partial_t + \pounds_v) v_a + 
	\nabla_a (h - \mbox{$\ds{\half v^2 + \Phi}$} ) ]   \nonumber \\
  &=& (\partial_t + \pounds_v) \mbox{$\ds{\Delta  v_a 
	+ \nabla_a [ \Delta (h - \half v^2 + \Phi ) ]}$}, 
\label{pert_euler}
\eea
and its curl, which expresses the conservation of circulation for an 
isentropic star, is
\be
0 = q^a \equiv i(\sigma + m \Omega) \epsilon^{abc} \nabla_b \Delta v_c, 
\label{form_of_q}
\ee
or
\be
0 = q^a = i(\sigma + m \Omega) \epsilon^{abc} \nabla_b \delta v_c 
	+ \Omega \epsilon^{abc} \nabla_b ( \pounds_{\delta v} \varphi_c ).
\ee


Using the spherical harmonic expansion (\ref{v_exp}) of $\delta v^a$ we can 
write the components of $q^a$ as

\be
0=q^r = \frac{1}{r^2} \sum_{l=m}^{\infty} 
	\Biggl\{ \ba[t]{l}
	[(\sigma+m\Omega)l(l+1)- 2m\Omega ]U_lY_l^m \\
	\\
	- 2 \Omega V_l[\sin\theta\partial_\theta Y_l^m 
				+ l(l+1)\cos\theta Y_l^m ] \\
	\\
	+ 2 \Omega W_l[\sin\theta\partial_\theta Y_l^m 
	+ 2 \cos\theta Y_l^m ] 
\Biggr\} \ e^{i\sigma t},   
\label{qr}
\ea
\ee

\be
0=q^{\theta} = \frac{1}{r^2 \sin\theta} \sum_{l=m}^{\infty} 
	\Biggl\{ \ba[t]{l}
	m (\sigma + m\Omega) 
	\ds{\left(\partial_rV_l-\frac{W_l}{r}\right) Y_l^m} \\
	\\
	- 2\Omega\partial_rV_l\cos\theta\sin\theta\partial_\theta Y_l^m 
	+ \ds{2\Omega m^2\frac{V_l}{r} Y_l^m} \nn \\
	\\
	- 2\Omega\partial_rW_l\sin^2\theta Y_l^m 
	- 2m\Omega\partial_rU_l\cos\theta Y_l^m  \nonumber \\
	\\
	+ \ds{(\sigma + m\Omega)\partial_rU_l \sin\theta\partial_\theta Y_l^m
	+ 2m\Omega\frac{U_l}{r} \sin\theta\partial_\theta Y_l^m}
\Biggr\} e^{i\sigma t},
\label{qtheta}
\ea
\ee
and
\be
0=q^{\varphi} = \frac{i}{r^2 \sin^2\theta} \sum_{l=m}^{\infty}
	\Biggl\{ \ba[t]{l}
	m (\sigma + m\Omega) \partial_rU_l Y_l^m
	- 2\Omega\partial_rU_l \cos\theta\sin\theta\partial_\theta Y_l^m 
	\nonumber\\
	\\
	+ \ds{2\Omega \frac{U_l}{r} [m^2-l(l+1)\sin^2\theta] Y_l^m   
	- 2m\Omega\partial_rV_l\cos\theta Y_l^m}
	\nonumber\\
	\\
	+ \left[ (\sigma + m\Omega)\left(
	\partial_rV_l-\ds{\frac{W_l}{r}}\right) 
	+ \ds{2m\Omega\frac{V_l}{r}}\right] \\
	\\
	\times \sin\theta\partial_\theta Y_l^m 
	\Biggr\} \ e^{i\sigma t}.    \label{qphi}
\ea
\ee
These components are not independent. The identity $\nabla_a q^a = 0$, 
which follows from equation (\ref{form_of_q}), serves as a check on the 
right-hand sides of (\ref{qr})-(\ref{qphi}).  


Let us rewrite these equations making use of the standard identities,
\begin{eqnarray}
\sin\theta\partial_\theta Y_l^m 
&=& l Q_{l+1} Y_{l+1}^m - (l+1) Q_l Y_{l-1}^m 
\label{identities1}  \\
\cos\theta Y_l^m &=& Q_{l+1} Y_{l+1}^m + Q_l Y_{l-1}^m 
\label{identities2} 
\end{eqnarray}
where
\be
Q_l \equiv \left[ \frac{(l+m)(l-m)}{(2l-1)(2l+1)} \right]^{\half}. 
\label{Q_l}
\ee
Defining a dimensionless comoving frequency 
\be
\kappa \equiv \frac{(\sigma + m\Omega)}{\Omega},  
\label{kappa}
\ee
we find that the equation $q^r=0$ becomes
\be
0  = \sum_{l=m}^{\infty} \Biggl\{ 
	\begin{array}[t]{l}
	[\half\kappa l(l+1) -m] U_lY_l^m \\
	\\
	+ (W_l-lV_l)(l+2)Q_{l+1}Y_{l+1}^m \\
	\\
	- [W_l+(l+1)V_l](l-1)Q_lY_{l-1}^m
	 \Biggr\},
\end{array}
\label{qr2}
\ee
$q^\theta =0$ becomes
\begin{eqnarray}
0 &=& \sum_{l=m}^{\infty} \left\{ \
- Q_{l+1}Q_{l+2} \left[ \vphantom{\frac{1}{2}}
lV'_l - W'_l \right] Y_{l+2}^m
	\right. \label{qtheta2} \\
	& & \nn \\
	& & \left.
\mbox{} - Q_{l+1} \left[ (m-\half\kappa l) U'_l 
	- m l \frac{U_l}{r}\right] Y_{l+1}^m
	\right. \nonumber\\
	& & \nn \\
	& & \left.
\mbox{} + \Biggl[ \ba[t]{l}
\ds{\left( \half\kappa m +(l+1)Q_l^2 - l Q_{l+1}^2  \right) V'_l} \\
\\
- \ds{\left(1- Q_l^2 - Q_{l+1}^2   \right) W'_l
- \half\kappa m \frac{W_l}{r} + m^2\frac{V_l}{r} }
\Biggr] Y_l^m 
\ea
	\right. \nonumber\\
	& & \nn \\
	& & \left.
\mbox{} - Q_l \left[ \left(m+\half\kappa (l+1)\vphantom{Q_l^2} \right) U'_l
+ m(l+1)\frac{U_l}{r} \right]  Y_{l-1}^m
	\right. \nonumber\\
	& & \nn \\
	& & \left.
+ Q_{l-1} Q_l \left[ \vphantom{\frac{1}{2}}
(l+1)V'_l + W'_l \right] Y_{l-2}^m
	 \ \right\} \nn                     
\end{eqnarray}
and
$q^\varphi =0$ becomes
\begin{eqnarray}
0 &=& \sum_{l=m}^{\infty} \left\{ \
- l Q_{l+1}Q_{l+2} \left[ U'_l-(l+1) \frac{U_l}{r}\right] Y_{l+2}^m
	\right. \label{qphi2} \\
	& & \nn \\
	& & \left. 
\mbox{} + Q_{l+1} \left[ (\half\kappa l-m) V'_l + m l \frac{V_l}{r}
-\half\kappa l \frac{W_l}{r} \right]  Y_{l+1}^m
	\right. \nonumber\\
	& & \nn \\
	& & \left. 
\mbox{} + \Biggl[ \ba[t]{l}
  	\ds{\left( \half\kappa m +(l+1)Q_l^2 - l Q_{l+1}^2  \right) U'_l} \\
	\\
	+ \ds{\left( m^2 - l(l+1)\left(1- Q_l^2 - Q_{l+1}^2\right)\right)
	\frac{U_l}{r}}
	\Biggr] Y_l^m
\ea
	\right. \nonumber\\
	& & \nn \\
	& & \left. 
\mbox{} - Q_l \left[ \left(\half\kappa (l+1)+m \vphantom{Q_l^2}\right)V'_l
+ m(l+1)\frac{V_l}{r} - \half\kappa (l+1) \frac{W_l}{r} \right] Y_{l-1}^m
	\right. \nonumber\\
	& & \nn \\
	& & \left. 
\mbox{} + (l+1)Q_{l-1}Q_l\left[ U'_l+l\frac{U_l}{r}\right] Y_{l-2}^m
	 \ \right\} \nn
\end{eqnarray}
where $'\equiv\frac{d}{dr}$.

Let us rewrite the equations one last time using the orthogonality 
relation for spherical harmonics,
\be
\int Y_{l'}^{m'} Y_l^{\ast m} d\Omega = \delta_{ll'} \delta_{mm'},
\ee
where $d\Omega$ is the usual solid angle element.

From equation (\ref{qr2}) we find that $\int q^r Y_l^{\ast m} d\Omega = 0$ 
gives
\be
0 = 
[\half\kappa l(l+1) -m] U_l + (l+1)Q_l [W_{l-1}-(l-1)V_{l-1}] 
- lQ_{l+1} [W_{l+1}+(l+2)V_{l+1}]   \label{eq2}
\ee

Similarly, $\int q^{\theta} Y_l^{\ast m} d\Omega = 0$ gives
\begin{eqnarray}
0 &=& Q_lQ_{l-1} \{ (l-2)V'_{l-2}-W'_{l-2}\} 
+ Q_l\left\{ [m-\half\kappa (l-1)]
  U'_{l-1}-m(l-1) \frac{U_{l-1}}{r}\right\}\nonumber\\
  &&\mbox{} + \left(1- Q_l^2 - Q_{l+1}^2   \right) W'_l
	- \left[\half\kappa m +(l+1)Q_l^2 - l Q_{l+1}^2  \right] V'_l 
	+ \half\kappa m \frac{W_l}{r} - m^2\frac{V_l}{r} \nonumber\\
  &&\mbox{} + Q_{l+1} \left\{ [m+\half\kappa (l+2)] U'_{l+1}+m(l+2)
  \frac{U_{l+1}}{r}\right\} \nonumber\\
  &&\mbox{} - Q_{l+2} Q_{l+1} \{ (l+3)V'_{l+2}+W'_{l+2}\} \label{eq3}
\end{eqnarray}
and $\int q^{\varphi} Y_l^{\ast m} d\Omega = 0$ gives
\begin{eqnarray}
0 &=& -(l-2)Q_lQ_{l-1} \left[ U'_{l-2}-(l-1) \frac{U_{l-2}}{r}\right] 
  + (l+3)Q_{l+2}Q_{l+1}\left[ U'_{l+2}+(l+2)\frac{U_{l+2}}{r}\right]
\nonumber\\
  &&\mbox{} + \left\{ 
  	\left[\half\kappa m +(l+1)Q_l^2 - l Q_{l+1}^2  \right] U'_l 
	+ \left[ m^2 - l(l+1)\left(1- Q_l^2 - Q_{l+1}^2\right)\right] 
	\frac{U_l}{r}
	\right\} \nonumber\\
  &&\mbox{} + Q_l\left\{ [\half\kappa (l-1)-m] V'_{l-1} 
		+m(l-1) \frac{V_{l-1}}{r}
  -\half\kappa (l-1) \frac{W_{l-1}}{r}\right\}\nonumber\\
  &&\mbox{} - Q_{l+1} \left\{ [\half\kappa (l+2)+m]V'_{l+1}+m(l+2)
  \frac{V_{l+1}}{r} - \half\kappa (l+2) \frac{W_{l+1}}{r}\right\}.  
\label{eq4}
\end{eqnarray}

\section{The Character of the Perturbation Modes}

From this last form of the equations it is clear that the rotation of
the star mixes the axial and polar contributions to $\delta v^a$.  That
is, rotation mixes those terms in (\ref{v_exp}) whose limit as $\Omega
\rightarrow 0$ is axial with those terms in (\ref{v_exp}) whose limit
as $\Omega \rightarrow 0$ is polar.  It is also evident that the axial
contributions to $\delta v^a$ with $l$ even mix only with the odd $l$
polar contributions, and that the axial contributions with $l$ odd mix
only with the even $l$ polar contributions.  In addition, we prove that 
for non-axisymmetric modes the lowest value of $l$ that appears in the 
expansion of $\delta v^a$ is always $l=m$. (When $m=0$ this lowest value 
of $l$ is either 0 or 1.)  

For an equilibrium model that is axisymmetric and invariant under parity,
one can resolve any degeneracy in the perturbation spectrum to make each 
discrete mode an eigenstate of parity with angular dependence 
$e^{im\varphi}$. The following theorem then holds.

\begin{thm}{Let $(\delta\rho, \delta v^a)$ with $\delta v^a\neq 0$ be a 
discrete normal mode of a uniformly rotating stellar model obeying a 
one-parameter equation of state. Then the decomposition of the mode into 
spherical harmonics $Y_l^m$ (i.e., into $(l, m)$ representations of the 
rotation group about the center of mass) has $l=m$ as the lowest 
contributing value of $l$, when $m\neq 0$; and has $0$ or $1$ as the 
lowest contributing value of $l$, when $m=0$.}
\label{thm1}
\end{thm}

Thus, we find two distinct classes of mixed, or hybrid, modes with 
definite behavior under parity.  Let us call a non-axisymmetric
mode an ``axial-led hybrid'' (or simply ``axial-hybrid'') if 
$\delta v^a$ receives contributions only from
\begin{center}
axial terms with $l \ = \ m, \ m+2, \ m+4,  \ \ldots$ and \\
polar terms with $l \ = \ m+1, \ m+3, \ m+5, \ \ldots$.
\end{center}
Such a mode has parity $(-1)^{m+1}$.

Similarly, we define a non-axisymmetric mode to be a ``polar-led 
hybrid'' (or ``polar-hybrid'') if $\delta v^a$ receives contributions 
only from
\begin{center}
polar terms with $l \ = \ m, \ m+2, \ m+4,  \ \ldots$ and \\
axial terms with $l \ = \ m+1, \ m+3, \ m+5,  \ \ldots$.
\end{center}
Such a mode has parity $(-1)^m$.

For the case $m=0$, there exists a set of axisymmetric modes with parity 
$+1$ that we call ``axial-led hybrids'' since $\delta v^a$ receives 
contributions only from 
\begin{center}
axial terms with $l \ = \ 1, \ 3, \ 5,  \ \ldots$ and \\
polar terms with $l \ = \ 2, \ 4, \ 6,  \ \ldots$;\phantom{and}
\end{center}
and there exist two sets of axisymmetric modes that may be designated as 
``polar-led hybrids.''  One set has parity $-1$ and $\delta v^a$ 
receives contributions only from 
\begin{center}
polar terms with $l \ = \ 1, \ 3, \ 5,  \ \ldots$ and \\
axial terms with $l \ = \ 2, \ 4, \ 6,  \ \ldots$.\phantom{and}
\end{center}
The other set has parity $+1$ and $\delta v^a$ receives contributions 
only from 
\begin{center}
polar terms with $l \ = \ 0,  \ 2, \ 4,  \ \ldots$ and \\
axial terms with $l \ = \ 1, \ 3, \ 5,  \ \ldots$.\phantom{and}
\end{center}

Note that the theorem holds for p-modes as well as for the rotational 
modes that are our main concern.  A p-mode is determined by its density 
perturbation and is therefore dominantly polar in character regardless of 
its parity.  For a rotational mode, however, the lowest $l$ term in its 
velocity perturbation is at least comparable in magnitude to the other 
contributing terms.  

We prove the theorem separately for each parity class in Appendix A.

\subsection{The purely axial solutions}

We have proved that the generic mode of a rotating isentropic 
newtonian star is a hybrid mixture of axial and polar terms.  
However, it is known that newtonian stars of this type do 
allow a set of purely axial modes (Provost et al. \cite{pea81}).  
To find these r-modes, let us assume that the only non-vanishing 
coefficient in the spherical harmonic expansion (\ref{v_exp}) 
of the perturbed velocity is $U_l(r)$, for some particular value of $l$.  
Eqs. (\ref{eq2})-(\ref{eq4}) must be satisfied for all $l$, 
but with our ansatz they reduce to the following set.

Eq. (\ref{eq2}) becomes,
\be
\left[\half\kappa l(l+1)-m\right]U_l=0,
\label{N:exact1}
\ee
and Eq. (\ref{eq4}) with $l\rightarrow l+2$, $l\rightarrow l$ 
and $l\rightarrow l-2$ gives the equations
\begin{eqnarray}
0 &=& 	-lQ_{l+1}Q_{l+2} \left[U'_l-(l+1)\frac{U_l}{r}\right]
	\label{N:exact2} \\
&& \nn \\
0 &=&	\ba[t]{l}
	\ds{\left(\half\kappa m+(l+1)Q_l^2-l Q_{l+1}^2\right)U'_l} \\
	\\
	+ \ds{\left[ m^2 - l(l+1)\left(1- Q_l^2 - Q_{l+1}^2\right)\right] 
	\frac{U_l}{r}}
	\ea \label{N:exact3} \\
&& \nn \\
0 &=& (l+1)Q_{l-1}Q_l \left[U'_l+l\frac{U_l}{r}\right],
	\label{N:exact4}
\end{eqnarray}
respectively.  Recall that we need only work with two of the three
equations (\ref{eq2})-(\ref{eq4}), since they are linearly
dependent as a result of the identity, $\nabla_a q^a = 0$.

We see immediately from Eq. (\ref{N:exact1}) that a non-trivial
solution to these equations exists if and only if 
\be
\kappa\Omega\equiv(\sigma+m\Omega)=\frac{2m\Omega}{l(l+1)},
\label{N:p&p_freq}
\ee
which is the r-mode frequency, Eq. (\ref{p&p_freq}), found by 
Papalouizou and Pringle \cite{pp78}.  

By Thm. (\ref{thm1}), we know that a non-axisymmetric $(m>0)$ 
mode must have $l=m$ as its lowest value of $l$ and that an 
axisymmetric $(m=0)$ mode must have $l=1$ as its lowest value 
of $l$.  Hence, in the present context of pure a spherical 
harmonic these are also the {\it only} allowed values of $l$.
{\it An r-mode with $l>m$ (or $l>1$ if $m=0$) cannot exist in 
isentropic newtonian models.}  We consider the axisymmetric and
non-axisymmetric cases separately.\\

\noindent
\textit{The case $m=0$ and $l=1$.}\\

\noindent
It is well known that uniform rotation is a purely axial
perturbation with $m=0$ and $l=1$, and we can see this
from our Eqs. (\ref{N:exact2})-(\ref{N:p&p_freq}) as follows.

\begin{figure}
\centerline{\resizebox{5in}{2.5in}{\includegraphics{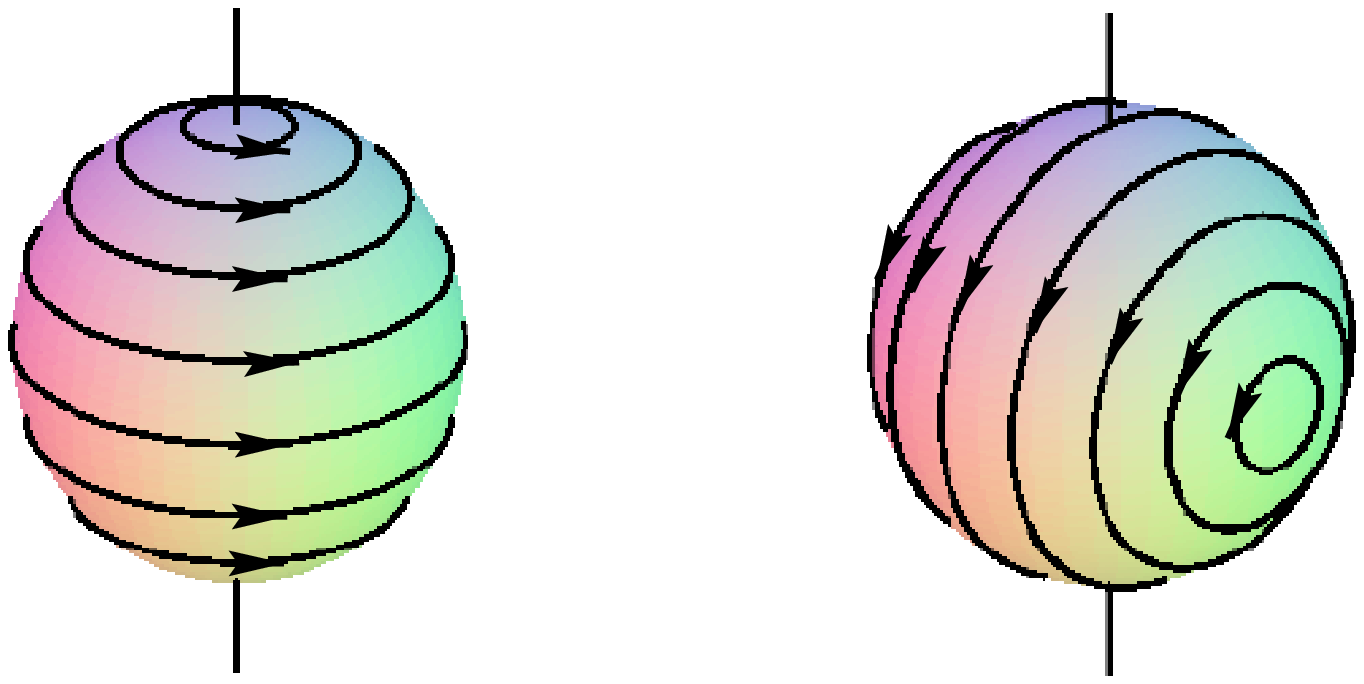}}}
\begin{tabular}{rcr}
\hphantom{123456789012345}$l=1$ & \hphantom{12345678901234567890123456} 
& $l=1$ \\
\hphantom{123456789012345}$m=0$ & \hphantom{12345678901234567890123456} 
& $m=1$
\end{tabular}
\centerline{\resizebox{5in}{2.5in}{\includegraphics{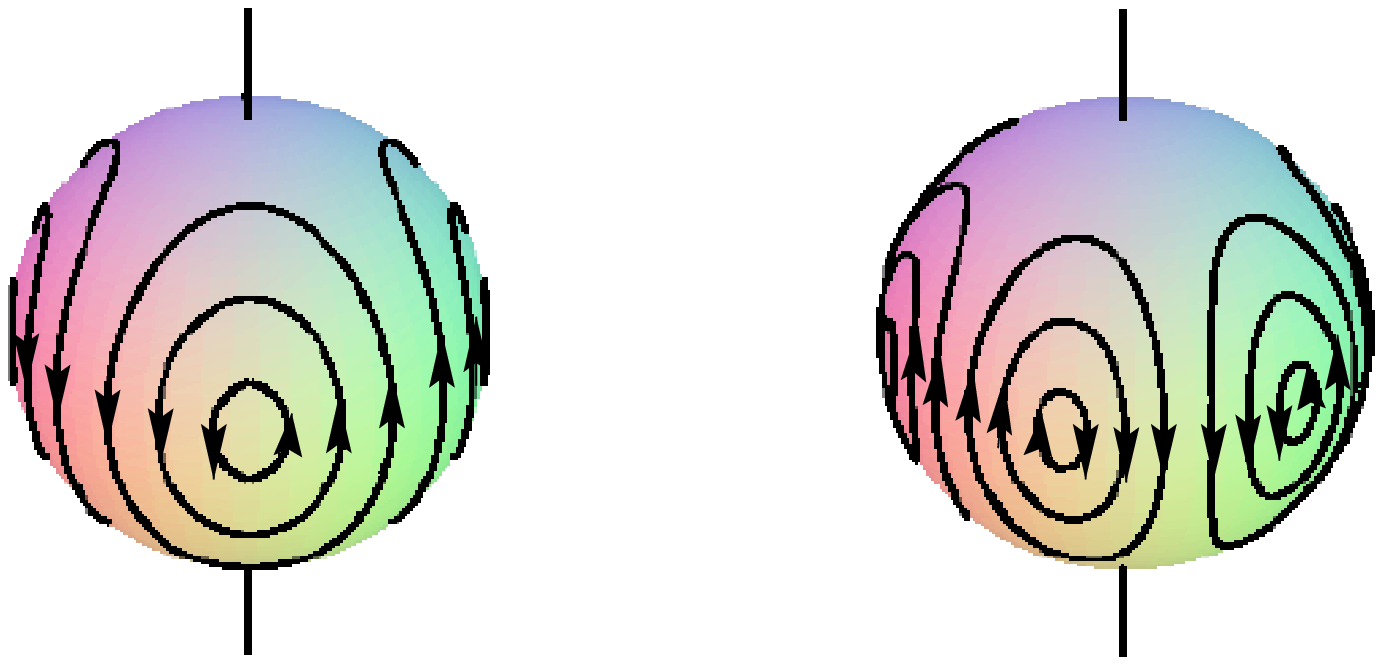}}}
\begin{tabular}{rcr}
\hphantom{123456789012345}$l=2$ & \hphantom{12345678901234567890123456} 
& $l=3$ \\
\hphantom{123456789012345}$m=2$ & \hphantom{12345678901234567890123456} 
& $m=3$
\end{tabular}
\caption{Images of the perturbed velocity field $\delta v^a$ for a few
of the newtonian r-modes at a fixed time.  The velocity profile shown 
rotates forward in the inertial frame with angular velocity
$(l-1)(l+2)\Omega/l(l+1)$ and backward in the rotating frame with
angular velocity $2\Omega/l(l+1)$.  Fluid elements oscillate in small
circles in the rotating frame as their velocity changes in accordance
with this rotating profile.}
\label{rmodes}
\end{figure}

With $m=0$ and $l=1$, Eq. (\ref{N:p&p_freq}) simply becomes
$\kappa\Omega=\sigma=0$.  The radial behaviour of this 
stationary solution is then determined by the other equations.  
The definition of $Q_l$, Eq. (\ref{Q_l}), gives,
\be
\ba[t]{ccccc}
Q^2_{l-1} = 0, & & Q^2_l = \ds{\frac{1}{3}} 
&\mbox{and} &  Q^2_{l+1} = \ds{\frac{4}{15}}.
\ea
\ee
These imply that Eq. (\ref{N:exact4}) is trivially satisfied, while
Eqs. (\ref{N:exact2}) and (\ref{N:exact3}) both become,
\be
0 = U'_l-\frac{2}{r}U_l
\ee
or
\be
U_l(r)=Kr^2,
\label{N:rad_dep}
\ee
for some constant $K$.  If we define the constant
\be
\hat\Omega = - i\left(\frac{3}{4\pi}\right)^\half K,
\ee
then our perturbed 3-velocity (\ref{v_exp}) simply 
becomes\footnote{We use the standard normalization for the 
spherical harmonic $Y_1^0 = \sqrt{3/4\pi} \cos\theta$. 
(See, e.g. Jackson \cite{j75}, p.99.)},
\be
\delta v^a = -iKr^2 \epsilon^{abc} \nabla_b Y_1^0 \nabla_c r 
= \hat\Omega \varphi^a,
\ee
which represents a small change $\hat\Omega$ in the uniform
angular velocity $\Omega$ of the star, as claimed.  This
perturbed velocity field is displayed in Fig. (\ref{rmodes}).\\

\noindent
\textit{The case $l=m>0$.}\\

\noindent
Eqs. (\ref{N:exact2})-(\ref{N:p&p_freq}) also have a simple 
solution when $l=m>0$.  The Papalouizou and Pringle frequency
(\ref{N:p&p_freq}) becomes,
\be
\kappa\Omega = (\sigma+m\Omega) = \frac{2\Omega}{(m+1)}.
\label{N:rmode_freq}
\ee
The definition of $Q_l$, Eq. (\ref{Q_l}), gives,
\be
\ba[t]{ccc}
Q^2_m = 0, & \mbox{and} & Q^2_{m+1} = \ds{\frac{1}{(2m+3)}}.
\ea
\ee
These, again, imply that Eq. (\ref{N:exact4}) is trivially satisfied, 
while Eqs. (\ref{N:exact2}) and (\ref{N:exact3}) both become,
\be
0 = U'_m -\frac{(m+1)}{r}U_m
\ee
with solution,
\be
U_m(r)=\left(\frac{r}{R}\right)^{m+1}.
\label{N:rmode_func}
\ee
We have chosen the normalization of this solution so that 
$U_m=1$ at the surface of the star, $r=R$.  Images of the 
perturbed velocity field $\delta v^a$ for the r-modes with
$1\leq l=m\leq 3$ are shown in Fig. (\ref{rmodes}). 
We note that the mode with $l=m=1$ has zero-frequency 
in the inertial frame,
\be
\sigma = (\kappa-1)\Omega = 0,
\ee
and represents rotation of the star about an axis 
perpendicular to its original axis of rotation.
In addition, we note that the r-mode with $l=m=2$ is the one 
expected to dominate the gravitational radiation driven 
instability in hot, young neutron stars.

\section{Numerical Method}

In our numerical solution, we restrict consideration to slowly 
rotating stars, finding axial- and polar-led hybrids to lowest 
order in the angular velocity $\Omega$.  That is, we assume that 
perturbed quantities introduced above obey the following ordering 
in $\Omega$:
\be
\begin{array}{rrrr}
W_l \sim O(1),      & V_l   \sim O(1), & U_l \sim O(1), & \\
\delta \rho \sim O(\Omega), & \delta p \sim O(\Omega), & 
\delta \Phi \sim O(\Omega), & \sigma \sim O(\Omega).           
\label{Newt:order}
\end{array}
\ee
The $\Omega \rightarrow 0$ limit of such a perturbation is a sum of 
the zero-frequency axial and polar perturbations considered in 
Sect. 2.2. Note that, although the relative orders of $\delta \rho$ 
and $\delta v^a$ are physically meaningful, there is an arbitrariness 
in their absolute order.  If $(\delta \rho, \delta v^a)$ is a solution 
to the linearized equations, so is $(\Omega\delta \rho, \Omega\delta v^a)$.  
We have chosen the order (\ref{Newt:order}) to reflect the existence of 
well-defined, nontrivial velocity perturbations of the spherical model.  
Other authors (e.g., Lindblom and Ipser \cite{li98}) adopt a convention in 
which $\delta v^a = O(\Omega)$ and $\delta\rho = O(\Omega^2)$.

To lowest order, the equations governing these perturbations are the 
perturbed Euler equations (\ref{eq2})-(\ref{eq4}) and the perturbed 
mass conservation equation, (\ref{continuity}), which becomes
\be
rW'_l + \left( 1 + r \frac{\rho '}{\rho}\right) W_l - l(l+1) V_l = 0.  
\label{eq1}
\ee

In addition, the perturbations must satisfy the boundary conditions of 
regularity at the center of the star, $r=0$, regularity at the surface of 
the star, $r=R$, and the vanishing of the lagrangian change in the pressure 
at the surface of the star,
\be
0 = \Delta p \equiv \delta p + \pounds_{\xi} p = \xi^r p' 
+ O\mbox{($\Omega$)}.
\ee
Equations (\ref{eq2})-(\ref{eq4}) and (\ref{eq1}) are a system of ordinary 
differential equations for $W_{l'}(r)$, $V_{l'}(r)$ and $U_{l'}(r)$ 
(for all $l'$). Together with the boundary conditions, these equations 
form a non-linear eigenvalue problem for the parameter $\kappa$, where 
$\kappa\Omega$ is the mode frequency in the rotating frame.

To solve for the eigenvalues we proceed as follows.  We first ensure 
that the boundary conditions are automatically satisfied by expanding 
$W_{l'}(r)$, $V_{l'}(r)$ and $U_{l'}(r)$ (for all $l'$) in regular 
power series about the surface and center of the star. Substituting 
these series into the differential equations results in a set of 
algebraic equations for the expansion coefficients. These algebraic 
equations may be solved for arbitrary values of $\kappa$ using standard 
matrix inversion methods. For arbitrary values of $\kappa$, however, the 
series solutions about the center of the star will not agree 
with those about the surface of the star.  The requirement that the series 
agree at some matching point, $0<r_0<R$, then becomes the condition that 
restricts the possible values of the eigenvalue, $\kappa_0$.  

The equilibrium solution $(\rho, \Phi)$ appears in the perturbation 
equations only through the quantity $(\rho'/\rho)$ in equation (\ref{eq1}). 
We begin by writing the series expansion for this quantity about $r=0$ as
\be \left( \frac{\rho '}{\rho}\right) = \frac{1}{R}
\sum^\infty_{\stackrel{i=1}{\mbox{\tiny $i$ odd}}} 
\pi_i \left(\frac{r}{R}\right)^i,                     
\label{rho_x}
\ee
and about $r=R$ as
\be \left(\frac{\rho '}{\rho}\right) = \frac{1}{R}
\sum^\infty_{k=-1} \tilde\pi_k \left(1-\frac{r}{R}\right)^k,  
\label{rho_y}
\ee
where the $\pi_i$ and $\tilde\pi_k$ are determined from the equilibrium 
solution.

Because (\ref{eq2}) relates $U_l(r)$ algebraically to $W_{l\pm 1}(r)$ and  
$V_{l\pm 1}(r)$, we may eliminate $U_{l'}(r)$ (all $l'$) from 
(\ref{eq3}) and (\ref{eq4}).  We then need only work with one of equations 
(\ref{eq3}) or (\ref{eq4}) since the equations (\ref{eq2}) through 
(\ref{eq4}) are related by $\nabla_a q^a=0$.  

We next replace $\rho'/\rho$, $W_{l'}$, and $V_{l'}$ in equations (\ref{eq3})
or (\ref{eq4}) by their series expansions.  We eliminate the $U_{l'}(r)$
from either (\ref{eq3}) or (\ref{eq4}) and, again, substitute for the
$W_{l'}(r)$ and $V_{l'}(r)$.  Finally, we write down the matching condition
at the point $r_0$ equating the series expansions about $r=0$ to the
series expansions about $r=R$.  The result is a linear algebraic system 
which we may represent schematically as
\be 
Ax=0. 
\label{linalg} 
\ee
In this equation, $A$ is a matrix which depends non-linearly on the
parameter $\kappa$, and $x$ is a vector whose components are the
unknown coefficients in the series expansions for the $W_{l'}(r)$ and
$V_{l'}(r)$. In Appendix B, we explicitly present the equations making up
this algebraic system as well as the forms of the regular series
expansions for $W_{l'}(r)$ and $V_{l'}(r)$.

To satisfy equation (\ref{linalg}) we must find those values of
$\kappa$ for which the matrix $A$ is singular, i.e., we must find the
zeroes of the determinant of $A$.  We truncate the spherical harmonic
expansion of $\delta v^a$ at some maximum index $l_{\mbox{\tiny max}}$
and we truncate the radial series expansions about $r=0$ and $r=R$ at
some maximum powers $i_{\mbox{\tiny max}}$ and $k_{\mbox{\tiny max}}$, 
respectively.

The resulting finite matrix is band diagonal. To find the zeroes of its
determinant we use standard root finding techniques combined with
routines from the LAPACK linear algebra libraries (Anderson et al. 
\cite{lapack}).  
We find that the eigenvalues, $\kappa_0$, computed in this manner converge 
quickly as we increase $l_{\mbox{\tiny max}}$, $i_{\mbox{\tiny max}}$ and
$k_{\mbox{\tiny max}}$.

The eigenfunctions associated with these eigenvalues are determined by
the perturbation equations only up to normalization.  Given a
particular eigenvalue, we find its eigenfunction by replacing one of
the equations in the system (\ref{linalg}) with the normalization
condition that
\be
\begin{array}{ll}
V_m(r=R) = 1 & \mbox{for polar-hybrids, or that} \\
V_{m+1}(r=R) = 1 & \mbox{for axial-hybrids.} 
\label{N:norm_cond})
\end{array}
\ee
Since we have eliminated one of the rows of the singular matrix $A$ in 
favor of this condition, the result is an algebraic system of the form
\be \tilde A x = b, \label{linalg2} \ee where $\tilde A$ is now a
non-singular matrix and $b$ is a known column vector.  We solve this
system for the vector $x$ using routines from LAPACK and reconstruct
the various series expansions from this solution vector of coefficients.

\section{Eigenvalues and Eigenfunctions}

We have computed the eigenvalues and eigenfunctions for uniform density 
stars and for $n=1$ polytropes, models obeying the polytropic equation of 
state $p=K\rho^2$, where $K$ is a constant.  Our numerical solutions for 
the uniform density star agree with the recent results of Lindblom and 
Ipser \cite{li98} who find analytic solutions for the hybrid modes in rigidly 
rotating uniform density stars with arbitrary angular velocity - the 
Maclaurin spheroids.  Their calculation uses the two-potential formalism 
(Ipser and Managan \cite{im85}; Ipser and Lindblom \cite{il90}) in 
which the equations for the perturbation 
modes are reformulated as coupled differential equations for a fluid 
potential, $\delta U$, and the gravitational potential, $\delta \Phi$. 
All of the perturbed fluid variables may be expressed in terms of these 
two potentials.  The analysis follows that of Bryan \cite{b1889} who 
found that the equations are separable in a non-standard spheroidal 
coordinate system.  

The Bryan/Lindblom-Ipser eigenfunctions $\delta U_0$ and $\delta \Phi_0$ 
turn out to be products of associated Legendre polynomials of their 
coordinates.  This simple form of their solutions leads us to expect that 
our series solutions might also have a simple form - even though their 
unusual spheroidal coordinates are rather complicated functions of $r$ 
and $\theta$.  In fact, we do find that the modes of the uniform density
star have a particularly simple structure.  For any particular mode, both 
the angular and radial series expansions terminate at some finite indices 
$l_0$ and $i_0$ (or $k_0$).  That is, the spherical harmonic expansion 
(\ref{v_exp}) of $\delta v^a$ contains only terms with $m\leq l\leq l_0$ 
for this mode, and the coefficients of this expansion - the $W_l(r)$, 
$V_l(r)$ and $U_l(r)$ - are polynomials of order $m+i_0$. For all 
$l_0\geq m$ there exist a number of modes terminating at $l_0$.

In Tables \ref{ef_m1a} to \ref{ef_m2p} we present the functions 
$W_l(r)$, $V_l(r)$ and $U_l(r)$ for all of the axial- and polar-led 
hybrids with $m=1$ and $m=2$ for a range of values of the terminating 
index $l_0$.  (See also Figure \ref{ef_mac}.)  For given values of 
$m>0$ and $l_0$ there are $l_0-m+1$ modes. (When $m=0$ there are $l_0$ 
modes. See Eq. (\ref{li_eq}) below.) We also find that the last 
term in the expansion (\ref{v_exp}), the term with $l=l_0$, is always 
axial for both types of hybrid modes.  This fact, together with the 
fact that the parity of the modes is,
\be
\pi = \Biggl\{ \begin{array}{lr}
(-1)^m     & \mbox{for polar-led hybrids\phantom{,}} \\
(-1)^{m+1} & \mbox{for axial-led hybrids,} 
\end{array}
\ee
(for $m>0$) implies that $l_0-m+1$ must be even for polar-led modes and 
odd for axial-led modes.

The fact that the various series terminate at $l_0$, $i_0$ and $k_0$ 
implies that 
Equations (\ref{linalg}) and (\ref{linalg2}) will be exact as long as we 
truncate 
the series at $l_{\mbox{\tiny max}}\geq l_0$, 
$i_{\mbox{\tiny max}}\geq i_0$ and $k_{\mbox{\tiny max}}\geq k_0$.

To find the eigenvalues of these modes we search the $\kappa$ axis for 
all of the zeroes of the determinant of the matrix $A$ in equation 
(\ref{linalg}). We begin by fixing $m$ and performing the search with 
$l_{\mbox{\tiny max}}=m$.  We then increase $l_{\mbox{\tiny max}}$ by 
1 and repeat the search (and so on).  At any given value of 
$l_{\mbox{\tiny max}}$, the search finds all of the eigenvalues 
associated with the eigenfunctions terminating at 
$l_0\leq l_{\mbox{\tiny max}}$.

In Table \ref{ev_mac}, we present the eigenvalues $\kappa_0$ found 
by this method for the axial- and polar-led hybrid modes of uniform 
density stars for a range of values of $l_0$ and $m$. Observe that 
many of the eigenvalues, (marked with a $\ast$) satisfy the CFS
instability condition $\sigma(\sigma+m\Omega)<0$ (see Sect. 1.1).
The modes whose frequencies satisfy this condition are subject to the 
non-axisymmetric gravitational radiation driven instability 
in the absence of viscosity.  The modes having $l_0=m>0$ 
(or $l_0=1$ for $m=0$) are the purely axial r-modes with eigenvalues
$\kappa_0 = 2/(m+1)$ (or $\kappa_0=0$ for $m=0$) discussed in Sect. 2.3.1.
We find that there are no purely polar modes satisfying our assumptions 
(\ref{Newt:order}) in these stellar models.

We have compared these eigenvalues with those of Lindblom and Ipser 
\cite{li98}.  To lowest non-trivial order in $\Omega$ their equation for 
the eigenvalue, $\kappa_0$, can be expressed in terms of associated
Legendre polynomials\footnote{The index $l$ used by Lindblom and Ipser 
is related to our $l_0$ by $l=l_0+1$.  Our convention agrees with the 
usual labelling of the $l_0=m$ pure axial modes.} (see Lindblom and 
Ipser's equation 6.4), as
\be
(4-\kappa_0^2) \frac{d}{d\kappa} P_{l_0+1}^m (\half\kappa_0) 
- 2m P_{l_0+1}^m (\half\kappa_0) = 0.                         
\label{li_eq}
\ee
For given values of $m>0$ and $l_0$ this equation has $l_0-m+1$ roots 
(corresponding to the number of distinct modes), which can easily be 
found numerically. (For $m=0$ there are $l_0$ roots.) For the range of 
values of $m$ and $l_0$ checked our eigenvalues agree with these to 
machine precision. (Compare our Table \ref{ev_mac} with Table 1 in 
Lindblom and Ipser \cite{li98}.)

We have also compared our eigenfunctions with those of Lindblom and 
Ipser. For a uniformly rotating, isentropic star, the fluid velocity 
perturbation, $\delta v^a$, is related (Ipser and Lindblom \cite{il90}) 
to their fluid potential $\delta U$ by 
\be
\nabla_a \delta U = - 
\left[ i\kappa\Omega g_{ab} + 2\nabla_b v_a \right] \delta v^b.
\ee
Since the $\varphi$ component of this equation is simply
\be
im \delta U = - \Omega r^2 \sin^2 \theta 
\left[  
\frac{2}{r}\delta v^r + 2\cot\theta\delta v^{\theta} 
+ i\kappa\delta v^{\varphi}
\right],
\ee
it is straightforward numerically to construct this quantity from 
the components of our $\delta v^a$ and compare it with the analytic 
solutions for $\delta U$ given by Lindblom and Ipser (see their 
Eq. 7.2).  We have compared these solutions on a $20\times 40$ 
grid in the ($r-\theta$) plane and found that they agree (up to 
normalization) to better than $1$ part in $10^9$ for all cases checked.

Because of the use of the two-potential formalism and the unusual
coordinate system used in their analysis, the axial- or polar-hybrid
character of the Bryan/Lindblom-Ipser solutions is not obvious.  Nor is
it evident that these solutions have, as their $\Omega\rightarrow 0$
limit, the zero-frequency convective modes described in Sect. 2.2.  The
comparison of their analytic results with our numerical work has served
the dual purpose of clarifying these properties of the solutions and of
testing the accuracy of our code.  The computational differences are
minor between the uniform density calculation and one in which the star
obeys a more realistic equation of state.  Thus, this testing gives us
confidence in the validity of our code for the polytrope calculation.
As a further check, we have written two independent codes and compared
the eigenvalues computed from each. One of these codes is based on the
set of equations described in Appendix B.  The other is based on the
set of second order equations that results from using the mass
conservation equation, (\ref{eq1}), to substitute for all the $V_l(r)$
in favor of the $W_l(r)$.

For the $n=1$ polytrope we will consider and, more generally, for any
isentropic equation of state, the purely axial r-modes are independent
of the equation of state.  In both isentropic and non-isentropic stars,
pure r-modes exist whose velocity field is, to lowest order in
$\Omega$, an axial vector field belonging to a single angular harmonic
(and restricted to harmonics with $l=m$ in the isentropic case).  The
frequency of such a mode is given (to order $\Omega$) by the Papalouizou
and Pringle \cite{pp78} expression, Eq. (\ref{p&p_freq}), and is
independent of the equation of state.  As we saw in Sect. 2.3.1, only 
those modes having $l=m$ (or $l=1$ for $m=0$) exist in isentropic stars, 
and for these modes the eigenfunctions are also independent of the 
(isentropic) equation of state.  This
independence of the equation of state occurs for the r-modes because
(to lowest order in $\Omega$) fluid elements move in surfaces of
constant $r$ (and thus in surfaces of constant density and pressure).
For the hybrid modes, however, fluid elements are not confined to
surfaces of constant $r$ and one would expect the eigenfrequencies and
eigenfunctions to depend on the equation of state.

Indeed, we find such a dependence.  The hybrid modes of the $n=1$ 
polytrope are not identical to those of the uniform density star. 
On the other hand, the modes do not appear to be very sensitive to 
the equation of state.  We have found that the character of the 
polytropic modes is similar to the modes of the uniform density star, 
except that the radial and angular series expansions do not terminate. 
For each eigenfunction in the uniform density star there is a 
corresponding eigenfunction in the polytrope with a slightly different 
eigenfrequency (See Table \ref{ev_poly}.)  For a given mode of the 
uniform density star, the series expansion (\ref{v_exp}) terminates 
at $l=l_0$.  For the corresponding polytrope mode, the expansion 
(\ref{v_exp}) does not terminate, but it does converge quickly.  
The largest terms in (\ref{v_exp}) with $l>l_0$ are more than an 
order of magnitude smaller than those with $l\leq l_0$ and they 
decrease rapidly as $l$ increases.  Thus, the terms that dominate 
the polytrope eigenfunctions are those that correspond to the non-zero 
terms in the corresponding uniform density eigenfunctions.  

In Figures \ref{ef_mac} and \ref{ef_poly} we display the coefficients 
$W_l(r)$, $V_l(r)$ and $U_l(r)$ of the expansion (\ref{v_exp}) for the 
same $m=2$ axial-led hybrid mode in each stellar model.  For the uniform 
density star (Figure \ref{ef_mac}) the only non-zero coefficients for 
this mode are those with $l\leq l_0=4$.  These coefficients are presented
explicitly in Table \ref{ef_m2a} and are low order polynomials in $r$.  
For the corresponding mode in the polytrope, we present in Figure 
\ref{ef_poly} the first seven coefficients of the expansion (\ref{v_exp}).  
Observe that those coefficients with $l\leq 4$ are similar to the 
corresponding functions in the uniform density mode and dominate the 
polytrope eigenfunction.  The coefficients with $4< l\leq 6$ are an 
order of magnitude smaller than the dominant coefficients and those 
with $l>6$ are smaller still. (Since they would be indistinguishable 
from the $(r/R)$ axis, we do not display the coefficients having $l>6$ 
for this mode.)  

Just as the angular series expansion fails to terminate for the polytrope 
modes, so too do the radial series expansions for the functions $W_l(r)$, 
$V_l(r)$ and $U_l(r)$.  We have seen that in the uniform density star 
these functions are polynomials in $r$ (Tables \ref{ef_m1a} through 
\ref{ef_m2p}).  In the polytropic star, the radial series do not 
terminate and we are required to work with both sets of radial series 
expansions - those about the center of the star and those about its 
surface - in order to represent the functions accurately everywhere 
inside the star.

In Figures \ref{fig3} through \ref{fig11} we compare corresponding 
functions from each type of star.  For example, Figures \ref{fig3}, 
\ref{fig4}, and \ref{fig5} show the functions $W_l(r)$, $V_l(r)$ and 
$U_l(r)$ (respectively) for $l\leq 6$ for a particular $m=1$ polar-led 
hybrid mode.  In the uniform density star this mode has eigenvalue 
$\kappa_0=1.509941$, and in the polytrope it has eigenvalue 
$\kappa_0=1.412999$.  The only non-zero functions in the uniform density 
mode are those with $l\leq l_0=2$ and they are simple polynomials in $r$ 
(see Table \ref{ef_m1p}).  Observe that these functions are similar, but 
not identical to, their counterparts in the polytrope mode, which have 
been constructed from their radial series expansions about $r=0$ and 
$r=R$ (with matching point $r_0=0.5R$).  Again, note the convergence with 
increasing $l$ of the polytrope eigenfunction.  The mode is dominated by 
the terms with $l\leq 2$ and those with $l>2$ decrease rapidly with $l$. 
(The $l=5$ and $l=6$ coefficients are virtually indistinguishable from 
the $(r/R)$ axis.)

Because the polytrope eigenfunctions are dominated by their $l\leq l_0$ 
terms, the eigenvalue search with $l_{\mbox{\tiny max}}=l_0$ will find 
the associated eigenvalues approximately.  We compute these approximate 
eigenvalues of the polytrope modes using the same search technique as 
for the uniform density star.  We then increase $l_{\mbox{\tiny max}}$ 
and search near one of the approximate eigenvalues for a corrected value, 
iterating this procedure until the eigenvalue converges to the desired 
accuracy. We present the eigenvalues found by this method in 
Table \ref{ev_poly}.

As a further comparison between the mode eigenvalues in the polytropic 
star and those in the uniform density star we have modelled a sequence 
of ``intermediate'' stars. By multiplying the expansions (\ref{rho_x}) 
and (\ref{rho_y}) for $(\rho'/\rho)$ by a scaling factor,
$\epsilon\in [0,1]$, we can simulate a continuous sequence of stellar
]models connecting the uniform density star ($\epsilon=0$) to the 
polytrope ($\epsilon=1$).  We find that an eigenvalue in the uniform 
density star varies smoothly as function of $\epsilon$ to the 
corresponding eigenvalue in the polytrope.

\section{Dissipation}

The effects of gravitational radiation and viscosity on the pure 
$l_0=m$ r-modes discussed in Sect. 2.3.1 have already been studied 
by a number of authors.  (Lindblom et al. \cite{lom98}, Owen et al. 
\cite{o98}, Andersson et al. \cite{aks98}, Kokkotas and Stergioulas 
\cite{ks98}, Lindblom et al. \cite{lmo99}) 
All of these modes are unstable to gravitational radiation 
reaction, and for some of them this instability strongly dominates 
viscous damping. We now consider the effects of dissipation on the 
axial- and polar-hybrid modes.

To estimate the timescales associated with viscous damping and 
gravitational radiation reaction we follow the methods used for 
the $l_0=m$ modes (Lindblom et al. \cite{lom98}, see also Ipser
and Lindblom \cite{il91}).  
When the energy radiated per cycle is small compared to the energy 
of the mode, the imaginary part of the mode frequency is accurately 
approximated by the expression
\be
\frac{1}{\tau} = - \frac{1}{2E} \frac{dE}{dt},   
\label{tau}
\ee
where $E$ is the energy of the mode as measured in the rotating frame,
\be
E = \frac{1}{2} \int \left[ 
\rho \delta v^a \delta v^{\ast}_a 
+ \left( \frac{\delta p}{\rho}+\delta\Phi\right) 
\delta\rho^{\ast}
\right] d^3 x .
\label{E}
\ee
The rate of change of this energy due to dissipation by viscosity 
and gravitational radiation is,
\begin{eqnarray}
\frac{dE}{dt} &=& - \int \left( 
2\eta\delta\sigma^{ab}\delta\sigma_{ab}^{\ast} 
+\zeta \delta\theta\delta\theta^{\ast}
\right) \nonumber \\
 & & -\sigma (\sigma + m\Omega) \sum_{l\geq 2} N_l \sigma^{2l} \left(
\left|\delta D_{lm}\right|^2   + \left|\delta J_{lm}\right|^2
\right). 
\label{dEdt}
\end{eqnarray}
The first term in (\ref{dEdt}) represents dissipation due to shear 
viscosity, where the shear, $\delta\sigma_{ab}$, of the perturbation is
\be
\delta\sigma_{ab} = \half\left( 
\nabla_a \delta v_b + \nabla_b \delta v_a 
- \frac{2}{3} g_{ab} \nabla_c \delta v^c
\right),
\ee
and the coefficient of shear viscosity for hot neutron-star matter is 
(Cutler and Lindblom \cite{cl87}; Sawyer \cite{s89})
\be
\eta = 2\times 10^{18} 
\left(\frac{\rho}{10^{15}\mbox{g}\!\cdot\!\mbox{cm}^{-3}}\right)^{\frac{9}{4}}
\left(\frac{10^9K}{T}\right)^2 \ 
\mbox{g}\!\cdot\!\mbox{cm}^{-1}\!\cdot\!\mbox{s}^{-1}.
\label{eta}
\ee

The second term in (\ref{dEdt}) represents dissipation due to bulk 
viscosity, where the expansion, $\delta\theta$, of the perturbation is
\be
\delta\theta = \nabla_c \delta v^c
\ee
and the bulk viscosity coefficient for hot neutron star matter is 
(Cutler and Lindblom \cite{cl87}; Sawyer \cite{s89})
\be
\zeta = 6\times 10^{25} 
\left(\frac{1\mbox{Hz}}{\sigma + m\Omega}\right)^2
\left(\frac{\rho}{10^{15}\mbox{g}\!\cdot\!\mbox{cm}^{-3}}\right)^2
\left(\frac{T}{10^9K}\right)^6 \ 
\mbox{g}\!\cdot\!\mbox{cm}^{-1}\!\cdot\!\mbox{s}^{-1}.
\label{zeta}
\ee

The third term in (\ref{dEdt}) represents dissipation due to 
gravitational radiation, with coupling constant
\be
N_l = \frac{4\pi G}{c^{2l+1}}\frac{(l+1)(l+2)}{l(l-1)[(2l+1)!!]^2}.
\ee
The mass, $\delta D_{lm}$, and current, $\delta J_{lm}$, multipole 
moments of the perturbation are given by (Thorne \cite{th80}, 
Lindblom et al. \cite{lom98})
\be
\delta D_{lm} = \int \delta\rho r^l Y_l^{\ast m} d^3 x,
\label{D}
\ee 
and
\be
\delta J_{lm} = \frac{2}{c} \left(\frac{l}{l+1}\right)^{\half} \int r^l 
\left( \rho\delta v_a + \delta\rho v_a \right) Y^{a,B\ast}_{lm} d^3 x 
\label{J}
\ee
where $Y^{a,B}_{lm}$ is the magnetic type vector spherical harmonic 
(Thorne \cite{th80}) given by,
\be
Y^{a,B}_{lm} = 
- \frac{r}{\sqrt{l(l+1)}} \epsilon^{abc} \nabla_b Y_l^{m} \nabla_c r.
\ee

To lowest order in $\Omega$, the energy (\ref{E}) of the hybrid modes 
is positive definite.  Their stability is therefore determined by the 
sign of the right hand side of equation (\ref{dEdt}). We have seen that 
many of the hybrid modes have frequencies satisfying the CFS instability
criterion $\sigma(\sigma+m\Omega)<0$.  
It is now clear that this makes the third term in Eq. (\ref{dEdt}) 
positive, implying that gravitational radiation 
reaction tends to drive these modes unstable.  As discussed in Sect. 1.1, 
however, to determine the actual stability of these modes, we must 
evaluate the various dissipative terms in (\ref{dEdt}).

We first substitute for $\delta v^a$ the spherical harmonic expansion 
(\ref{v_exp}) and use the orthogonality relations for vector spherical 
harmonics (Thorne \cite{th80}) to perform the angular integrals.  The energy 
of the modes in the rotating frame then becomes
\be
E = \sum^{\infty}_{l=m} \half \int_0^R \rho 
\left[ W_l^2 + l(l+1)V_l^2 + l(l+1)U_l^2 \right] dr.
\ee

To calculate the dissipation due to gravitational radiation reaction 
we must evaluate the multipole moments (\ref{D}) and (\ref{J}).  
To lowest order in $\Omega$ the mass multipole moments (\ref{D}) vanish 
and the current multipole moments are given by 
\be
\delta J_{lm} = \frac{2il}{c} \int_0^R \rho r^{l+1} U_l dr.  
\label{J_int}
\ee

To calculate the dissipation due to bulk viscosity we must evaluate 
the expansion, $\delta\theta = \nabla_c \delta v^c$, of the 
perturbation.  For uniform density stars this quantity vanishes 
identically by the mass conservation equation (\ref{continuity}).  
For the $l_0=m$, pure axial modes the expansion, again, vanishes 
identically, regardless of the equation of state.  To compute the 
bulk viscosity of these modes it is necessary to work to higher order 
in $\Omega$ (Andersson et al. \cite{aks98}, Lindblom et al. \cite{lmo99}).  
On the other hand, for the 
new hybrid modes in which we are interested, the expansion of the 
fluid perturbation is non-zero in the slowly rotating polytropic stars. 
After substituting for $\delta v^a$ its series expansion and performing 
the angular integrals, the bulk viscosity contribution to (\ref{dEdt}) 
becomes
\be
\left( \frac{dE}{dt} \right)_B = - \sum^{\infty}_{l=m} \int_0^R 
\frac{\zeta}{r^2} \left[ r W'_l + W_l - l(l+1)V_l \right]^2 dr
\ee
In a similar manner, the contribution to (\ref{dEdt}) from shear 
viscosity becomes
\be
\left( \frac{dE}{dt} \right)_S = 
- \sum^{\infty}_{l=m} \int_0^R  
\begin{array}[t]{l}
\ds{\frac{2\eta}{r^2}} \Biggl\{
\ds{\frac{2}{3} \left[ r^3 \left(\frac{W_l}{r^2}\right)'\right]^2 
+ \half l(l+1) W_l^2} \\
\\
+ \ds{\half l(l+1)\left[ r^3 \left(\frac{V_l}{r^2}\right)'\right]^2
+ \frac{1}{3} l(l+1)(2l^2+2l-3) V_l^2} \\
\\
+  \ds{l(l+1) W_l \left[ r^5 \left(\frac{V_l}{r^4}\right)'\right] 
+ \frac{2}{3} l(l+1) V_l \left(rW_l\right)'} \\
\\
+ \ds{\half l(l+1) \left[ r^3 \left(\frac{U_l}{r^2}\right)'\right]^2 
+ \half l(l+1)(l^2+l-2) U_l^2 }
\Biggr\} dr.
\end{array}
\ee

Given a numerical solution for one of the hybrid mode eigenfunctions, 
these radial integrals can be performed numerically. The resulting 
contributions to (\ref{dEdt}) also depend on the angular velocity and 
temperature of the star.  Let us express the imaginary part of the 
hybrid mode frequency (\ref{tau}) as,
\be
\frac{1}{\tau} = \frac{1}{\tilde \tau_S} \left( \frac{10^9 K}{T} \right)^2
+ \frac{1}{\tilde \tau_B} \left( \frac{T}{10^9 K} \right)^6 
\left( \frac{\pi G \bar{\rho}}{\Omega^2} \right) 
+ \sum_{l\geq 2} \frac{1}{\tilde \tau_l}
\left( \frac{\Omega^2}{\pi G \bar{\rho}} \right)^{l+1}, \label{tau2}
\ee
where $\bar{\rho}$ is average density of the star. (Compare this 
expression to the corresponding expression in Lindblom et al. \cite{lom98} 
- their equation (22) - for the $l_0=m$ pure axial modes.)  

The bulk viscosity term in this equation is stronger by a factor 
$\Omega^{-4}$ than that for the $l_0=m$ pure axial modes. This is 
because the expansion $\delta\theta$ of the hybrid mode is nonzero 
to lowest order in $\Omega$ for the polytropic star, whereas it is 
order $\Omega^2$ for the pure axial modes.  This implies that the 
damping due to bulk viscosity will be much stronger for the hybrid 
modes than for the pure axial modes in slowly rotating stars.

Note that the contribution to (\ref{tau2}) from gravitational
radiation reaction consists of a sum over all the values of $l$ with
a non-vanishing current multipole.  This sum is, of course, dominated by 
the lowest contributing multipole.

In Tables \ref{times_m1a} to \ref{times_m2p} we present the timescales 
for these various dissipative effects in the uniform density and 
polytropic stellar models that we have been considering with 
$R=12.57\mbox{km}$ and $M=1.4M_{\odot}$.  For the reasons discussed 
above, we do not present bulk viscosity timescales for the uniform 
density star.

\begin{figure}
\centerline{\includegraphics[3in,3.25in][6in,8.25in]{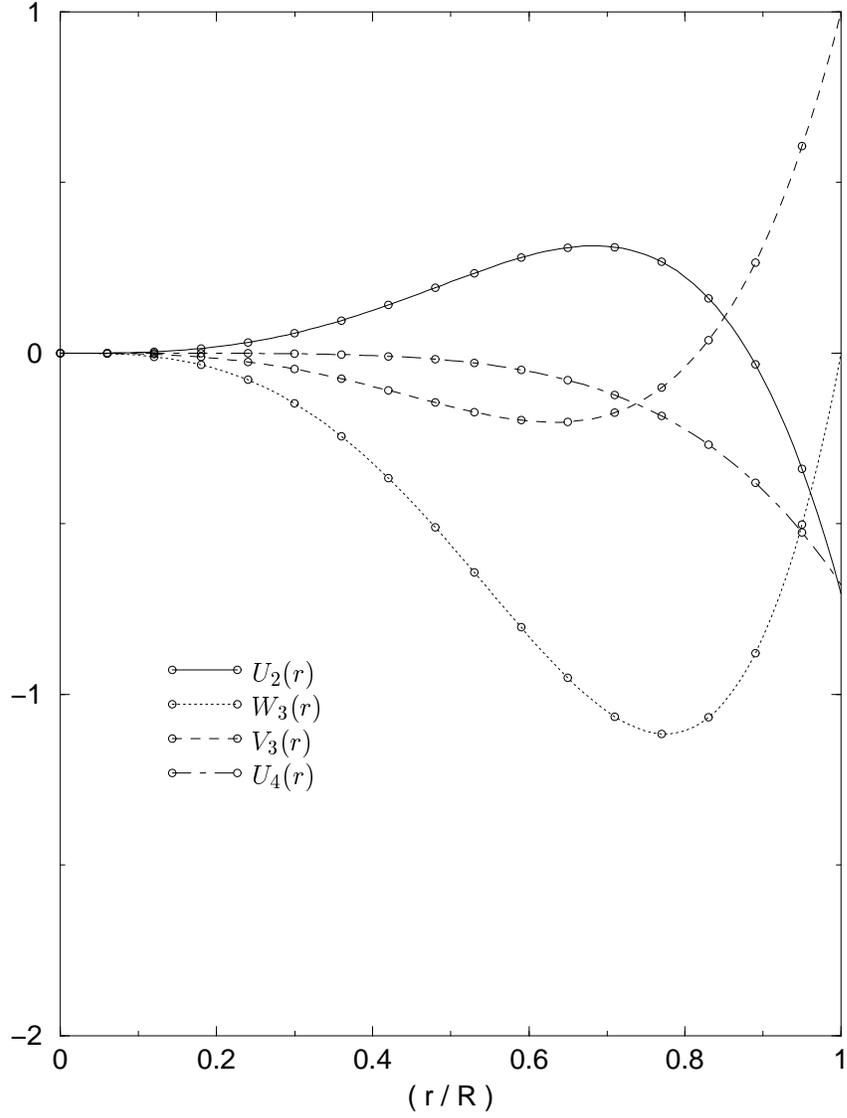}}
\caption{All of the non-zero coefficients $W_l(r)$, $V_l(r)$, 
$U_l(r)$ of the spherical harmonic expansion (\ref{v_exp}) 
for a particular $m=2$ axial-led hybrid mode of the uniform 
density star.  The mode has eigenvalue $\kappa_0 = -0.763337$.  
Note that the largest value of $l$ that appears in the expansion 
(\ref{v_exp}) is $l_0=4$ and that the functions $W_l(r)$, $V_l(r)$ 
and $U_l(r)$ are low order polynomials in $(r/R)$. (See Table 
\ref{ef_m2a}.) The mode is normalized so that $V_3(r=R)=1$.}
\label{ef_mac}
\end{figure}

\begin{figure}
\centerline{\includegraphics[3in,3.5in][6in,8.25in]{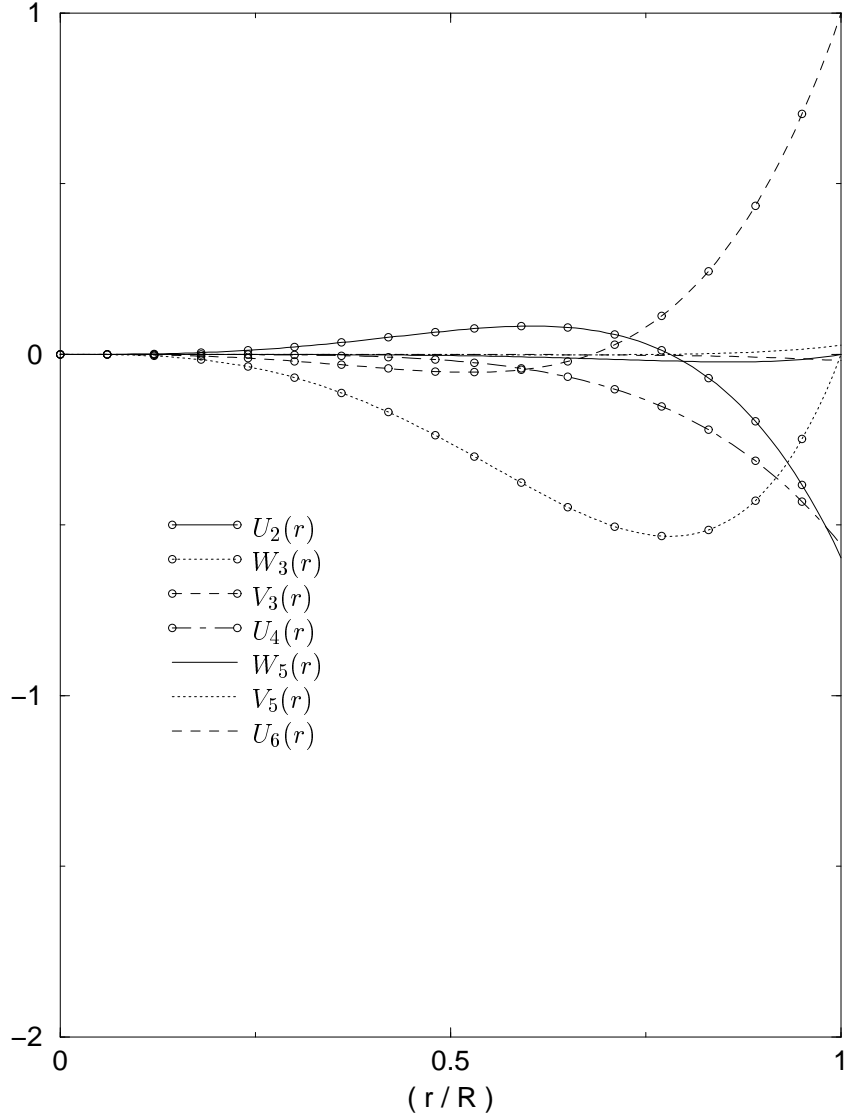}}
\caption{The coefficients $W_l(r)$, $V_l(r)$, $U_l(r)$ with 
$l\leq 6$ of the spherical harmonic expansion (\ref{v_exp}) 
for a particular $m=2$ axial-led hybrid mode of the polytropic 
star.  This is the polytrope mode that corresponds to the 
uniform density mode displayed in Figure \ref{ef_mac}. For the 
polytrope the mode has eigenvalue $\kappa_0 = -1.025883$.  
The expansion (\ref{v_exp}) converges rapidly with increasing 
$l$ and is dominated by the terms with $2\leq l\leq 4$, i.e., 
by the terms corresponding to those which are non-zero for the 
uniform density mode.  Observe that the coefficients shown with 
$4<l\leq 6$ are an order of magnitude smaller than those with 
$2\leq l\leq 4$. Those with $l>6$ are smaller still and are not 
displayed here. The mode is, again, normalized so that $V_3(r=R)=1$.}
\label{ef_poly}
\end{figure}

\begin{figure}
\centerline{\includegraphics[3in,2.5in][6in,8in]{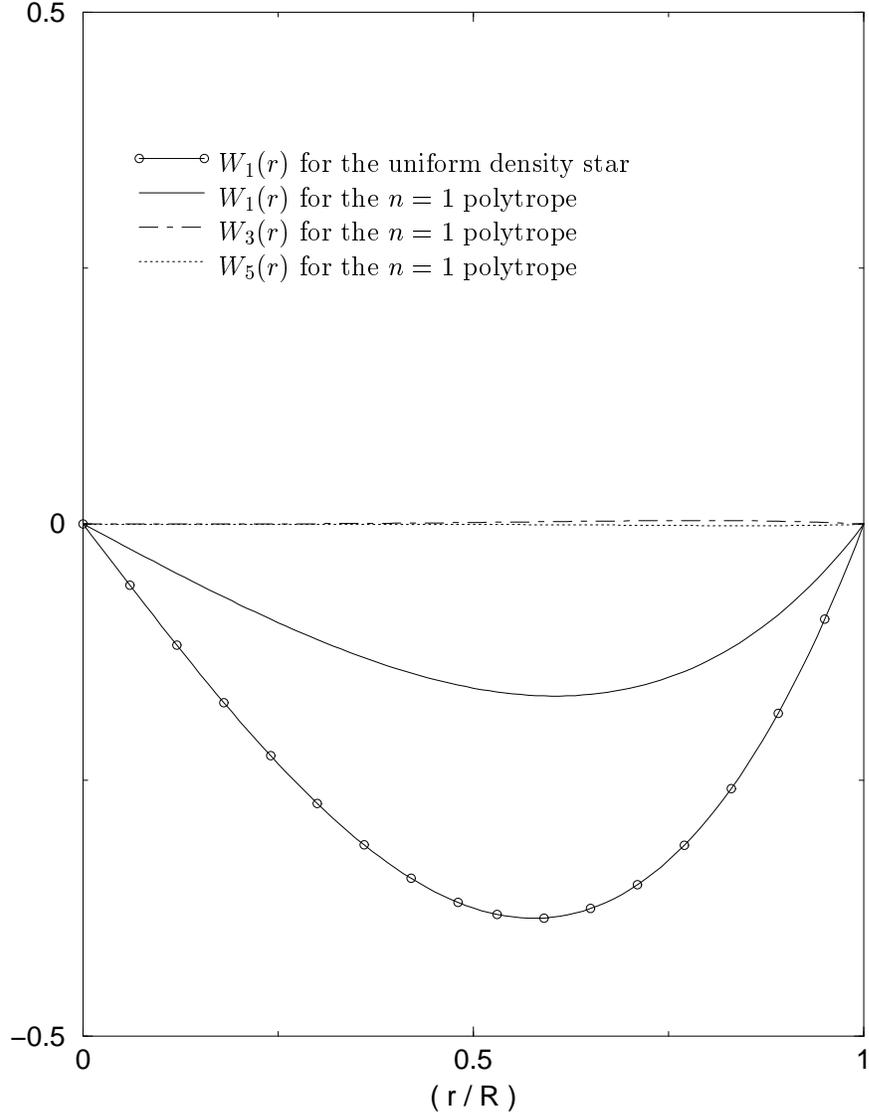}}
\caption{The functions $W_l(r)$ with $l\leq 6$ for a particular 
$m=1$ polar-led hybrid mode. For the uniform density star this 
mode has eigenvalue $\kappa_0=1.509941$ and $W_1=-x+x^3$ ($x=r/R$) 
is the only non-vanishing $W_l(r)$ (see Table \ref{ef_m1p}). The 
corresponding mode of the polytropic star has eigenvalue 
$\kappa_0=1.412999$.  Observe that $W_1(r)$ for the polytrope, 
which has been constructed from its power series expansions about 
$r=0$ and $r=R$, is similar, though not identical, to the 
corresponding $W_1(r)$ for the uniform density star. Observe also 
that the functions $W_l(r)$ with $l>1$ for the polytrope are more 
than an order of magnitude smaller than $W_1(r)$ and become smaller 
with increasing $l$. ($W_5(r)$ is virtually indistinguishable from 
the $(r/R)$ axis.)}
\label{fig3}
\end{figure}

\begin{figure}
\centerline{\includegraphics[3in,2.5in][6in,8in]{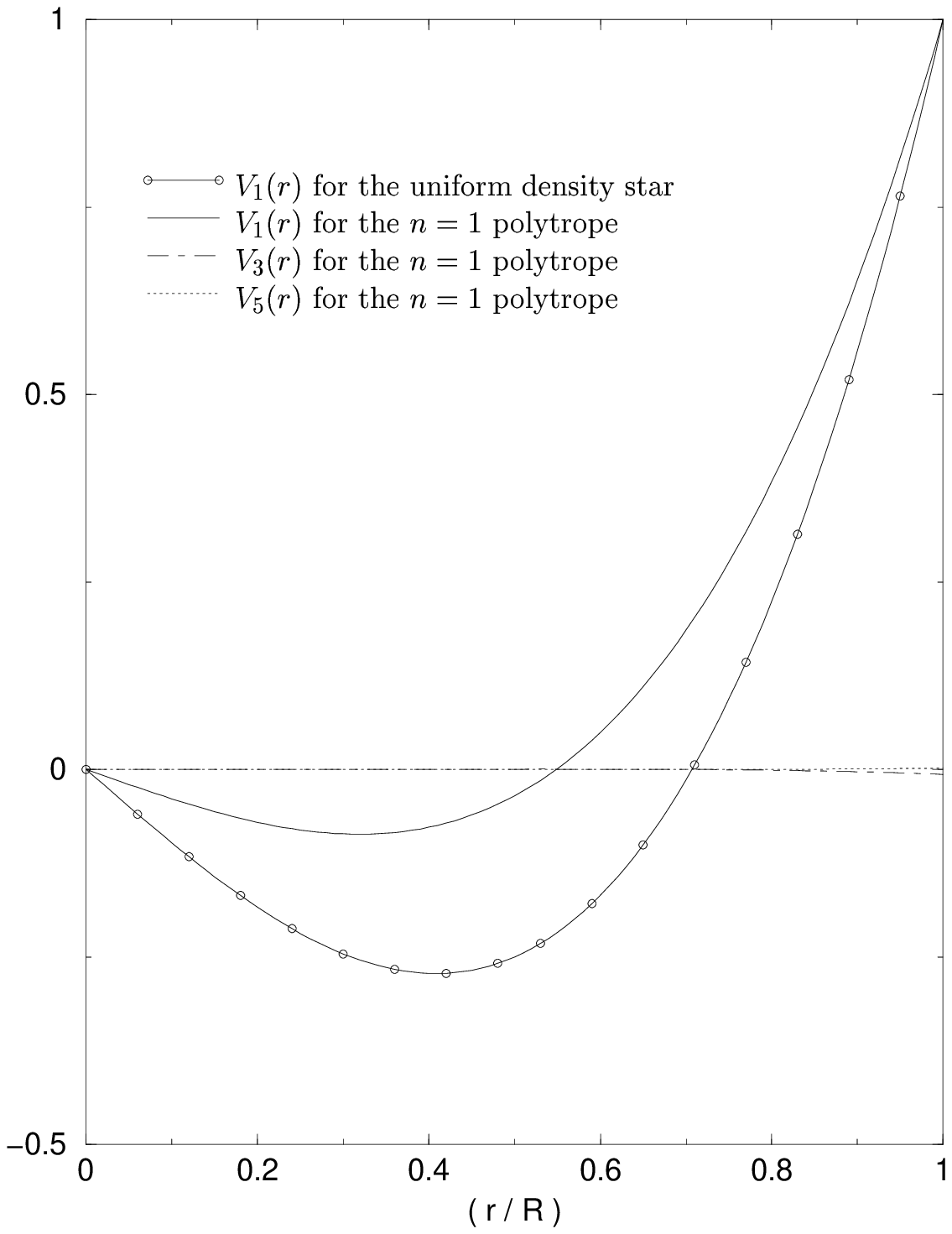}}
\caption{The functions $V_l(r)$ with $l\leq 6$ for the same mode as 
in Figure \ref{fig3}.}
\label{fig4}
\end{figure}

\begin{figure}
\centerline{\includegraphics[3in,2.5in][6in,8in]{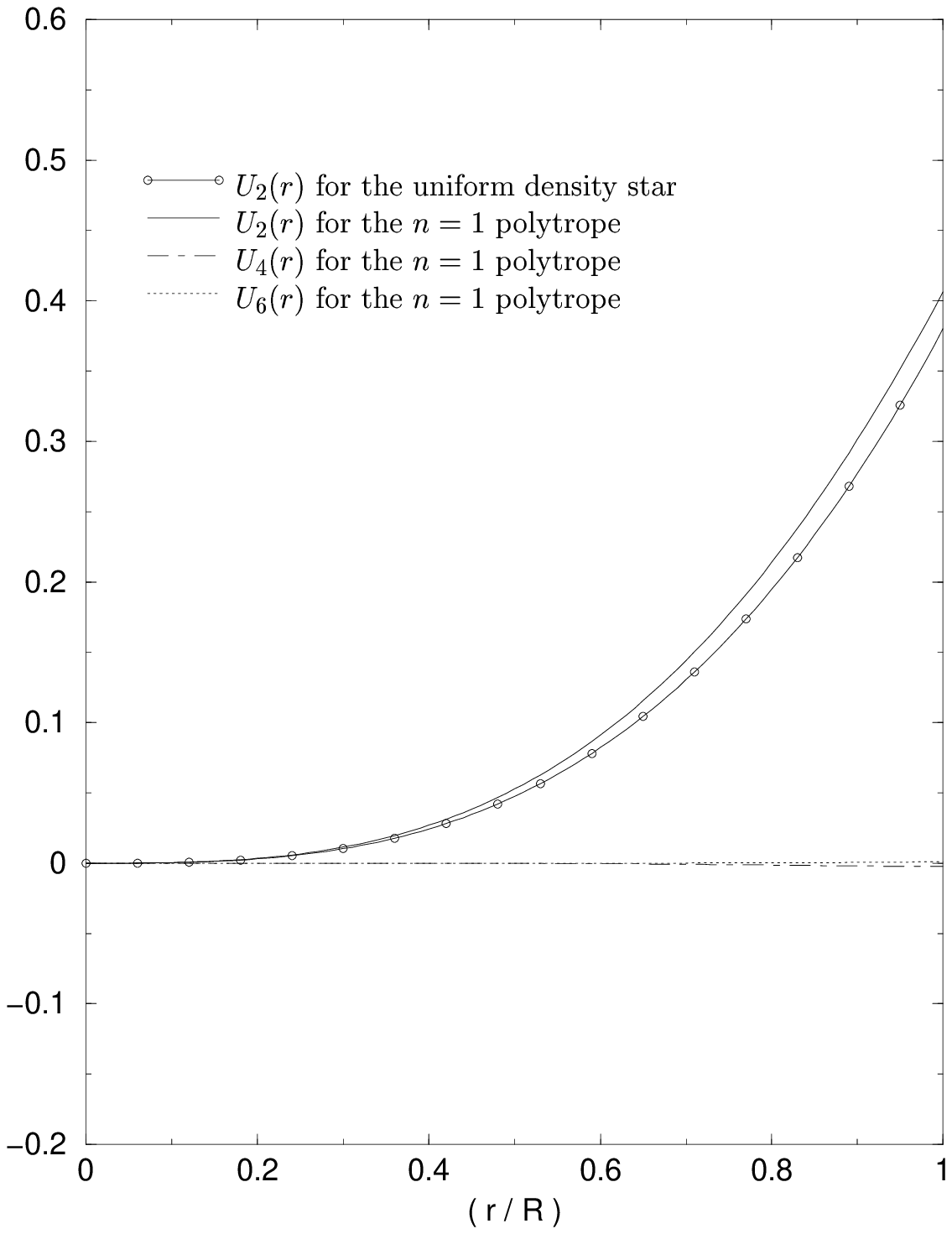}}
\caption{The functions $U_l(r)$ with $l\leq 6$ for the same mode as 
in Figure \ref{fig3}.}
\label{fig5}
\end{figure}

Given the form of their eigenfunctions, it seems reasonable to expect 
that some of the unstable hybrid modes might grow on a timescale which 
is comparable to that of the pure $l_0=m$ r-modes. For example, the 
$m=2$ axial-led hybrids all have $U_2(r)\neq 0$ (see, for example, 
Figures \ref{ef_mac} and \ref{ef_poly}). By equation (\ref{J_int}), 
this leads one to expect a non-zero current quadrupole moment 
$\delta J_{22}$, and this is the multipole moment that dominates the 
gravitational radiation in the r-modes. Upon closer inspection, however, 
one finds that this is not the case.  In fact, we find that all of the 
multipoles $\delta J_{lm}$ vanish (or nearly vanish) for $l<l_0$, 
where $l_0$ is the largest value of $l$ contributing a dominant term 
to the expansion (\ref{v_exp}) of $\delta v^a$.  

\begin{figure}
\centerline{\includegraphics[3in,2.5in][6in,8in]{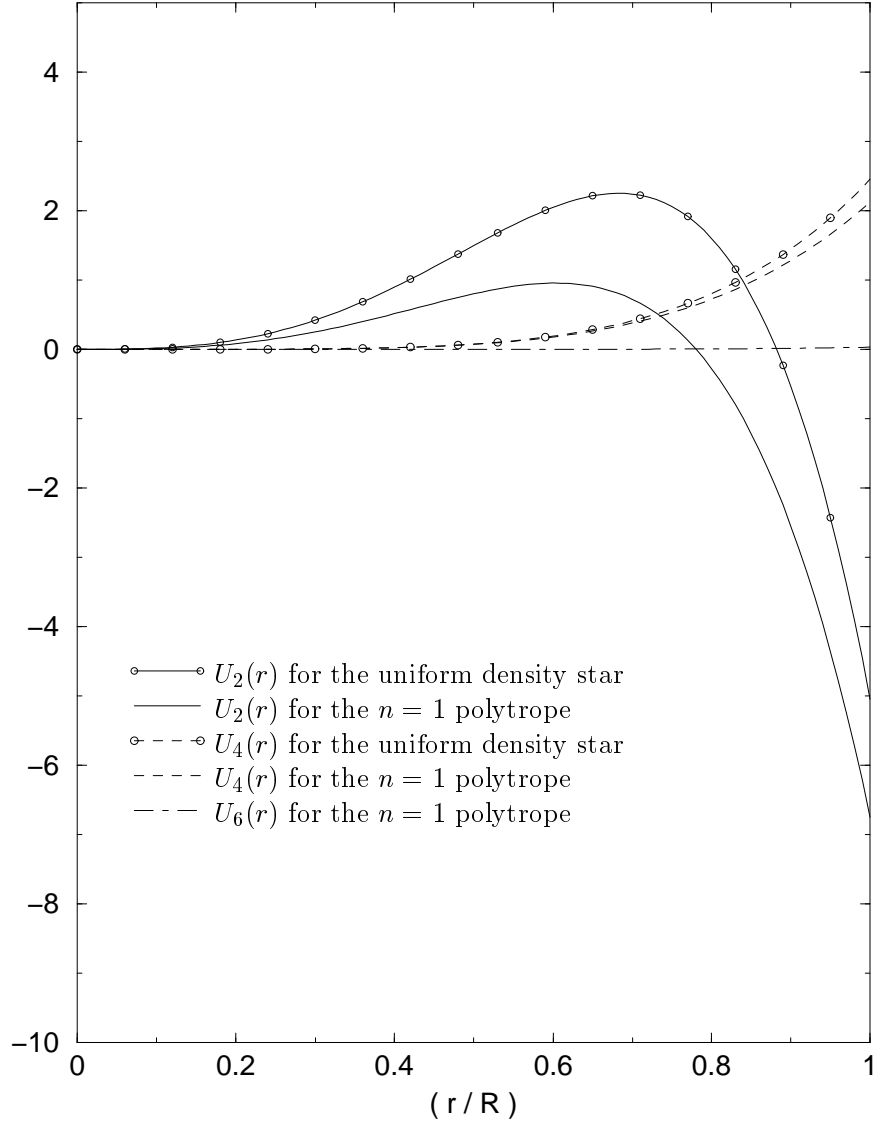}}
\caption{The functions $U_l(r)$ with $l\leq 7$ for a particular $m=2$ 
axial-led hybrid mode.  For the uniform density star this mode has 
eigenvalue $\kappa_0=0.466901$ and $U_2(r)$ and $U_4(r)$ are the only 
non-vanishing $U_l(r)$. (See Table \ref{ef_m2a} for their explicit 
forms.) The corresponding mode of the polytropic star has eigenvalue 
$\kappa_0=0.517337$. Observe that $U_2(r)$ and $U_4(r)$ for the 
polytrope, which have been constructed from their power series 
expansions about $r=0$ and $r=R$, are similar, though not identical, 
to the corresponding functions for the uniform density star.  Observe 
also that $U_6(r)$ is more than an order of magnitude smaller than 
$U_2(r)$ and $U_4(r)$.}
\label{fig6}
\end{figure}

\begin{figure}
\centerline{\includegraphics[3in,2.5in][6in,8in]{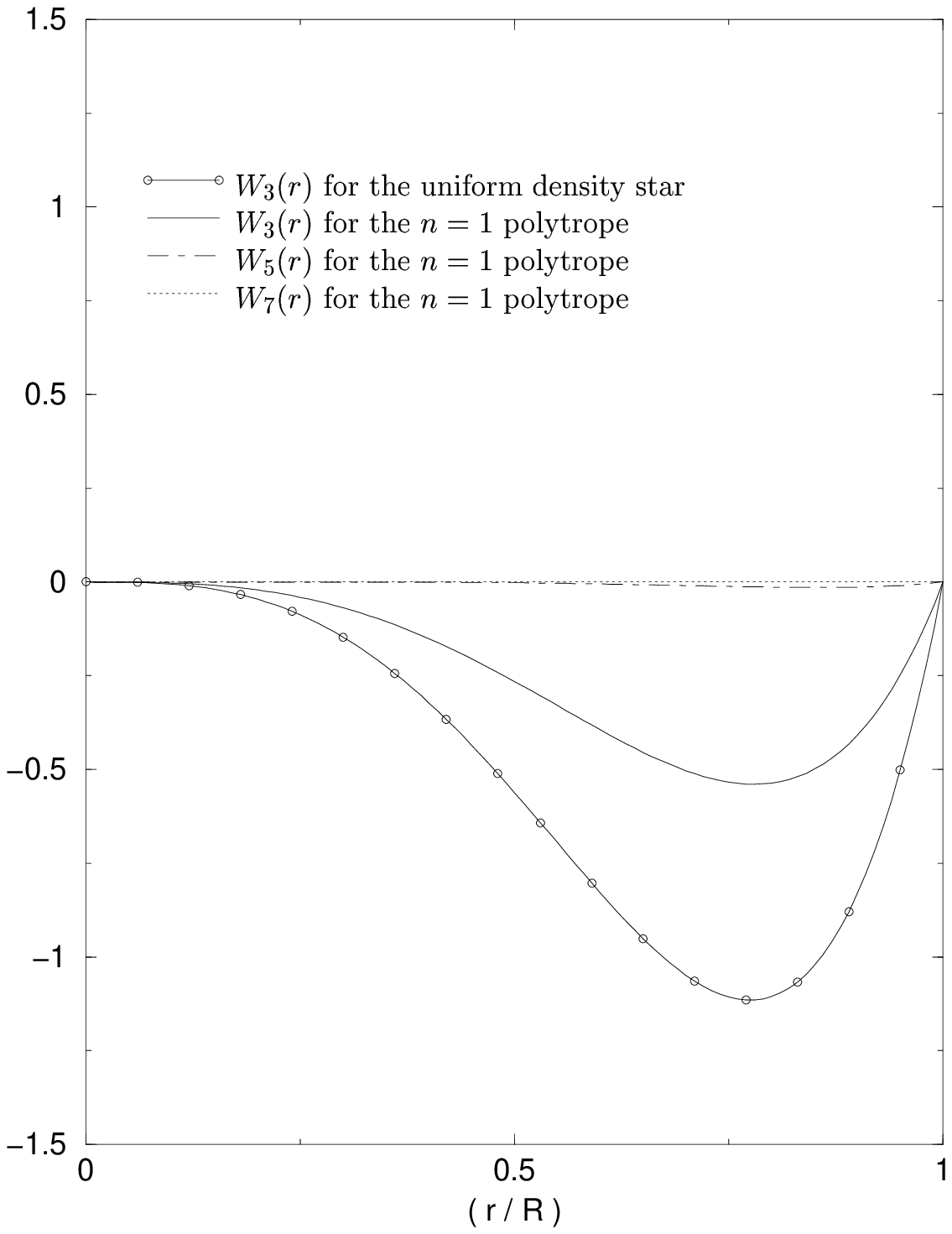}}
\caption{The functions $W_l(r)$ with $l\leq 7$ for the same mode as 
in Figure \ref{fig6}.}
\label{fig7}
\end{figure}

\begin{figure}
\centerline{\includegraphics[3in,2.5in][6in,8in]{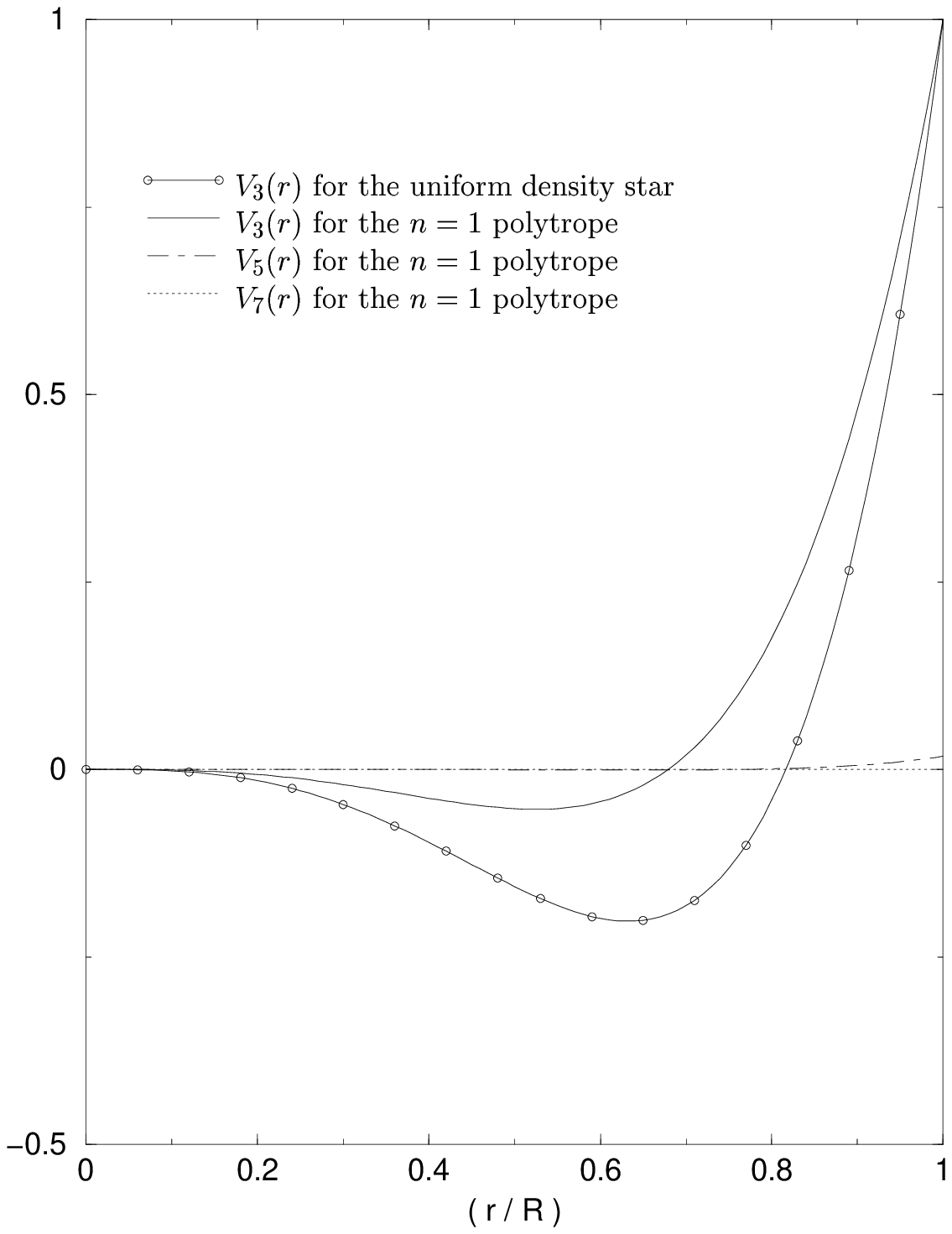}}
\caption{The functions $V_l(r)$ with $l\leq 7$ for the same mode as 
in Figure \ref{fig6}.}
\label{fig8}
\end{figure}

\begin{figure}
\centerline{\includegraphics[3in,2.5in][6in,8in]{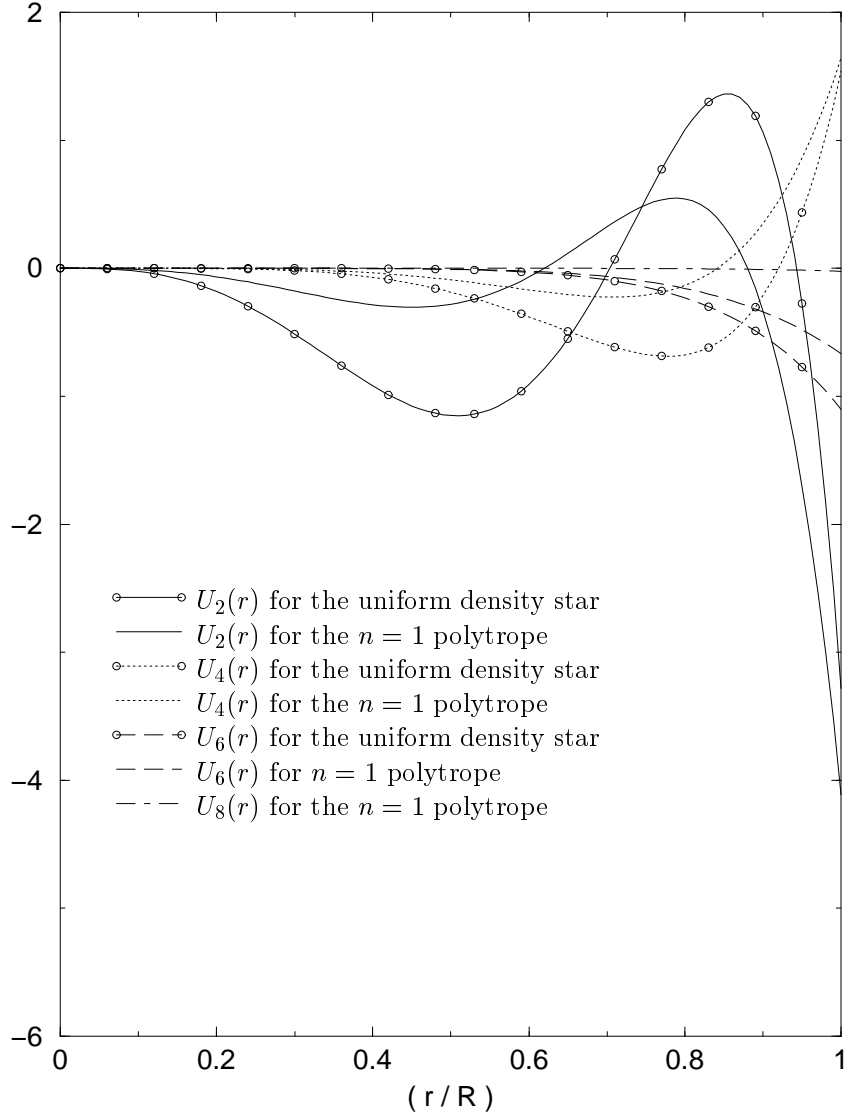}}
\caption{The functions $U_l(r)$ with $l\leq 8$ for a particular 
$m=2$ axial-led hybrid mode.  For the uniform density star this mode 
has eigenvalue $\kappa_0=0.359536$ and $U_2(r)$, $U_4(r)$ and $U_6(r)$ 
are the only non-vanishing $U_l(r)$. (See Table \ref{ef_m2a} for their 
explicit forms.) The corresponding mode of the polytropic star has 
eigenvalue $\kappa_0=0.421678$.  Observe that $U_2(r)$, $U_4(r)$ and 
$U_6(r)$ for the polytrope, which have been constructed from their 
power series expansions about $r=0$ and $r=R$, are similar, though not 
identical, to the corresponding functions for the uniform density star.  
Observe also that $U_8(r)$ is more than an order of magnitude smaller 
than $U_2(r)$, $U_4(r)$ and $U_6(r)$.}
\label{fig9}
\end{figure}

\begin{figure}
\centerline{\includegraphics[3in,2.5in][6in,8in]{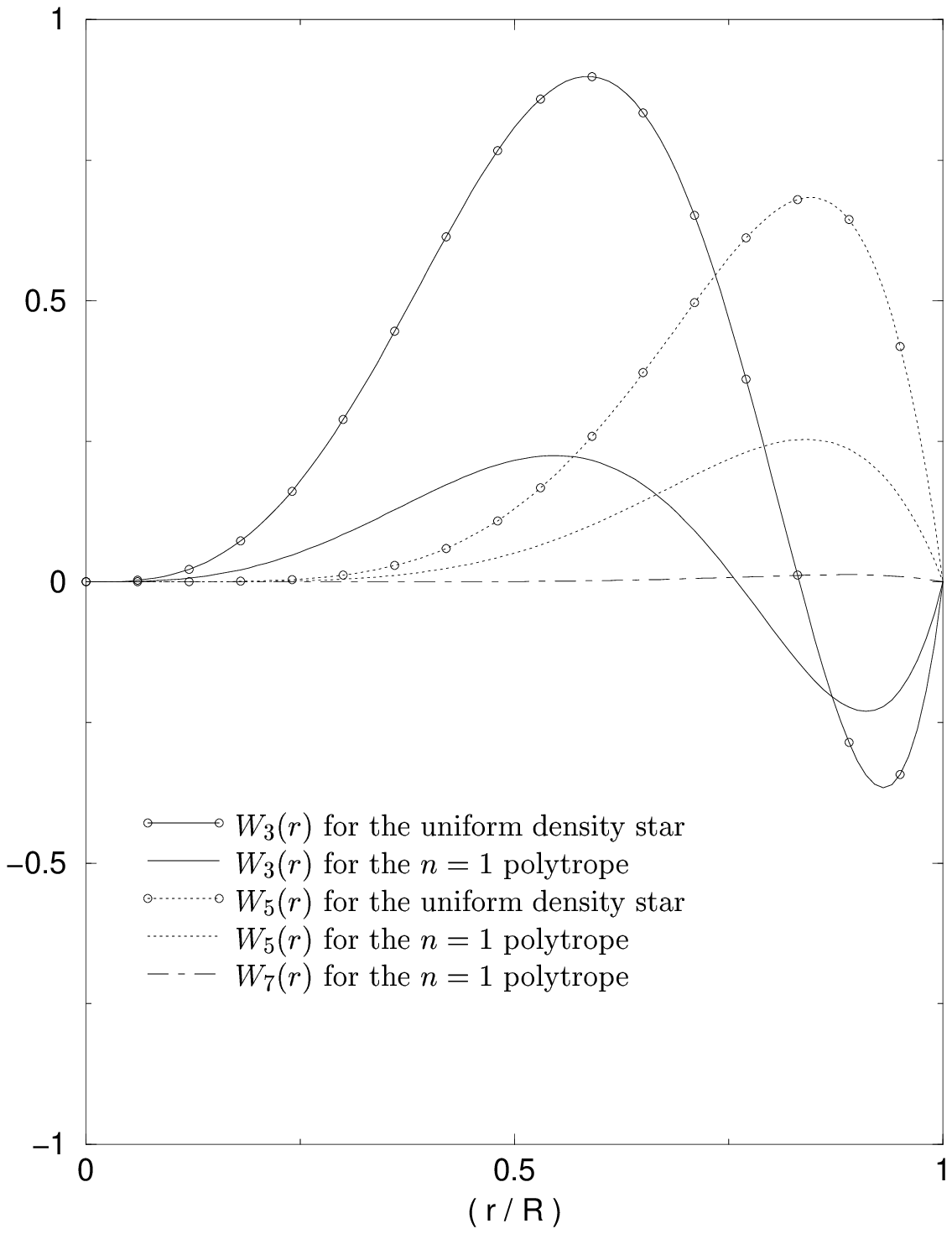}}
\caption{The functions $W_l(r)$ with $l\leq 8$ for the same mode 
as in Figure \ref{fig9}.}
\label{fig10}
\end{figure}

\begin{figure}
\centerline{\includegraphics[3in,2.5in][6in,8in]{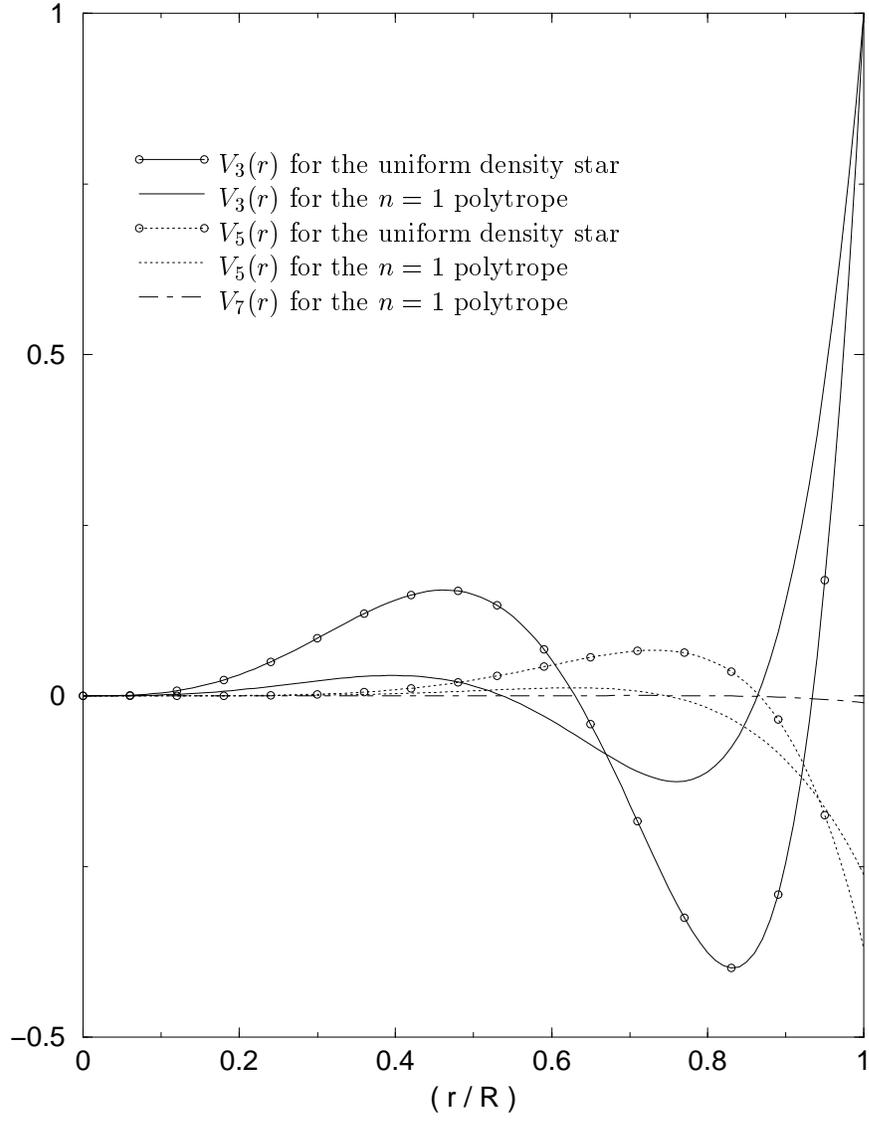}}
\caption{The functions $V_l(r)$ with $l\leq 8$ for the same mode 
as in Figure \ref{fig9}.}
\label{fig11}
\end{figure}

In the uniform density star, these multipoles vanish identically.  
Consider, for example, the $m=2$, $l_0=4$ axial-hybrid with 
eigenvalue $\kappa=0.466901$. (See Table \ref{ef_m2a} and Figure 
\ref{fig6}) For this mode, $U_2\propto (7x^3-9x^5)$, where $x=(r/R)$. 
By equation (\ref{J_int}), we then find that
\be
\delta J_{22} \propto \int_0^1 x^3 (7x^3-9x^5) dx \equiv 0,
\ee
and that $\delta J_{42}$ is the only non-zero radiation multipole.  
In general, the only non-zero multipole for an axial- or polar-hybrid 
mode in the uniform density star is $\delta J_{l_0\, m}$. 

That this should be the case is not obvious from the form of our 
eigenfunctions.  However, Lindblom and Ipser's \cite{li98} analytic 
solutions provide an explanation.  Their equations (7.1) and (7.3) 
reveal that the perturbed gravitational potential, $\delta \Phi$, 
is a pure spherical harmonic to lowest order in $\Omega$. 
In particular,
\be
\delta \Phi \propto Y_{l_0+1}^m.
\ee
This implies that the only non-zero current multipole is 
$\delta J_{l_0\, m}$. 

We find a similar result for the polytropic star. Because of the 
similarity between the modes in the polytrope and the modes in the 
uniform density star, we find that although the lower $l$ current 
multipoles do not vanish identically, they very nearly vanish and
the radiation is dominated by higher $l$ multipoles.

The fastest growth times we find in the hybrid modes are of order 
$10^4$ seconds (at $10^9K$ and $\Omega=\sqrt{\pi G\bar{\rho}}$).  
Thus, the spin-down of a newly formed neutron star will be dominated 
by the $l_0=m=2$ mode with small contributions from the $l_0=m$ pure 
axial modes with $2\leq m\lesssim 10$ and from the fastest growing 
hybrid modes.


\clearpage
\begin{table}
\centerline{\rotatebox{180}{\resizebox{2in}{7.25in}
{\includegraphics[2in,3.3in][4in,10.8in]{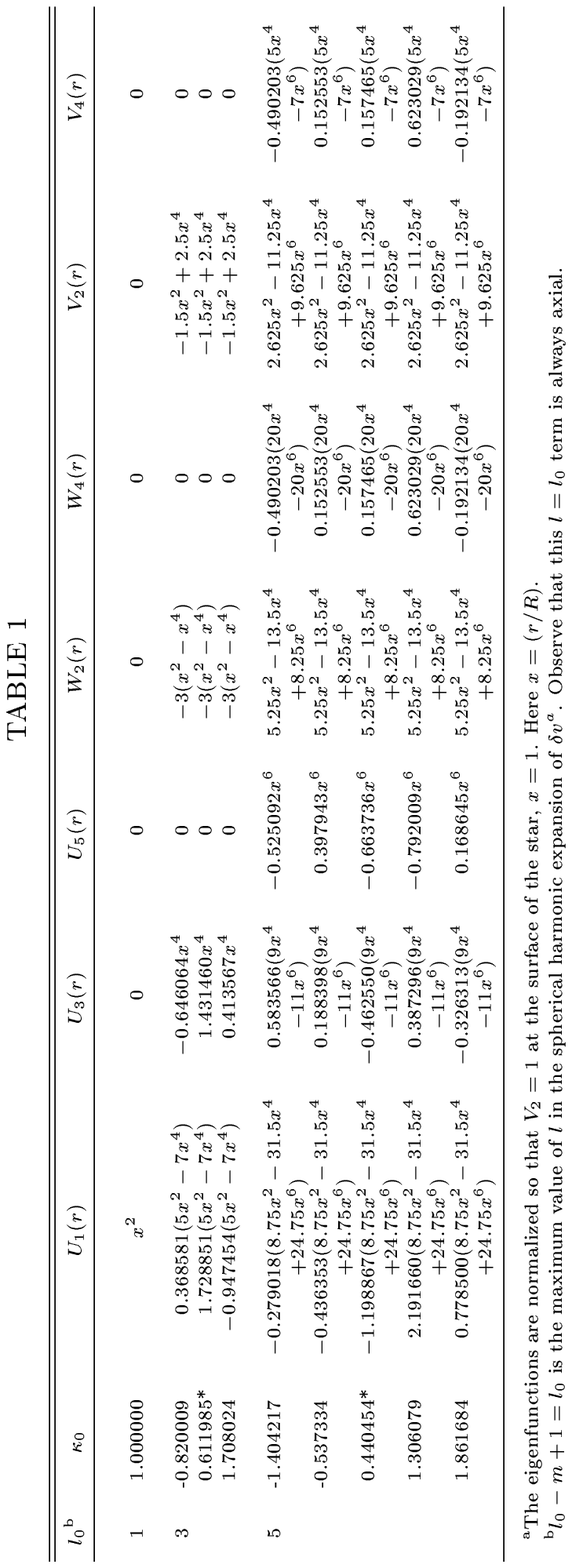}}}}
\caption{Axial-Hybrid Eigenfunctions with $m=1$ for 
Uniform Density Stars.}
\label{ef_m1a}
\end{table}

\clearpage
\begin{table}
\centerline{\rotatebox{180}{\resizebox{1.5in}{7.25in}
{\includegraphics[3.5in,3.3in][5in,10.7in]{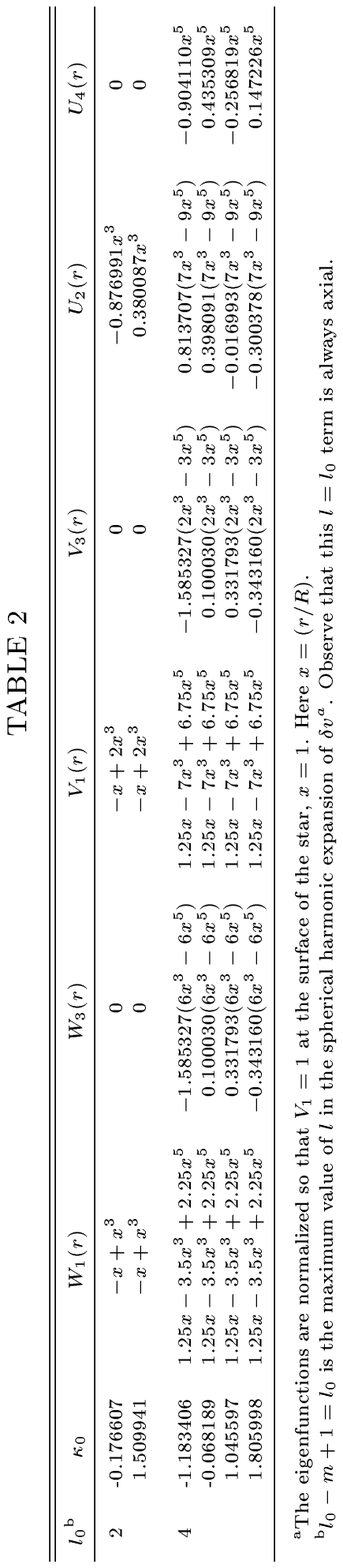}}}}
\caption{Polar-Hybrid Eigenfunctions with $m=1$ for 
Uniform Density Stars.}
\label{ef_m1p}
\end{table}

\clearpage
\begin{table}
\centerline{\rotatebox{180}{\resizebox{1.5in}{7in}
{\includegraphics[3.5in,3.3in][5in,10.7in]{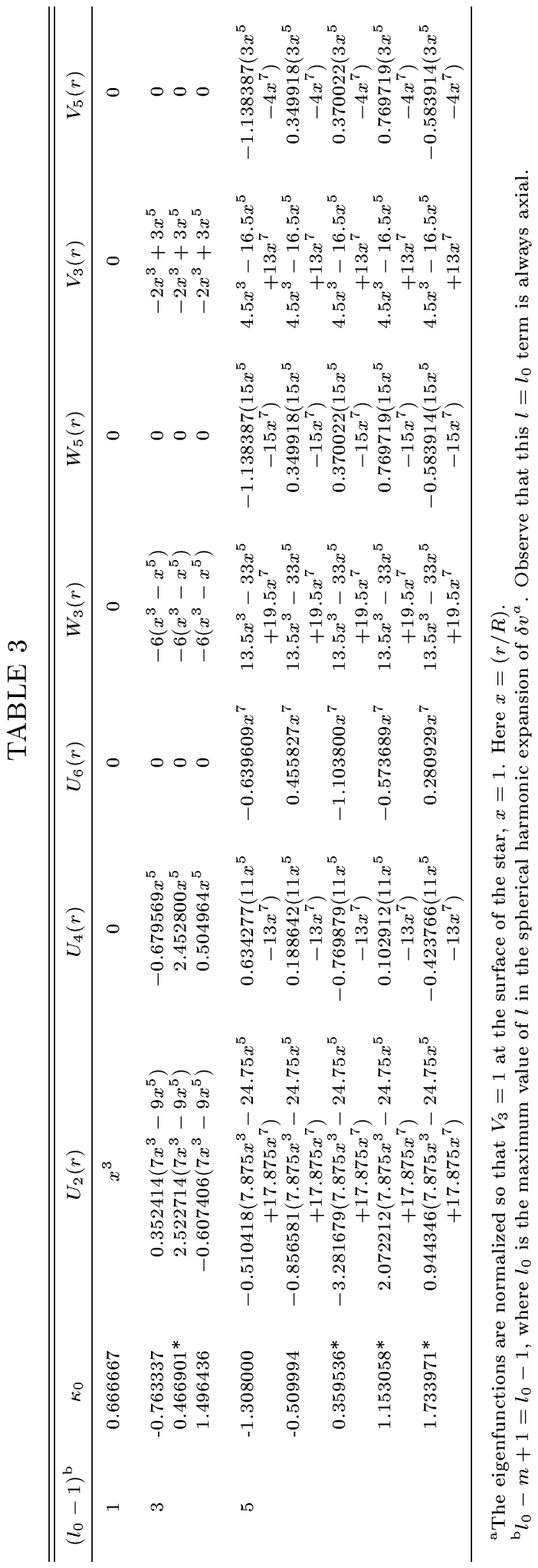}}}}
\caption{Axial-Hybrid Eigenfunctions with $m=2$ for 
Uniform Density Stars.}
\label{ef_m2a}
\end{table}

\clearpage
\begin{table}
\centerline{\rotatebox{180}
{\includegraphics[3.5in,3.5in][5in,10.8in]{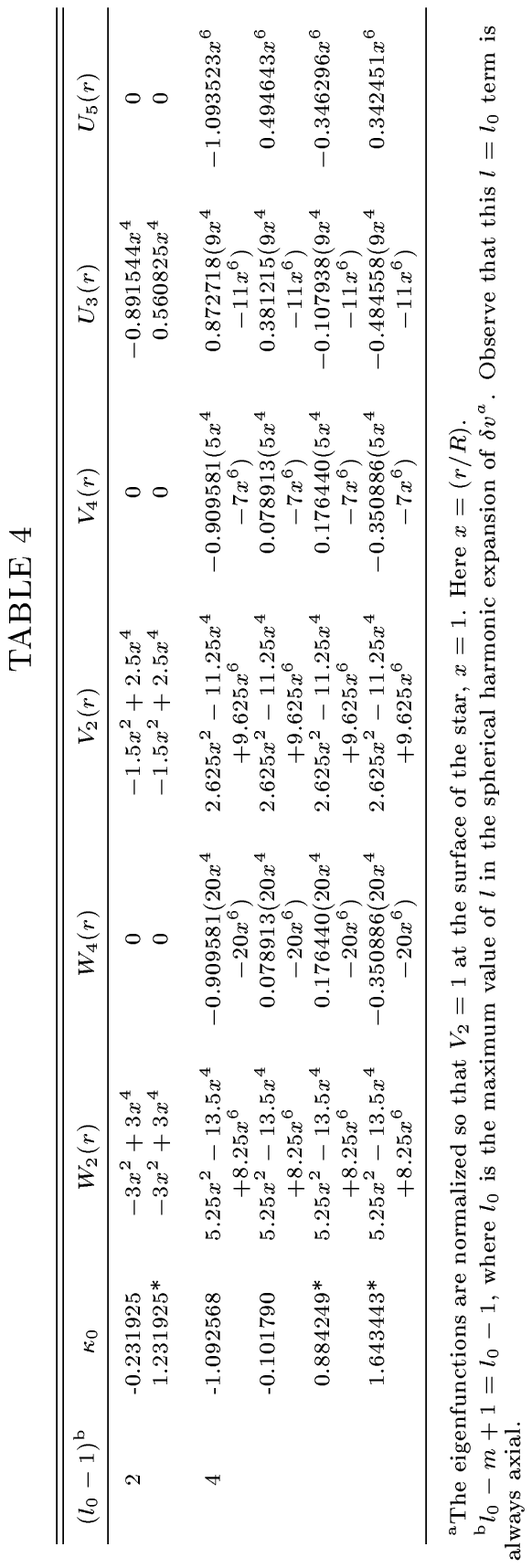}}}
\caption{Polar-Hybrid Eigenfunctions with $m=2$ for 
Uniform Density Stars.}
\label{ef_m2p}
\end{table}

\clearpage
\begin{table}
\centerline{\includegraphics[0.75in,2in][7.75in,6in]{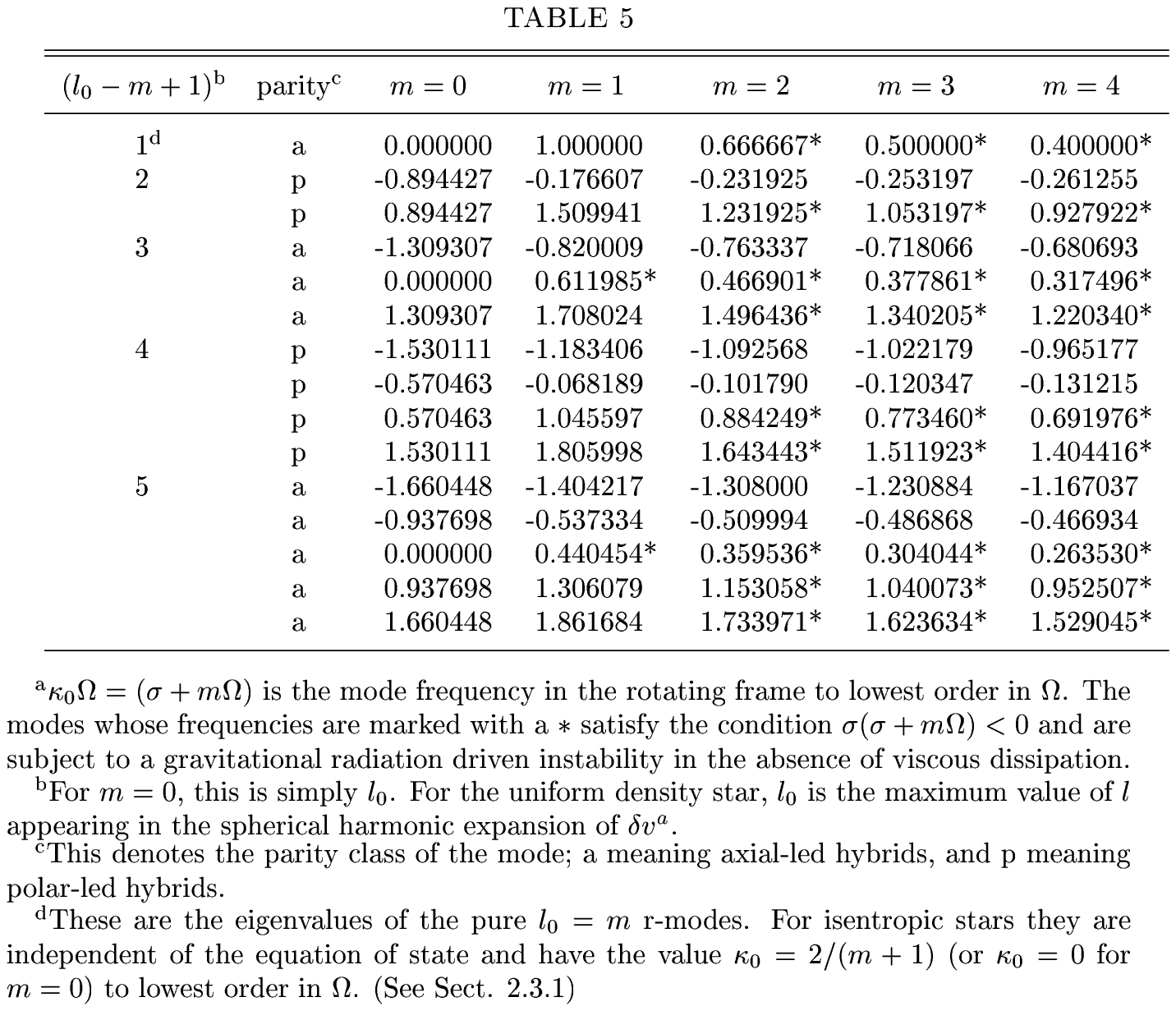}}
\caption{Eigenvalues $\kappa_0$ for Uniform Density Stars.}
\label{ev_mac}
\end{table}

\clearpage
\begin{table}
\centerline{\includegraphics[0.75in,3.5in][7.75in,7in]{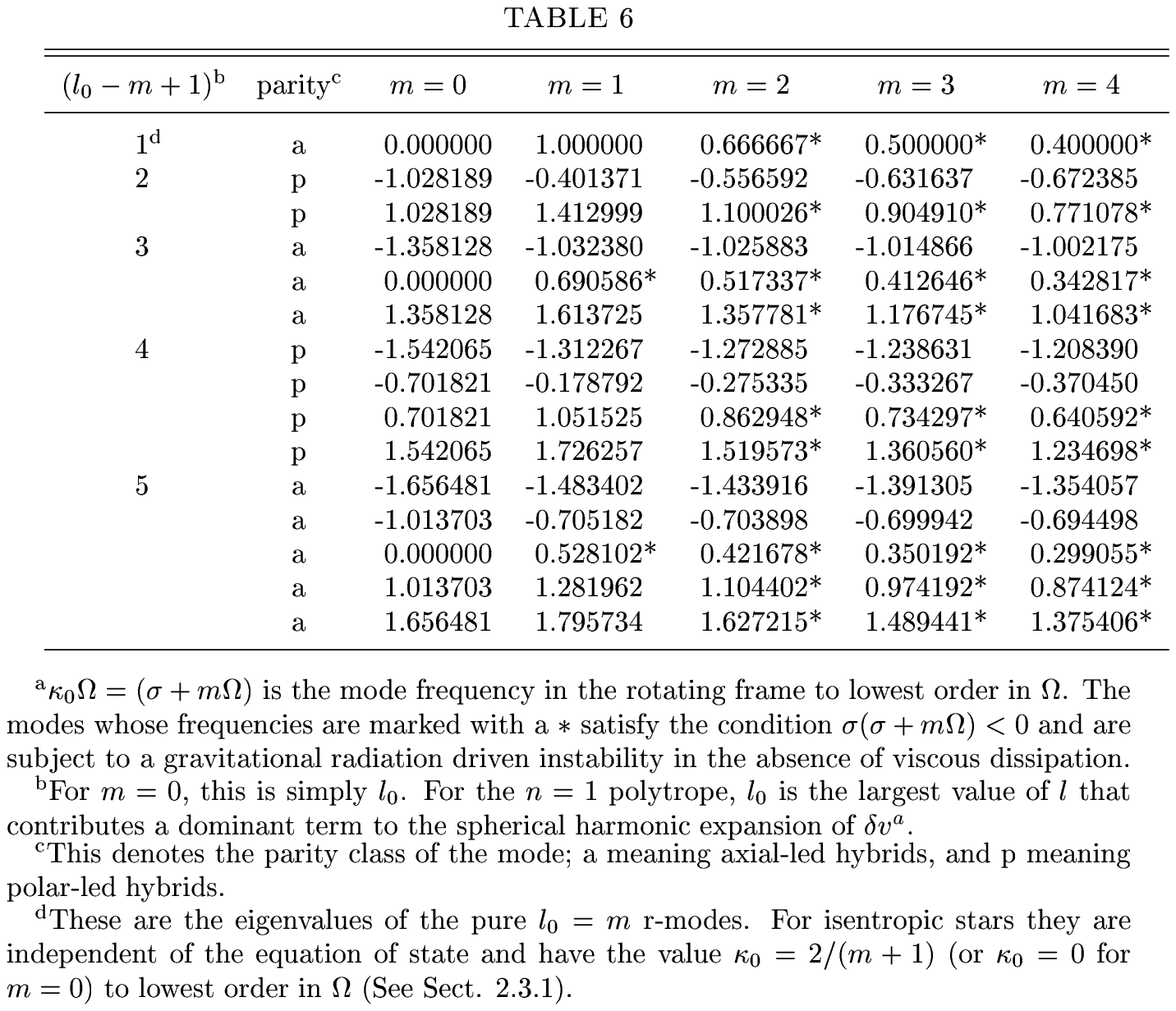}}
\caption{Eigenvalues $\kappa_0$ for the $p=K\rho^2$ Polytrope.}
\label{ev_poly}
\end{table}

\clearpage
\begin{table}
\centerline{\includegraphics[1in,4.5in][8in,8in]{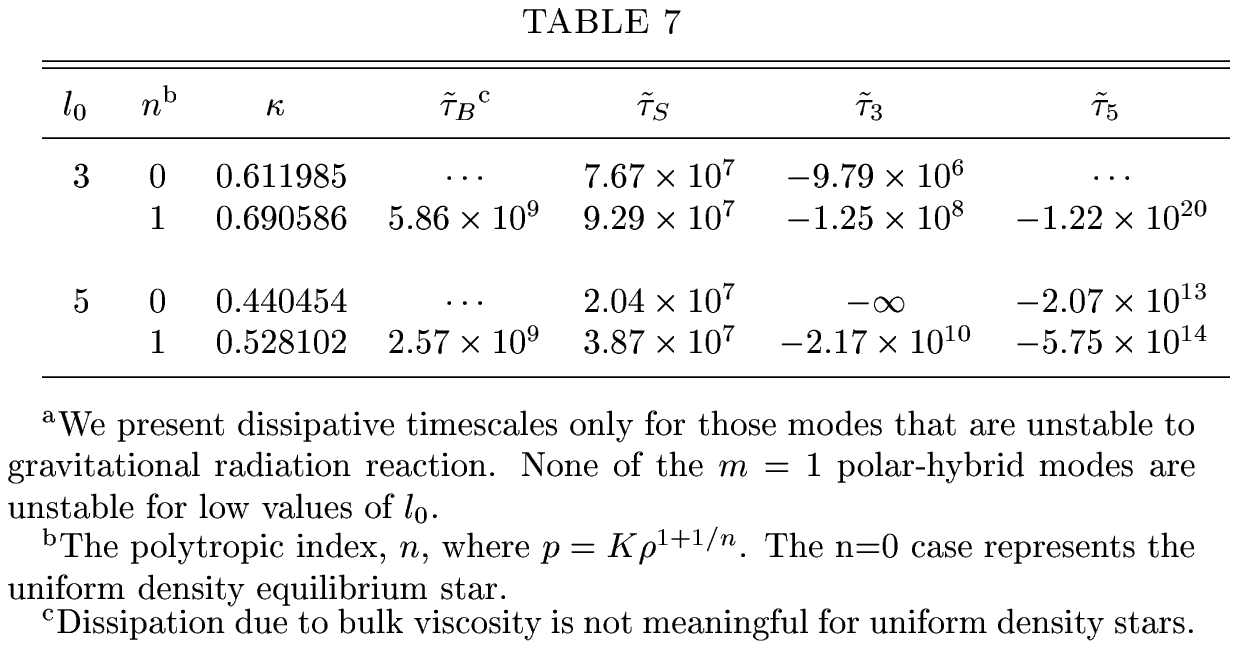}}
\caption{Dissipative timescales (in seconds) for $m=1$ axial-hybrid 
modes at $T=10^9K$ and $\Omega=\sqrt{\pi G \bar{\rho}}$.}
\label{times_m1a}
\end{table}

\clearpage
\begin{table}
\centerline{\resizebox{6.5in}{3in}
{\includegraphics[0.75in,3.5in][7.75in,7in]{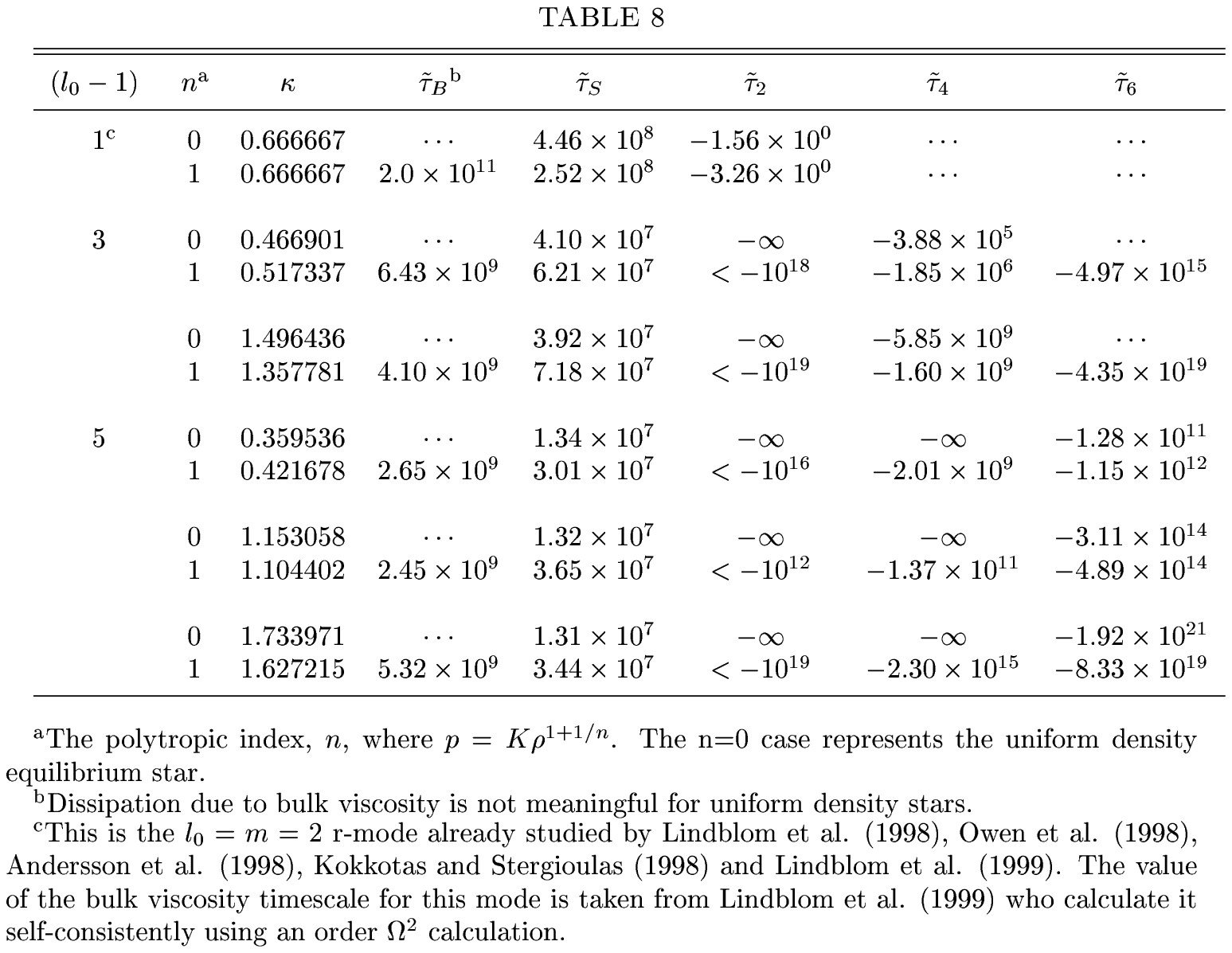}}}
\caption{Dissipative timescales (in seconds) for $m=2$ axial-hybrid 
modes at $T=10^9K$ and $\Omega=\sqrt{\pi G \bar{\rho}}$.}
\label{times_m2a}
\end{table}

\clearpage
\begin{table}
\centerline{\includegraphics[1in,4.5in][8in,8in]{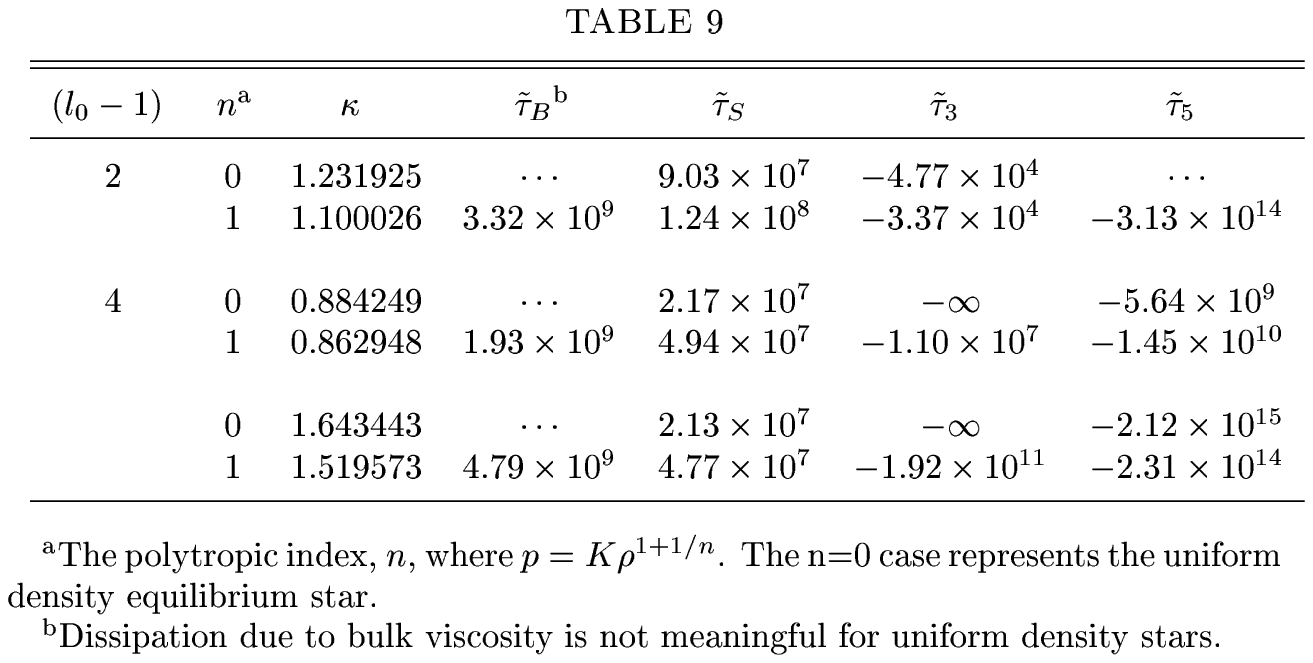}}
\caption{Dissipative timescales (in seconds) for $m=2$ polar-hybrid 
modes at $T=10^9K$ and $\Omega=\sqrt{\pi G \bar{\rho}}$.}
\label{times_m2p}
\end{table}


\clearpage

\clearpage

\clearpage

\clearpage

\clearpage

\clearpage

\clearpage

\clearpage

\clearpage

\clearpage

\clearpage

\clearpage


\chapter{Relativistic Stars: Analytic Results}

In Ch. 2 we examined the rotationally restored hybrid modes
of isentropic newtonian stars.  We now consider the 
corresponding modes in relativistic stars.  As in the 
newtonian case, we must begin by examining the perturbations 
of the non-rotating star, and finding all of the modes 
belonging to its degenerate zero-frequency subspace.

\section{Stationary Perturbations of Spherical Stars}

The equilibrium of a spherical perfect fluid star is described
by a static, spherically symmetric spacetime with metric 
$g_{\alpha\beta}$ of the form,
\be
ds^2 = -e^{2\nu(r)} dt^2 + e^{2\lambda(r)} dr^2 + r^2 d \theta^2 
	+ r^2 sin^2 \theta d \varphi^2,
\ee
and with energy-momentum tensor,
\be
T_{\alpha\beta} = (\epsilon+p)u_\alpha u_\beta + p g_{\alpha\beta},
\ee
where $\ep(r)$ is the total fluid energy density, $p(r)$ is the fluid 
pressure and
\be
u^\alpha = e^{- \nu} t^\alpha
\ee
is the fluid 4-velocity - with $t^\alpha=(\partial_t)^\alpha$ 
the timelike Killing vector of the spacetime.

These satisfy an equation of state of the form
\be
p = p(\ep)
\label{GR_sph:eos}
\ee
as well as the Einstein equations,
$G_{\alpha\beta}=8\pi T_{\alpha\beta}$, which are equivalent to
\be
\frac{dp}{dr} = - \frac{(\ep+p)(M+4\pi r^3p)}{r(r-2M)},
\label{GR_sph:tov}
\ee
\be
\frac{dM}{dr} = 4\pi r^2\ep
\label{GR_sph:dMdr}
\ee
and
\be
\frac{d\nu}{dr} = - \frac{1}{(\ep+p)}\frac{dp}{dr}
\label{GR_sph:dnudr}
\ee
where
\be
M(r) \equiv \half r(1-e^{-2\lambda}).
\label{GR_sph:mass_def}
\ee
(See, e.g., Wald \cite{wald}, Ch.6.)

We are, again, interested in the space of zero-frequency modes, 
the linearized, time-independent perturbations of this static
equilibrium.  As in the newtonian case, we find that this 
zero-frequency subspace is spanned by two classes of perturbations.
To identify these classes explicitly, we must examine the 
equations governing the perturbed configuration.

Writing the change in the metric as 
$h_{\alpha\beta}\equiv\delta g_{\alpha\beta}$, we express the 
perturbed configuration in terms of the set
$(h_{\alpha\beta}, \delta u^\alpha, \delta \ep, \delta p)$. These
must satisfy the perturbed Einstein equations
$\delta G_\alpha^{\ \beta} = 8\pi\delta T_\alpha^{\ \beta}$,
together with an equation of state (which may, in general, differ 
from that of the equilibrium configuration).  

The perturbed Einstein tensor is given by
\be
\delta G_\alpha^{\ \beta}  = -\half \biggl\{ \ba[t]{l}
  \nabla_\gamma \nabla^\gamma h_\alpha^{\ \beta}
- \nabla_\gamma \nabla^\beta h_\alpha^{\ \gamma}
- \nabla^\gamma \nabla_\alpha h_\gamma^{\ \beta}
+ \nabla_\alpha \nabla^\beta h \\
\\
+ 2 R_\alpha^{\ \gamma} h_\gamma^{\ \beta}
+( \nabla^\alpha \nabla^\beta h_{\alpha\beta}
  -\nabla_\gamma \nabla^\gamma h
  -R^{\alpha\beta} h_{\alpha\beta} )g_\alpha^{\ \beta}
\biggr\}
\ea
\label{pert_Gmunu}
\ee
where $h\equiv h_\alpha^{\ \alpha}$, $\nabla_\alpha$ is the 
covariant derivative associated with the equilibrium metric and
\be
R_\alpha^{\ \beta}
= 8\pi(T_\alpha^{\ \beta} - \half T g_\alpha^{\ \beta})
= 8\pi\left[(\ep+p)u_\alpha u^\beta 
+ \half (\ep-p)g_\alpha^{\ \beta}\right]
\ee
is the equilibrium Ricci tensor.
The perturbed energy-momentum tensor is given by
\be
\delta T_\alpha^{\ \beta}  = 
(\delta\ep+\delta p)u_\alpha u^\beta + \delta p \delta_\alpha^{\ \beta}
+(\ep+p)\delta u_\alpha u^\beta+(\ep+p)u_\alpha \delta u^\beta.
\label{pert_Tmunu}
\ee

Following Thorne and Campolattaro \cite{tc67}, we expand our
perturbed variables in scalar, vector and tensor spherical harmonics.
The perturbed energy density and pressure are scalars and therefore 
must have polar parity, $(-1)^l$.
\be
\delta\ep = \delta\ep(r) Y_l^m,
\label{GR:del_ep}
\ee
\be
\delta p = \delta p(r) Y_l^m.
\label{GR:del_p}
\ee

The perturbed 4-velocity for a polar-parity mode can be written
\be
\delta u_P^{\alpha} = \biggl\{
\half H_0(r) Y_l^m t^\alpha + \frac{1}{r} W(r) Y_l^m r^{\alpha} 
+ V(r) \nabla^{\alpha} Y_l^m 
\biggr\} e^{-\nu}
\label{GR:del_u_po}
\ee
while that of an axial-parity mode can be written
\be
\delta u_A^{\alpha} = 
- U(r) e^{(\lambda-\nu)} \epsilon^{\alpha\beta\gamma\delta} 
\nabla_{\beta} Y_l^m u_\gamma \nabla_{\delta}\, r.
\label{GR:del_u_ax}
\ee
(We have chosen the exact form of these expressions for later convenience.)

To simplify the form of the metric perturbation we will, again, follow 
Thorne and Campolattaro \cite{tc67} and work in the Regge-Wheeler \cite{rw57}
gauge.  The metric perturbation for a polar-parity mode can be 
written
\be
h_{\mu\nu}^P = \left[
\ba[c]{cccc}
H_0(r) e^{2\nu} & H_1(r) & 0 & 0 \\
H_1(r) & H_2(r) e^{2\lambda} & 0 & 0 \\
0 & 0 & r^2 K(r) & 0\\
0 & 0 & 0 & r^2 sin^2\theta K(r)
\ea
\right] Y_l^m,
\label{GR:h_po}
\ee
while that of an axial-parity mode can be written
\be
h_{\mu\nu}^A = \left[
\ba[c]{cccc}
0 & 0 & h_0(r) \,(\frac{-1}{sin\theta})\partial_\varphi Y_l^m & h_0(r)\,\sinY\\
0 & 0 & h_1(r) \,(\frac{-1}{sin\theta})\partial_\varphi Y_l^m & h_1(r)\,\sinY\\
\mbox{\scriptsize symm} & \mbox{\scriptsize symm} & 0 & 0 \\
\mbox{\scriptsize symm} & \mbox{\scriptsize symm} & 0 & 0 \\
\ea
\right]
\label{GR:h_ax}
\ee

The Regge-Wheeler gauge is unique for perturbations having spherical
harmonic index $l\geq 2$.  However, when $l=1$ or $l=0$, there remain
additional gauge degrees of
freedom\footnote{Letting $e_{AB}$ be the metric on a two-sphere with
$\ep_{AB}$ and $D_A$ the associated volume element and covariant
derivative, respectively, one finds the following.  When $l\geq 2$ the 
polar tensors $D_{\! A}D_{\! B} Y_l^m$ and $e_{AB} Y_l^m$ are linearly 
independent, but when $l=1$, they coincide.  In addition, the axial tensor 
$\ep_{\left(A\right.}^{\ \ B}D_{\left.\! C\right)}D_{\! B} Y_l^m$ vanishes 
identically for $l=1$ and, of course, $D_{\! A} Y_l^m$ vanishes for $l=0$.}.  
In addition, the components of the perturbed Einstein equation acquire a 
slightly different form in each of these three cases 
(cf. Campolattaro and Thorne \cite{ct70}) and will be presented separately
below.

We have derived these components
using the Maple tensor package by substituting expressions 
(\ref{GR:del_ep})-(\ref{GR:h_ax}) into Eqs. (\ref{pert_Gmunu}) and 
(\ref{pert_Tmunu}) (making liberal use of the equilibrium equations
(\ref{GR_sph:tov}) through (\ref{GR_sph:mass_def}) to simplify the 
resulting expressions).  
The resulting set of equations for the case $l\geq 2$
are equivalent to those presented in Thorne and Campolattaro \cite{tc67}
upon specializing their equations to the case of stationary perturbations 
and making the necessary changes of 
notation\footnote{In particular, their fluid variables (denoted by the 
subscript ${\scriptstyle TC}$) are related to 
ours as follows: $U_{TC}(r,t)=Ut$, $W_{TC}(r,t)=-re^{\lambda}Wt$,
$V_{TC}(r,t)=Vt$ and $\delta\ep/(\ep+p)=\delta p/\gamma p=-(K+\half H_0)$.
 Their equilibrium metric has the opposite signature and differs 
in the definitions of the metric potentials
$\nu_{TC}=\half\nu$ and $\lambda_{TC}=\half\lambda$.}.  
Similarly, the set of 
equations for the case $l=1$ are equivalent to those presented 
in Campolattaro and Thorne \cite{ct70}.  

\subsection{The case $l\geq 2$}
The non-vanishing components of the perturbed Einstein equation for 
$l\geq 2$ are as follows.  We will use Eq. (\ref{thth-_l2}) below, to 
replace $H_2$ by $H_0$.  From $\delta G_t^{\ t}=8\pi\delta T_t^{\ t}$ 
we have 
\bea
0 &=& e^{-2\lambda} r^2 K'' 
+ e^{-2\lambda} (3-r\lambda')rK' - \left[\half l(l+1)-1\right] K \nn\\
&& \nn \\
&& - e^{-2\lambda} rH_0'
-\left[\half l(l+1)+1-8\pi r^2\ep\right]H_0
+8\pi r^2\delta\ep.
\label{tt_l2}
\eea
From $\delta G_r^{\ r}=8\pi\delta T_r^{\ r}$ we similarly have
\bea
0 &=& e^{-2\lambda}(1+r\nu')rK' - \left[\half l(l+1)-1\right] K \nn\\
&& \nn \\
&& - e^{-2\lambda}rH_0'
+\left[\half l(l+1)-1-8\pi r^2 p\right]H_0
-8\pi r^2\delta p.
\label{rr_l2}
\eea
From 
$\delta G_\theta^{\ \theta}+\delta G_\varphi^{\ \varphi}
=8\pi\left(\delta T_\theta^{\ \theta}+\delta T_\varphi^{\ \varphi}\right)$ 
we have
\bea
0 &=& e^{-2\lambda} r^2 K'' 
+ e^{-2\lambda} \left[r(\nu'-\lambda')+2\right]rK' 
-16\pi r^2\delta p \nn\\
&& \nn \\
&& - e^{-2\lambda}r^2H_0''
- e^{-2\lambda}(3r\nu'-r\lambda'+2)rH_0'
-16\pi r^2 p H_0.
\label{thth+_l2}
\eea
From 
$\delta G_\theta^{\ \theta}-\delta G_\varphi^{\ \varphi}
=8\pi\left(\delta T_\theta^{\ \theta}-\delta T_\varphi^{\ \varphi}\right)$ 
we have
\be
H_2 = H_0.
\label{thth-_l2}
\ee
From $\delta G_r^{\ \theta}=8\pi\delta T_r^{\ \theta}$ we have
\be
K' = e^{-2\nu} \left[e^{2\nu} H_0 \right]'.
\label{rth_l2}
\ee
From $\delta G_t^{\ r}=8\pi\delta T_t^{\ r}$ we have
\be
0 = H_1 + \frac{16\pi(\epsilon+p)}{l(l+1)} e^{2\lambda} r W.
\label{tr_l2}
\ee
From $\delta G_t^{\ \theta}=8\pi\delta T_t^{\ \theta}$ we have
\be
0 = e^{-(\nu-\lambda)}
\left[e^{(\nu-\lambda)}  H_1\right]'
+ 16\pi(\ep+p)e^{2\lambda}V.
\label{tth_l2}
\ee
From $\delta G_t^{\ \varphi}=8\pi\delta T_t^{\ \varphi}$ we have
\be
h_0^{''} - (\nu'+\lambda') h_0' 
+ \left[ \frac{(2-l^2-l)}{r^2}e^{2\lambda} 
- \frac{2}{r}(\nu'+\lambda') - \frac{2}{r^2} \right] h_0 
= \frac{4}{r}(\nu'+\lambda') U.
\label{tph_l2}
\ee
From $\delta G_r^{\ \varphi}=8\pi\delta T_r^{\ \varphi}$ we have
\be
(l-1)(l+2) h_1 = 0.
\label{rph_l2}
\ee
Finally, from $\delta G_\theta^{\ \varphi}=8\pi\delta T_\theta^{\ \varphi}$ 
we have
\be
e^{-(\nu-\lambda)}
\left[e^{(\nu-\lambda)}  h_1\right]' = 0.
\label{thph_l2}
\ee

\subsection{The case $l=1$}
The $l=1$ case differs from $l\geq 2$ in two respects (Campolattaro
and Thorne \cite{ct70}).  Firstly, $H_2(r)\neq H_0(r)$, because the 
equation 
$\delta G_\theta^{\ \theta}-\delta G_\varphi^{\ \varphi}
=8\pi\left(\delta T_\theta^{\ \theta}-\delta T_\varphi^{\ \varphi}\right)$ 
vanishes identically.  Secondly, we may exploit the aforementioned gauge
freedom for this case to eliminate the metric functions $K(r)$ and $h_1(r)$. 
(We note that Eq.(\ref{rph_l2}) implies $h_1(r)=0$ for $l\geq 2$ anyway.)
With these two differences taken into account the non-vanishing components 
of the perturbed Einstein equation for $l=1$ are as follows. From 
$\delta G_t^{\ t}=8\pi\delta T_t^{\ t}$ we have
\be
0 = e^{-2\lambda} rH_2' + \left(2-8\pi r^2\ep\right)H_2
-8\pi r^2\delta\ep.
\label{tt_l1}
\ee
From $\delta G_r^{\ r}=8\pi\delta T_r^{\ r}$ we have
\be
0 = e^{-2\lambda}rH_0' - H_0 + \left(1+8\pi r^2 p\right)H_2
+8\pi r^2\delta p.
\label{rr_l1}
\ee
From 
$\delta G_\theta^{\ \theta}+\delta G_\varphi^{\ \varphi}
=8\pi\left(\delta T_\theta^{\ \theta}+\delta T_\varphi^{\ \varphi}\right)$ 
we have
\bea
0 &=& e^{-2\lambda}r^2H_0''
+ e^{-2\lambda}(2r\nu'-r\lambda'+1)rH_0' - H_0 \nn\\
&& + e^{-2\lambda}(1+r\nu')rH_2' + (1+16\pi r^2 p)H_2
+16\pi r^2\delta p
\label{thth+_l1}
\eea
From $\delta G_r^{\ \theta}=8\pi\delta T_r^{\ \theta}$ we have
\be
0 = rH_0' + (r\nu'-1)H_0 + (r\nu'+1)H_2
\label{rth_l1}
\ee
From $\delta G_t^{\ r}=8\pi\delta T_t^{\ r}$ we, again, have
\be
0 = H_1 + 8\pi(\epsilon+p) e^{2\lambda} r W.
\label{tr_l1}
\ee
From $\delta G_t^{\ \theta}=8\pi\delta T_t^{\ \theta}$ we, again, have
\be
0 = e^{-(\nu-\lambda)}
\left[e^{(\nu-\lambda)}  H_1\right]'
+ 16\pi(\ep+p)e^{2\lambda}V.
\label{tth_l1}
\ee
Finally, from $\delta G_t^{\ \varphi}=8\pi\delta T_t^{\ \varphi}$ 
we have
\be
h_0^{''} - (\nu'+\lambda') h_0' 
- \left[ \frac{2}{r}(\nu'+\lambda') + \frac{2}{r^2} \right] h_0 
= \frac{4}{r}(\nu'+\lambda') U.
\label{tph_l1}
\ee

\subsection{The case $l=0$}
The $l=0$ case differs yet again from the previous two, being the 
case of stationary, spherically symmetric perturbations of a static,
spherical equilibrium.  To maximize the similarity to the preceding
two cases we will use the same form for the perturbed metric except
that we may now exploit the gauge freedom for this case to eliminate 
the functions $K(r)$, $H_1(r)$ and $h_1(r)$.  The non-vanishing 
components of the perturbed Einstein equation for $l=0$ are as follows. 
From $\delta G_t^{\ t}=8\pi\delta T_t^{\ t}$ we have
\be
0 = e^{-2\lambda} rH_2' + \left(1-8\pi r^2\ep\right)H_2
-8\pi r^2\delta\ep.
\label{tt_l0}
\ee
From $\delta G_r^{\ r}=8\pi\delta T_r^{\ r}$ we have
\be
0 = e^{-2\lambda}rH_0' + \left(1+8\pi r^2 p\right)H_2
+8\pi r^2\delta p.
\label{rr_l0}
\ee
From 
$\delta G_\theta^{\ \theta}+\delta G_\varphi^{\ \varphi}
=8\pi\left(\delta T_\theta^{\ \theta}+\delta T_\varphi^{\ \varphi}\right)$ 
we have
\bea
0 &=& e^{-2\lambda}r^2H_0''
+ e^{-2\lambda}(2r\nu'-r\lambda'+1)rH_0' \nn\\
&& + e^{-2\lambda}(1+r\nu')rH_2' + 16\pi r^2 pH_2 +16\pi r^2\delta p
\label{thth+_l0}
\eea
Finally, from $\delta G_t^{\ r}=8\pi\delta T_t^{\ r}$ we have
\be
0 = 16\pi(\epsilon+p) W.
\label{tr_l0}
\ee

\subsection{Decomposition of the zero-frequency subspace.}
By inspection of the above three sets of equations, it is evident 
that they decouple into two independent classes. For $l\geq 2$ Eqs.
(\ref{tt_l2})-(\ref{rth_l2}) involve only the variables
$(H_0, H_2, K, \delta\ep, \delta p)$ while Eqs. 
(\ref{tr_l2})-(\ref{tph_l2}) involve only the variables 
$(H_1, h_0, W, V, U)$
(with $h_1(r)\equiv0$ implied by Eq. (\ref{rph_l2})).  
Similarly for $l=1$ Eqs. (\ref{tt_l1})-(\ref{rth_l1}) 
involve only $(H_0, H_2, \delta\ep, \delta p)$ 
while Eqs. (\ref{tr_l1})-(\ref{tph_l1}) involve only 
$(H_1, h_0, W, V, U)$ and, in fact, are identical to Eqs.
(\ref{tr_l2})-(\ref{tph_l2}).  Finally, for $l=0$ Eqs. 
(\ref{tt_l0})-(\ref{thth+_l0}) involve only 
$(H_0, H_2, \delta\ep, \delta p)$ while Eq. (\ref{tr_l0}) simply 
implies that $W(r)\equiv 0$.

Thus, any solution, 
\be
(H_0, H_1, H_2, K, h_0, W, V, U, \delta\ep, \delta p),
\ee
to the equations governing the time-independent perturbations of 
a static, spherical star is a superposition of (i) a solution
\be
(0, H_1, 0, 0, h_0, W, V, U, 0, 0)
\ee
to Eqs. (\ref{tr_l2})-(\ref{tph_l2}) or (\ref{tr_l0}) and 
(ii) a solution
\be
(H_0, 0, H_2, K, 0, 0, 0, 0, \delta\ep, \delta p)
\ee
to Eqs. (\ref{tt_l2})-(\ref{rth_l2}), (\ref{tt_l1})-(\ref{rth_l1})
or (\ref{tt_l0})-(\ref{thth+_l0}).

For the solutions of type (ii), the vanishing of the $(tr)$, 
$(t\theta)$ and $(t\varphi)$ components of the perturbed metric 
in our coordinate system implies that these solutions are static.
If, as in the newtonian case, one assumes the linearization 
stability\footnote{Again, we are aware of a proof of this linearization
stability property under assumptions on the equation of state that are 
satisfied by uniform density stars, but would not allow polytropes 
(K\"unzle and Savage \cite{ks80}).}
of these solutions, i.e., that any solution to the static perturbation 
equations is tangent to a family of exact static solutions, then the 
theorem that any static self-gravitating
perfect fluid is spherical implies that any solution of type (ii) is 
simply a neighboring spherical equilibrium. 

Thus, under the assumption of linearization stability we have shown that
all stationary non-radial ($l>0$) perturbations of a spherical star have
\[H_0=H_2=K=\delta\ep=\delta p=0\]
and satisfy Eqs. (\ref{tr_l2})-(\ref{tph_l2}); that is,
\bea
0 &=& H_1 +\frac{16\pi(\epsilon+p)}{l(l+1)} e^{2\lambda} r W,  
\label{GR:sph_H1} \\
&& \nn \\
0 &=& e^{-(\nu-\lambda)}\left[e^{(\nu-\lambda)}  H_1\right]'
+ 16\pi(\ep+p)e^{2\lambda}V,
\label{GR:sph_H1_2} 
\eea
\be
h_0^{''} - (\nu'+\lambda') h_0' 
+ \left[ \frac{(2-l^2-l)}{r^2}e^{2\lambda} 
- \frac{2}{r}(\nu'+\lambda') - \frac{2}{r^2} \right] h_0 
= \frac{4}{r}(\nu'+\lambda') U.
\label{GR:sph_h_0''}
\ee
Observe that if we use Eq. (\ref{GR:sph_H1}) to eliminate $H_1(r)$ from 
Eq. (\ref{GR:sph_H1_2}) we obtain
\be
V = \frac{e^{-(\nu+\lambda)}}{l(l+1)(\epsilon+p)} 
\left[(\epsilon+p)e^{\nu+\lambda} r W \right]'.
\label{GR:sph_V}
\ee
This equation is clearly the generalization to relativistic stars
of the conservation of mass equation in newtonian gravity, Eq. 
(\ref{N:sph_cont}) from Sect. 2.1.
The other two equations relate the perturbation of the spacetime 
metric to the perturbation of the fluid 4-velocity and vanish
in the newtonian limit.

These perturbations must be regular everywhere and satisfy the boundary
condition that the lagrangian change in the pressure vanish at the 
surface of the star. (See Sect. 3.4 below.)  As with newtonian stars, 
this boundary condition requires only that 
\be
W(R) = 0.
\ee
leaving $W(r)$ and $U(r)$ otherwise undetermined.  If $W(r)$ and 
$U(r)$ are specified, then the functions $H_1(r)$, $h_0(r)$ and 
$V(r)$ are determined by the above equations. (These solutions are
subject to matching conditions to the solutions in the exterior
spacetime, which must also be regular at infinity. See Sect. 3.4.)

Finally, we consider the equation of state of the perturbed star.
For an adiabatic oscillation of a barotropic star (i.e., a star
that satisfies a one-parameter equation of state, $p=p(\ep)$) 
Eq. (\ref{Del_etc}) implies that the
perturbed pressure and energy density are related by
\be
\frac{\delta p}{\gamma p} = \frac{\delta \ep}{(\ep+p)} 
+ \xi^r \left[\frac{\ep'}{(\ep+p)} - \frac{p'}{\gamma p}
\right]
\label{ad_osc}
\ee
for some adiabatic index $\gamma(r)$ which need not be the function
\be
\Gamma(r) \equiv \frac{(\ep+p)}{p}\frac{dp}{d\ep}
\ee
associated with the equilibrium equation of state.  Here, 
$\xi^\alpha$ is the lagrangian displacement vector and is related 
to our perturbation variables by
\be
q^\alpha_{\ \beta} \pounds_u\xi^\beta = 
\delta u^\alpha - \mbox{$\ds{\half}$} u^\alpha u^\beta u^\gamma 
h_{\beta\gamma}
\ee
Thus, we have
\be
e^{-\nu}\partial_t\xi^r = \delta u^r
\ee
or (taking the initial displacement (at $t=0$) to be zero)
\be
\xi^r = t e^\nu \delta u^r.
\label{form_xi_r}
\ee

For the class of perturbations under consideration, we have
seen that $\delta p = \delta\ep = 0$, thus Eqs. (\ref{ad_osc}) and
(\ref{form_xi_r}) require that
\be
\delta u^r \left[\frac{\ep'}{(\ep+p)} 
- \frac{p'}{\gamma p}\right] = 0.
\label{eos_req}
\ee
For axial-parity perturbations this equation is automatically
satisfied, since $\delta u_A^\alpha$ has no $r$-component (Eq.
(\ref{GR:del_u_ax})).  Thus, a spherical barotropic star always 
admits a class of zero-frequency r-modes.

For polar-parity perturbations, $\delta u^r_P = e^{-\nu} W(r)/r \neq 0$,
and Eq. (\ref{eos_req}) will be satisfied if and only if 
\be
\gamma(r)\equiv\Gamma(r)=\frac{(\ep+p)}{p}\frac{dp}{d\ep}.
\ee
Thus, a spherical barotropic star admits a class of zero-frequency
g-modes if and only if the perturbed star obeys the same 
one-parameter equation of state as the equilibrium star. Once again, 
we will call such a star isentropic, because isentropic models and 
their adiabatic perturbations obey the same one-parameter equation
of state. (That all axial-parity fluid perturbations of a spherical
relativistic star are time-independent was shown by Thorne and 
Campolattaro \cite{tc67}.  The time-independent g-modes in spherical,
isentropic, relativistic stars were found by Thorne \cite{t69a}.)

Summarizing our results, we have shown the following.  A spherical
barotropic star always admits a class of zero-frequency r-modes 
(stationary fluid currents with axial parity); but admits
zero-frequency g-modes (stationary fluid currents with polar parity)
if and only if the star is isentropic.  Conversely, the zero-frequency
subspace of non-radial perturbations of a spherical isentropic 
star is spanned by the r- and g-modes - that is, by convective fluid 
motions having both axial and polar parity and with vanishing perturbed 
pressure and density.  Being stationary, these r- and g-modes do not 
couple to gravitational radiation, although the r-modes do induce a 
nontrivial metric perturbation ($h_{t\theta}, h_{t\varphi}\neq 0$) 
in the spacetime exterior to the
star (frame-dragging).  One would expect this large subspace of 
modes, which is degenerate at zero-frequency, to be split by 
rotation, as it is in newtonian stars, so let us now 
consider the perturbations of slowly rotating relativistic stars.


\section{Perturbations of Slowly Rotating Stars}

The equilibrium of an isentropic, perfect fluid star that is
rotating slowly with uniform angular velocity $\Omega$ is
described (Hartle \cite{h67}, Chandrasekhar and Miller \cite{cm74}) 
by a stationary, axisymmetric 
spacetime with metric, $g_{\alpha\beta}$, of the form 
\be
ds^2 = -e^{2\nu(r)} dt^2 + e^{2\lambda(r)} dr^2 + r^2 d \theta^2 + 
r^2 sin^2 \theta d \varphi^2  - 2 \omega(r) r^2 sin^2\theta dt d\varphi
\label{equil_metric}
\ee
(accurate to order $\Omega$) and with energy-momentum tensor
\be
T_{\alpha\beta} = (\epsilon+p)u_\alpha u_\beta + p g_{\alpha\beta},
\label{equil_stress}
\ee
where $\ep(r)$ is the total fluid energy density, $p(r)$ is the fluid 
pressure and
\be
u^\alpha = e^{-\nu} ( t^\alpha + \Omega \varphi^\alpha)
\label{equil_4v}
\ee
is the fluid 4-velocity to order $\Omega$ - with 
$t^\alpha=(\partial_t)^\alpha$ and
$\varphi^\alpha=(\partial_\varphi)^\alpha$, respectively the
timelike and rotational Killing vectors of the spacetime.

That the star is rotating slowly is the assumption that $\Omega$
is small compared to the Kepler velocity, 
$\Omega_K\sim\sqrt{M/R^3}$, the 
angular velocity at which the star is unstable to mass shedding
at its equator.  In particular, we neglect all quantities of order 
$\Omega^2$ or higher. To this order, the star retains
its spherical shape, because the centrifugal deformation of its figure
is an order $\Omega^2$ effect (Hartle \cite{h67}).

In constructing such an equilibrium configuration, the equations 
(\ref{GR_sph:eos})-(\ref{GR_sph:mass_def}) governing a spherical star,
\be
p = p(\ep),
\label{GR:eos}
\ee

\be
\frac{dp}{dr} = - \frac{(\ep+p)(M+4\pi r^3p)}{r(r-2M)},
\label{tov}
\ee
\be
\frac{dM}{dr} = 4\pi r^2\ep,
\label{GR:dMdr}
\ee
and
\be
\frac{d\nu}{dr} = - \frac{1}{(\ep+p)}\frac{dp}{dr},
\label{dnudr}
\ee
with
\be
M(r) \equiv \half r(1-e^{-2\lambda}),
\label{GR:mass_def}
\ee
are joined by an equation (Hartle \cite{h67})
that determines the new metric function $\om(r)$ in terms of the
spherical metric functions $\nu(r)$ and $\lambda(r)$,
\be
\frac{e^{(\nu+\lambda)}}{r^4}\frac{d}{dr}
\left( r^4 e^{-(\nu+\lambda)} \frac{d\bom}{dr} \right)
- \frac{4}{r} 
\left(\frac{d\nu}{dr}+\frac{d\lambda}{dr}\right) \bom = 0
\label{hartle}
\ee
where
\be
\bom(r)\equiv\Omega-\om.
\label{bom}
\ee
Outside the star, Eq. (\ref{hartle}) has the solution,
\be
\bom = \Omega-\frac{2J}{r^3}
\label{hartle_ext_soln}
\ee
where $J$ is the angular momentum of the spacetime.
This new metric variable is a quantity of order
$\Omega$ and governs the dragging of inertial frames
induced by the rotation of the star (Hartle \cite{h67}).
Apart from the frame-dragging effect, however, the 
spacetime is unchanged from the spherical configuration.


Since the equilibrium spacetime is stationary and axisymmetric, 
we may decompose our perturbations into modes of the 
form\footnote{We will, again, always choose $m \geq 0$ since the 
complex conjugate of an $m<0$ mode with frequency $\sigma$ is an 
$m>0$ mode with frequency $-\sigma$.  Note that $\sigma$ is the 
frequency measured by an inertial observer at infinity.} 
$e^{i(\sigma t + m \varphi)}$.  
We will use the lagrangian perturbation formalism reviewed in 
Sect. 1.2.2 and begin by expanding the displacement vector 
$\xi^\alpha$ and metric perturbation $h_{\alpha\beta}$ in 
tensor spherical harmonics.

The lagrangian displacement vector can be written
\be
\xi^{\alpha} \equiv \frac{1}{i\kappa\Omega} \sum_{l=m}^\infty 
\biggl\{  \ba[t]{l}
\ds{\frac{1}{r} W_l(r) Y_l^m r^{\alpha} 
+ V_l(r) \nabla^{\alpha} Y_l^m } \\
\\
- \ds{i U_l(r) P^\alpha_{\ \mu} \epsilon^{\mu\beta\gamma\delta} 
\nabla_{\beta} Y_l^m \nabla_{\!\gamma} \, t \nabla_{\!\delta} \, r}
\biggr\} e^{i\sigma t},
\ea
\label{xi_exp}
\ee
where we have defined,
\be
P^\alpha_{\ \mu} \equiv e^{(\nu+\lambda)} 
\left( \delta^\alpha_{\ \mu}
- t_\mu \nabla^\alpha t
\right)
\ee
and the comoving frequency $\kappa\Omega \equiv \sigma+m\Omega$.
The exact form of this expression has been chosen for later convenience.
In particular, we have chosen a gauge in which $\xi_t\equiv 0$.  
Note also the choice of phase between the terms in (\ref{xi_exp}) with 
polar parity (those with coefficients $W_l$ and $V_l$) and the terms 
with axial parity (those with coefficients $U_l$).


Working again in the Regge-Wheeler gauge, we express our metric 
perturbation as
\be
h_{\mu\nu} = \sum_{l=m}^\infty \left[
\ba[c]{cccc}
H_{0,l}(r)e^{2\nu}Y_l^m & H_{1,l}(r)Y_l^m & 
h_{0,l}(r) \,(\frac{m}{sin\theta})Y_l^m & i h_{0,l}(r)\,\sinY\\
H_{1,l}(r)Y_l^m & H_{2,l}(r)e^{2\lambda}Y_l^m & 
h_{1,l}(r) \,(\frac{m}{sin\theta})Y_l^m & i h_{1,l}(r)\,\sinY\\
\mbox{\scriptsize symm} & \mbox{\scriptsize symm} &
r^2K_l(r)Y_l^m & 0\\
\mbox{\scriptsize symm} & \mbox{\scriptsize symm} &
0 & r^2sin^2\theta K_l(r)Y_l^m
\ea
\right]e^{i\sigma t}.
\label{h_components}
\ee
Again, note the choice of phase between the polar-parity components
(those with coefficients $H_{0,l}$, $H_{1,l}$, $H_{2,l}$ and $K_l$) 
and the axial-parity components (those with coefficients
$h_{0,l}$ and $h_{1,l}$).

Based on our knowledge of the newtonian spectrum, we expect to find 
a class of hybrid modes whose spherical limit ($\Omega\rightarrow 0$) 
is a sum of the relativistic zero-frequency r- and g-modes found in the 
preceding section.  We will, therefore, assume that our perturbation 
variables obey an ordering in powers of $\Omega$ that reflects this 
spherical limit.
\be
\ba{lll}
W_l, V_l, U_l, H_{1,l}, h_{0,l} & \sim  & O(1) \\
H_{0,l}, H_{2,l}, K_l, h_{1,l}, \delta \epsilon, \delta p, \sigma & 
\sim & O(\Omega).
\ea
\label{ordering}
\ee
In addition, we assume that the $O(1)$ quantities obey the 
$O(1)$ perturbation equations (\ref{GR:sph_H1}), 
(\ref{GR:sph_h_0''}) and (\ref{GR:sph_V}) for all $l$.

Eq. (\ref{DelQ}) relates the Eulerian change in the 4-velocity to 
$\xi^\alpha$ and $h_{\alpha\beta}$,
\be
\delta u^\alpha = q^\alpha_{\ \beta}\pounds_u\xi^\beta 
+ \mbox{$\half$} u^\alpha u^\beta u^\gamma h_{\beta\gamma}.
\ee
The ordering (\ref{ordering}) implies that $\delta u^\alpha$
is given, to zeroth order in $\Omega$, by
\be
\delta u^\alpha = i\kappa\Omega e^{-\nu} q^\alpha_{\ \beta}\xi^\beta,
\ee
and the spherical limit of this expression reveals that the mode
$(\xi^\alpha, h_{\alpha\beta})$ is manifestly a sum of the 
axial and polar perturbations considered in Sect. 3.1. 
(Eqs. (\ref{GR:del_u_po}), (\ref{GR:del_u_ax}), (\ref{GR:h_po})
and (\ref{GR:h_ax}).)

In the newtonian calculation, it was the conservation of circulation 
in isentropic stars that brought about the rotational splitting of the 
zero-frequency modes at zeroth order in the angular velocity $\Omega$ 
(see Sect. 2.2).  The equation that enforces this conservation law 
is the curl of the perturbed Euler equation (\ref{form_of_q}) and in general
relativity it takes the form (Friedman \cite{f78}; see also Friedman
and Ipser \cite{fi92}),
\be
0  = \Delta \pounds_u \omega_{\alpha\beta} = 
\pounds_u \mbox{$\Delta$} \omega_{\alpha\beta} = 
i\kappa\mbox{$\Omega$} e^{-\nu} \mbox{$\Delta$} \omega_{\alpha\beta}
\ee
or simply
\be
\Delta \omega_{\alpha\beta} = 0,   \label{Del_om}
\ee
where
\be
\omega_{\alpha\beta} \equiv 
2 \nabla_{\left[\alpha\right.} 
\Biggl(\frac{\epsilon+p}{n} u_{\left.\beta\right]}\Biggr)
\ee
is the relativistic vorticity tensor.

We begin by expressing Eq. (\ref{Del_om}),
\be
0 = \Delta \omega_{\alpha\beta} = 
\nabla_\alpha\left[\Delta\left(\frac{\epsilon+p}{n} u_\beta 
\right)\right]
- \nabla_\beta\left[\Delta\left(\frac{\epsilon+p}{n} u_\alpha 
\right)\right],
\ee
in terms of $\xi^\alpha$ and $h_{\alpha\beta}$.

Making use of Eq. (\ref{Del_etc}) we have
\be
\Delta \left( \frac{\epsilon+p}{n} u_\alpha \right) = 
\frac{\epsilon+p}{n} \left[ \Delta u_\alpha 
- \half q^{\alpha\beta}\Delta g_{\alpha\beta}\left( 
	\frac{\gamma p}{\epsilon+p}\right)u_\alpha 
\right],
\label{Del_hu}
\ee
where
\be
\Delta u_\alpha\equiv \Delta(g_{\alpha\beta}u^\beta)
=\Delta g_{\alpha\beta}u^\beta+g_{\alpha\beta}\Delta u^\beta
\ee

The ordering (\ref{ordering}) implies that 
$u^\alpha u^\beta h_{\alpha\beta}$ and 
$g^{\alpha\beta}h_{\alpha\beta}$ vanish to zeroth order 
in $\Omega$, since the only zeroth order metric components are
$h_{tr}$, $h_{t\theta}$ and $h_{t\varphi}$.
Therefore,
\begin{eqnarray}
\half u^\alpha u^\beta \Delta g_{\alpha\beta} 
	&=& u^\alpha u^\beta \nabla_\alpha \xi_\beta  \\
\half q^{\alpha\beta}\Delta g_{\alpha\beta} 
	&=& q^{\alpha\beta}\nabla_\alpha \xi_\beta
\label{Del_q}  \\
\Delta u^\alpha &=& 
u^\alpha u^\beta u^\gamma \nabla_\beta\xi_\gamma \\
\Delta u_\alpha &=& 
h_{\alpha\beta}u^\beta + 
	u^\beta\nabla_\beta\xi_\alpha 
	+ u^\beta\nabla_\alpha\xi_\beta 
	+ u_\alpha u^\beta u^\gamma\nabla_\beta\xi_\gamma.
\label{Del_u}
\end{eqnarray}

From Eqs. (\ref{Del_etc}) and (\ref{dnudr}) and the relation,
\bea
u^\alpha u^\beta \nabla_\alpha\xi_\beta 
&=& 
- \xi^\beta u^\alpha\nabla_\alpha u_\beta 
+ u^\alpha \nabla_\alpha(u^\beta \xi_\beta) \nn \\
&=& - \xi^\beta \nabla_\beta \nu + O(\Omega),
\eea
we obtain,
\begin{eqnarray}
\half q^{\alpha\beta}\Delta g_{\alpha\beta} 
	&=& \left(\frac{\epsilon+p}{\gamma p}\right)\nu' 
		e^{-2\lambda}\xi_r 
\label{form_Dp}  \\
u^\alpha u^\beta \nabla_\alpha\xi_\beta
	&=& - \nu' e^{-2\lambda}\xi_r
\end{eqnarray}
to zeroth order in $\Omega$.
We will also use the explicit form of $u_\varphi$ from 
Eq. (\ref{equil_4v}),
\be
u_\varphi = e^{-\nu}{\bar\omega}r^2\sin^2\theta,
\label{u_phi}
\ee
and the components of $\Delta u_\alpha$ to zeroth order in 
$\Omega$,
\begin{eqnarray}
\Delta u_r       &=& e^{-\nu} \biggl[
h_{tr}+i\kappa\Omega\xi_r
+\Omega r^2\partial_r\left(\frac{1}{r^2}\xi_\varphi\right)
+\frac{e^{2\nu}}{r^2}\partial_r\left(r^2\omega 
e^{-2\nu}\right)\xi_\varphi
\biggr] \\
\Delta u_\theta  &=& e^{-\nu} \biggl[
h_{t\theta}+i\kappa\Omega\xi_\theta
+\Omega\partial_\theta\xi_\varphi
-2{\bar\omega}\cot\theta\xi_\varphi
\biggr] \\
\Delta u_\varphi &=& e^{-\nu} \biggl[
\ba[t]{l}
h_{t\varphi}+i\kappa\Omega\xi_\varphi
+\Omega\partial_\varphi\xi_\varphi
+2{\bar\omega}\sin\theta\cos\theta\xi_\theta 
\vphantom{\frac{x^a_b}{y^a_b}} \\
+e^\nu\partial_r\left(r^2{\bar\omega}e^{-\nu} 
\right)\sin^2\theta e^{-2\lambda}\xi_r
\vphantom{\frac{x^a_b}{y^a_b}}
\biggr]. 
\ea
\label{Del_u_cov_compts}
\end{eqnarray}
For reference, we explicitly write the components of 
$i\kappa\Omega{\vec\xi}$ to zeroth order in $\Omega$,
\be
\label{xi_components}
\ba[t]{ll}
i\kappa\Omega\xi^t = O(\Omega)  &
i\kappa\Omega\xi^\theta = \ds{\sum_l \frac{1}{r^2\sin\theta} 
			\left[V_l \sinY + m U_l Y_l^m \right]e^{i\sigma t}} \\
 & \\
i\kappa\Omega\xi^r = \ds{\sum_l \frac{1}{r} W_l Y_l^m e^{i\sigma t}} &
i\kappa\Omega\xi^\varphi = \ds{\sum_l \frac{i}{r^2\sin^2\theta}    
			\left[m V_l Y_l^m + U_l \sinY \right]e^{i\sigma t}} \\
 & \\
i\kappa\Omega\xi_t = 0 &
i\kappa\Omega\xi_\theta = \ds{\sum_l \frac{1}{\sin\theta} 
			\left[V_l \sinY + m U_l Y_l^m \right]e^{i\sigma t}}  \\
 & \\
i\kappa\Omega\xi_r = \ds{\sum_l \frac{e^{2\lambda}}{r} W_l Y_l^m e^{i\sigma t}} &
i\kappa\Omega\xi_\varphi = \ds{\sum_l i \left[m V_l Y_l^m 
+ U_l \sinY \right]e^{i\sigma t}}.
\ea
\ee


By making use of Eqs. (\ref{Del_hu}) through 
(\ref{Del_u_cov_compts}) and the expressions (\ref{xi_components}) 
and (\ref{h_components}) for the components of 
$i\kappa\Omega{\vec\xi}$ and $h_{\alpha\beta}$, we may now 
write the spatial components of $\Delta\omega_{\alpha\beta}$. 
We will use Eq. (\ref{GR:sph_H1}) to eliminate $H_{1,l}$ (for all $l$)
from the resulting expressions and drop the ``0'' subscript on the metric
functions $h_{0,l}$, writing $h_{0,l}\equiv h_l$.  


\begin{eqnarray}
\Delta\omega_{\theta\varphi}
 &=& \left(\frac{\epsilon+p}{n}\right) \biggl\{
\partial_\theta\Delta u_\varphi 
- \partial_\varphi\Delta u_\theta
- \partial_\theta\left[
\half q^{\alpha\beta}\Delta g_{\alpha\beta}\left( 
\frac{\gamma p}{\epsilon+p}\right)u_\varphi
\right]\biggr\} 
\nonumber \\
 &=& \left(\frac{\epsilon+p}{n}\right) 
\frac{e^{-\nu}\sin\theta}{i\kappa\Omega} 
\label{om_th_ph_form1} \\
 & & \times\sum_l\biggl\{
	\ba[t]{l}
	\left[ l(l+1)\kappa\Omega(h_l+U_l)
	-2m{\bar\omega}U_l\right] Y_l^m \\
	\\
	-2{\bar\omega}V_l\left[\sinY+l(l+1)\cosY\right] \\
	\\
	+\frac{e^{2\nu}}{r}\partial_r\left(r^2{\bom}e^{-2\nu}
	\right)W_l\left[\sinY+2\cosY\right]
	\biggr\}e^{i\sigma t}
	\ea
\nonumber 
\end{eqnarray}


\begin{eqnarray}
\Delta\omega_{r\theta}
 &=& \left(\frac{\epsilon+p}{n}\right) e^\nu \biggl[
\partial_r\left(e^{-\nu}\Delta u_\theta\right) 
- \partial_\theta\left(e^{-\nu}\Delta u_r\right)
\biggr]
\nonumber \\
 &=& \left(\frac{\epsilon+p}{n}\right) 
\frac{e^\nu}{\kappa\Omega\sin\theta}
\label{om_r_th_form1} \\
 & & \times\sum_l\Biggl\{
\ba[t]{l}
	m\kappa\Omega\partial_r\left[e^{-2\nu}(h_l+U_l)\right]Y_l^m 
	-2\partial_r\left({\bar\omega}e^{-2\nu}U_l\right)\cos\theta\sinY \\
	\\
	+\frac{1}{r^2}\partial_r\left(r^2{\bar\omega}e^{-2\nu}\right)U_l
	\left[m^2+l(l+1)(\cos^2\theta-1)\right]Y_l^m \\
	\\
	-2m\partial_r\left({\bar\omega}e^{-2\nu}V_l\right)\cos\theta Y_l^m
	+\frac{m}{r^2}\partial_r\left(r^2{\bar\omega}e^{-2\nu}\right)V_l\sinY \\
	\\
	+\kappa\Omega\left[\partial_r\left(e^{-2\nu}V_l\right)
	+e^{-2\nu}\left(\frac{16\pi r(\epsilon+p)}{l(l+1)}
	-\frac{1}{r}\right)e^{2\lambda}W_l
	\right]\sinY
	\Biggr\}e^{i\sigma t}
	\ea
\nonumber
\end{eqnarray}


\begin{eqnarray}
\Delta\omega_{\varphi r}
 &=& \left(\frac{\epsilon+p}{n}\right) e^\nu 
\biggl\{ \ba[t]{l}
	\partial_\varphi\left(e^{-\nu}\Delta u_r\right) 
	- \partial_r\left(e^{-\nu}\Delta u_\varphi\right) \\
	\\
	+ \partial_r\left[
	\half q^{\alpha\beta}\Delta g_{\alpha\beta}
	\left( \frac{\gamma p}{\epsilon+p}\right) e^{-\nu}u_\varphi
\right]\biggr\}
\ea
\nonumber \\
 &=& \left(\frac{\epsilon+p}{n}\right) \frac{e^\nu}{i\kappa\Omega}
\label{om_ph_r_form1} \\
 & & \times\sum_l\Biggl\{
\ba[t]{l}
	m\kappa\Omega\left[\partial_r\left(e^{-2\nu}V_l\right)
	+e^{-2\nu}\left(\frac{16\pi r(\epsilon+p)}{l(l+1)}
	-\frac{1}{r}\right)e^{2\lambda}W_l
	\right]Y_l^m \\
	\\
	-2\partial_r\left({\bar\omega}e^{-2\nu}V_l\right)\cos\theta\sinY
	+\frac{m^2}{r^2}\partial_r\left(r^2{\bar\omega}e^{-2\nu}\right)V_lY_l^m \\
	\\
	+\partial_r\left[\frac{1}{r}\partial_r\left(r^2{\bom}e^{-2\nu}\right)W_l
	\right](\cos^2\theta-1)Y_l^m \\
	\\
	-2m\partial_r\left({\bar\omega}e^{-2\nu}U_l\right)\cosY
	+\kappa\Omega\partial_r\left[e^{-2\nu}(h_l+U_l)\right]\sinY \\
	\\
	+\frac{m}{r^2}\partial_r\left(r^2{\bar\omega}e^{-2\nu}\right)U_l\sinY
	\Biggr\}e^{i\sigma t}
	\ea
\nonumber
\end{eqnarray}

Note that the three spatial components of $\Delta\om_{\alpha\beta}=0$ 
are not independent. They are related by the identity
\be
\nabla_{\left[\alpha\right.}\Delta\omega_{\left.\beta\gamma\right]} = 0,
\label{not_ind}
\ee
which, therefore, serves as a check on the above expressions.


Let us write these three equations making use of the identities 
(\ref{identities1})-(\ref{identities2}),
\begin{eqnarray}
\sin\theta\partial_\theta Y_l^m &=& l Q_{l+1} Y_{l+1}^m 
	- (l+1) Q_l Y_{l-1}^m \\
\cos\theta Y_l^m &=& Q_{l+1} Y_{l+1}^m + Q_l Y_{l-1}^m 
\end{eqnarray}
where $Q_l$ was defined in Eq. (\ref{Q_l}) to be
\be
Q_l \equiv \left[ \frac{(l+m)(l-m)}{(2l-1)(2l+1)} \right]^{\half}. 
\ee

\noindent
From $\Delta\omega_{\theta\varphi}=0$ we have,
\be
0 = \sum_l\Biggl\{
\ba[t]{l}
\left[ l(l+1)\kappa\Omega(h_l+U_l)
-2m{\bar\omega}U_l\right] Y_l^m \\
 \\
+\left[\frac{e^{2\nu}}{r}\partial_r\left(r^2{\bar\omega}e^{-2\nu}\right)W_l
-2l{\bar\omega}V_l\right](l+2)Q_{l+1}Y_{l+1}^m  \\
 \\
-\left[\frac{e^{2\nu}}{r}\partial_r\left(r^2{\bar\omega}e^{-2\nu}\right)W_l
+2(l+1){\bar\omega}V_l\right](l-1)Q_lY_{l-1}^m
\Biggr\}
\ea
\label{om_th_ph_form2}
\ee


\noindent
From $\Delta\omega_{r\theta}=0$ we have,
\be
0 = \sum_l\Biggl\{
\ba[t]{l}
\left[-2\partial_r\left({\bar\omega}e^{-2\nu}U_l\right)
+\frac{(l+1)}{r^2}\partial_r\left(r^2{\bar\omega}e^{-2\nu}\right)U_l
\right] l Q_{l+1}Q_{l+2}Y_{l+2}^m \\
 \\
+\biggl[
	\ba[t]{l}
	l\kappa\Omega\partial_r\left(e^{-2\nu}V_l\right)
	-2m\partial_r\left({\bar\omega}e^{-2\nu}V_l\right) \\
	 \\
	+\frac{lm}{r^2}\partial_r\left(r^2{\bar\omega}e^{-2\nu}\right)V_l
	+l\kappa\Omega e^{-2\nu}\left(\frac{16\pi r(\epsilon+p)}{l(l+1)}
					-\frac{1}{r}\right)e^{2\lambda}W_l
	\biggr]Q_{l+1}Y_{l+1}^m 
	\ea \\
 \\
+\biggl[
	\ba[t]{l}
	m\kappa\Omega\partial_r\left[e^{-2\nu}(h_l+U_l)\right]
	+2\partial_r\left({\bar\omega}e^{-2\nu}U_l\right)
	\left((l+1)Q_l^2-l Q_{l+1}^2\right) \\
	 \\
	+\frac{1}{r^2}\partial_r\left(r^2{\bar\omega}e^{-2\nu}\right)U_l
	\left[m^2+l(l+1)\left(Q_{l+1}^2+Q_l^2-1\right)\right]
	\biggr] Y_l^m 
	\ea \\
 \\
-\biggl[
	\ba[t]{l}
	(l+1)\kappa\Omega\partial_r\left(e^{-2\nu}V_l\right)
	+2m\partial_r\left({\bar\omega}e^{-2\nu}V_l\right) \\
	 \\
	+\frac{m(l+1)}{r^2}\partial_r\left(r^2{\bar\omega}e^{-2\nu}\right)V_l
	+(l+1)\kappa\Omega e^{-2\nu}\left(\frac{16\pi r(\epsilon+p)}{l(l+1)}
					-\frac{1}{r}\right)e^{2\lambda}W_l
	\biggr]Q_l Y_{l-1}^m 
	\ea \\
 \\
+\left[2\partial_r\left({\bar\omega}e^{-2\nu}U_l\right)
+\frac{l}{r^2}\partial_r\left(r^2{\bar\omega}e^{-2\nu}\right)U_l
\right] (l+1) Q_{l-1}Q_lY_{l-2}^m 
\Biggr\}
\ea
\label{om_r_th_form2}
\ee

\noindent
From $\Delta\omega_{\varphi r}=0$ we have,
\be
0 = \sum_l\Biggl\{
\ba[t]{l}
\biggl[
\partial_r\left[\frac{1}{r}\partial_r\left(r^2{\bar\omega}e^{-2\nu}\right)W_l\right]
-2l\partial_r\left({\bar\omega}e^{-2\nu}V_l\right)
\biggr] Q_{l+2}Q_{l+1}Y_{l+2}^m \\
 \\
+\biggl[
l\kappa\Omega\partial_r\left[e^{-2\nu}(h_l+U_l)\right]
-2m\partial_r\left({\bar\omega}e^{-2\nu}U_l\right)
+\frac{ml}{r^2}\partial_r\left(r^2{\bar\omega}e^{-2\nu}\right)U_l
\biggr] Q_{l+1} Y_{l+1}^m  \\
 \\
+\biggl[
	\ba[t]{l}
	m\kappa\Omega\partial_r\left(e^{-2\nu}V_l\right)
	+2\partial_r\left({\bar\omega}e^{-2\nu}V_l\right)
	\left((l+1)Q_l^2-l Q_{l+1}^2\right) \\
	\\
	+\frac{m^2}{r^2}\partial_r\left(r^2{\bar\omega}e^{-2\nu}\right)V_l
	+\partial_r\left[\frac{1}{r}\partial_r\left(r^2{\bar\omega}e^{-2\nu}
	\right)W_l\right] \left(Q_{l+1}^2+Q_l^2-1\right) \\
	\\
	+m\kappa\Omega e^{-2\nu}\left(\frac{16\pi r(\epsilon+p)}{l(l+1)}
					-\frac{1}{r}\right)e^{2\lambda}W_l
	\biggr] Y_l^m 
	\ea \\
 \\
-\biggl[
	\ba[t]{l}
	(l+1)\kappa\Omega\partial_r\left[e^{-2\nu}(h_l+U_l)\right] \\
	\\
	+2m\partial_r\left({\bar\omega}e^{-2\nu}U_l\right)
	+\frac{m(l+1)}{r^2}\partial_r\left(r^2{\bar\omega}e^{-2\nu}\right)U_l
	\biggr] Q_l Y_{l-1}^m  
	\ea \\
 \\
+\biggl[
\partial_r\left[\frac{1}{r}\partial_r\left(r^2{\bar\omega}e^{-2\nu}\right)W_l\right]
+2(l+1)\partial_r\left({\bar\omega}e^{-2\nu}V_l\right)
\biggr] Q_{l-1}Q_l Y_{l-2}^m 
\Biggr\}
\ea
\label{om_ph_r_form2}
\ee


Let us rewrite the equations one last time using the orthogonality 
relation for spherical harmonics,
\be
\int Y_{l'}^{m'} Y_l^{m\ast} d\Omega = \delta_{ll'} \delta_{mm'},
\ee
where $d\Omega$ is the usual solid angle element on the unit 2-sphere.

\noindent
From $\Delta\omega_{\theta\varphi}=0$ we have, for all allowed $l$,
\be
0 = \ba[t]{l}
\left[ l(l+1)\kappa\Omega(h_l+U_l)-2m{\bar\omega}U_l\right] \\
 \\
+(l+1)Q_l \left[
\frac{e^{2\nu}}{r}\partial_r\left(r^2{\bar\omega}e^{-2\nu}\right)W_{l-1}
-2(l-1){\bar\omega}V_{l-1}
\right]  \\
 \\
-l Q_{l+1} \left[\frac{e^{2\nu}}{r}\partial_r
\left(r^2{\bar\omega}e^{-2\nu}\right)W_{l+1}
+2(l+2){\bar\omega}V_{l+1} \right]
\ea
\label{om_th_ph}
\ee

\noindent
From $\Delta\omega_{r\theta}=0$ we have, for all allowed $l$,
\bea
0 &=&
(l-2)Q_{l-1}Q_l \left[
-2\partial_r\left({\bar\omega}e^{-2\nu}U_{l-2}\right)
+\frac{(l-1)}{r^2}\partial_r\left(r^2{\bar\omega}e^{-2\nu}\right)U_{l-2}
\right] 
\label{om_r_th} \\
&&\nn \\
&&+ Q_l \biggl[
	\ba[t]{l}
	(l-1)\kappa\Omega\partial_r\left(e^{-2\nu}V_{l-1}\right)
	-2m\partial_r\left({\bar\omega}e^{-2\nu}V_{l-1}\right) \\
	 \\
	+\frac{m(l-1)}{r^2}\partial_r\left(r^2{\bar\omega}e^{-2\nu}\right)V_{l-1}
	+(l-1)\kappa\Omega e^{-2\nu}\left(\frac{16\pi r(\epsilon+p)}{(l-1)l}
					-\frac{1}{r}\right)e^{2\lambda}W_{l-1}
	\biggr]
	\ea \nn \\
&&\nn \\
&&+\biggl[
	\ba[t]{l}
	m\kappa\Omega\partial_r\left[e^{-2\nu}(h_l+U_l)\right]
	+2\partial_r\left({\bar\omega}e^{-2\nu}U_l\right)
	\left((l+1)Q_l^2-l Q_{l+1}^2\right) \\
	 \\
	+\frac{1}{r^2}\partial_r\left(r^2{\bar\omega}e^{-2\nu}\right)U_l
	\left[m^2+l(l+1)\left(Q_{l+1}^2+Q_l^2-1\right)\right]
	\biggr]
	\ea \nn \\
&&\nn \\
&&- Q_{l+1} \biggl[
	\ba[t]{l}
	(l+2)\kappa\Omega\partial_r\left(e^{-2\nu}V_{l+1}\right)
	+2m\partial_r\left({\bar\omega}e^{-2\nu}V_{l+1}\right) \\
	 \\
	+\frac{m(l+2)}{r^2}\partial_r\left(r^2{\bar\omega}e^{-2\nu}\right)V_{l+1}
	+(l+2)\kappa\Omega e^{-2\nu}\left(\frac{16\pi r(\epsilon+p)}{(l+1)(l+2)}
					-\frac{1}{r}\right)e^{2\lambda}W_{l+1}
	\biggr]
	\ea \nn \\
&&\nn \\
&&+(l+3)Q_{l+1}Q_{l+2} \left[
2\partial_r\left({\bar\omega}e^{-2\nu}U_{l+2}\right)
+\frac{(l+2)}{r^2}\partial_r\left(r^2{\bar\omega}e^{-2\nu}\right)U_{l+2}
\right]
\nn
\eea

\noindent
From $\Delta\omega_{\varphi r}=0$ we have, for all allowed $l$,
\bea
0 &=& Q_{l-1}Q_l \biggl[
\partial_r\left[\frac{1}{r}\partial_r
\left(r^2{\bar\omega}e^{-2\nu}\right)W_{l-2}\right]
-2(l-2)\partial_r\left({\bar\omega}e^{-2\nu}V_{l-2}\right)
\biggr] 
\label{om_ph_r} \\
&&\nn \\
&&+ Q_l \biggl[
	\ba[t]{l}
	(l-1)\kappa\Omega\partial_r\left[e^{-2\nu}(h_{l-1}+U_{l-1})\right] \\
	\\
	-2m\partial_r\left({\bar\omega}e^{-2\nu}U_{l-1}\right)
	+\frac{m(l-1)}{r^2}\partial_r\left(r^2{\bar\omega}e^{-2\nu}\right)U_{l-1}
	\biggr] 
	\ea \nn \\
&&\nn  \\
&&+\biggl[
	\ba[t]{l}
	m\kappa\Omega\partial_r\left(e^{-2\nu}V_l\right)
	+2\partial_r\left({\bar\omega}e^{-2\nu}V_l\right)
	\left((l+1)Q_l^2-l Q_{l+1}^2\right) \\
	\\
	+\frac{m^2}{r^2}\partial_r\left(r^2{\bar\omega}e^{-2\nu}\right)V_l
	+\partial_r\left[\frac{1}{r}\partial_r\left(r^2{\bar\omega}e^{-2\nu}
	\right)W_l\right] \left(Q_{l+1}^2+Q_l^2-1\right) \\
	\\
	+m\kappa\Omega e^{-2\nu}\left(\frac{16\pi r(\epsilon+p)}{l(l+1)}
					-\frac{1}{r}\right)e^{2\lambda}W_l
	\biggr]
	\ea \nn \\
&&\nn  \\
&&- Q_{l+1} \biggl[
	\ba[t]{l}
	(l+2)\kappa\Omega\partial_r\left[e^{-2\nu}(h_{l+1}+U_{l+1})\right] \\
	\\
	+2m\partial_r\left({\bar\omega}e^{-2\nu}U_{l+1}\right)
	+\frac{m(l+2)}{r^2}\partial_r\left(r^2{\bar\omega}e^{-2\nu}\right)U_{l+1}
	\biggr] 
	\ea \nn \\
&&\nn  \\
&&+ Q_{l+1}Q_{l+2} \biggl[
\partial_r\left[\frac{1}{r}\partial_r
\left(r^2{\bar\omega}e^{-2\nu}\right)W_{l+2}\right]
+2(l+3)\partial_r\left({\bar\omega}e^{-2\nu}V_{l+2}\right)
\biggr]
\nn
\eea

It is instructive to consider the newtonian limit of these equations,
\be
\om(r), \nu(r), \lambda(r), h_l(r) \rightarrow 0.
\label{newt_lim}
\ee
We have already seen that Eq. (\ref{GR:sph_V}) is the relativistic 
generalization of the newtonian mass conservation equation (\ref{N:sph_cont})
(or Eq. (\ref{eq1})), and that the other zeroth order perturbation 
equations, (\ref{GR:sph_H1}) and
(\ref{GR:sph_h_0''}), simply vanish in the newtonian limit.
Similarly, one can readily observe that the conservation of circulation 
equations have as their newtonian limit the corresponding equations 
presented in Sect. 2.2,
\[
\ba{lcl}
\mbox{Eq. (\ref{om_th_ph})} & \rightarrow  & \mbox{Eq. (\ref{eq2})}, \\
\mbox{Eq. (\ref{om_r_th})}  & \rightarrow  & \mbox{Eq. (\ref{eq4})}, \\
\mbox{Eq. (\ref{om_ph_r})} & \rightarrow  & \mbox{Eq. (\ref{eq3})}
\ea
\]
(and similarly for the other forms of these equations).

This correspondence leads us to expect the same structure for the 
relativistic modes as was found in the newtonian case: we expect to 
find a discrete set of axial- and polar-led hybrid modes with opposite
behaviour under parity.  Further, we expect a one-to-one correspondence 
between these relativistic hybrid modes and the newtonian modes, to
which the relativistic hybrids should approach in the newtonian limit.

\section{Character of the Perturbation Modes}


In deriving the components of the curl of the perturbed Euler equation 
in newtonian gravity (\ref{eq2})-(\ref{eq4}), we required no 
assumptions about the ordering of the
perturbation variables $(\delta\rho,\delta v^a)$ in powers of the
angular velocity $\Omega$.  Thus, our theorem concerning the 
character of the newtonian modes, Thm. (\ref{thm1}) applied to any 
discrete normal mode of a uniformly rotating barotropic star with
arbitrary angular velocity.  

We conjecture that the perturbations of relativistic stars obey
the same principle: If $(\xi^\alpha, h_{\alpha\beta})$ with
$\xi^\alpha\neq 0$ is a discrete normal mode of a uniformly rotating 
stellar model obeying a one-parameter equation of state, then the 
decomposition of the mode into spherical harmonics $Y_l^m$ has 
$l=m$ as the lowest contributing value of $l$, when $m\neq 0$; and 
has 0 or 1 as the lowest contributing value of $l$, when $m=0$.

However, in deriving the curl of the perturbed Euler equation for 
relativistic models we have imposed assumptions that restrict its
generality.  We have derived Eqs. (\ref{om_th_ph})-(\ref{om_ph_r})
in a form that requires a slowly rotating equilibrium model, assumes 
the ordering (\ref{ordering}) and neglects terms of order $\Omega^2$
and higher.  Under these more restrictive assumptions, the following 
theorem holds.

\begin{thm}{Let $(g_{\alpha\beta}(\Omega), T_{\alpha\beta}(\Omega))$
be a family of stationary, axisymmetric spacetimes describing a 
sequence of stellar models in uniform rotation with angular velocity
$\Omega$ and obeying a one-parameter equation of state, where
$(g_{\alpha\beta}(0), T_{\alpha\beta}(0))$ is a static spherically
symmetric spacetime describing the non-rotating model.
Let $(\xi^\alpha(\Omega), h_{\alpha\beta}(\Omega))$ with 
$\xi^\alpha\neq 0$ be a family of discrete normal modes of these 
spacetimes obeying the same one-parameter equation of state, where
$(\xi^\alpha(0), h_{\alpha\beta}(0))$ is a stationary non-radial 
perturbation of the static spherical model. 
Let $(\xi^\alpha(\Omega_0), h_{\alpha\beta}(\Omega_0)$ be a member
of this family with $\Omega_0\ll \Omega_K$, the angular velocity of a 
particle in orbit at the star's equator. Then the decomposition of 
$(\xi^\alpha(\Omega_0), h_{\alpha\beta}(\Omega_0)$ into 
spherical harmonics $Y_l^m$ (i.e., into $(l, m)$ representations of 
the rotation group about its center of mass) has $l=m$ as the lowest 
contributing value of $l$, when $m\neq 0$; and $l=1$ as the lowest 
contributing value of $l$, when $m=0$.}
\label{thm2}
\end{thm}


We designate a non-axisymmetric mode with parity $(-1)^{m+1}$ an
``axial-led hybrid'' if $\xi^\alpha$ and $h_{\alpha\beta}$
receive contributions only from
\begin{center}
axial terms with $l \ = \ m, \ m+2, \ m+4,  \ \ldots$ and \\
polar terms with $l \ = \ m+1, \ m+3, \ m+5, \ \ldots$.
\end{center}
Similarly, we designate a non-axisymmetric mode with parity $(-1)^m$ a
``polar-led hybrid'' if $\xi^\alpha$ and $h_{\alpha\beta}$
receive contributions only from
\begin{center}
polar terms with $l \ = \ m, \ m+2, \ m+4,  \ \ldots$ and \\
axial terms with $l \ = \ m+1, \ m+3, \ m+5,  \ \ldots$.
\end{center}

For the case $m=0$, we designate axisymmetric modes with 
parity\footnote{The family of modes for which $\xi^\alpha$ and 
$h_{\alpha\beta}$ receive contributions only from polar terms 
with $l=0,2,4,\ldots$ and axial terms with $l=1,3,5,\ldots$
would also have parity $+1$ and could be designated 
``polar-led hybrids.''  However, these modes require a more 
general theorem to establish their character.} 
$+1$ ``axial-led hybrids'' if $\xi^\alpha$ and $h_{\alpha\beta}$
receive contributions only from
\begin{center}
axial terms with $l \ = \ 1, \ 3, \ 5,  \ \ldots$ and \\
polar terms with $l \ = \ 2, \ 4, \ 6, \ \ldots$,\hphantom{and}
\end{center}
and we designate axisymmetric modes with parity $-1$ 
``polar-led hybrids'' if $\xi^\alpha$ and $h_{\alpha\beta}$
receive contributions only from
\begin{center}
polar terms with $l \ = \ 1, \ 3, \ 5, \ \ldots$ and \\
axial terms with $l \ = \ 2, \ 4, \ 6, \ \ldots$.\hphantom{and }
\end{center}
We prove the theorem separately for each parity class in Appendix C.


\section{Boundary Conditions and Explicit Solutions}

A physically reasonable solution $(\xi^\alpha, h_{\alpha\beta})$
to the perturbation equations (\ref{GR:sph_H1}), (\ref{GR:sph_h_0''}),
(\ref{GR:sph_V}) and (\ref{om_th_ph})-(\ref{om_ph_r}), 
must be regular everywhere in the spacetime.  Of course, the fluid 
variables $W_l(r)$, $V_l(r)$ and $U_l(r)$ (for all $l$) have support 
only inside the star, $r\in [0,R]$.  The metric functions $H_{1,l}(r)$ 
will also have support only inside the star (for all $l$), since they 
are directly proportional to $W_l(r)$ by Eq. (\ref{GR:sph_H1}).
The metric functions $h_l(r)$, on the other hand, satisfy a nontrivial 
differential equation, (\ref{GR:sph_h_0''}), in the exterior spacetime
and will, therefore, have support on the whole domain $r\in [0,\infty]$.
Let us now consider the boundary and matching conditions that our
solutions must satisfy.

At the surface of the star, $r=R$, the perturbed pressure, $\Delta p$, 
must vanish. (This is how one defines the surface of the perturbed star.)
The lagrangian change in the pressure is given by 
Eq. (\ref{Del_etc}),
\be
\Delta p = -\half \, \gamma \, p \, 
q^{\alpha\beta}\Delta g_{\alpha\beta}.
\ee
Making use of Eq. (\ref{form_Dp}) and the equilibrium equations
(\ref{tov}) and (\ref{dnudr}), we find that at $r=R$
\be
0 = \Delta p = \frac{-\ep M_0}{R^2\left(R-2M_0\right)}
\sum_l W_l(R) Y_l^m e^{i\sigma t}
\label{bc}
\ee
where $M_0=M(R)$ is the gravitational mass of the equilibrium star and
satisfies $2M_0<R$.

For the equations of state we 
consider\footnote{This restriction can be dropped if the boundary 
condition $\Delta p(r=R)=0$ is replaced by $\Delta h(r=R)=0$, with $h$ the
comoving enthalpy defined below in Sect. 4.1.1, Eq. (\ref{enth_def}).}
the energy density $\epsilon(r)$ either goes to a constant or vanishes 
at the surface of the star in the manner,
\[
\ep(r) \sim \left(1-\frac{r}{R}\right)^k
\]
(for some constant $k$). 
If $\epsilon(R)\neq 0$, then Eq. (\ref{bc}) requires that 
$W_l(R)=0$ for all $l$.
Otherwise, one finds from Eq. (\ref{GR:sph_V}) that 
$V_l(r)$ will diverge at the surface unless $W_l(R)=0$.  
Thus, the boundary condition,
\be
W_l(R)=0 \ \ \ \ \mbox{(all $l$)}
\label{bc_on_W}
\ee
must be satisfied at the surface of the star.  (By Eq. 
(\ref{GR:sph_H1}), this also implies that $H_{1,l}(r)$ vanishes 
at the surface of the star).

In the exterior vacuum spacetime, $r>R$, we have only to satisfy the 
single 
equation\footnote{This equation was first written down by Regge and 
Wheeler \cite{rw57} in the context of Schwarzschild perturbations.}
(\ref{GR:sph_h_0''}) for all $l$, which becomes
\be
h_l^{''} + \left[ \frac{(2-l^2-l)}{r^2}e^{2\lambda} 
	- \frac{2}{r^2} \right] h_l  = 0,
\ee
or
\be
(1-\frac{2M_0}{r}) h_l^{''} - \left[ \frac{l(l+1)}{r^2} 
	- \frac{4M_0}{r^3} \right] h_l  = 0,
\label{h_l''_ext}
\ee
where we have used $e^{-2\lambda}=(1-2M_0/r)$ for $r>R$.

Since this exterior equation does not couple $h_l(r)$ having different 
values of $l$, we can find its solution explicitly.  The solution that 
is regular at spatial infinity can be written
\be
h_l(r) = \sum_{s=0}^\infty {\hat h}_{l,s} 
\left(\frac{R}{r}\right)^{l+s}.
\label{h_ext}
\ee
If we substitute this series expansion into Eq. (\ref{h_l''_ext}), 
we find the following recursion relation for the expansion 
coefficients,
\be
{\hat h}_{l,s} = \left(\frac{2M_0}{R}\right) 
\frac{(l+s-2)(l+s+1)}{s(2l+s+1)} {\hat h}_{l,s-1}
\label{h_ext_soln}
\ee
with ${\hat h}_{l,0}$ an arbitrary normalization constant.  We, 
therefore, have the full solution to zeroth order in $\Omega$ of the 
perturbation equations in the exterior spacetime.

This exterior solution must be matched at the surface of the star to 
the interior solution for $h_l(r)$.  One requires that the solutions
be continuous at the surface,
\be
\lim_{\varepsilon\rightarrow 0}  \left[
h_l(R-\varepsilon) - h_l(R+\varepsilon) \right] = 0,
\label{cont_cond}
\ee
for all $l$, and that the Wronskian of the interior and exterior 
solutions vanish at $r=R$, i.e. that
\be
\lim_{\varepsilon\rightarrow 0} \left[
h_l(R-\varepsilon) h'_l(R+\varepsilon) 
- h'_l(R-\varepsilon) h_l(R+\varepsilon)
\right] = 0,
\label{match_cond}
\ee
for all $l$.

Thus, in solving the perturbation equations to zeroth order in $\Omega$ 
we need only work in the interior of the star (as in the newtonian case).  
Inside the star, the perturbation $(\xi^\alpha, h_{\alpha\beta})$ must 
satisfy the full set of coupled equations (\ref{GR:sph_H1}), 
(\ref{GR:sph_h_0''}), (\ref{GR:sph_V}) and (\ref{om_th_ph})-(\ref{om_ph_r}) 
for all $l$, subject to the boundary and matching conditions 
(\ref{bc_on_W}), (\ref{cont_cond}) and (\ref{match_cond}).

Finally, we note that since we are working in linearized perturbation
theory there is a scale invariance to the equations.  If 
$(\xi^\alpha, h_{\alpha\beta})$ is a solution to the perturbation
equations then $(K\xi^\alpha, Kh_{\alpha\beta})$ is also a 
solution, for constant $K$.  Thus, in order to find a particular
mode of oscillation we must impose some normalization condition in 
addition to the boundary and matching conditions just discussed.
We choose the condition that 
\be
\begin{array}{ll}
U_m(r=R) = 1 & \mbox{for axial-hybrids, or that} \\
U_{m+1}(r=R) = 1 & \mbox{for polar-hybrids.} 
\end{array}
\label{norm_cond}
\ee

\subsection{The purely axial solutions}

In Sect. 3.3 we saw that if a mode of a slowly rotating star has 
a stationary non-radial perturbation as its spherical limit, then 
it is generically a hybrid mode with mixed axial and polar angular 
behaviour.  However, we have also seen in Sect. 2.3.1 that newtonian stars 
retain a vestigial set of purely axial modes  - the so-called 
``classical r-modes'' - whose angular behaviour is a purely axial 
harmonic having $l=m$.  Let us now address the question of whether 
or not such r-mode solutions exist in the relativistic models.  

For relativistic stars, Kojima \cite{k98} has recently derived an 
equation governing purely axial perturbations to lowest order in the 
star's angular velocity.  Based on this equation, he has argued for the 
existence of a continuous spectrum of modes, and his argument has been 
made precise in a recent paper of Beyer and Kokkotas \cite{bk99}.
Beyer and Kokkotas, however, also point out that the continuous spectrum 
they find may be an artifact of the vanishing of the imaginary part of the 
frequency in the slow rotation limit.  (Or, more broadly, it may be
an artifact of the slow rotation approximation.)

In addition, Kojima and Hosonuma \cite{kh99} have studied the mixing of
axial and polar perturbations to order $\Omega^2$ in rotating relativistic 
stars, again finding a continuous mode spectrum.  Their calculation uses 
the Cowling approximation (which ignores all metric perturbations) and 
assumes an ordering of the perturbation variables in powers of $\Omega$ 
which forbids the mixing of axial and polar terms at zeroth order.

In contrast to these results, we do not find a continuous spectrum of 
purely axial modes for isentropic stars, but rather, a discrete spectrum of 
axial-led hybrids.  Indeed, we show below that there are only two purely 
axial modes (and their complex conjugates) in isentropic rotating 
relativistic stars.  Both are discrete stationary modes having spherical 
harmonic index $l=1$.  (One mode has $m=0$ and the other has $m=1$.)
These modes are the generalization to slowly rotating stars of the $l=1$ 
axial modes discussed by Campolattaro and Thorne \cite{ct70} and the 
generalization to relativistic stars of the $l=1$ axial modes discussed
in Sect. 2.3.1. In particular, none of the newtonian r-modes having 
$l=m\geq 2$ retain their purely axial character in isentropic relativistic 
stars.

As in Sect 2.3.1, let us write down the equations governing an axial mode 
belonging to a pure spherical harmonic of index $l$.  In other words, 
let us assume that $h_l(r)$ and $U_l(r)$ (for some particular value of $l$)
are the only non-vanishing 
coefficients in the spherical harmonic expansions of the lagrangian 
displacement (\ref{xi_exp}) and the perturbed metric (\ref{h_components}).  
The set of equations that have to be satisfied are the zeroth order 
(spherical) equations (\ref{GR:sph_H1}), (\ref{GR:sph_h_0''}) and 
(\ref{GR:sph_V}); the order $\Omega$ conservation of circulation
equations (\ref{om_th_ph})-(\ref{om_ph_r}) and the matching conditions
at the surface of the star (\ref{cont_cond}) and (\ref{match_cond}).

With $h_l(r)$ and $U_l(r)$ the only non-vanishing perturbation variables,
Eqs. (\ref{GR:sph_H1}) and (\ref{GR:sph_V}) vanish identically, while Eq.
(\ref{GR:sph_h_0''}) remains unchanged,
\be
h_l^{''} - (\nu'+\lambda') h_l' 
+ \left[ \frac{(2-l^2-l)}{r^2}e^{2\lambda} 
- \frac{2}{r}(\nu'+\lambda') - \frac{2}{r^2} \right] h_l 
= \frac{4}{r}(\nu'+\lambda') U_l.
\label{koji1}
\ee
Eq. (\ref{om_th_ph}) becomes,
\be
0 = \left[ l(l+1)\kappa\Omega(h_l+U_l)-2m{\bar\omega}U_l\right],
\label{koji2}
\ee
and Eq. (\ref{om_r_th}) with $l\rightarrow l+2$, $l\rightarrow l$ and
$l\rightarrow l-2$ gives the 
equations\footnote{Alternatively, one can get these equations from the 
coefficients of $Y_l^m$ and $Y_{l\pm 2}^m$ in  Eq. (\ref{om_r_th_form2}).}
\begin{eqnarray}
0 &=& 	lQ_{l+1}Q_{l+2} \left[
	-2\partial_r\left({\bar\omega}e^{-2\nu}U_l\right)
	+\frac{(l+1)}{r^2}\partial_r\left(r^2{\bar\omega}e^{-2\nu}\right)U_l
	\right], \label{koji3} \\
0 &=&	\ba[t]{l}
	m\kappa\Omega\partial_r\left[e^{-2\nu}(h_l+U_l)\right] \\
	\\
	+ 2\partial_r\left({\bar\omega}e^{-2\nu}U_l\right)
	\left((l+1)Q_l^2-l Q_{l+1}^2\right) \\
	 \\
	+\frac{1}{r^2}\partial_r\left(r^2{\bar\omega}e^{-2\nu}\right)U_l
	\left[m^2+l(l+1)\left(Q_{l+1}^2+Q_l^2-1\right)\right],
	\ea \label{koji4} \\
0 &=& (l+1)Q_{l-1}Q_l \left[
	2\partial_r\left({\bar\omega}e^{-2\nu}U_l\right)
	+\frac{l}{r^2}\partial_r\left(r^2{\bar\omega}e^{-2\nu}\right)U_l
	\right], \label{koji5}
\end{eqnarray}
respectively.  Recall that we need only work with two of the three 
equations (\ref{om_th_ph})-(\ref{om_ph_r}) since they are linearly 
dependent as a result of Eq. (\ref{not_ind}).  

By Thm. (\ref{thm2}), we know that a non-axisymmetric $(m>0)$ mode must
have $l=m$ as its lowest value of $l$ and that an axisymmetric $(m=0)$ 
mode must have $l=1$ as its lowest value of $l$.  (Hence, in the present 
context of pure a spherical harmonic these are also the {\it only} allowed 
values of $l$.) We consider each case separately.\\

\noindent
\textit{The case $m=0$ and $l=1$.}\\

\noindent
We seek a solution to Eqs. (\ref{koji1})-(\ref{koji5}) with $m=0$ and 
$l=1$. The fluid perturbation turns out to have a particularly simple form 
in this case.  From the definition of $Q_l$, Eq. (\ref{Q_l}), we find that
\be
\ba[t]{ccccc}
Q^2_{l-1} = 0, & & Q^2_l = \ds{\frac{1}{3}} 
&\mbox{and} &  Q^2_{l+1} = \ds{\frac{4}{15}}.
\ea
\ee
These imply that Eq. (\ref{koji5}) is trivially satisfied, while
Eqs. (\ref{koji3}) and (\ref{koji4}) both become,
\be
0 = \partial_r\left({\bar\omega}e^{-2\nu}U_l\right)
-\frac{1}{r^2}\partial_r\left(r^2{\bar\omega}e^{-2\nu}\right)U_l
\ee
with solution
\be
U_l(r)=K_1r^2,
\label{rad_dep}
\ee
for some constant $K_1$.

The exterior solution (\ref{h_ext}) also takes on a particularly simple
form in the present case.  When $l=1$, the recursion relation 
(\ref{h_ext_soln}) terminates and the exterior solution simply becomes
\be
h_l(r) = \frac{K_2}{r},
\label{h_ext_l=1}
\ee
for some constant $K_2$. 

Eq. (\ref{koji2}) can be satisfied for $m=0$ if either 
$\kappa=0$ or $h_l\equiv -U_l = -K_1r^2$.  
In the latter case, however, the matching conditions (\ref{cont_cond}) 
and (\ref{match_cond}) at the surface of the star would require that
$K_2=-K_1R^3= 0$. Thus, a non-trivial solution will exist if and only if
\be
\kappa=0.
\ee
(Since $m=0$, this implies that the frequency $\sigma$ also vanishes.)

Finally, we consider Eq. (\ref{koji1}).  Defining $f(r)=h_l(r)/r^2$, 
it is not difficult to show that
\be
\frac{1}{r^2} \Biggl\{
h_l^{''} - (\nu'+\lambda') h_l' 
\ba[t]{l}
- \ds{\left[\frac{2}{r}(\nu'+\lambda') + \frac{2}{r^2} \right] h_l }
\Biggr\} = \\
\\
\ds{\frac{e^{(\nu+\lambda)}}{r^4}\left( r^4 e^{-(\nu+\lambda)} f' \right)'
- \frac{4}{r} (\nu'+\lambda') f.}
\label{hartle_op}
\ea
\ee
But this is simply the operator appearing in Hartle's \cite{h67} 
equation, Eq. (\ref{hartle})!  Indeed, 
defining\footnote{The exact form of these expressions assumes a
standard definition of the spherical harmonics $Y_l^m$. (See, e.g. 
Jackson \cite{j75}, p.99.) In particular, we are using 
$Y_1^0 = \sqrt{3/4\pi} \cos\theta$.}
\bea
\hat\Omega &=& - i\left(\frac{3}{4\pi}\right)^\half K_1 \\
\hat J &=& \frac{i}{2}\left(\frac{3}{4\pi}\right)^\half K_2 \\
\hat\omega(r) &=& i\left(\frac{3}{4\pi}\right)^\half f(r) \ \ 
\mbox{$= \ \ds{ i\left(\frac{3}{4\pi}\right)^\half \frac{h_l(r)}{r^2}}$} \\
\tilde\om(r) &=& \hat\Omega - \hat\omega(r)
\eea
Eq. (\ref{koji1}) becomes precisely Hartle's equation - here
governing a change $\hat\Omega$ in the uniform angular velocity 
$\Omega$ of the star and the associated changes $\hat J$ and 
$\hat\om$ in the angular momentum $J$ of the spacetime and the 
frame-dragging metric variable $\om$, respectively.

Just as in the newtonian case (Sect 2.3.1), we find that the only 
allowed purely axial mode with $m=0$ is a stationary perturbation 
in the fluid velocity of the form,
\be
\delta u^\alpha = e^{-\nu}\hat{\Omega}\varphi^\alpha.
\ee
Associated with this fluid perturbation is a change in the frame-dragging 
metric variable $\om\rightarrow\om+\hat\om$ given by a solution to Eq. 
(\ref{hartle}) for $\tilde\om=\hat\Omega-\hat\om$. Observe that this change, 
$\hat\Omega$, in the angular velocity of the star is uniform - a fact 
which follows from the radial dependence of the perturbed velocity 
field, Eq. (\ref{rad_dep}).  In a star that is already rotating uniformly, 
a perturbation inducing differential rotation would violate conservation of
circulation.  Thus, the equation that enforces conservation of circulation
 - the curl of the perturbed Euler equation - with components 
(\ref{om_th_ph})-(\ref{om_ph_r}), restricts the radial behaviour of
$\delta u^\alpha$.  In the non-rotating star, however, such a restriction
does not apply (See Sect. 3.1.4 and Campolattaro and Thorne \cite{ct70}).
Since Eqs. (\ref{om_th_ph})-(\ref{om_ph_r}) are of order $\Omega$, they
vanish in the spherical limit and impose no restriction on the
form of $U_l(r)$.  In freely specifying $U_l(r)$, in this case, one 
is specifying the form of differential rotation about the $z$-axis.
The corresponding metric perturbation (the frame dragging term) is 
then determined by Hartle's equation in the form of Eq. (\ref{koji1}).\\

\noindent
\textit{The case $l=m>0$.}\\

\noindent
We now seek solutions to Eqs. (\ref{koji1})-(\ref{koji5}) with $l=m$.
The fluid perturbation, again, turns out to have a particularly simple 
form.  In this case, Eq. (\ref{koji2}) becomes,
\be
(m+1)\kappa\Omega(h_m + U_m) = 2\bom U_m.
\label{koji6}
\ee
From the definition of $Q_l$, Eq. (\ref{Q_l}), we find that
\be
\ba[t]{ccc}
Q^2_m = 0, & \mbox{and} & Q^2_{m+1} = \ds{\frac{1}{(2m+3)}}.
\ea
\ee
These, again, imply that Eq. (\ref{koji5}) is trivially satisfied, 
while Eqs. (\ref{koji3}) and (\ref{koji4}) both become,
\be
0 = 2 \partial_r\left({\bar\omega}e^{-2\nu}U_m\right)
-\frac{(m+1)}{r^2}\partial_r\left(r^2{\bar\omega}e^{-2\nu}\right)U_m
\ee
where we have used Eq. (\ref{koji6}) to substitute for 
$\kappa\Omega(h_m + U_m)$ in Eq. (\ref{koji4}).

\noindent
This last equation has solution
\be
U_m(r) = K_1 r^{m+1} \bom^{\half(m-1)} e^{-(m-1)\nu}
\label{l=m_formU}
\ee
for some constant $K_1$.
Then, Eq. (\ref{koji6}) implies,
\be
h_m(r) = K_1 \left[\frac{2\bom}{(m+1)\kappa\Omega}-1\right]
r^{m+1}\bom^{\half(m-1)} e^{-(m-1)\nu}
\label{l=m_formh}
\ee

\noindent
Finally, consider Eq. (\ref{koji1}).  Letting $f=\bom^\half e^{-\nu}$,
and substituting for $U_m(r)$ and $h_m(r)$ the expressions
(\ref{l=m_formU}) and (\ref{l=m_formh}), respectively, one can 
show that,
\bea
0 &=& \frac{1}{U_m(r)} 
\Biggl\{ \ba[t]{l}
	\ds{h_m^{''} - (\nu'+\lambda') h_m' 
	- \frac{4}{r}(\nu'+\lambda') U_m} \\
	\\
	+ \ds{\left[ \frac{(2-m^2-m)}{r^2}e^{2\lambda} 
	- \frac{2}{r}(\nu'+\lambda') - \frac{2}{r^2} \right] h_m }
	\Biggr\}
	\ea \nn \\
&& \nn \\
&=& (m-1) \Biggl\{ \ba[t]{l}
	\ds{\frac{4\bom'}{(m+1)\kappa\Omega}
	\left(\frac{f'}{f}+\frac{1}{r}\right)} \\
	\\
	+ \ds{\left[\frac{2\bom}{(m+1)\kappa\Omega}-1\right]} 
	\Biggl[ \ds{\frac{f''}{f} + (m-2)\left(\frac{f'}{f}\right)^2
	+\frac{2(m+1)}{r}\left(\frac{f'}{f}\right)} \\
	\\
	- \ds{(\nu'+\lambda')\left(\frac{f'}{f}+\frac{1}{r}\right)
	+ \frac{(m+2)}{r^2}\left(1-e^{2\lambda}\right)} \Biggr]
	\Biggr\}
	\ea
\eea
If $m=1$ this equation is obviously satisfied.  However, with some 
effort, one can also show that it is satisfied only if $m=1$.
For $m>1$, therefore, the system of Eqs. (\ref{koji1})-(\ref{koji5})
is overdetermined and no solutions exist.  {\it The purely axial modes
with $l=m\geq 2$ do not exist in isentropic relativistic stars.}

This result contradicts the claims by Kojima \cite{k98} and Kojima
and Hosonuma \cite{kh99} that a continuous spectrum of purely axial 
modes exists in isentropic relativistic stars.

Kojima \cite{k98} bases his conclusion on Eqs. (\ref{koji1}) and
(\ref{koji2}) (which he derives using slightly different notation)
and he does not distinguish between the isentropic and non-isentropic
cases.  In non-isentropic stars, Eqs. (\ref{koji1}) and (\ref{koji2})
are, indeed, the only 
equations\footnote{In isentropic stars Eqs. 
(\ref{koji3})-(\ref{koji5}) have the form shown and involve only
$h_l$ and $U_l$.  In non-isentropic stars, however, these equations are 
modified in such a way that they give a coupling to order $\Omega$
polar variables such as the perturbed pressure and density.}
governing axial perturbations to order $\Omega$.  Kojima \cite{k98} 
derives the striking result that these equations can be combined into 
a single ``master equation'' whose highest derivative term has a
frequency-dependent coefficient.  The vanishing of this coefficient at 
a particular radius when the frequency is real gives rise to the
continuous spectrum.  However, because the frequency is complex (with
imaginary part higher order in $\Omega$) we conjecture that a higher-order
calculation for non-isentropic models will find only a discrete set of 
modes - those modes that generalize to relativistic stars the newtonian 
r-modes first studied by Papalouizou and Pringle \cite{pp78}
(see also Provost et al. \cite{pea81}, Saio \cite{s82} and 
Smeyers and Martens \cite{sm83}).


For the case of isentropic stars, on the other hand, it is no longer true
that Eqs. (\ref{koji1}) and (\ref{koji2}) are the only equations to be
satisfied.  As we have just seen, these are joined by Eqs. 
(\ref{koji3})-(\ref{koji5}) and comprise an overdetermined system for
modes with $m\geq 2$.  The result is not a continuous spectrum of 
purely axial modes, but no such modes at all!

Kojima and Hosonuma \cite{kh99} perform a higher order calculation and
they do distinguish between what we are calling the isentropic and
non-isentropic cases. (They use the terms ``barotropic'' and 
``non-barotropic''.)  However, they also make the simplifying 
assumption that the metric perturbation is small and can be ignored 
altogether (the Cowling approximation).  Based on this assumption they,
again, claim to find a continuous spectrum of pure r-modes.  However, 
we have seen in the isentropic case that the full perturbation equations 
(which include contributions from the perturbed metric) forbid the 
existence of such r-modes.  Again, in the non-isentropic case, we expect 
that a higher order calculation that includes the metric perturbation will 
find only a discrete r-mode spectrum.

Finally, let us return to the solution (\ref{l=m_formU})-(\ref{l=m_formh})
with $l=m=1$,
\bea
U_m(r) &=& K_1r^2 \\
h_m(r) &=& K_1\left(\frac{\bom}{\kappa\Omega}-1\right)r^2.
\eea
This solution must satisfy the matching conditions (\ref{cont_cond}) and 
(\ref{match_cond}) to the exterior solution with $l=1$, 
Eq. (\ref{h_ext_l=1}),
\be
h_m(r) = \frac{K_2}{r}.
\ee
Eq. (\ref{cont_cond}) gives,
\be
0 = K_1R^3\left(\frac{\bom(R)}{\kappa\Omega}-1\right)-K_2,
\ee
while (\ref{match_cond}) gives,
\be
0 = K_1R^3 \left[2\left(\frac{\bom(R)}{\kappa\Omega}-1\right)
+ R\frac{\bom'(R)}{\kappa\Omega}
\right] +K_2,
\ee
Adding these equations, we find,
\be
0 = \bom(R) + \frac{1}{3}R\bom'(R) - \kappa\Omega
\ee
and using Eq. (\ref{hartle_ext_soln}) to substitute for $\bom(R)$ and
$\bom'(R)$ we find,
\be
0 = \Omega - \kappa\Omega
\ee
or
\be
\kappa = 1.
\ee
Thus, the purely axial solution with $l=m=1$ has a co-rotating frequency 
equal to the angular velocity of the star.  Since 
$\kappa\Omega=\sigma+m\Omega$, this implies that $\sigma=0$, i.e., that 
the mode is stationary as seen by an inertial observer.  This mode is the
generalization to slowly rotating stars of the $l=m=1$ mode found in
spherical stars by Campolattaro and Thorne \cite{ct70}, and the 
generalization to relativistic stars of the $l=m=1$ mode found in
Sect 2.3.1.  It represents uniform rotation about an axis perpendicular 
to the rotational axis of the star.  We note, again, that only uniform 
rotation is an acceptable perturbation, as was the case with the $m=0$, 
$l=1$ mode.

In newtonian isentropic stars there remained a large set of purely
axial modes with $l=m$; the $l=m=2$ mode being the one expected to
dominate the gravitational wave-driven spin-down of a hot, young
neutron star.  In relativistic stars, however, we see that all such
pure r-modes with $l=m\geq 2$ are forbidden by the perturbation
equations, and instead must be replaced by axial-led hybrids. 
Let us then turn to the problem of finding these important hybrid
modes.

\subsection{Relativistic corrections to the ``classical'' r-modes
in uniform density stars}

Before turning to the general problem of numerically solving for the
hybrid modes of fully relativistic stars, let us look directly for the
post-newtonian corrections to the $l=m$ newtonian r-modes.  The 
equilibrium structure of a slowly rotating star with uniform density 
is particularly simple (see Chandrasekhar and Miller \cite{cm74}) and
lends itself readily to such a post-newtonian analysis.


For a spherically symmetric star with constant density,
\be
\ep(r)=\frac{3M_0}{4\pi R^3},
\label{GR:equil_ep}
\ee
the equilibrium equations (\ref{GR:eos})-(\ref{GR:mass_def}) have 
the well-known exact solution inside the star ($r\leq R$),
\bea
p(r) &=& \ep\left\{
\frac{\left(1-\tmr\right)^\half 
- \left[1-\tmr\left(\rx\right)^2\right]^\half}
{3\left[1-\tmr\left(\rx\right)^2\right]^\half 
- \left(1-\tmr\right)^\half}
\right\}
\label{GR:equil_p} \\
M(r) &=& M_0 \left(\frac{r}{R}\right)^3 \\
e^{2\nu(r)} &=& \left\{
\frac{3}{2}\left[1-\tmr\left(\rx\right)^2\right]^\half 
- \half\left(1-\tmr\right)^\half
\right\}^{2} \\
e^{-2\lambda(r)} &=& 1 - \frac{2M_0}{R}\left(\frac{r}{R}\right)^2
\eea
where $M_0$ is the gravitational mass of the star and $R$ is its radius.
(See, e.g., Wald \cite{wald} Ch. 6.)

To find the equilibrium solution corresponding to the slowly rotating
star, we must also solve Hartle's \cite{h67} equation (\ref{hartle}),
\be
0 = r^2\bom'' + [4-r(\nu'+\lambda')]r\bom'-4r(\nu'+\lambda')\bom
\label{hartle2}
\ee 
where we may use the spherical solution to write 
\bea
r(\nu'+\lambda')
 &=& 4\pi r^2(\ep+p)e^{2\lambda} \\
 &=& \frac{3\left(\tmr\right)\left(\rx\right)^2 
\left(1-\tmr\right)^\half}
{\left[1-\tmr\left(\rx\right)^2\right]\left\{
3\left[1-\tmr\left(\rx\right)^2\right]^\half 
- \left(1-\tmr\right)^\half
\right\}}. \nn
\eea

To simplify the problem, we expand our equilibrium solution in powers
of $(2M_0/R)$ and work only to linear 
order\footnote{This expansion will give us the first post-newtonian
(1PN) corrections to the $l=m$ newtonian r-modes.}. 
We will need the expressions,
\be
r(\nu'+\lambda') = \frac{3}{2}\left(\rx\right)^2\left(\tmr\right)
+ O\left(\tmr\right)^2
\label{GR:equil_nulam}
\ee
and
\be
e^{-2\nu} = 1+\left[\frac{3}{2}
-\half\left(\rx\right)^2\right]\left(\tmr\right)
+ O\left(\tmr\right)^2
\label{GR:equil_gtt}
\ee

Since we are also working to linear order in the star's angular 
velocity, we may set $\Omega=1$ without loss of generality. 
We write,
\be
\bar\omega =
\sum^\infty_{i=0} \omega_i \left(\frac{r}{R}\right)^{2i}
\ee
and solve Eq. (\ref{hartle2}) subject to the following boundary 
condition (Hartle \cite{h67}) at the surface of the star,
\be
1 = \Omega = \left[\bom + \frac{1}{3}R\ \bom'\right]_{r=R}
\label{hartleBC}
\ee
To order $(2M_0/R)$ the solution is,
\be
\bom(r) = 1 - \left(1-\frac{3r^2}{5R^2}\right)\left(\tmr\right)
+ O\left(\tmr\right)^2.
\label{GR:equil_bom}
\ee

With this explicit equilibrium solution in hand, we now consider the 
perturbation equations.  We are required to solve Eqs. (\ref{GR:sph_H1}), 
(\ref{GR:sph_h_0''}), (\ref{GR:sph_V}) and (\ref{om_th_ph})-(\ref{om_ph_r})
subject to the boundary, matching and normalization conditions
(\ref{bc_on_W}), (\ref{cont_cond}), (\ref{match_cond}) and (\ref{norm_cond}). 
We seek the post-newtonian corrections to the $l=m$ newtonian r-modes
discussed in Sect. 2.3.1, Eqs. (\ref{N:rmode_freq}) and (\ref{N:rmode_func}),
\bea
\kappa &=& \frac{2}{(m+1)} \\ \nn \\
U_m &=& \left(\frac{r}{R}\right)^{m+1}.
\eea
Therefore, let us make the following ansatz for our perturbed solution
inside the star,
\bea
\kappa &=& \frac{2}{(m+1)}\left[1+\kappa_1\left(\tmr\right)
+ O\left(\tmr\right)^2 \right]
\label{ans:freq} \\ \nn \\
U_m(r) &=& \left(\rx\right)^{m+1}\left[
1+u_{m,0}\left(1-\frac{r^2}{R^2}\right)\left(\tmr\right)
+ O\left(\tmr\right)^2\right]
\label{ans:u_m} \\ \nn \\
h_m(r) &=& \left(\rx\right)^{m+1}
\left[h_{m,0} + h_{m,1}\left(\rx\right)^2\right]\left(\tmr\right)
+ O\left(\tmr\right)^2
\label{ans:v_m+1} \\ \nn \\
W_{m+1}(r) &=& w_{m,0}\left(\rx\right)^{m+1}
\left(1-\frac{r^2}{R^2}\right)\left(\tmr\right) + O\left(\tmr\right)^2
\label{ans:h_m} \\ \nn \\
V_{m+1}(r) &=& \left(\rx\right)^{m+1}
\left[v_{m,0} + v_{m,1}\left(\rx\right)^2\right]\left(\tmr\right)
+ O\left(\tmr\right)^2
\label{ans:w_m+1} \\ \nn \\
U_{m+2}(r) &=& u_{m+2,0}\left(\rx\right)^{m+3}\left(\tmr\right)
+ O\left(\tmr\right)^2
\label{ans:u_m+2}
\eea
where we have chosen the form of $U_m(r)$ so as to automatically
satisfy the normalization condition (\ref{norm_cond}) and we have 
chosen the form of $W_{m+1}(r)$ so as to automatically satisfy the boundary
condition (\ref{bc_on_W}).  Note that we have assumed that 
$h_l$, $V_{l'}$, $W_{l'}$ and $U_{l''}$ are of order $(2M_0/R)^2$ 
or higher for all $l>m$, $l'>m+1$ and $l''>m+2$.  We will justify 
this ansatz by finding a self-consistent solution to the perturbation 
equations.

Observe that the exterior solution (\ref{h_ext}) for $h_m(r)$ 
already has a natural expansion in powers of $(2M_0/R)$ as a 
result of the recursion relation (\ref{h_ext_soln}),
\be
h_m(r) = {\hat h}_{m,0} \left(\frac{R}{r}\right)^m\left(\tmr\right)
+ O\left(\tmr\right)^2.
\ee
The normalization constant, ${\hat h}_{m,0}$, is determined by the 
matching condition (\ref{cont_cond}),
\be
{\hat h}_{m,0} = h_{m,0} + h_{m,1} 
\ee
while (\ref{match_cond}) imposes the following condition on the interior
solution,
\be
0 = {\hat h}_{m,0} \biggl\{-m(h_{m,0}+h_{m,1}) 
- \left[ (m+1)h_{m,0} + (m+3)h_{m,1} \right]
\biggr\}
\ee
or,
\be
0 = (2m+1)h_{m,0} + (2m+3)h_{m,1}
\label{GR:ex_h_BC}
\ee

Turning now to the perturbation equations, we will write Eqs. 
(\ref{GR:sph_h_0''}), (\ref{GR:sph_V}), (\ref{om_th_ph}) and
(\ref{om_r_th}) to order $(2M_0/R)$.  
Eq. (\ref{GR:sph_H1}) merely expresses $H_{1,l}(r)$ in terms of
$W_l(r)$, and we need not work with Eq. (\ref{om_ph_r}) since
(\ref{om_th_ph})-(\ref{om_ph_r}) are related by Eq. (\ref{not_ind}).
Hence, a complete set of perturbation equations, accurate to first order 
in $(2M_0/R)$, is as follows.

Eq. (\ref{GR:sph_V}) with $l=m+1$ is,
\be
0 = (m+1)(m+2)V_{m+1} - \left(rW_{m+1}\right)'
\label{GR:ex1}
\ee

Eq. (\ref{GR:sph_h_0''}) with $l=m$ is,
\be
0 = r^2 h_m'' - m(m+1)h_m-4r(\nu'+\lambda')U_m
\label{GR:ex2}
\ee

Eq. (\ref{om_th_ph}) with $l=m$ is,
\be
0 = \ba[t]{l}
m(m+1)\kappa h_m + m\left[(m+1)\kappa-2\bom\right]U_m \\
 \\
-2mQ_{m+1} \left[W_{m+1}+(m+2)V_{m+1}\right]
\ea
\label{GR:ex3}
\ee

Eq. (\ref{om_th_ph}) with $l=m+2$ is,
\be
0 = \ba[t]{l}
\left[(m+2)(m+3)\kappa-2m\right]U_{m+2} \\
 \\
+2(m+3)Q_{m+2}\left[W_{m+1}-(m+1)V_{m+1}\right]
\ea
\label{GR:ex4}
\ee

Eq. (\ref{om_r_th}) with $l=m$ is,
\be
0 = \ba[t]{l}
m\kappa rh'_m + m\kappa e^{2\nu}r\left(e^{-2\nu}U_m\right)' 
-\ds{\frac{2me^{2\nu}r}{(2m+3)}\left(\bom e^{-2\nu}U_m\right)' }\\
 \\
-\ds{\frac{m(m+2)e^{2\nu}}{(2m+3)r}}
\left(r^2\bom e^{-2\nu}\right)'U_m 
+2(m+3)Q_{m+1}Q_{m+2}\left[rU'_{m+2}+(m+2)U_{m+2}\right]\\
 \\
-Q_{m+1}\left\{[(m+2)\kappa+2m]rV'_{m+1}
+2m(m+2)V_{m+1}-(m+2)\kappa W_{m+1}\right\} 
\ea
\label{GR:ex5}
\ee
where we have used the fact that $Q^2_m\equiv 0$ and
$Q^2_{m+1}=1/(2m+3)$, and we have set $\Omega=1$.
One can readily verify that all other non-trivial equations
are satisfied by a solution to these; for example,
Eq. (\ref{om_r_th}) with $l=m+2$.  

We now substitute for the equilibrium quantities $(\nu'+\lambda')$,
$e^{-2\nu}$ and $\bom$ using Eqs. (\ref{GR:equil_nulam}), 
(\ref{GR:equil_gtt}) and (\ref{GR:equil_bom}), respectively.  
We also substitute for 
the perturbation variables $\kappa$, $U_m$, $h_m$, $W_{m+1}$, 
$V_{m+1}$ and $U_{m+2}$ using our ansatz, Eqs. (\ref{ans:freq}) 
to (\ref{ans:u_m+2}). 
Collecting powers of $(2M_0/R)$, one finds that the 
zeroth order terms vanish identically as a consequence of the 
newtonian solution.  At order $(2M_0/R)$ we find the following 
set of equations.

Eq. (\ref{GR:ex1}) becomes,
\be
0 = \ba[t]{l}
\ds{(m+2)\biggl[(m+1)v_{m+1,0}-w_{m+1,0}\biggr]\left(\rx\right)^{m+1}} \\
 \\
+\ds{\biggl[(m+1)(m+2)v_{m+1,1}+(m+4)w_{m+1,0}\biggr]\left(\rx\right)^{m+3}}
\ea
\ee

Eq. (\ref{GR:ex2}) becomes,
\be
0 = \biggl[ 2(2m+3)h_{m,1}-6\biggr]\left(\rx\right)^{m+3}
\ee

Eq. (\ref{GR:ex3}) becomes,
\be
0 = \ba[t]{l}
\ds{2m\Biggl\{ 
h_{m,0}+\kappa_1+1-Q_{m+1}\biggl[w_{m+1,0}+(m+2)v_{m+1,0}\biggr]
\Biggr\}\left(\rx\right)^{m+1}} \\
 \\
+\ds{2m\Biggl\{
h_{m,1}-\frac{3}{5}+Q_{m+1}\biggl[w_{m+1,0}-(m+2)v_{m+1,1}\biggr]
\Biggr\}\left(\rx\right)^{m+3}}
\ea
\ee

Eq. (\ref{GR:ex4}) becomes,
\be
0 = \ba[t]{l}
\ds{2(m+3)Q_{m+2}\biggl[w_{m+1,0}-(m+1)v_{m+1,0}\biggr]
\left(\rx\right)^{m+1}} \\ 
\\
+\ds{\Biggl\{
\frac{4(2m+3)}{(m+1)}u_{m+2,0}
-2(m+3)Q_{m+2}\biggl[w_{m+1,0}+(m+1)v_{m+1,1}\biggr]
\Biggr\}\left(\rx\right)^{m+3}}
\ea
\ee

Eq. (\ref{GR:ex5}) becomes,
\be
0 = \ba[t]{l}
\Biggl\{ 
\ba[t]{l}
2mh_{m,0}+2m\kappa_1+2m \\
 \\
-\ds{2Q_{m+1}\biggl[[(m+2)+m(2m+3)]v_{m+1,0}-\frac{(m+2)}{(m+1)}w_{m+1,0}\biggr]
\Biggr\}\left(\rx\right)^{m+1}} \\
\ea
 \\ \\
+\Biggl\{ \ba[t]{l}
\ds{\frac{2m(m+3)}{(m+1)}h_{m,1} - \frac{4m(m+2)}{(m+1)(2m+3)}u_{m,0}} 
- \ds{\frac{m(m+3)}{(m+1)}} \\
 \\
- \ds{\frac{m(m+3)}{5(2m+3)} - \frac{2m(m+2)}{5(2m+3)}
+2(m+3)(2m+5)Q_{m+1}Q_{m+2}u_{m+2,0}} \\
 \\
-\ds{\frac{2Q_{m+1}}{(m+1)}}
\biggl[ \ba[t]{l}
	\ds{[(m+2)(m+3)+m(m+1)(2m+5)]v_{m+1,1}} \\
	\\
	+(m+2)w_{m+1,0}\biggr]
	\Biggr\}\ds{\left(\rx\right)^{m+3}}
\ea
\ea
\ea
\ee

We must now solve these algebraic equations for the eight constants 
$\kappa_1$, $u_{m,0}$, $h_{m,0}$, $h_{m,1}$, $w_{m+1,0}$, $v_{m+1,0}$, 
$v_{m+1,1}$ and $u_{m+2,0}$ defined by our ansatz. Observe that of the 
four equations obtained from the coefficients of $(r/R)^{m+1}$, only 
two are linearly independent.  (This is an example of the linear 
dependence in the perturbation equations expanded about $r=0$ discussed 
in detail in Appendix D.) Thus, we have seven independent equations 
together with our matching condition (\ref{GR:ex_h_BC}) for the eight 
unknown quantities.  We find the following solution.


\be
\kappa = \frac{2}{(m+1)}\left[1
-\frac{4(m-1)(2m+11)}{5(2m+1)(2m+5)}\left(\tmr\right)
+ O\left(\tmr\right)^2 \right]
\label{GR:ex_sol:freq}
\ee

\be
U_m(r) = \left(\rx\right)^{m+1}\left[
1+u_{m,0}\left(1-\frac{r^2}{R^2}\right)\left(\tmr\right)
+ O\left(\tmr\right)^2\right]
\label{GR:ex_sol:u_m}
\ee

\be
h_m(r) = \left(\rx\right)^{m+1}\left[-\frac{3}{(2m+1)} 
+ \frac{3}{(2m+3)}\left(\rx\right)^2\right]\left(\tmr\right)
+ O\left(\tmr\right)^2
\label{GR:ex_sol:v_m+1}
\ee

\be
W_{m+1}(r) = (m+1)(m+2)K\left(\rx\right)^{m+1}
\left(1-\frac{r^2}{R^2}\right)\left(\tmr\right) + O\left(\tmr\right)^2
\label{GR:ex_sol:h_m}
\ee

\be
V_{m+1}(r) = K\left(\rx\right)^{m+1}
\left[(m+2) - (m+4)\left(\rx\right)^2\right]\left(\tmr\right)
+ O\left(\tmr\right)^2
\label{GR:ex_sol:w_m+1}
\ee

\be
U_{m+2}(r) = -K\, Q_{m+2}\frac{(m+1)^2(m+3)}{(2m+3)}
\left(\rx\right)^{m+3}\left(\tmr\right)
+ O\left(\tmr\right)^2
\label{GR:ex_sol:u_m+2}
\ee
where we have defined
\be
K \equiv \frac{6(m-1)Q_{m+1}}{5(m+2)(2m+5)}
\ee
and where
\be
u_{m,0} = - \frac{K\, Q_{m+1}}{24m(m+2)(2m+3)}
\ba[t]{l}
\biggl\{48(m+1)^4(m+3)^2 \\
 \\
+(2m+3)^2(2m+5)\biggl[m(m+2)^2-48\biggr]
\biggr\}
\ea
\ee

Since our solution satisfies the full perturbation equations 
to order $(2M_0/R)$, our ansatz was self-consistent.  Thus, 
we have explicitly found the first post-newtonian corrections 
to the $l=m$ newtonian r-modes of uniform density stars.

The solution reveals the expected mixing of axial and polar 
terms in the spherical harmonic expansion of $\delta u^\alpha$.
All of the newtonian r-modes with $m\geq 2$ pick up both axial
and polar corrections of order $(2M_0/R)$. (When $m=1$, the 
constant $K$ vanishes and we recover the purely axial solution 
with $l=m=1$ described in Sect. 3.4.1.)  In addition, we see
that the newtonian r-mode frequency also picks up a small
relativistic correction.  To order $(2M_0/R)$ the frequency 
decreases, but it is not clear whether this represents a tendency
to stabilize or destabilize the mode.  Finally, we note that the 
metric perturbation (whose radial behaviour is determined by the 
function $h_m$) is of the same order as the post-newtonian 
corrections to the fluid perturbation. Thus, there is no 
justification for the Cowling approximation in constructing
the hybrid mode solutions.

This analytic solution provides some useful insights into the 
effect of general relativity on the rotational modes of
neutron stars. However, it is limited to uniform density stars, 
it is accurate only to first post-newtonian order and it applies
only to those modes that are purely axial in the newtonian star.
Thus, let us now examine the numerical methods that will allow
us, in principle, to relax all three of these restrictions.  


\chapter{Relativistic Stars: Numerical Results}

\section{Method of Solution}

We now consider the problem of solving numerically for the 
hybrid modes of slowly rotating relativistic stars. We 
follow the same approach used for the newtonian problem
(Sect. 2.4). That is, we expand all quantities in regular
power series about the center and surface of the star and 
solve an algebraic system for the coefficients of these 
expansions.

In the newtonian case, the equilibrium solution did not play
a large role in the perturbation equations, entering only into
the perturbed mass conservation equation (\ref{eq1}). For the 
relativistic problem, however, equilibrium variables appear
in all of the perturbation equations. Thus, we begin with a 
discussion of our numerical solution to the equilibrium equations 
(\ref{GR:eos})-(\ref{hartle}).

\subsection{Numerical solution of the equilibrium equations}

As discussed in Sect. 3.2, we require our equilibrium solution to 
be that of a slowly rotating perfect fluid obeying a barotropic 
(one-parameter) equation of state,
\be
p = p(\ep).
\ee
For such models, it is convenient to define a comoving enthalpy,
\be
h(p) \equiv\int_0^p \frac{dp'}{\ep(p')+p'}
\label{enth_def}
\ee
and to re-express the equilibrium equations (\ref{GR:eos})-(\ref{hartle}) 
such that $h$ is the independent integration variable rather than $r$
(Lindblom \cite{l92}).  We write these modified equations below.

To integrate the equations in their usual form (for a given equation of 
state), one begins by fixing a quantity at the center of the star such 
as the central energy density $\ep_c$ or the central pressure $p_c$.  
One then integrates Eqs. (\ref{GR:eos})-(\ref{hartle}) from $r=0$ out 
to the surface of the star - i.e., to the radius, $r=R$, at which the 
pressure drops to zero.

With the equations re-expressed so that $h$ is the independent variable
one proceeds in a similar manner. One begins by fixing the central 
enthalpy $h_c$. (This also fixes the central pressure and energy density 
as a result of the equation of state and Eq. (\ref{enth_def})). One then 
integrates from $h=h_c$ out to the surface of the star, $h=0$. This 
method, unlike the usual approach, has the advantage that the domain of 
integration is known from the outset.

Using 
\be
\frac{dp}{dh}=(\ep+p)
\label{dpdh}
\ee
one can write Eq. (\ref{dnudr}) as
\be
\frac{d\nu}{dh} = -1,
\label{dnudh}
\ee
or
\be
\nu(h)-\nu_c = h_c - h
\ee
where $\nu_c$ is an arbitrary integration constant.

Lindblom \cite{l92} has re-expressed Eqs. (\ref{tov}) and 
(\ref{GR:dMdr}) as,
\be
\frac{dr}{dh} = -\frac{r(r-2M)}{(M+4\pi r^3p)}
\label{drdh}
\ee
and
\be
\frac{dM}{dh} = 4\pi r^2\ep \frac{dr}{dh},
\label{dMdh}
\ee
respectively.  Similarly, we write Eq. (\ref{hartle}) as the 
pair of first order equations
\be
\frac{d\bom}{dh} = e^{(\nu-\nu_c+\lambda)} f \frac{dr}{dh}
\label{dbomdh}
\ee
and
\be
\frac{df}{dh} = \left[ 16\pi(\ep+p)e^{-(\nu-\nu_c-\lambda)}\bom
-\frac{4}{r}f\right] \frac{dr}{dh}
\label{dfdh}
\ee
where we have defined
\be
f \equiv e^{-(\nu-\nu_c+\lambda)} \frac{d\bom}{dr}.
\ee

As in the usual approach, the equations (\ref{drdh})-(\ref{dfdh})
are singular at the center of the star.  Therefore,
we start the numerical integration near $h=h_c$ using the following
truncated power series solutions (see Lindblom \cite{l92} for the
first two of these expressions),
\be
r(h) = \left[ \frac{3(h_c-h)}{2\pi (\ep_c+p_c)}\right]^\half
\left\{
1 - \frac{1}{4}\left[\ep_c-3p_c+\frac{3}{5}\ep_1\right]
\frac{(h_c-h)}{(\ep_c+3p_c)}
\right\},
\ee

\be
M(h) = \frac{4\pi}{3}\ep_cr^3(h)\left\{
1 + \frac{3\ep_1}{5\ep_c}(h_c-h)\right\},
\ee

\be
\bom(h) = \om_c \left\{
1 + \frac{12(\ep_c+p_c)}{5(\ep_c+3p_c)}(h_c-h)\right\},
\ee

\be
f(h) =  \frac{16\pi}{5}(\ep_c+p_c)\om_cr(h)\left\{
1 + \frac{5}{7}\left[\frac{6(2\ep_c-3p_c)}{5(\ep_c+3p_c)}
+\frac{\ep_1}{(\ep_c+p_c)}\right](h_c-h)
\right\},
\ee
The central energy density $\ep_c$, the central pressure $p_c$ 
and the constant
\be
\ep_1 = - \left. \frac{d\ep}{dh}\right|_{h=h_c},
\ee
are determined from the equation of state, while $\om_c$ is a
normalization constant which must be fixed arbitrarily to
begin the integration. (We simply set $\om_c =1$.)

Beginning with these initial conditions, we integrate Eqs. 
(\ref{drdh})-(\ref{dfdh}) to the surface of the star, $h=0$, 
using a standard Runge-Kutta algorithm (Press et al. \cite{nr}). 
The mass and radius of the star are given at this point by 
$M_0=M(0)$ and $R=r(0)$. The constant $\nu_c$ is then given
in terms of these quantities by matching to the exterior 
solution at the surface of the star,
\be
\nu_c = - h_c + \nu(0) = - h_c + \half \ln\left(1-\frac{2M_0}{R}\right).
\ee
Finally, the angular velocity $\Omega$ and the angular momentum
$J$ of the star are determined by matching the interior solution to 
the exterior solution (\ref{hartle_ext_soln}).  (See, e.g., Hartle 
and Thorne \cite{ht68}.) At the surface of the star, $h=0$, we have,
\be
\bom(0) = \Omega - \frac{2J}{R^3}
\ee
and
\be
\frac{d\bom}{dr} = \frac{6J}{R^4} = e^{-\nu_c} f(0),
\ee
which gives
\be
\Omega = \left[\bom + \frac{1}{3}R\frac{d\bom}{dr}\right]_{r=R}
 = \bom(0) + \frac{1}{3}Re^{-\nu_c} f(0)
\ee
and
\be
J = \frac{1}{6}R^4 e^{-\nu_c} f(0).
\ee
Once these quantities are determined, we may then set $\Omega=1$
by renormalizing the solutions $\bom(h)$ and $f(h)$ 
(Hartle and Thorne \cite{ht68}).  

For the special case of polytropic stellar models, one can 
explicitly relate the energy density and pressure to the enthalpy 
using Eq. (\ref{enth_def}).  The equation of state for a polytrope
is given by
\bea
p &=& K \rho^\frac{n+1}{n} \\
\ep &=& \rho + np
\eea
where $n$ is the polytropic index, $\rho$ is the rest-mass-energy 
density and $K$ is a constant.  Since $K$ has dimensions
$(\mbox{length})^{2/n}$, it is convenient to work only with 
dimensionless quantities by making the redefinitions,
\bea
p &\rightarrow& K^{-n} p \nn \\
\rho &\rightarrow& K^{-n} \rho \nn \\
\ep &\rightarrow& K^{-n} \ep. \nn
\eea
The equation of state then becomes
\bea
p &=& \rho^\frac{n+1}{n} \\
\ep &=& \rho + np
\eea
and Eq. (\ref{enth_def}) gives
\bea
h(p) &=& \int_0^p \frac{dp'}{\ep(p')+p'} \\
     &=& ln\left[1+(n+1)p^\frac{1}{n+1}\right]
\eea
or
\bea
\rho(h) &=& \left(\frac{e^h-1}{n+1}\right)^n \\
p(h)    &=& \left(\frac{e^h-1}{n+1}\right)^{n+1} \\
\ep(h)  &=& \left(\frac{e^h-1}{n+1}\right)^n 
\left[ 1+ n\left(\frac{e^h-1}{n+1}\right) \right].
\eea
These expressions may then be used in the integration of Eqs. 
(\ref{drdh})-(\ref{dfdh}) with the initial values
\bea
\ep_c  &=& \left(\frac{e^{h_c}-1}{n+1}\right)^n 
\left[ 1+ n \left(\frac{e^{h_c}-1}{n+1}\right)\right] \\
p_c    &=& \left(\frac{e^{h_c}-1}{n+1}\right)^{n+1} \\
\ep_1  &=& - \frac{ne^{2h_c}}{(e^{h_c}-1)}
\left(\frac{e^{h_c}-1}{n+1}\right)^n.
\eea

Finally, we comment on one last aspect of the numerical
solution of the equilibrium equations.  In order to implement
our numerical approach to the perturbation equations, we need 
the power series expansions of the equilibrium quantities that
appear in these equations.  The equilibrium quantities we will 
require (see Sect. 4.1.2 below) and the forms of their series 
expansions are as follows.  
About the center of the star, $r=0$, we write
\begin{eqnarray}
\frac{r(\epsilon'+p')}{(\epsilon+p)} &=&
\sum^\infty_{i=1} \pi_i \left(\frac{r}{R}\right)^{2i},  
\label{pi_i}
\\
e^{2\lambda} &=& 1 + \sum^\infty_{i=1} 
E_i \left(\frac{r}{R}\right)^{2i}, 
\label{E_i}
\\ 
r\nu' &=& \sum^\infty_{i=1} \nu_i \left(\frac{r}{R}\right)^{2i}, 
\label{nu_i}
\\ 
r\lambda' &=& \sum^\infty_{i=1} 
\lambda_i \left(\frac{r}{R}\right)^{2i}, 
\label{la_i}
\\ 
\frac{\bar\omega}{\Omega} &=& 
\sum^\infty_{i=0} \omega_i \left(\frac{r}{R}\right)^{2i}, 
\label{om_i}
\\
\frac{\mu}{\Omega} &=& \sum^\infty_{i=1} 
\mu_i \left(\frac{r}{R}\right)^{2i}, 
\label{mu_i}
\end{eqnarray}
and about the surface of the star, $r=R$, we write
\begin{eqnarray}
\frac{r(\epsilon'+p')}{(\epsilon+p)} &=&
\sum^\infty_{k=-1} \tilde\pi_k \left(1-\frac{r}{R}\right)^k,  
\label{pi_k}
\\
e^{2\lambda} &=& 
\sum^\infty_{k=0} \tilde E_k \left(1-\frac{r}{R}\right)^k,  
\label{E_k}
\\ 
r\nu' &=& 
\sum^\infty_{k=0} \tilde\nu_k \left(1-\frac{r}{R}\right)^k, 
\label{nu_k}
\\ 
r\lambda' &=& 
\sum^\infty_{k=0} \tilde\lambda_k \left(1-\frac{r}{R}\right)^k,  
\label{la_k}
\\ 
\frac{\bar\omega}{\Omega} &=& 
\sum^\infty_{k=0} \tilde\omega_k \left(1-\frac{r}{R}\right)^k,  
\label{om_k}
\\
\frac{\mu}{\Omega} &=& \sum^\infty_{k=0} 
\tilde\mu_k \left(1-\frac{r}{R}\right)^k,
\label{mu_k}
\end{eqnarray}
where $'\equiv d/dr$ and where we have defined the 
function
\be
\mu(r) \equiv re^{2\nu}({\bar\omega}e^{-2\nu})'.  
\label{mu_def}
\ee

All of these quantities are expressible in terms of the 
functions $r(h)$, $M(h)$, $\nu(h)$, $\bom(h)$, $f(h)$,
$p(h)$ and $\ep(h)$ defined by our Runge-Kutta solution to 
the equilibrium equations.  The coefficients of these series,
$\pi_i$, $\tilde\pi_k$ etc., are extracted from our numerical 
solution using a polynomial fitting algorithm (Press et al. 
\cite{nr}).  We note that although the polynomial fit accurately 
reproduces the various equilibrium functions defined above, it does 
not find the coefficients of the higher order terms in the fitting 
polynomials with great accuracy.  Since it is the coefficients, 
themselves, that are used in the numerical solution of the 
perturbation equations, this limits the accuracy of our solutions.

\subsection{Numerical solution of the perturbation equations}

Having described our numerical method for the solution of the 
equilibrium equations, we now turn to the perturbation equations 
(\ref{GR:sph_H1}), (\ref{GR:sph_h_0''}), (\ref{GR:sph_V}) and 
(\ref{om_th_ph})-(\ref{om_ph_r}).  These must be solved 
numerically subject to the boundary, matching and normalization 
conditions (\ref{bc_on_W}), (\ref{cont_cond}), (\ref{match_cond}) 
and (\ref{norm_cond}) at the surface of the star.

Since Eq. (\ref{GR:sph_H1}) merely expresses $H_{1,l}(r)$ in terms 
of $W_l(r)$, we may ignore this metric variable and drop 
Eq. (\ref{GR:sph_H1}) from the system of equations to be solved, 
for all $l$.  In addition, we need not work with Eq. (\ref{om_ph_r}) 
since (\ref{om_th_ph})-(\ref{om_ph_r}) are related by 
Eq. (\ref{not_ind}).  Thus, a complete set of perturbation equations 
is provided by Eqs. (\ref{GR:sph_h_0''}), (\ref{GR:sph_V}), 
(\ref{om_th_ph}) and (\ref{om_r_th}) which we will now re-express
in a form more suitable for our numerical approach.

We write Eq. (\ref{GR:sph_h_0''}) as
\be
0 = \ba[t]{l}
r^2 h_l^{''} - r(\nu'+\lambda')rh_l' 
- \left[2 + (2-l^2-l) e^{2\lambda} + r(\nu'+\lambda')\right] h_l \\
\\
- 4r(\nu'+\lambda') U_l,
\label{h_0''}
\ea
\ee
Eq. (\ref{GR:sph_V}) as
\be
0 = rW_l' 
+\left[1+r(\nu'+\lambda')+\frac{r(\ep'+p')}{(\ep+p)}\right]W_l
- l(l+1)V_l,
\label{Vsub}
\ee
Eq. (\ref{om_th_ph}) as
\be
0 = \ba[t]{l}
\left[ l(l+1)\kappa\Omega(h_l+U_l)-2m{\bar\omega}U_l\right] \\
 \\
+(l+1)Q_l \left[\left(2\bom+\mu\right)W_{l-1} 
- 2(l-1){\bar\omega}V_{l-1}\right]  \\
 \\
-l Q_{l+1} \left[\left(2\bom+\mu\right)W_{l+1} 
+2(l+2){\bar\omega}V_{l+1} \right]
\ea
\label{th_ph}
\ee
and Eq. (\ref{om_r_th}) as
\bea
0 &=&
(l-2)l(l+1)Q_{l-1}Q_l \left[
-2\bom rU'_{l-2}+2(l-1)\bom U_{l-2} +(l-3)\mu U_{l-2}
\right] 
\label{r_th} \\
&&\nn \\
&&+ (l+1) Q_l \biggl[
	\ba[t]{l}
	(l-1)l\kappa\Omega rV'_{l-1} -2ml\bom rV'_{l-1}
	-2(l-1)l\kappa\Omega r\nu' V_{l-1} \\
	 \\
	+2m(l-1)l\bom V_{l-1} + m(l-3)l\mu V_{l-1} \\
	\\
	-(l-1)l\kappa\Omega e^{2\lambda} W_{l-1}
	+4\kappa\Omega r(\nu'+\lambda') W_{l-1}
	\biggr]
	\ea \nn \\
&&\nn \\
&&+l(l+1)\biggl\{
	\ba[t]{l}
	m\kappa\Omega r\left(h'_l+U'_l\right)
	-2m\kappa\Omega r\nu' \left(h_l+U_l\right) \\
	\\
	+ \left((l+1)Q_l^2-l Q_{l+1}^2\right) 
	\left(2\bom rU'_l + 2\mu U_l\right) \\
	 \\
	+\left[m^2+l(l+1)\left(Q_{l+1}^2+Q_l^2-1\right)\right]
	(2\bom+\mu) U_l
	\biggr\}
	\ea \nn \\
&&\nn \\
&&-l Q_{l+1} \biggl[
	\ba[t]{l}
	(l+1)(l+2)\kappa\Omega rV'_{l+1}
	+2m(l+1)\bom rV'_{l+1}
	-2(l+1)(l+2)\kappa\Omega r\nu' V_{l+1} \\
	 \\
	+2m(l+1)(l+2)\bom V_{l+1}
	+m(l+1)(l+4)\mu V_{l+1} \\
	\\
	-(l+1)(l+2)\kappa\Omega e^{2\lambda} W_{l+1}
	+4\kappa\Omega r(\nu'+\lambda') W_{l+1}
	\biggr]
	\ea \nn \\
&&\nn \\
&&+l(l+1)(l+3)Q_{l+1}Q_{l+2} \left[
2\bom rU'_{l+2} +2(l+2)\bom U_{l+2} + (l+4)\mu U_{l+2}
\right].
\nn
\eea
where we have used the definition (\ref{mu_def}) of the function
$\mu(r)$.

Eqs. (\ref{h_0''})-(\ref{r_th}) comprise a system of ordinary 
differential equations for the variables $h_{l'}$, $U_{l'}$, 
$W_{l'}$ and $V_{l'}$ (for all $l'$).  Together with the boundary
and matching conditions at the surface of the star, these equations
form a non-linear eigenvalue problem for the parameter $\kappa$, where
$\kappa\Omega$ is the mode frequency in the rotating frame.

To solve for the eigenvalues we proceed exactly as in the newtonian
case (see Sect. 2.4).  We first ensure that the boundary and matching 
conditions are satisfied by expanding $h_{l'}(r)$, $U_{l'}(r)$, 
$W_{l'}(r)$ and $V_{l'}(r)$ (for all $l'$) in regular power series 
about the surface and center of the star.  (We present these expansions 
explicitly in Appendix D, Eqs. (\ref{h_i})-(\ref{V_k})).  
Substituting these series into the 
differential equations (\ref{h_0''})-(\ref{r_th}) results in a set 
of algebraic equations for the expansion coefficients.  These 
algebraic equations may be solved for arbitrary values of $\kappa$ 
using standard matrix inversion methods.  For arbitrary values of 
$\kappa$, however, the series solutions about the center of the star 
will not agree with those about the surface of the star.  
The requirement that the series agree at some matching point, 
$0<r_0<R$, then becomes the condition that restricts the possible 
values of the eigenvalue, $\kappa_0$.  

We begin by replacing all of the equilibrium quantities in
Eqs. (\ref{h_0''})-(\ref{r_th}) with their series expansions,
(\ref{pi_i})-(\ref{mu_k}) and by replacing our perturbation
variables $h_{l'}(r)$, $U_{l'}(r)$, $W_{l'}(r)$ and $V_{l'}(r)$ 
with their series expansions, (\ref{h_i})-(\ref{V_k}).  The result
is a set of algebraic equations for the expansion coefficients
$h_i$, $\tilde h_k$ etc., which we present explicitly in Appendix D,
Eqs. (\ref{alg1})-(\ref{alg8}).  We then write down the matching conditions
at the point $r_0$ equating the series expansions about $r=0$ to the
series expansions about $r=R$.  (These matching conditions are also
presented explicitly in Appendix D, Eqs. (\ref{match_h1})-(\ref{match_h2}).)
The result is a linear algebraic system which we may represent 
schematically as
\be Ax=0. 
\label{GR:linalg} 
\ee
In this equation, $A$ is a matrix which depends non-linearly on the
parameter $\kappa$, and $x$ is a vector whose components are the
unknown coefficients in the series expansions for $h_{l'}(r)$, 
$U_{l'}(r)$, $W_{l'}(r)$ and $V_{l'}(r)$.

To satisfy equation (\ref{GR:linalg}) we must find those values of
$\kappa$ for which the matrix $A$ is singular, i.e., we must find the
zeroes of the determinant of $A$.  We truncate the spherical harmonic
expansions of $\delta u^\alpha$ and $h_{\alpha\beta}$ at some maximum 
index $l_{\mbox{\tiny max}}$ and we truncate the radial series 
expansions about $r=0$ and $r=R$ at some maximum powers 
$i_{\mbox{\tiny max}}$ and $k_{\mbox{\tiny max}}$, respectively.

The resulting finite matrix is band diagonal. To find the zeroes of 
its determinant we use the same routines from the LAPACK linear 
algebra libraries (Anderson et al. \cite{lapack}) and root finding 
techniques that we used in the newtonian calculation.

The eigenfunctions associated with these eigenvalues are determined 
by the perturbation equations only up to normalization.  Given a
particular eigenvalue, we find its eigenfunction by replacing one of
the equations in the system (\ref{GR:linalg}) with the normalization
condition (\ref{norm_cond}).
Since this eliminates one of the rows of the singular matrix $A$ 
in favor of the normalization equation, the result is an algebraic 
system of the form
\be 
\tilde A x = b, 
\label{GR:linalg2} 
\ee 
where $\tilde A$ is now a non-singular matrix and $b$ is a known 
column vector.  We solve this system for the vector $x$ using 
routines from LAPACK and reconstruct the various series expansions 
from this solution vector of coefficients.

\section{Eigenvalues and Eigenfunctions}

We have computed the eigenvalues and eigenfunctions for 
uniform density stars in full general relativity.  If $M_0$ 
is the gravitational mass of the spacetime and $R$ the coordinate 
radius of the star, then the dimensionless constant $(2M_0/R)$
is a useful measure of the strength of relativistic effects. 
We have studied a set of axial- and polar-hybrids over a range
of values of this constant from the nearly newtonian to the 
relativistic regimes, $10^{-6}\lesssim(2M_0/R)\lesssim 0.2$.

Although the code is fully relativistic and was written to handle
any polytropic equation of state, it does not yet accurately
find the modes for compressible stars (polytropic index $n\neq 0$)
or for highly relativistic uniform density stars 
$(2M_0/R \grsim 0.2)$.  The problem appears to be related to
the fitting of the equilibrium variables to their power series 
expansions, and the difficulty in accurately computing the 
coefficients of the higher order terms in the fitting polynomials 
(see Sect. 4.1.1).  Work is in progress to address this problem 
and extend this study to the broader class of stellar models.  
However, we do not expect qualitative differences between our 
current results and the results of such a study.

The structure of the hybrid mode spectrum in relativistic stars
appears to be identical to that in newtonian stars.  We find 
that for each of the hybrid modes considered, there exists a 
family of relativistic modes parametrized by $(2M_0/R)$,
whose limit as $(2M_0/R)\rightarrow 0$ is the corresponding
newtonian mode.

\begin{figure}
\centerline{\includegraphics[3in,3in][6in,8in]{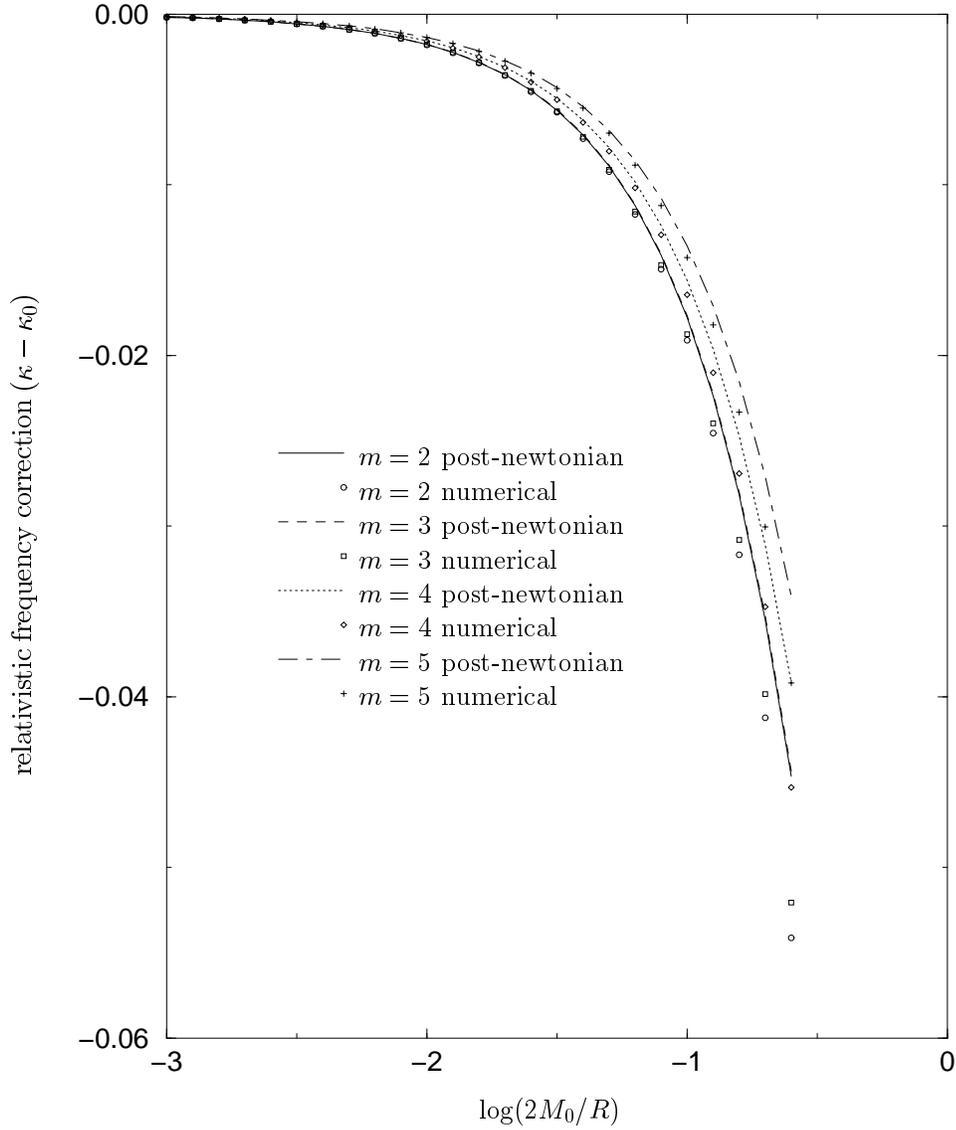}}
\caption{The difference between the relativistic and newtonian
eigenvalues $(\kappa-\kappa_0)$ as the uniform density star becomes
increasingly relativistic.  These frequency corrections are shown
for the modes whose newtonian limit is a pure r-mode with
$2\leq l=m\leq 5$.  The curves were calculated using the 
analytic post-newtonian expression (\ref{GR:ex_sol:freq}), while
the symbols were calculated numerically.}
\label{freqs1}
\end{figure}

By means of the post-newtonian solution of Sect. 3.4.2, we have
already verified our claim that the newtonian r-modes with
$l=m\geq 2$ do not exist in relativistic stars as purely axial
modes.  We may now make further use of this analytic solution 
to test the accuracy of our relativistic code in the small 
$(2M_0/R)$ regime.  We do so by examining the difference 
between the frequencies of corresponding modes in the newtonian 
and relativistic stars.  If $\sigma_0$ is the frequency of a 
mode in the newtonian star with eigenvalue $\kappa_0$ and 
$\sigma$ is the frequency of the corresponding relativistic
mode with eigenvalue $\kappa$, then this difference is given by,
\be
\sigma-\sigma_0 = (\kappa-\kappa_0)\Omega.
\ee
In Fig. \ref{freqs1} we display $\kappa-\kappa_0$ as a function 
of $(2M_0/R)$ for the modes whose newtonian limit is a pure
r-mode with $2\leq l=m\leq 5$.  The curves show the 
post-newtonian solution given by Eq. (\ref{GR:ex_sol:freq}), 
while the symbols show the fully relativistic numerical 
solution.  At $(2M_0/R)=10\%$ the fractional change in the 
frequency from the newtonian value is approximately $4\%$
for these modes.  Observe that not only do the analytic and 
numerical solutions agree in the small $(2M_0/R)$ regime, 
but that the code scales correctly with $(2M_0/R)$.  

\begin{figure}
\centerline{\includegraphics[3in,3in][6in,8in]{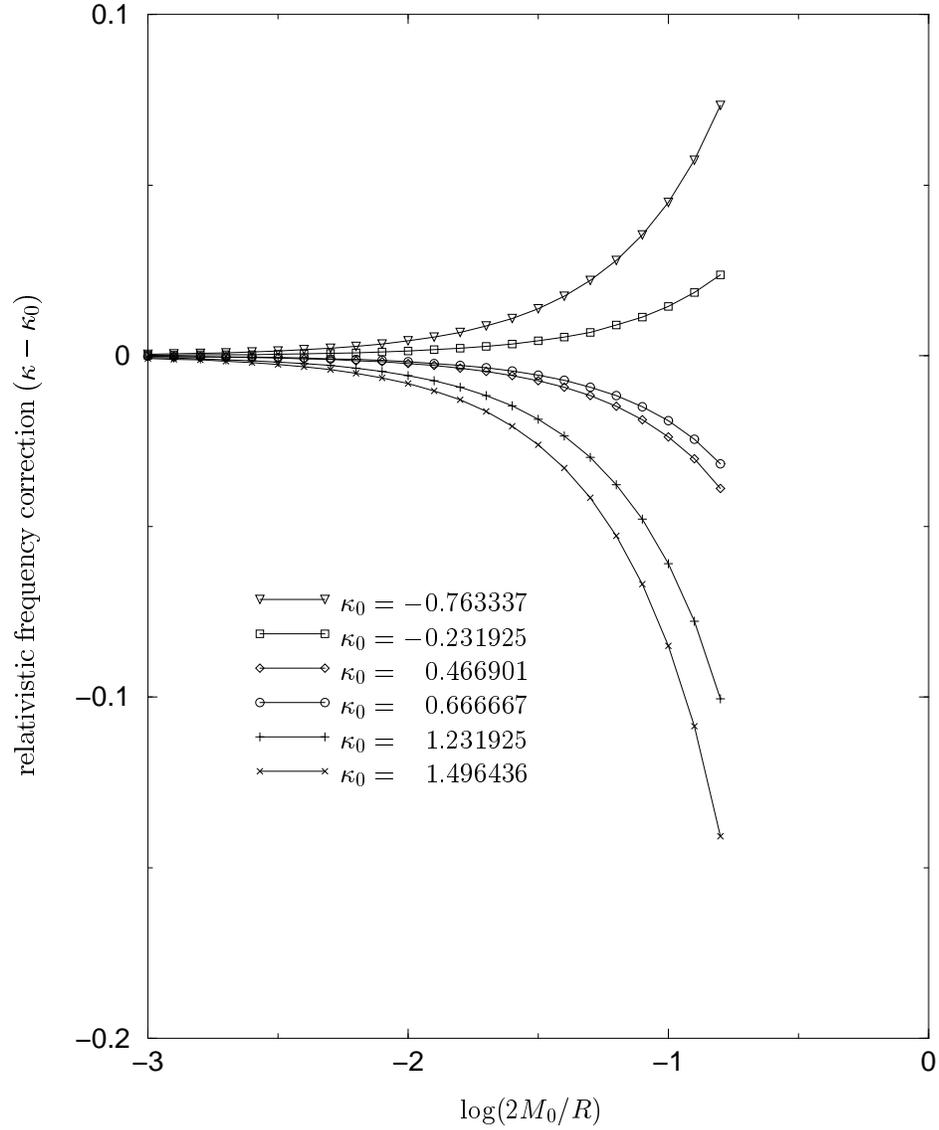}}
\caption{The difference between the relativistic and newtonian
eigenvalues $(\kappa-\kappa_0)$ as the uniform density star becomes
increasingly relativistic.  These frequency corrections are shown
for a number of both axial- and polar-led hybrid modes with $m=2$
(see Table \ref{ev_mac}).  All of the data points were calculated
numerically.}
\label{freqs2}
\end{figure}

This agreement with our analytic solution gives us confidence 
that our code is able to find the relativistic modes.  Thus,
we may now explore those modes for which we have not worked
out a post-newtonian solution.  In Fig. \ref{freqs2} we 
display $\kappa-\kappa_0$ as a function of $(2M_0/R)$ for a
set of modes whose newtonian limits are axial- and polar-led 
hybrids with $m=2$.  The modes whose frequency corrections are
shown correspond to the first six entries in the $m=2$ column 
of Table \ref{ev_mac}.  At $(2M_0/R)=10\%$ the fractional 
change in the frequency from the newtonian value is, again,
approximately $4\%$.  Fig. \ref{freqs2} also reveals the
feature that the frequencies of modes with $\kappa_0<0$ increase
in the relativistic star, while the frequencies of modes with 
$\kappa_0>0$ decrease.  So the magnitude of the frequency 
always decreases as the star becomes more relativistic. 
It is natural that general relativity will have such an 
effect for two reasons.  Gravitational redshift will tend to
decrease the fluid oscillation frequencies measured by an 
inertial observer at infinity (i.e., frequencies measured with 
respect to the Killing time parameter, $t$).  Also,
``the magnitude of the centrifugal force is determined not
by the angular velocity $\Omega$ of the fluid relative to a
distant observer but by its angular velocity relative to the
local inertial frame, $\bar\om(r)$.'' (Hartle and Thorne 
\cite{ht68}.)  Thus, the centrifugal and Coriolis forces 
diminish - and the modes oscillate less rapidly - as the 
dragging of inertial frames becomes more pronounced.  
It is not clear what effect these small frequency shifts will 
have (if any) on the stability of the modes.  

\begin{figure}
\centerline{\includegraphics[3in,3.25in][6in,8.25in]{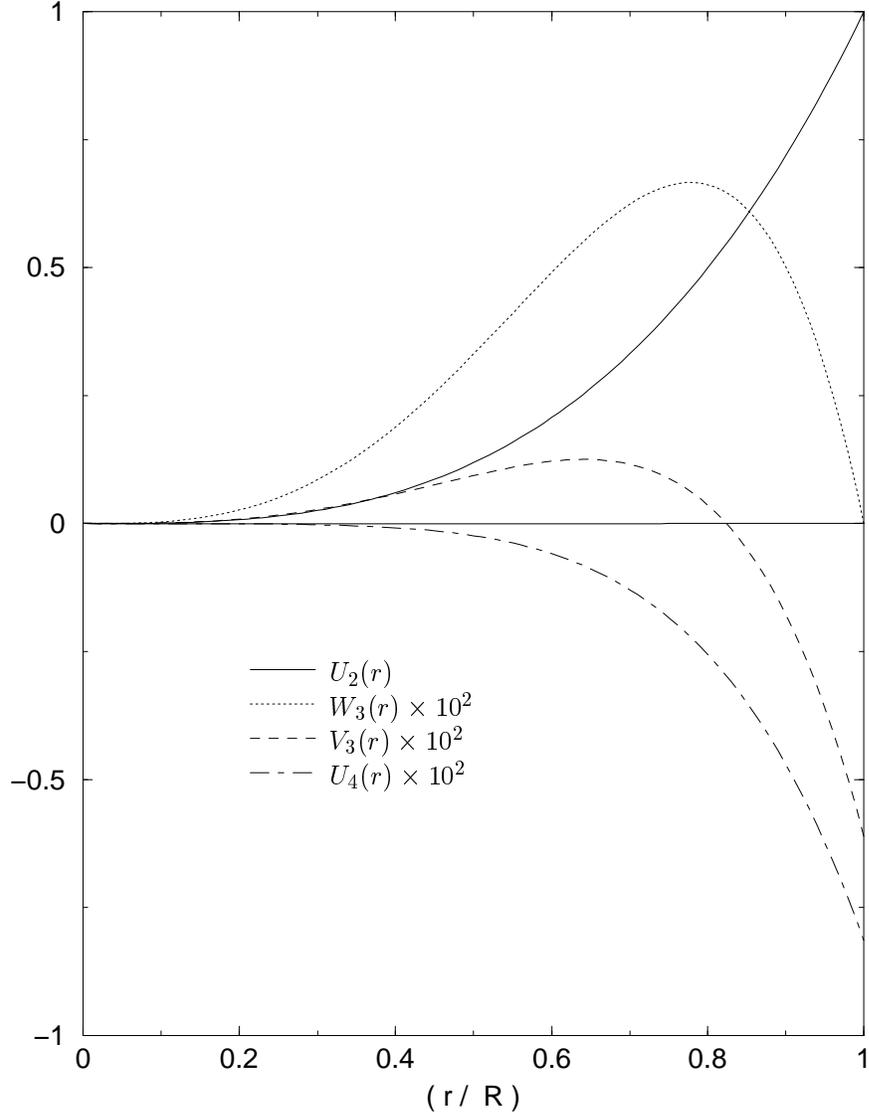}}
\caption{Coefficients $U_l(r)$, $W_l(r)$, and $V_l(r)$ with $l\leq 4$
of the spherical harmonic expansion (\ref{xi_exp}) for the $m=2$
axial-led hybrid mode whose newtonian limit is the pure axial r-mode 
with $l=2$ and comoving frequency $\kappa_0\Omega=2\Omega/3$.  
The mode is shown in the uniform density star with $(2M_0/R)=0.2$, 
for which its comoving frequency has shifted to
$\kappa\Omega=0.625\Omega$.  The vertical scale is set by the 
normalization of $U_2(r)$ to unity at the surface of the star. 
Observe that there are both axial and 
polar relativistic corrections of order $1\%$.  The coefficients
of expansion (\ref{xi_exp}) with $l>4$ are of order $0.01\%$ and
smaller and are not shown.}
\label{GR:fig1}
\end{figure}

\begin{figure}
\centerline{\includegraphics[3in,3in][6in,8in]{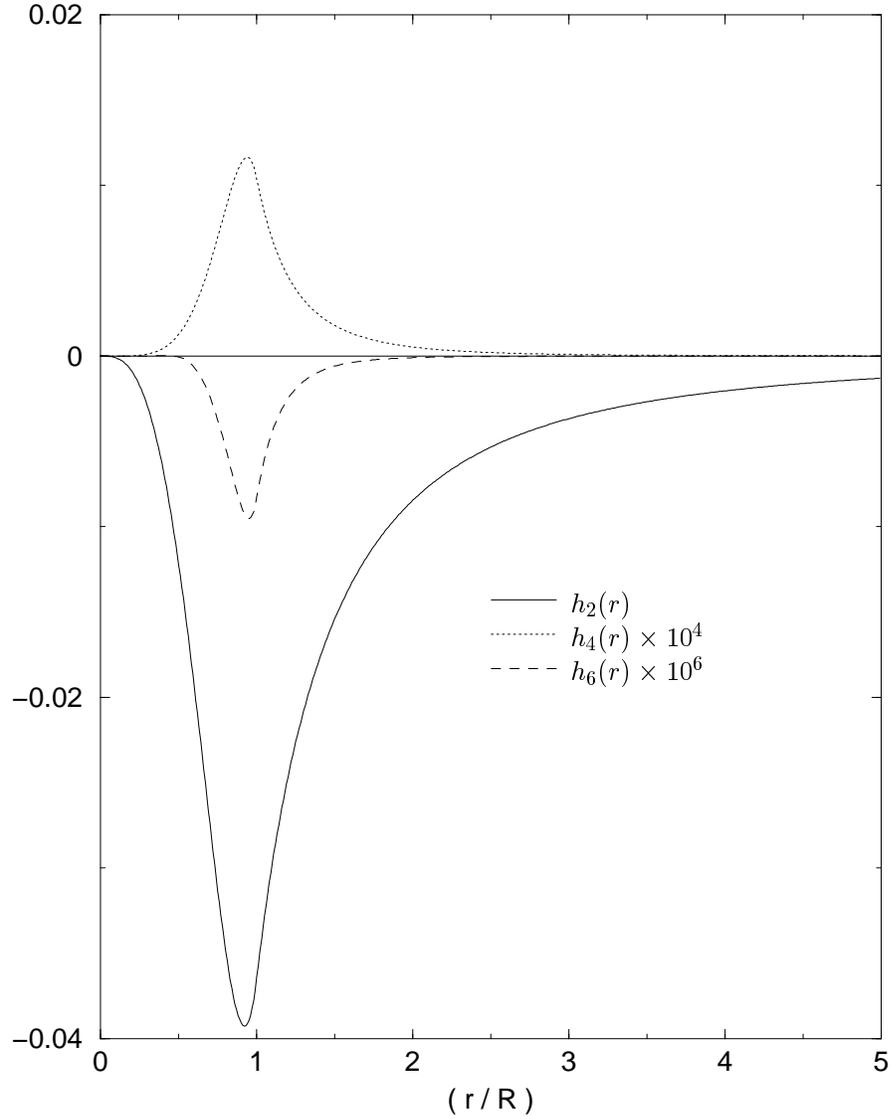}}
\caption{Coefficients $h_l(r)(\equiv h_{0,l}(r))$ with $l\leq 6$
of the spherical harmonic expansion (\ref{h_components}) of the 
perturbed metric for the same mode shown in Fig. \ref{GR:fig1}.
The vertical scale is the same as that of Fig. \ref{GR:fig1}
and is set by the normalization of $U_2(r)$.
Observe that, as expected, $h_2(r)$ dominates the perturbed metric, 
which implies that this mode couples strongly to current quadrupole 
radiation.}
\label{GR:fig2}
\end{figure}

We now turn to a discussion of the eigenfunctions.  
From Sect. 2.3.1, we know that the radial behaviour of the $l=m$
newtonian r-modes is given by $U_m(r) = (r/R)^{m+1}$, and in
Sect. 3.4.2 we computed their post-newtonian corrections.
The mode expected to dominate the gravitational-wave 
instability in a hot, young neutron star is the newtonian
r-mode with $l=m=2$.  In Fig. \ref{GR:fig1}, we display
the functions $U_l(r)$, $W_l(r)$, and $V_l(r)$ for $l\leq 4$ 
associated with this mode in a uniform density star with
$(2M_0/R)=0.2$.  These components of the eigenfunction were 
computed numerically using the fully relativistic code.  
The $20\%$ corrections to the structure of the equilibrium 
star induce only $1\%$ corrections to the character of this 
mode; thus, the function $U_2(r)$ is barely distinguishable 
from its newtonian form, $(r/R)^3$, and the coefficients shown 
of the other axial and polar terms have been multiplied by a 
factor of 100 to make them visible on the scale of the plot.  
These functions are normalized so that $U_2(r)=1$ at the 
surface of the star, $r=R$.  In Fig. \ref{GR:fig2} we show 
the metric functions $h_l(r)$ for $l\leq 6$ for the same mode.
The vertical scale is set by the normalization of $U_2(r)$,
and this implies that the metric perturbation is at most a 
$4\%$ correction to the equilibrium metric.  The fact that
$h_2(r)$ dominates the perturbed metric is the statement that
this mode couples strongly to current quadrupole radiation.

\begin{figure}
\centerline{\includegraphics[3in,3.25in][6in,8.25in]{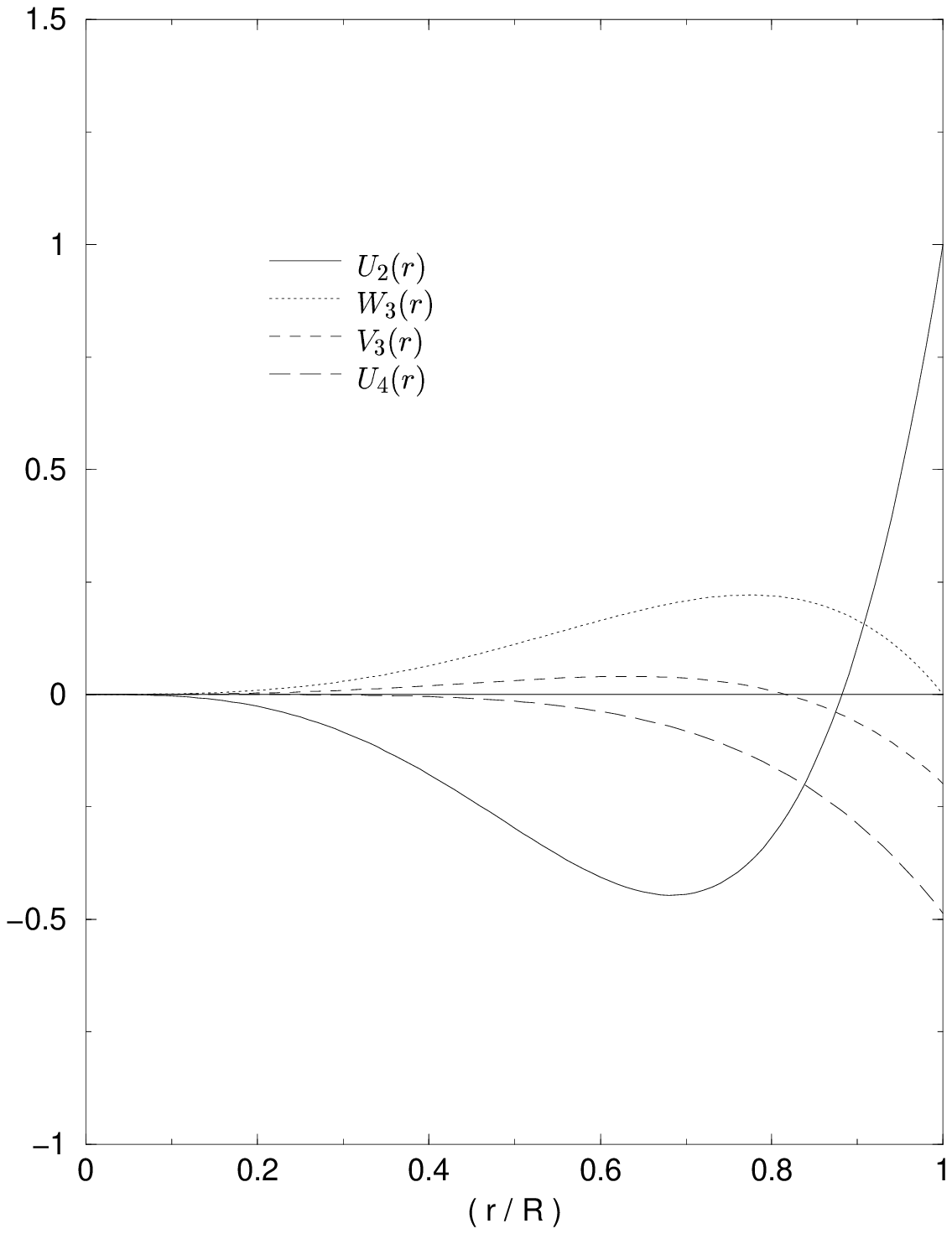}}
\caption{Coefficients $U_l(r)$, $W_l(r)$, and $V_l(r)$ with $l\leq 4$
of the spherical harmonic expansion (\ref{xi_exp}) for the $m=2$
axial-led hybrid mode whose newtonian limit has comoving frequency 
$\kappa_0\Omega=0.466901\Omega$.  The mode is shown in the uniform 
density star with $(2M_0/R)=10^{-6}$ to check for agreement with
the newtonian calculation.  As expected, these functions agree with 
the newtonian forms displayed in Figs. \ref{fig6}-\ref{fig8}, 
up to the difference in the normalization conditions used in the 
newtonian (\ref{N:norm_cond}) and relativistic (\ref{norm_cond}) 
calculations and corrections of order $10^{-6}$.}
\label{GR:fig3}
\end{figure}

\begin{figure}
\centerline{\includegraphics[3in,3.25in][6in,8.25in]{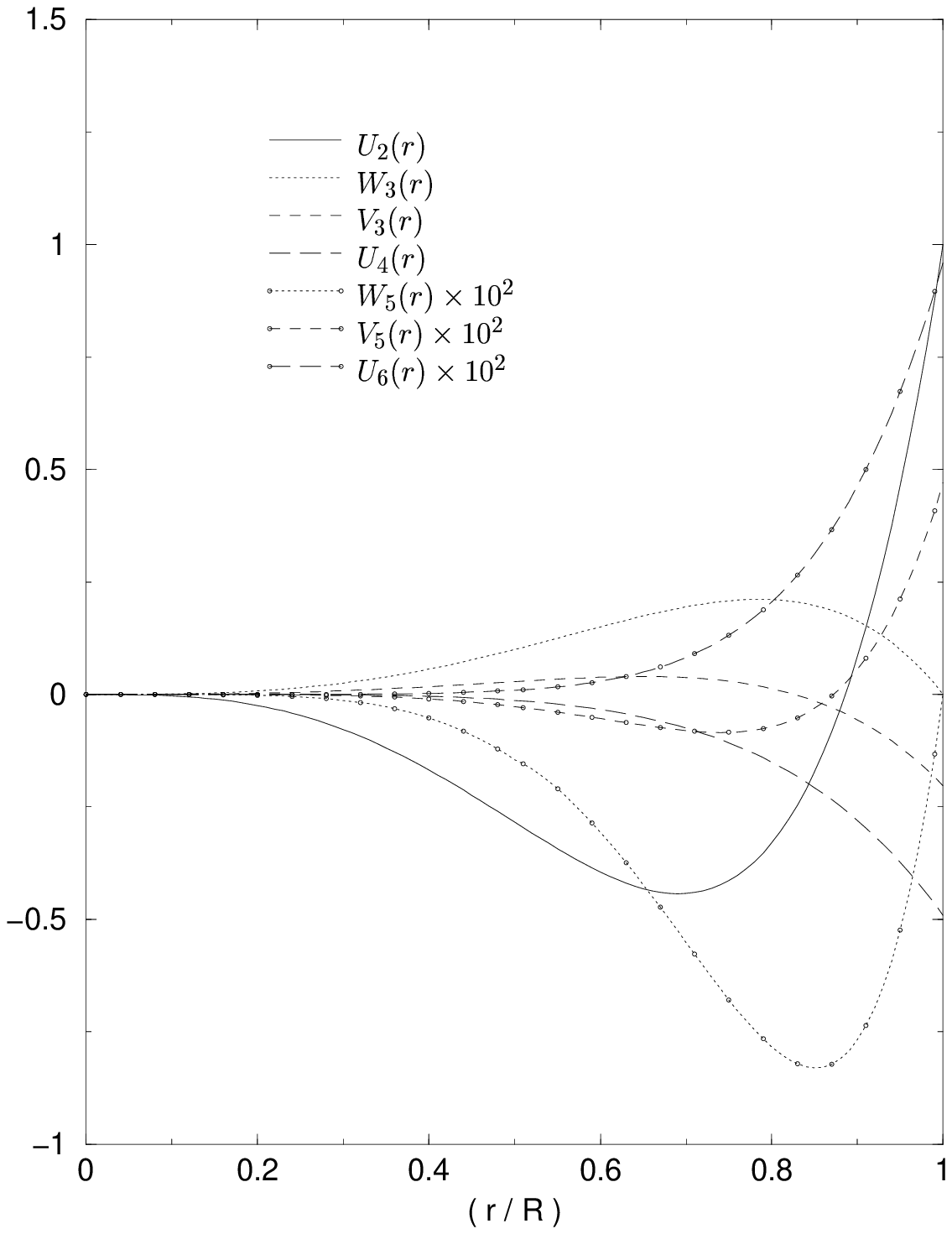}}
\caption{Coefficients $U_l(r)$, $W_l(r)$, and $V_l(r)$ with $l\leq 6$
of the spherical harmonic expansion (\ref{xi_exp}) for the same mode
shown in Fig. \ref{GR:fig3} but now with $(2M_0/R)=0.1$.  The 
comoving frequency has shifted to $\kappa\Omega=0.443\Omega$, and
the mode has acquired relativistic corrections of order $1\%$.
Accordingly, the functions $U_2$, $W_3$, $V_3$ and $U_4$ are 
indistinguishable from those shown in Fig. \ref{GR:fig3}.
The coefficients of expansion (\ref{xi_exp}) with $l>6$ are of order 
$0.01\%$ and smaller and are not shown.}
\label{GR:fig4}
\end{figure}

\begin{figure}
\centerline{\includegraphics[3in,3.25in][6in,8.25in]{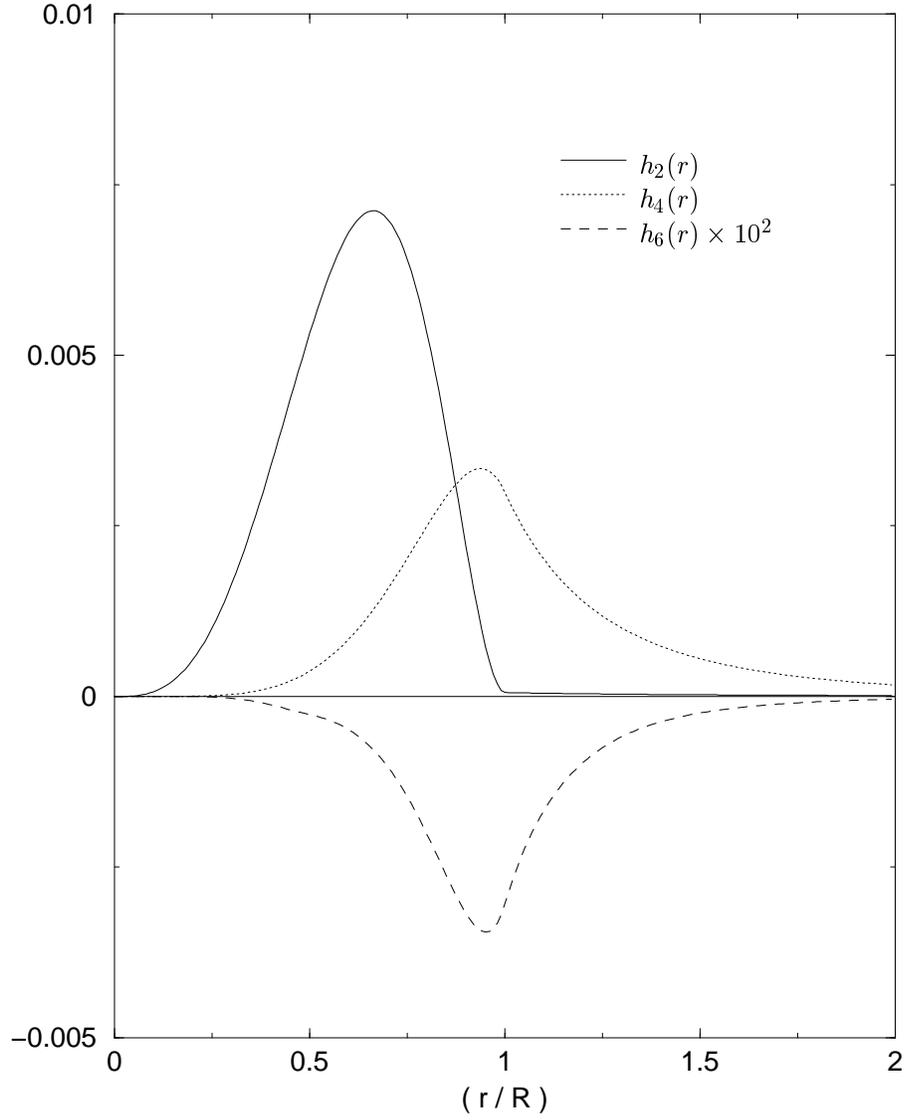}}
\caption{Coefficients $h_l(r)(\equiv h_{0,l}(r))$ with $l\leq 6$
of the spherical harmonic expansion (\ref{h_components}) for the 
same mode shown in Fig. \ref{GR:fig4} in the uniform density star 
with $(2M_0/R)=0.1$.  The vertical scale is the same as that of 
Fig. \ref{GR:fig4} and is set by the normalization of $U_2(r)$.  
Observe that $h_2(r)$ nearly vanishes in the exterior spacetime and
that the metric perturbation is dominated by $h_4(r)$.  This is
the expected result that the mode couples strongly to the
$l=4$ current multipole only (see Sect 2.6).}
\label{GR:fig5}
\end{figure}

Figs. \ref{GR:fig3}-\ref{GR:fig5} are a series of plots 
displaying the $m=2$ axial-led hybrid mode whose newtonian 
eigenvalue in the uniform density star is $\kappa_0=0.466901$.  
(See Figs. \ref{fig6}-\ref{fig8} and Table \ref{ef_m2a}.)
Fig. \ref{GR:fig3} shows the functions $U_l(r)$, $W_l(r)$, 
and $V_l(r)$ for this mode as calculated by the relativistic 
code in the newtonian regime, $(2M_0/R)=10^{-6}$.  As expected,
these functions agree with the newtonian forms displayed 
in Figs. \ref{fig6}-\ref{fig8}, up to the difference in the 
normalization conditions used in the newtonian (\ref{N:norm_cond}) and 
relativistic (\ref{norm_cond}) calculations and corrections of order 
$10^{-6}$.  In Fig. \ref{GR:fig4} we display the functions 
$U_l(r)$, $W_l(r)$, and $V_l(r)$ for $l\leq 6$ for the same 
mode, but now with $(2M_0/R)=0.1$, and in Fig. \ref{GR:fig5} 
we display the corresponding metric functions $h_l(r)$ for 
$l\leq 6$.  Observe that $U_2$, $W_3$, $V_3$ and $U_4$ are
barely distinguishable from their newtonian forms and that
the mode has acquired relativistic corrections of order $1\%$. 
Fig. \ref{GR:fig5} reveals the interesting feature that the
metric function $h_2(r)$ nearly vanishes in the exterior 
spacetime.  Recall the discussion at the end of Sect. 2.6 of
the vanishing of the $l=2$ current multipole for this
particular mode in the uniform density newtonian star; we
saw that the only nonzero current multipole is that with $l=4$.
Here, the relativistic calculation reveals explicitly that
the perturbed metric in the exterior spacetime is dominated
by $h_4(r)$, and that $h_2(r)$ is smaller by two orders of
magnitude in the exterior spacetime.

\begin{figure}
\centerline{\includegraphics[3in,3.25in][6in,8.25in]{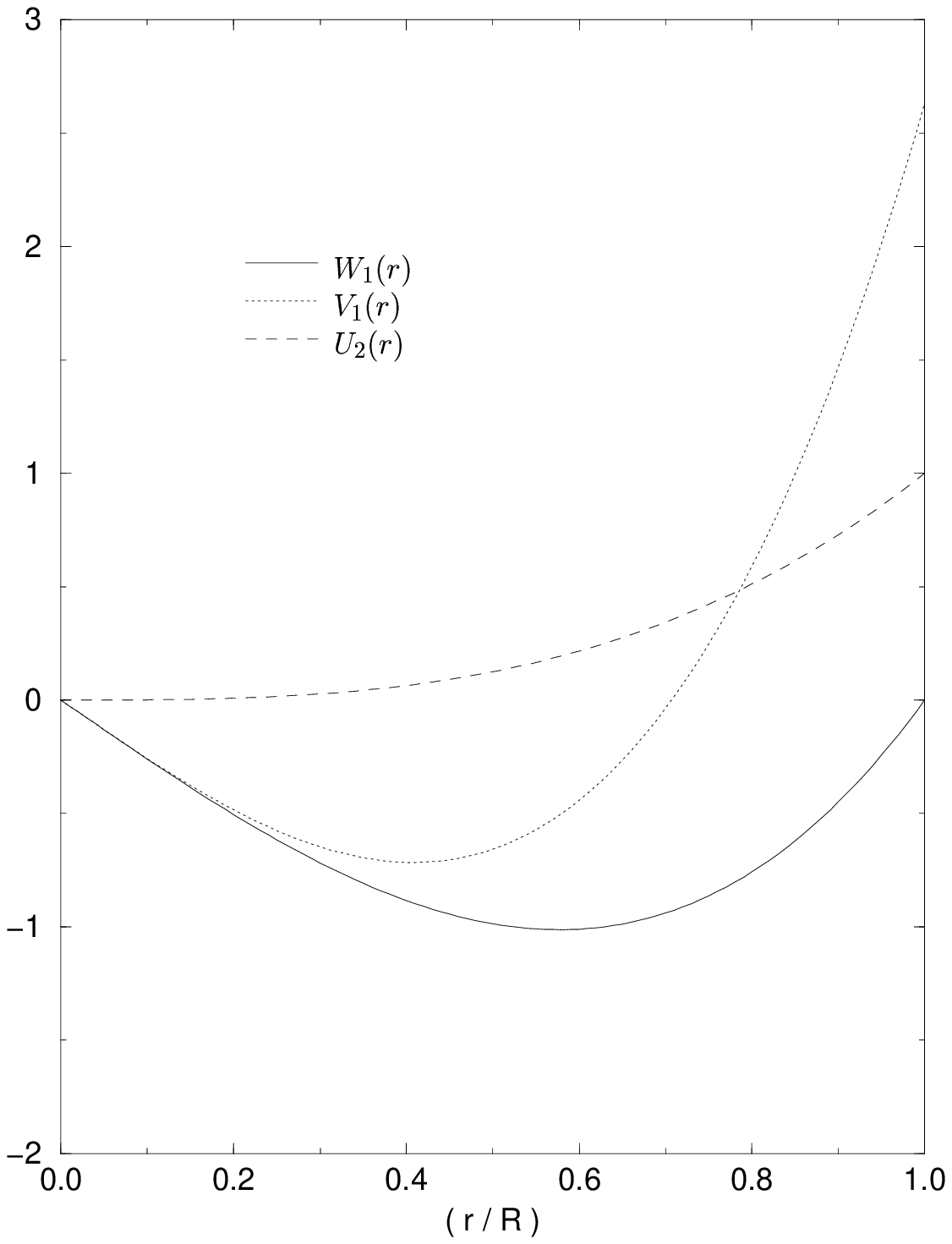}}
\caption{Coefficients $U_l(r)$, $W_l(r)$, and $V_l(r)$ with $l\leq 2$
of the spherical harmonic expansion (\ref{xi_exp}) for the $m=1$
polar-led hybrid mode whose newtonian limit has comoving frequency 
$\kappa_0\Omega=1.509941\Omega$.  The mode is shown in the uniform 
density star with $(2M_0/R)=10^{-6}$ to check for agreement with
the newtonian calculation.  As expected, these functions agree with 
the newtonian forms displayed in Figs. \ref{fig3}-\ref{fig5}, 
up to the difference in the normalization conditions used in the 
newtonian (\ref{N:norm_cond}) and relativistic (\ref{norm_cond})
 calculations and corrections of order $10^{-6}$.}
\label{GR:fig6}
\end{figure}

\begin{figure}
\centerline{\includegraphics[3in,3.25in][6in,8.25in]{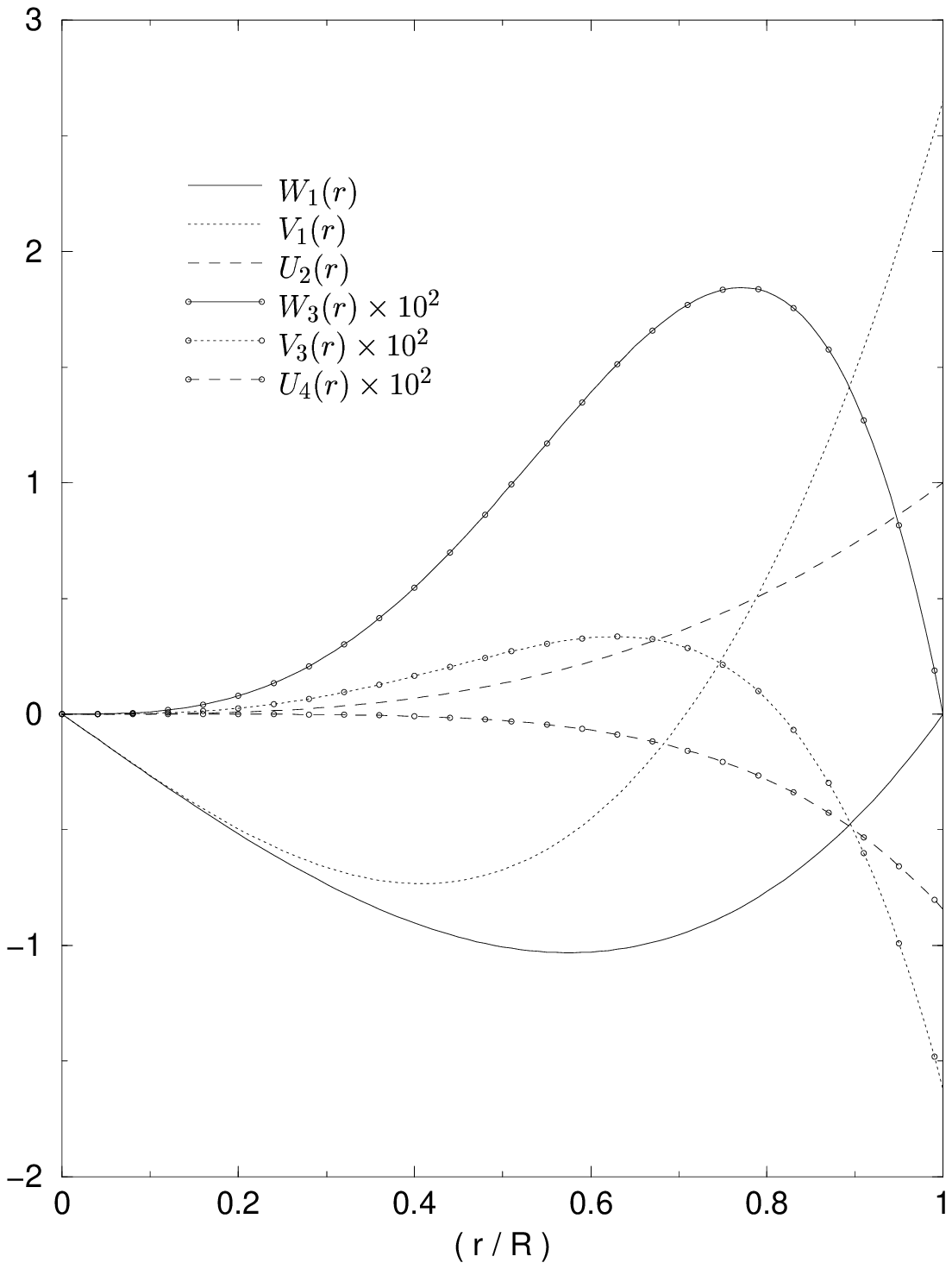}}
\caption{Coefficients $U_l(r)$, $W_l(r)$, and $V_l(r)$ with $l\leq 4$
of the spherical harmonic expansion (\ref{xi_exp}) for the same mode
shown in Fig. \ref{GR:fig6} but now with $(2M_0/R)=0.05$.  The 
comoving frequency has shifted to $\kappa\Omega=1.4802\Omega$, and
the mode has acquired relativistic corrections of order $1\%$.
Accordingly, the functions $W_1$, $V_1$ and $U_2$ are 
indistinguishable from those shown in Fig. \ref{GR:fig6}.
The coefficients of expansion (\ref{xi_exp}) with $l>6$ are of order 
$0.01\%$ and smaller and are not shown.}
\label{GR:fig7}
\end{figure}

\begin{figure}
\centerline{\includegraphics[3in,3.25in][6in,8.25in]{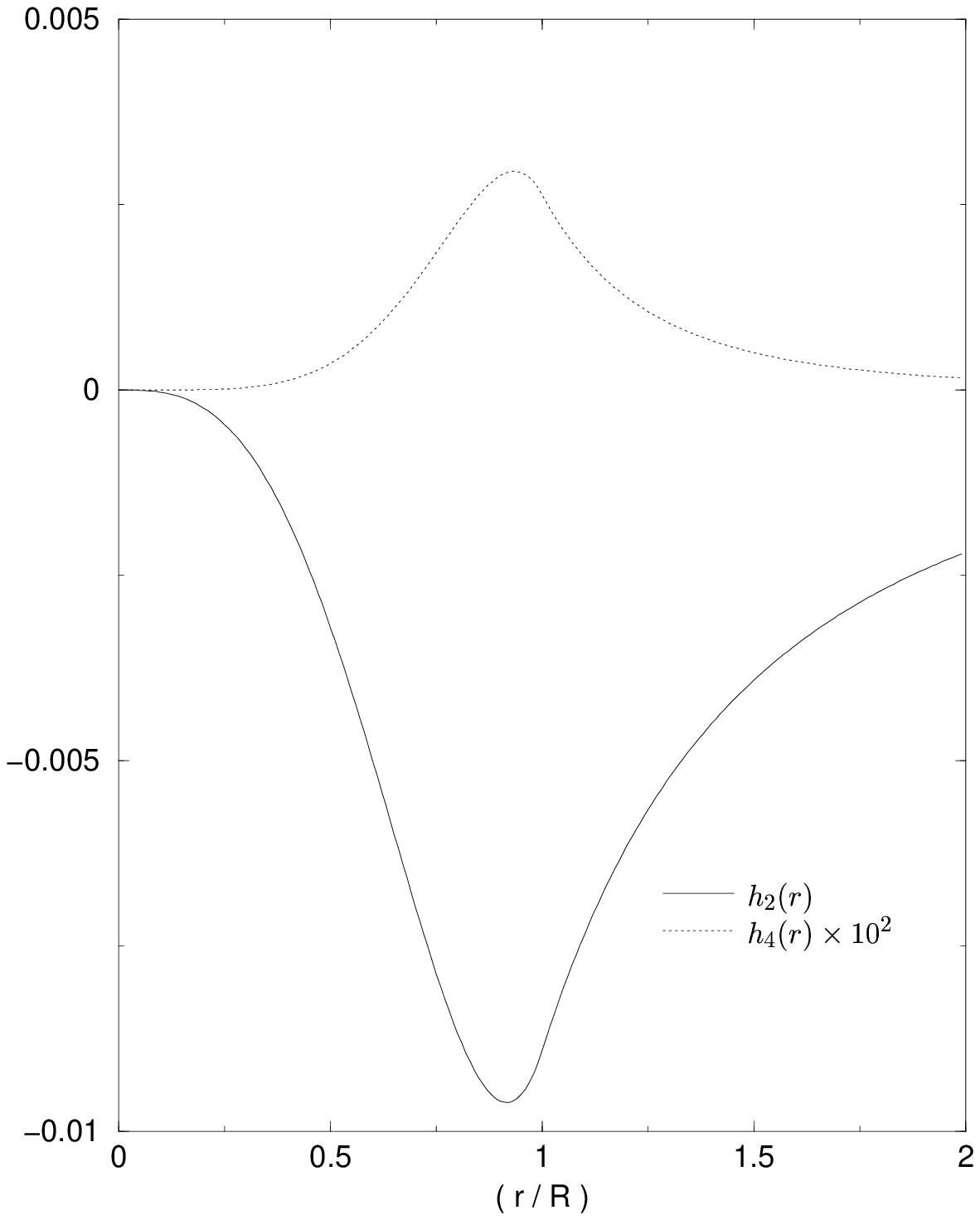}}
\caption{Coefficients $h_l(r)(\equiv h_{0,l}(r))$ with $l\leq 4$
of the spherical harmonic expansion (\ref{h_components}) for the 
same mode shown in Fig. \ref{GR:fig7} in the uniform density star 
with $(2M_0/R)=0.05$.  The vertical scale is the same as that of 
Fig. \ref{GR:fig7} and is set by the normalization of $U_2(r)$.  
The metric perturbation is dominated by $h_2(r)$, which implies
that this stable mode will be rapidly damped by current quadrupole 
radiation.}
\label{GR:fig8}
\end{figure}

Finally, we present a series of plots displaying a polar-led
hybrid mode and its relativistic corrections.  
Figs. \ref{GR:fig6}-\ref{GR:fig8} show the $m=1$ polar-led 
hybrid mode whose newtonian eigenvalue in the uniform density 
star is $\kappa_0=1.509941$. (See Figs. \ref{fig3}-\ref{fig5} 
and Table \ref{ef_m1p}  Fig. \ref{GR:fig6} shows the 
functions $U_l(r)$, $W_l(r)$, and $V_l(r)$ for this mode as 
calculated by the relativistic code in the newtonian regime, 
$(2M_0/R)=10^{-6}$.  As expected, these functions agree with 
the newtonian forms displayed in Figs. \ref{fig3}-\ref{fig5}, 
up to the difference in the normalization conditions used in 
the newtonian (\ref{N:norm_cond}) and relativistic (\ref{norm_cond}) 
calculations and corrections of order $10^{-6}$.  In Fig. \ref{GR:fig7} 
we display the functions $U_l(r)$, $W_l(r)$, and $V_l(r)$ for 
$l\leq 4$ for the same mode, but now with $(2M_0/R)=0.05$, 
and in Fig. \ref{GR:fig8} we display the corresponding 
metric functions $h_l(r)$ for $l\leq 4$.  Observe that 
$W_1$, $V_1$ and $U_2$ are barely distinguishable from their 
newtonian forms and that the mode has acquired relativistic 
corrections of order $1\%$.  Fig. \ref{GR:fig8} shows that
$h_2(r)$ dominates the metric perturbation.  However, since 
the mode is stable, this strong coupling to current quadrupole 
radiation will serve only to damp the mode rapidly.


\clearpage

\clearpage

\clearpage

\clearpage

\clearpage

\clearpage

\clearpage

\clearpage

\clearpage

\clearpage

%
\clearpage
\addcontentsline{toc}{chapter}{\numberline {Appendices}}
\appendix
\vspace*{\fill}
\begin{center}
\Huge\bf
Appendices
\end{center}
\vspace*{\fill}
\vspace*{\fill}

\chapter{Proof of Theorem \ref{thm1}}

\section{Axial-Led Hybrids with $m>0$}

Let $l$ be the smallest value of $l'$ for which $U_{l'}\neq 0$ in the 
spherical harmonic expansion (\ref{v_exp}) of the perturbed velocity 
field $\delta v^a$. The axial parity of $\delta v^a$, $(-1)^{l+1}$, 
and the vanishing of $Y_l^m$ for $l<m$ implies $l\geq m$.  That the 
mode is axial-led means $W_{l'}=0$ and $V_{l'}=0$ for $l'\leq l$.   
We show by contradiction that $l=m$.

Suppose $l\geq m+1$. From equation (\ref{eq2}), 
$\int q^r Y_l^{\ast m} d\Omega = 0$, we have
\be
\left[\half\kappa l(l+1) -m\right] U_l = lQ_{l+1} [W_{l+1}+(l+2)V_{l+1}],
\ee
and from equation (\ref{eq3}) with $l$ replaced by $l-1$, 
$\int q^{\theta} Y_{l-1}^{\ast m}d\Omega = 0$, we have
\be
Q_{l+1} \left[ (l+2)V'_{l+1} + W'_{l+1} \right] = 
\left\{ \left[m+\half\kappa (l+1)\right] U'_l 
+ m(l+1) \frac{U_l}{r} \right\}.
\ee
These two equations, together imply that 
\[
U'_l + \frac{l}{r} U_l = 0,
\]
or 
\[
U_l = K r^{-l},
\]
which is singular at $r=0$.

\section{Axial-Led Hybrids with $m=0$}

Let $m=0$ and let $l$ be the smallest value of $l'$ for which 
$U_{l'}\neq 0$ in the spherical harmonic expansion (\ref{v_exp}) of 
the perturbed velocity field $\delta v^a$.  Since $\nabla_a Y_0^0=0$, 
the mode vanishes unless $l\geq 1$. That the mode is axial-led means 
$W_{l'}=0$ and $V_{l'}=0$ for $l'\leq l$.  We show by contradiction 
that $l=1$.

Suppose $l \geq 2$. Then $\int q^{\varphi} Y_{l-2}^{\ast 0} d\Omega = 0$ 
becomes,
\be
U'_l + \frac{l}{r} U_l = 0,
\ee
or 
\[
U_l = K r^{-l},
\]
which is singular at $r=0$.

\section{Polar-Led Hybrids with $m>0$}

Let $l$ be the smallest value of $l'$ for which $W_{l'}\neq 0$ or 
$V_{l'}\neq 0$ in the spherical harmonic expansion (\ref{v_exp}) 
of the perturbed velocity field $\delta v^a$.  The polar parity of 
$\delta v^a$, $(-1)^l$, and the vanishing of $Y_l^m$ for $l<m$ 
implies $l\geq m$.  That the mode is polar-led means $U_{l'}=0$ 
for $l'\leq l$.   We show by contradiction that $l=m$.

Suppose $l \geq m+1$. Then $\int q^r Y_{l-1}^{\ast m} d\Omega = 0$ 
becomes
\be
W_l+(l+1)V_l = 0,
\ee
and $\int q^{\varphi} Y_{l-1}^{\ast m} d\Omega = 0$ becomes,
\be
0 = 
\begin{array}[t]{l}
- \ds{\left\{\left[\half\kappa (l+1)+m\right] V'_l 
+ m(l+1)\frac{V_l}{r} 
- \half\kappa (l+1)\frac{W_l}{r} 
\right\}} \\
\\
+\ds{ (l+2) Q_{l+1} \left[ U'_{l+1} + (l+1) \frac{U_{l+1}}{r} \right] }
\end{array}
\ee
These two equations, together imply that
\[
- \left[\half\kappa (l+1)+m\right] \left[V'_l  + (l+1) \frac{V_l}{r} \right]
+ (l+2) Q_{l+1} \left[ U'_{l+1} + (l+1) \frac{U_{l+1}}{r} \right]
= 0,
\]
or
\[
- \left[\half\kappa (l+1)+m\right] V_l 
+ (l+2) Q_{l+1} U_{l+1} = K r^{-{(l+1)}},
\]
which is singular at $r=0$.

\section{Polar-Led Hybrids with $m=0$}

Let $m=0$ and let $l$ be the smallest value of $l'$ for which 
$W_{l'}\neq 0$ or $V_{l'}\neq 0$ in the spherical harmonic expansion 
(\ref{v_exp}) of the perturbed velocity field $\delta v^a$. When $l=0$ 
the mode is automatically polar-led; thus we need only consider the 
case $l\geq 1$. That the mode is polar-led means $U_{l'}=0$ for 
$l'\leq l$.  We show by contradiction that $l=1$.

Suppose $l \geq 2$. Then $\int q^r Y_{l-1}^{\ast 0} d\Omega = 0$ becomes
\be
W_l+(l+1)V_l = 0,
\ee
and $\int q^{\varphi} Y_{l-1}^{\ast 0} d\Omega = 0$ becomes,
\be
- \half\kappa (l+1)\left[V'_l - \frac{W_l}{r} \right]
+ (l+2) Q_{l+1} \left[ U'_{l+1} + (l+1) \frac{U_{l+1}}{r} \right].
= 0
\ee
These two equations, together imply that
\[
- \half\kappa (l+1) \left[V'_l  + (l+1) \frac{V_l}{r} \right]
+ (l+2) Q_{l+1} \left[ U'_{l+1} + (l+1) \frac{U_{l+1}}{r} \right] = 0,
\]
or
\[
- \half\kappa (l+1) V_l + (l+2) Q_{l+1} U_{l+1} = K r^{-{(l+1)}},
\]
which is singular at $r=0$.


\chapter{Algebraic Equations: Newtonian}
In this appendix, we make use of the following definitions:
\begin{eqnarray}
a_l &\equiv& \half\kappa m +(l+1)Q_l^2 - l Q_{l+1}^2 \\
b_l &\equiv& m^2 - l(l+1)\left(1- Q_l^2 - Q_{l+1}^2\right) \\
c_l &\equiv& \half\kappa l(l+1) - m 
\end{eqnarray}
For reference, we repeat the definitions (\ref{Q_l}) and (\ref{kappa}):
\begin{eqnarray}
\kappa  &\equiv& \frac{(\sigma + m\Omega)}{\Omega} \\
Q_l     &\equiv& \left[ \frac{(l+m)(l-m)}{(2l-1)(2l+1)} \right]^{\half}
\end{eqnarray}

\section{Axial-Led Hybrids}

\noindent For $l=m,\:m+2,\:m+4,\ldots$ the regular series 
expansions\footnote{We present 
the form of the series expansions for $U_l(r)$ for reference; 
however, we do not need 
these series since we eliminate the $U_l(r)$ using equation 
(\ref{eq2}).} about
the center of the star, $r=0$, are

\begin{eqnarray}
W_{m+j+1}(r) &=& \left(\frac{r}{R}\right)^{m+j} 
\sum^\infty_{\stackrel{i=1}{\mbox{\tiny $i$ odd}}} 
w_{j+1,i} \left(\frac{r}{R}\right)^i                    
\label{ax_x1}\\
V_{m+j+1}(r) &=& \left(\frac{r}{R}\right)^{m+j} 
\sum^\infty_{\stackrel{i=1}{\mbox{\tiny $i$ odd}}} 
v_{j+1,i} \left(\frac{r}{R}\right)^i                     
\label{ax_x3}\\
U_{m+j}(r) &=& \left(\frac{r}{R}\right)^{m+j} 
\sum^\infty_{\stackrel{i=1}{\mbox{\tiny $i$ odd}}} 
u_{j,i} \left(\frac{r}{R}\right)^i
\end{eqnarray} 

where $j=0,2,4,\ldots$.

\noindent The regular series expansions about $r=R$, which 
satisfy the boundary 
condition $\Delta p = 0$ are
\begin{eqnarray}
W_{m+j+1}(r) &=& \sum^\infty_{k=1} \tilde w_{j+1,k} 
\left(1-\frac{r}{R}\right)^k
\label{ax_y1}\\
V_{m+j+1}(r) &=& \sum^\infty_{k=0} \tilde v_{j+1,k} 
\left(1-\frac{r}{R}\right)^k
\label{ax_y3}\\
U_{m+j}(r) &=& \sum^\infty_{k=0} \tilde u_{j,k} 
\left(1-\frac{r}{R}\right)^k
\end{eqnarray} 
where $j=0,2,4,\ldots$.

\noindent
These series expansions must agree in the interior of the star. 
We impose the matching condition that the series 
(\ref{ax_x1})-(\ref{ax_x3}) truncated at $i_{\mbox{\tiny max}}$ 
be equal at the point $r=r_0$ to the corresponding series 
(\ref{ax_y1})-(\ref{ax_y3}) truncated at $k_{\mbox{\tiny max}}$. 
That is,
\begin{eqnarray}
0 &=& \left(\frac{r_0}{R}\right)^{m+j} 
\sum^{i_{\mbox{\tiny max}}}_{\stackrel{i=1}{\mbox{\tiny $i$ odd}}} 
w_{j+1,i} \left(\frac{r_0}{R}\right)^i                    
- \sum^{k_{\mbox{\tiny max}}}_{k=1} \tilde w_{j+1,k} 
\left(1-\frac{r_0}{R}\right)^k
\label{ax_match_a}
\\
0 &=& \left(\frac{r_0}{R}\right)^{m+j} 
\sum^{i_{\mbox{\tiny max}}}_{\stackrel{i=1}{\mbox{\tiny $i$ odd}}} 
v_{j+1,i} \left(\frac{r_0}{R}\right)^i                     
- \sum^{k_{\mbox{\tiny max}}}_{k=0} \tilde v_{j+1,k} 
\left(1-\frac{r_0}{R}\right)^k
\label{ax_match_b}
\end{eqnarray}

\noindent 
When we substitute (\ref{ax_x1})-(\ref{ax_x3}) and (\ref{rho_x}) 
into (\ref{eq1}), the 
coefficient of $(r/R)^{m+j+i}$ in the resulting equation is
\be
0 = \ba[t]{l} \ds{(m+j+i+1)w_{j+1,i} 
+ \sum^{i-2}_{\stackrel{s=1}{\mbox{\tiny $s$ odd}}} \pi_s 
\, w_{j+1,i-s-1}} \\
\\
- \ds{(m+j+1)(m+j+2)v_{j+1,i}}
\label{ax_1}
\ea
\ee

\noindent Similarly, when we substitute (\ref{ax_y1})-(\ref{ax_y3}) 
and (\ref{rho_y}) into 
(\ref{eq1}), the coefficient of $[1-(r/R)]^k$ in the resulting equation is
\be
0 = \ba[t]{l} 
\ds{(k+1) \left[\tilde w_{j+1,k} - \tilde w_{j+1,k+1} \right]
+ \sum^k_{s=0} \left(\tilde\pi_{s-1} - \tilde\pi_{s-2} \right) 
\tilde w_{j+1,k-s+1}} \\
\\
- \ds{(m+j+2)(m+j+1)\tilde v_{j+1,k}}
\label{ax_2}
\ea
\ee
where we have defined $\tilde \pi_{-2} \equiv 0 \equiv \tilde w_{j+1,0}$.

\noindent When we use (\ref{eq2}) to eliminate the $U_l(r)$ from (\ref{eq4}) 
and then substitute for the $W_{l\pm 1}(r)$ and $V_{l\pm 1}(r)$ using 
(\ref{ax_x1})-(\ref{ax_x3}), the coefficient of $(r/R)^{m+j+i}$ in the 
resulting equation is
\begin{eqnarray}
0 &=& (i+1)(m+j-2)(m+j-1)Q_{m+j}Q_{m+j-1}Q_{m+j-2}c_{m+j}c_{m+j+2}
\label{ax_3} \\
&& \nonumber\\
&& \hspace{2.0in}
\mbox{} \times \Biggl[ w_{j-3,i+4} - (m+j-3) v_{j-3,i+4} \Biggr]
\nonumber\\
&& \nonumber\\
&&\mbox{} - \Biggl\{ \ba[t]{l}
	(i+1)(m+j-2)^2Q^2_{m+j-1}c_{m+j}
	+\half\kappa (m+j-1)c_{m+j-2}c_{m+j} \\
	\\
	+ (m+j+1) \left[(m+j+i)a_{m+j}+b_{m+j}\right] c_{m+j-2}
	\Biggr\} \\
	\\
	\times Q_{m+j} c_{m+j+2} \ w_{j-1,i+2} \vphantom{\Biggl\{}
\ea \nonumber \\
&& \nonumber \\
&&\mbox{} + \Biggl\{  \ba[t]{l}
	\left[\half\kappa (m+j-1)(m+j+i)-(i+1)m\right] c_{m+j-2}c_{m+j} \\
	\\
	+ (m+j+1)(m+j-1) \left[(m+j+i)a_{m+j}+b_{m+j} \right] c_{m+j-2} \\
	\\
	-  (i+1)(m+j)(m+j-2)^2Q^2_{m+j-1}c_{m+j} \Biggr\} \\
	\\
	\times Q_{m+j}c_{m+j+2} \ v_{j-1,i+2}\vphantom{\Biggl\{}
\ea \nonumber\\
&& \nonumber \\
&&\mbox{} + \Biggl\{ \ba[t]{l} 
	\half\kappa (m+j+2) c_{m+j}c_{m+j+2} \\
	\\
	+ (m+j)\left[(m+j+i)a_{m+j}+b_{m+j}\right] c_{m+j+2} \\
	\\
	- (2m+2j+i+2)(m+j+3)^2Q^2_{m+j+2}c_{m+j}
	\Biggr\} \\ 
	\\ \times Q_{m+j+1}c_{m+j-2} \ w_{j+1,i} \vphantom{\Biggl\{}
\ea \nonumber\\
&& \nonumber \\
&&\mbox{} + \Biggl\{ \ba[t]{l} 
	(m+j+2)(m+j)\left[(m+j+i)a_{m+j}+b_{m+j}\right]c_{m+j+2} \\
	\\
	- \left[\half\kappa (m+j+2)(m+j+i)
	+m(2m+2j+i+2)\right] c_{m+j}c_{m+j+2} \\
	\\
	+ (2m+2j+i+2)(m+j+3)^2(m+j+1)Q^2_{m+j+2}c_{m+j}
	\Biggr\} \\
	\\
	\times Q_{m+j+1}c_{m+j-2} \ v_{j+1,i} \vphantom{\Biggl\{}
\ea \nonumber\\
&& \nonumber \\
&&\mbox{} + (2m+2j+i+2)(m+j+3)(m+j+2)Q_{m+j+3}Q_{m+j+2}Q_{m+j+1} 
\nonumber\\
&& \nonumber \\
&& \hspace{1.0in}
\mbox{} \times c_{m+j-2}c_{m+j} 
\Biggl[ w_{j+3,i-2} + (m+j+4) v_{j+3,i-2} \Biggr] \nn
\end{eqnarray}

\noindent When we use (\ref{eq2}) to eliminate the $U_l(r)$ 
from (\ref{eq4}) and 
then substitute for the $W_{l\pm 1}(r)$ and $V_{l\pm 1}(r)$ 
using (\ref{ax_y1})-(\ref{ax_y3}), 
the coefficient of $[1-(r/R)]^k$ in the resulting equation is
\begin{eqnarray}
0 &=& - (m+j-k-1)(m+j-1)(m+j-2)Q_{m+j}Q_{m+j-1}Q_{m+j-2}
\label{ax_4}\\
&& \nonumber \\
&& \hspace{1.0in}
\mbox{} \times c_{m+j}c_{m+j+2} \Biggl[ \tilde w_{j-3,k} 
- (m+j-3) \tilde v_{j-3,k} \Biggr]
\nonumber\\
&& \nonumber \\
&&\mbox{} - (k+1)(m+j-1)(m+j-2)Q_{m+j}Q_{m+j-1}Q_{m+j-2}c_{m+j}c_{m+j+2}
\nonumber\\
&& \nonumber \\
&& \hspace{2.0in}
\mbox{} \times \Biggl[ \tilde w_{j-3,k+1} 
- (m+j-3) \tilde v_{j-3,k+1} \Biggr]
\nonumber\\
&& \nonumber \\
&&\mbox{} + \Biggl\{ \ba[t]{l}
	(m+j-k-1)(m+j-2)^2Q^2_{m+j-1}c_{m+j} \\
	\\
	- \half\kappa(m+j-1)c_{m+j-2}c_{m+j} \\
	\\
	-(m+j+1)\left(b_{m+j}+ka_{m+j}\right)c_{m+j-2}
	\Biggr\} \\
	\\
	\times Q_{m+j}c_{m+j+2} \ \tilde w_{j-1,k} \vphantom{\Biggl\{}
\ea
\nonumber\\
&& \nonumber \\
&&\mbox{} + \Biggl\{ \ba[t]{l}
(m+j-2)^2Q^2_{m+j-1}c_{m+j} + (m+j+1)a_{m+j}c_{m+j-2}
\Biggr\} \\
\\
\times (k+1)Q_{m+j}c_{m+j+2} \ \tilde w_{j-1,k+1} \vphantom{\Biggl\{}
\ea
\nonumber\\
&& \nonumber \\
&&\mbox{} + \Biggl\{ \ba[t]{l}
	(m+j-k-1)(m+j-2)^2(m+j)Q^2_{m+j-1}c_{m+j} \\
	\\
	+ \left[\half\kappa k(m+j-1)+m(m+j-k-1)\right] c_{m+j-2}c_{m+j} \\
	\\
	+ (m+j+1)(m+j-1)\left(b_{m+j}+ka_{m+j}\right) c_{m+j-2}
	\Biggr\} \\
	\\
	\times Q_{m+j}c_{m+j+2} \ \tilde v_{j-1,k} \vphantom{\Biggl\{}
\ea
\nonumber\\
&& \nonumber \\
&&\mbox{} + \Biggl\{ \ba[t]{l}
	(m+j)(m+j-2)^2Q^2_{m+j-1}c_{m+j} \\
	\\
	+ \left[m-\half\kappa (m+j-1)\right] c_{m+j-2}c_{m+j} \\
	\\
	-(m+j+1)(m+j-1)a_{m+j}c_{m+j-2}
	\Biggr\} \\
	\\
	\times (k+1)Q_{m+j}c_{m+j+2} \ \tilde v_{j-1,k+1}\vphantom{\Biggl\{}
\ea
\nonumber\\
&& \nonumber \\
&&\mbox{} + \Biggl\{ \ba[t]{l}
	(m+j)\left(b_{m+j}+ka_{m+j}\right)c_{m+j+2} \\
	\\
	+ \half\kappa (m+j+2)c_{m+j}c_{m+j+2} \\
	\\
	- (m+j+k+2)(m+j+3)^2Q^2_{m+j+2}c_{m+j}
	\Biggr\} \\
	\\
	\times Q_{m+j+1}c_{m+j-2} \ \tilde w_{j+1,k} \vphantom{\Biggl\{}
\ea
\nonumber\\
&& \nonumber \\
&&\mbox{} + \Biggl\{ \ba[t]{l}
-(m+j)a_{m+j}c_{m+j+2}+(m+j+3)^2Q^2_{m+j+2}c_{m+j}
\Biggr\} \\
\\
(k+1)Q_{m+j+1}c_{m+j-2} \ \tilde w_{j+1,k+1} \vphantom{\Biggl\{}
\ea
\nonumber\\
&& \nonumber \\
&&\mbox{} + \Biggl\{ \ba[t]{l}
	(m+j+2)(m+j)\left(b_{m+j}+ka_{m+j}\right)c_{m+j+2} \\
	\\
	- \left[m(m+j+k+2)+\half\kappa k(m+j+2)\right] c_{m+j}c_{m+j+2} \\
	\\
	+(m+j+k+2)(m+j+3)^2(m+j+1)Q^2_{m+j+2}c_{m+j}
	\Biggr\} \\
	\\
	\times Q_{m+j+1}c_{m+j-2} \ \tilde v_{j+1,k} \vphantom{\Biggl\{}
\ea
\nonumber\\
&& \nonumber \\
&&\mbox{} + \Biggl\{ \ba[t]{l}
	-(m+j+2)(m+j)a_{m+j}c_{m+j+2} \\
	\\
	+\left[\half\kappa (m+j+2)+m\right] c_{m+j}c_{m+j+2} \\
	\\
	-(m+j+3)^2(m+j+1)Q^2_{m+j+2}c_{m+j}
	\Biggr\} \\
	\\
	\times (k+1)Q_{m+j+1}c_{m+j-2} \ \tilde v_{j+1,k+1}
	\vphantom{\Biggl\{}
\ea
\nonumber\\
&& \nonumber \\
&&\mbox{} + (m+j+k+2)(m+j+3)(m+j+2)Q_{m+j+3}Q_{m+j+2}Q_{m+j+1}
\nonumber\\
&& \nonumber \\
&& \hspace{1.0in}
\mbox{} \times c_{m+j-2}c_{m+j} \Biggl[ \tilde w_{j+3,k} 
+ (m+j+4) \tilde v_{j+3,k} \Biggr]
\nonumber\\
&& \nonumber \\
&&\mbox{} - (k+1)(m+j+3)(m+j+2)Q_{m+j+3}Q_{m+j+2}Q_{m+j+1}c_{m+j-2}c_{m+j}
\nonumber\\
&& \nonumber \\
&& \hspace{2.0in}
\mbox{} \times \Biggl[ \tilde w_{j+3,k+1} + (m+j+4) \tilde v_{j+3,k+1} \Biggr]
\nonumber 
\end{eqnarray}

The equations (\ref{ax_match_a}) through (\ref{ax_4}) make up the 
algebraic system 
(\ref{linalg}) for eigenvalues of the axial-led hybrid modes.  
One truncates the angular 
and radial series expansions at indices $j_{\mbox{\tiny max}}$, 
$i_{\mbox{\tiny max}}$ and 
$k_{\mbox{\tiny max}}$ and constructs the matrix $A$ by keeping 
the appropriate number of 
equations for the number of unknown coefficients $w_{j+1,i}$, 
$v_{j+1,i}$, $\tilde w_{j+1,k}$ and $\tilde v_{j+1,k}$. 
In following this procedure, however, one must 
be aware of the following subtlety in the equations.

For each $q\equiv j+i$ the set of equations
\[
\begin{array}{ll}
\mbox{(\ref{ax_1})} & \mbox{with} \ \ i=1 \ \ 
\mbox{and} \ \ j=q-1, \ \ \mbox{and} \\
\mbox{(\ref{ax_3})} & \mbox{for all} \ \ i=1,3,\ldots,q \ \ 
\mbox{with} \ \ j=q-i
\end{array}
\]
can be shown to be linearly dependent for arbitrary $\kappa$ 
and for any equilibrium
stellar model. For example, taking the simplest case of $q=1$, 
one can show that equation 
(\ref{ax_1}) with $i=1$ and $j=0$ becomes
\[
0 = (m+2) \left[w_{1,1} - (m+1) v_{1,1} \right] 
\]
while equation (\ref{ax_3}) with $i=1$ and $j=0$ becomes
\[
0 = \Biggl\{ \begin{array}[t]{l}
\half\kappa(m+2)c_m c_{m+2} 
+ m\left[(m+1)a_m + b_m \right] \\
\\
- (2m+3)(m+3)^2 Q^2_{m+2}c_m \Biggr\}
Q_{m+1}c_{m-2} \left[w_{1,1} - (m+1) v_{1,1} \right].
\end{array}
\]

This problem can be solved by eliminating one of these equations 
from the subset for 
each $q$ (for example, equation (\ref{ax_3}) with $i=1$).  Thus, 
to properly construct the 
algebraic system (\ref{linalg}) we use, for all 
$j=0,2,\ldots,j_{\mbox{\tiny max}}$, 
the equations
\[
\begin{array}{ll}
\mbox{(\ref{ax_match_a})} & \\
\mbox{(\ref{ax_match_b})} & \\
\mbox{(\ref{ax_1})} & \mbox{with} \ \ i=1,3,\ldots,i_{\mbox{\tiny max}} \\ 
\mbox{(\ref{ax_2})} & \mbox{with} \ \ k=0,1,\ldots,k_{\mbox{\tiny max}}-1 \\ 
\mbox{(\ref{ax_3})} & \mbox{with} \ \ i=3,5,\ldots,i_{\mbox{\tiny max}} \\
\mbox{(\ref{ax_4})} & \mbox{with} \ \ k=0,1,\ldots,k_{\mbox{\tiny max}}-1. \\ 
\end{array}
\]

\section{Polar-Led Hybrids}

\noindent For $l=m,\:m+2,\:m+4,\ldots$ the regular series 
expansions\footnote{We present 
the form of the series expansions for $U_l(r)$ for reference; 
however, we do not need 
these series since we eliminate the $U_l(r)$ using equation 
(\ref{eq2}).} about
the center of the star, $r=0$, are

\begin{eqnarray}
W_{m+j}(r) &=& \left(\frac{r}{R}\right)^{m+j} 
\sum^\infty_{\stackrel{i=0}{\mbox{\tiny $i$ even}}} 
w_{j,i} \left(\frac{r}{R}\right)^i
\label{po_x1}\\
V_{m+j}(r) &=& \left(\frac{r}{R}\right)^{m+j} 
\sum^\infty_{\stackrel{i=0}{\mbox{\tiny $i$ even}}} 
v_{j,i} \left(\frac{r}{R}\right)^i                    
\label{po_x2}\\
U_{m+j+1}(r) &=& \left(\frac{r}{R}\right)^{m+j} 
\sum^\infty_{\stackrel{i=2}{\mbox{\tiny $i$ even}}} 
u_{j+1,i} \left(\frac{r}{R}\right)^i
\end{eqnarray}

where $j=0,2,4,\ldots$.

\noindent The regular series expansions about $r=R$, which 
satisfy the boundary 
condition $\Delta p = 0$ are

\begin{eqnarray}
W_{m+j}(r) &=& 
\sum^\infty_{k=1} \tilde w_{j,k} \left(1-\frac{r}{R}\right)^k
\label{po_y1}\\
V_{m+j}(r) &=& 
\sum^\infty_{k=0} \tilde v_{j,k} \left(1-\frac{r}{R}\right)^k
\label{po_y2}\\
U_{m+j+1}(r) &=& 
\sum^\infty_{k=0} \tilde u_{j+1,k} \left(1-\frac{r}{R}\right)^k
\end{eqnarray}

where $j=0,2,4,\ldots$.

\noindent
These series expansions must agree in the interior of the star. 
We impose the matching 
condition that the series (\ref{po_x1})-(\ref{po_x2}) truncated at 
$i_{\mbox{\tiny max}}$ be equal at the 
point $r=r_0$ to the corresponding series (\ref{po_y1})-(\ref{po_y2})
truncated at $k_{\mbox{\tiny max}}$. 
That is,

\begin{eqnarray}
0 &=& \left(\frac{r_0}{R}\right)^{m+j} 
\sum^{i_{\mbox{\tiny max}}}_{\stackrel{i=0}{\mbox{\tiny $i$ even}}} 
w_{j,i} \left(\frac{r_0}{R}\right)^i                    
- \sum^{k_{\mbox{\tiny max}}}_{k=1} \tilde w_{j,k} 
\left(1-\frac{r_0}{R}\right)^k
\label{po_match_a}
\\
0 &=& \left(\frac{r_0}{R}\right)^{m+j} 
\sum^{i_{\mbox{\tiny max}}}_{\stackrel{i=0}{\mbox{\tiny $i$ even}}} 
v_{j,i} \left(\frac{r_0}{R}\right)^i                     
- \sum^{k_{\mbox{\tiny max}}}_{k=0} \tilde v_{j,k} 
\left(1-\frac{r_0}{R}\right)^k
\label{po_match_b}
\end{eqnarray}

\noindent 
When we substitute (\ref{po_x1})-(\ref{po_x2}) and (\ref{rho_x}) 
into (\ref{eq1}), the 
coefficient of $(r/R)^{m+j+i}$ in the resulting equation is
\be
0 = (m+j+i+1)w_{j,i} + \sum^{i-2}_{\stackrel{s=0}{\mbox{\tiny $s$ even}}} 
\pi_{s+1} \, w_{j,i-s-2} - (m+j)(m+j+1)v_{j,i}        \label{po_1}
\ee

\noindent Similarly, when we substitute (\ref{po_y1})-(\ref{po_y2}) and 
(\ref{rho_y}) into 
(\ref{eq1}), the coefficient of $[1-(r/R)]^k$ in the resulting equation is
\be
0 = \ba[t]{l}
\ds{(k+1) \left[\tilde w_{j,k} - \tilde w_{j,k+1} \right]
+ \sum^k_{s=0} \left(\tilde\pi_{s-1} 
- \tilde\pi_{s-2} \right) \tilde w_{j,k-s+1}} \\
\\
- (m+j)(m+j+1)\tilde v_{j,k}
\ea
\label{po_2}
\ee
where we have defined $\tilde \pi_{-2} \equiv 0 \equiv \tilde w_{j,0}$.

\noindent When we use (\ref{eq2}) to eliminate the $U_l(r)$ from (\ref{eq3}) 
and then substitute for the $W_{l\pm 1}(r)$ and $V_{l\pm 1}(r)$ using 
(\ref{po_x1})-(\ref{po_x2}), the coefficient of $(r/R)^{m+j+i}$ in the 
resulting equation is
\begin{eqnarray}
0 &=& - im(m+j-1)Q_{m+j}Q_{m+j-1} c_{m+j+1} 
\label{po_3}\\
&&\nonumber\\
&& \hspace{1.0in}
\mbox{} \times \Biggl[ w_{j-2, i+2} - (m+j-2)v_{j-2, i+2} \Biggr]
\nonumber\\
&&\nonumber\\
&&\mbox{} + \Biggl\{ 
[(i+1)m-\half\kappa(m+j-1)(m+j+i)](m+j-1)Q^2_{m+j} c_{m+j+1}
\nonumber\\
&&\nonumber\\
& & \hspace{0.5in}
\mbox{} + \left[
(m+j+i)\left(1- Q_{m+j}^2 - Q_{m+j+1}^2\right)+\half\kappa m
\right] c_{m+j-1} c_{m+j+1} 
\nonumber\\
&&\nonumber\\
& & \hspace{0.5in}
\mbox{} - [m(2m+2j+i+2)+\half\kappa (m+j+2)(m+j+i)] 
\nonumber\\
&&\nonumber\\
& & \hspace{1.0in}
\mbox{} \times (m+j+2)Q^2_{m+j+1} c_{m+j-1} \Biggr\}  w_{j,i} 
\nonumber\\
&&\nonumber\\
&&\mbox{} + \Biggl\{ [(i+1)m-\half\kappa (m+j-1)(m+j+i)]
\nonumber\\
&&\nonumber\\
&& \hspace{1.0in}
\mbox{} \times (m+j-1)(m+j+1)Q^2_{m+j} c_{m+j+1}
\nonumber\\
&&\nonumber\\
&& \hspace{0.5in}
\mbox{} - \left[ m^2 + (m+j+i)a_{m+j} \right]  c_{m+j-1} c_{m+j+1} 
\nonumber\\
&&\nonumber\\
&& \hspace{0.5in}
\mbox{} + \left[m(2m+2j+i+2)+ \half\kappa (m+j+2)(m+j+i)\right] 
\nonumber\\
&&\nonumber\\
&& \hspace{1.0in}
\mbox{} \times (m+j)(m+j+2)Q^2_{m+j+1}c_{m+j-1} \Biggr\}  v_{j,i}
\nonumber\\
&&\nonumber\\
&&\mbox{} + Q_{m+j+2}Q_{m+j+1}\left[m(m+j+i)+m(m+j+1)(2m+2j+i+2)\right] 
\nonumber\\
&&\nonumber\\
&& \hspace{1.0in}
\mbox{} \times c_{m+j-1} \Biggl[ w_{j+2, i-2} + (m+j+3)v_{j+2, i-2} \Biggr]
\nonumber
\end{eqnarray}

\noindent When we use (\ref{eq2}) to eliminate the $U_l(r)$ 
from (\ref{eq3}) and 
then substitute for the $W_{l\pm 1}(r)$ and $V_{l\pm 1}(r)$ 
using (\ref{po_y1})-(\ref{po_y2}), 
the coefficient of $[1-(r/R)]^k$ in the resulting equation is
\begin{eqnarray}
0 &=& m(m+j-1)(m+j-k) Q_{m+j}Q_{m+j-1} c_{m+j+1} 
\label{po_4}\\
&&\nonumber\\
&& \hspace{1.0in}
\mbox{} \times \Biggl[ \tilde w_{j-2,k} 
- (m+j-2) \tilde v_{j-2, k} \Biggr]
\nonumber\\
&&\nonumber\\
&&\mbox{} + (k+1)m(m+j-1) Q_{m+j}Q_{m+j-1} c_{m+j+1}
\nonumber\\
&&\nonumber\\
&& \hspace{1.0in}
\mbox{} \times \Biggl[ \tilde w_{j-2,k+1} 
- (m+j-2) \tilde v_{j-2, k+1} \Biggr]
\nonumber\\
&&\nonumber\\
&&\mbox{} + \Biggl\{ 
- [(\half\kappa k + m)(m+j-1)-km] (m+j-1) Q^2_{m+j} c_{m+j+1}
\nonumber\\
&&\nonumber\\
&& \hspace{0.5in}
\mbox{} + \left[
\half\kappa m + k\left(1- Q_{m+j}^2 - Q_{m+j+1}^2\right) \right] 
c_{m+j-1} c_{m+j+1}
\nonumber\\
&&\nonumber\\
&& \hspace{0.5in}
\mbox{} - [(\half\kappa k+m)(m+j+2)+km] 
\nonumber\\
&&\nonumber\\
&& \hspace{1.0in}
\mbox{} \times (m+j+2) Q^2_{m+j+1} c_{m+j-1} \Biggr\} \tilde w_{j,k} 
\nonumber\\
&&\nonumber\\
&&\mbox{} - (k+1) \Biggl\{ 
[m-\half\kappa (m+j-1)] (m+j-1)Q^2_{m+j} c_{m+j+1}
\nonumber\\
&&\nonumber\\
&& \hspace{1.0in}
\mbox{} + \left(1- Q_{m+j}^2 - Q_{m+j+1}^2\right) c_{m+j-1} c_{m+j+1}
\nonumber\\
&&\nonumber\\
&& \hspace{1.0in}
\mbox{} - [m+\half\kappa (m+j+2)] 
\nonumber\\
&&\nonumber\\
&& \hspace{1.0in}
\mbox{} \times (m+j+2) Q^2_{m+j+1} c_{m+j-1} \Biggr\} \tilde w_{j,k+1} 
\nonumber\\
&&\nonumber\\
&&\mbox{} + \Biggl\{ 
-[(\half\kappa k +m)(m+j-1)-km] 
\nonumber\\
&&\nonumber\\
&& \hspace{1.0in}
\mbox{} \times (m+j-1)(m+j+1) Q^2_{m+j} c_{m+j+1}
\nonumber\\
&&\nonumber\\
&& \hspace{0.5in}
\mbox{} - \left( m^2+ka_{m+j}\right) c_{m+j-1} c_{m+j+1}
\nonumber\\
&&\nonumber\\
&& \hspace{0.5in}
\mbox{} + [(\half\kappa k +m)(m+j+2)+km] 
\nonumber\\
&&\nonumber\\
&& \hspace{1.0in}
\mbox{} \times (m+j)(m+j+2) Q^2_{m+j+1} c_{m+j-1} \Biggr\} \tilde v_{j,k} 
\nonumber\\
&&\nonumber\\
&&\mbox{} + (k+1)\Biggl\{ 
- [m-\half\kappa (m+j-1)] 
\nonumber\\
&&\nonumber\\
&& \hspace{1.0in}
\mbox{} \times (m+j-1)(m+j+1) Q^2_{m+j} c_{m+j+1}
\nonumber\\
&&\nonumber\\
&& \hspace{1.0in}
\mbox{} + a_{m+j} c_{m+j-1} c_{m+j+1}
\nonumber\\
&&\nonumber\\
&& \hspace{1.0in}
\mbox{} - [m+\half\kappa (m+j+2)] 
\nonumber\\
&&\nonumber\\
&& \hspace{1.0in}
\mbox{} \times (m+j)(m+j+2) Q^2_{m+j+1} c_{m+j-1} \Biggr\} \tilde v_{j,k+1}
\nonumber\\
&&\nonumber\\
&&\mbox{} + m(m+j+2)(m+j+k+1) Q_{m+j+2}Q_{m+j+1} c_{m+j-1}
\nonumber\\
&&\nonumber\\
&& \hspace{1.0in}
\mbox{} \times \Biggl[ \tilde w_{j+2,k} 
+ (m+j+3) \tilde v_{j+2,k} \Biggr]
\nonumber\\
&&\nonumber\\
&&\mbox{} - (k+1)m(m+j+2) Q_{m+j+2}Q_{m+j+1} c_{m+j-1}
\nonumber\\
&&\nonumber\\
&& \hspace{1.0in}
\mbox{} \times \Biggl[ \tilde w_{j+2,k+1} 
+ (m+j+3) \tilde v_{j+2,k+1} \Biggr]
\nonumber
\end{eqnarray}

The equations (\ref{po_match_a}) through (\ref{po_4}) make up the 
algebraic system 
(\ref{linalg}) for eigenvalues of the polar-led hybrid modes.  
As in the case of the axial-led
hybrids, one truncates the angular and radial series expansions 
at indices $j_{\mbox{\tiny max}}$, $i_{\mbox{\tiny max}}$ and 
$k_{\mbox{\tiny max}}$ and constructs the 
matrix $A$ by keeping the appropriate number of equations for 
the number of unknown 
coefficients $w_{j,i}$, $v_{j,i}$, $\tilde w_{j,k}$ and 
$\tilde v_{j,k}$. 

We, again, find that certain subsets of these equations are 
linearly dependent for
arbitrary $\kappa$ and for any equilibrium stellar model.  
For all $j$, it can be shown that 
both equation (\ref{po_1}) with $i=0$ and equation (\ref{po_3}) 
with $i=0$ are proportional to 
\[
0 = \left[w_{j,0} - (m+j) v_{j,0} \right].
\]

This problem can, again, be solved by eliminating, for example, 
equation (\ref{po_3}) 
with $i=0$ for all $j$.  Thus, to properly construct the 
algebraic system (\ref{linalg}) 
we use, for all $j=0,2,\ldots,j_{\mbox{\tiny max}}$, 
the equations
\[
\begin{array}{ll}
\mbox{(\ref{po_match_a})} & \\
\mbox{(\ref{po_match_b})} & \\
\mbox{(\ref{po_1})} & \mbox{with} \ \ i=0,2,\ldots,i_{\mbox{\tiny max}} \\ 
\mbox{(\ref{po_2})} & \mbox{with} \ \ k=0,1,\ldots,k_{\mbox{\tiny max}}-1 \\ 
\mbox{(\ref{po_3})} & \mbox{with} \ \ i=2,4,\ldots,i_{\mbox{\tiny max}} \\
\mbox{(\ref{po_4})} & \mbox{with} \ \ k=0,1,\ldots,k_{\mbox{\tiny max}}-1. \\ 
\end{array}
\]


\chapter{Proof of Theorem \ref{thm2}}

\section{Axial-led hybrids with $m>0$.}

Let $l$ be the smallest value of $l'$ for which $U_{l'}\neq 0$ in 
the spherical harmonic expansion (\ref{xi_exp}) of the displacement
vector $\xi^\alpha$, or for which $h_{l'}\equiv h_{0,l'}\neq 0$ in 
the spherical harmonic expansion (\ref{h_components}) of the metric
perturbation $h_{\alpha\beta}$.  The axial parity of 
$(\xi^\alpha, h_{\alpha\beta})$, $(-1)^{l+1}$, and the vanishing of 
$Y_l^m$ for $l<m$ implies $l\geq m$.  That the mode is axial-led means 
$W_{l'}=0$, $V_{l'}=0$ and $H_{1,l'}=0$ for $l'\leq l$.  
We show by contradiction that $l=m$.

Suppose $l\geq m+1$.  From Eq. (\ref{om_th_ph}),
$\int \Delta\om_{\theta\varphi} Y_l^{\ast m} d\Omega = 0$, 
we have
\be
0 = \ba[t]{l}
l(l+1)\kappa\Omega(h_l+U_l)-2m{\bar\omega}U_l
\\ \\
-l Q_{l+1} \left[\frac{e^{2\nu}}{r}\partial_r
\left(r^2{\bar\omega}e^{-2\nu}\right)W_{l+1}
+2(l+2){\bar\omega}V_{l+1} \right],
\ea
\ee
and from Eq. (\ref{om_ph_r}) with $l\rightarrow l-1$,
$\int \Delta\om_{\varphi r} Y_{l-1}^{\ast m} d\Omega = 0$, 
we have
\be
0 = \ba[t]{l}
- \biggl\{
(l+1)\kappa\Omega\partial_r\left[e^{-2\nu}(h_l+U_l)\right]
+2m\partial_r\left({\bar\omega}e^{-2\nu}U_l\right)
+\frac{m(l+1)}{r^2}\partial_r\left(r^2{\bar\omega}e^{-2\nu}\right)U_l
\biggr\}
 \\ \\
+ Q_{l+1} \biggl[
\partial_r\left[\frac{1}{r}\partial_r
\left(r^2{\bar\omega}e^{-2\nu}\right)W_{l+1}\right]
+2(l+2)\partial_r\left({\bar\omega}e^{-2\nu}V_{l+1}\right)
\biggr]
\ea
\ee
Together these give,
\bea
0 &=& 2\partial_r\left({\bar\omega}e^{-2\nu}U_l\right)
+\frac{l}{r^2}\partial_r\left(r^2{\bar\omega}e^{-2\nu}\right)U_l
\nn \\ \\
  &=& 2\left(r^2\bom e^{-2\nu}\right)^{-\frac{l}{2}}
\partial_r\left[r^l\left(\bom e^{-2\nu}\right)^{\half(l+2)}U_l
\right]
\eea
or,
\be
U_l = K \left(\bom e^{-2\nu}\right)^{-\half(l+2)}r^{-l}
\ee
(for some constant $K$) which is singular as $r\rightarrow 0$.

\section{Axial-led hybrids with $m=0$.}

Let $m=0$ and let $l$ be the smallest value of $l'$ for which 
$U_{l'}\neq 0$ in the spherical harmonic expansion (\ref{xi_exp}) of 
the displacement vector $\xi^\alpha$, or for which 
$h_{l'}\equiv h_{0,l'}\neq 0$ in the spherical harmonic expansion 
(\ref{h_components}) of the metric perturbation $h_{\alpha\beta}$.  
Since $\nabla_a Y_0^0=0$, the mode vanishes unless $l\geq 1$. 
That the mode is axial-led means $W_{l'}=0$, $V_{l'}=0$ and 
$H_{1,l'}=0$ for $l'\leq l$.  We show by contradiction that $l=1$.

Suppose $l\geq 2$.  Then Eq. (\ref{om_r_th}) with $l\rightarrow l-2$,
$\int \Delta\om_{r\theta} Y_{l-2}^{\ast 0} d\Omega = 0$, 
becomes
\bea
0 &=& 2\partial_r\left({\bar\omega}e^{-2\nu}U_l\right)
+\frac{l}{r^2}\partial_r\left(r^2{\bar\omega}e^{-2\nu}\right)U_l
\nn \\ \\
  &=& 2\left(r^2\bom e^{-2\nu}\right)^{-\frac{l}{2}}
\partial_r\left[r^l\left(\bom e^{-2\nu}\right)^{\half(l+2)}U_l
\right]
\eea
or,
\be
U_l = K \left(\bom e^{-2\nu}\right)^{-\half(l+2)}r^{-l}
\ee
(for some constant $K$) which is singular as $r\rightarrow 0$.

\section{Polar-led hybrids with $m\geq 0$.}

Let $l$ be the smallest value of $l'$ for which $W_{l'}\neq 0$ or 
$V_{l'}\neq 0$ in the spherical harmonic expansion (\ref{xi_exp}) 
of the displacement vector $\xi^\alpha$, or for which $H_{1,l'}\neq 0$ 
in the spherical harmonic expansion (\ref{h_components}) of the metric
perturbation $h_{\alpha\beta}$.  The polar parity of 
$(\xi^\alpha, h_{\alpha\beta})$, $(-1)^l$, and the vanishing of $Y_l^m$ 
for $l<m$ implies $l\geq m$.  That the mode is polar-led means 
$U_{l'}=0$ and $h_{l'}=0$ for $l'\leq l$.  
We show by contradiction that $l=m$ when $m>0$ and that $l=1$ when $m=0$.

Suppose $l\geq m+1$.  From Eq. (\ref{om_th_ph}) with 
$l\rightarrow l-1$,
$\int \Delta\om_{\theta\varphi} Y_{l-1}^{\ast m} d\Omega = 0$, 
we have
\be
0 = (l-1) Q_l \left[\frac{e^{2\nu}}{r}\partial_r
\left(r^2{\bar\omega}e^{-2\nu}\right)W_l
+2(l+1){\bar\omega}V_l \right].
\label{polar_proof}
\ee

Substituting for $V_l$ using Eq. (\ref{GR:sph_V}), we find
\bea
0 &=& \frac{l}{r}\partial_r\left(r^2{\bar\omega}e^{-2\nu}\right)W_l
+ 2\bom e^{-2\nu}\frac{e^{-(\nu+\lambda)}}{(\ep+p)}
\partial_r\left[(\ep+p)e^{(\nu+\lambda)}rW_l\right]
\nn \\ \\
 &=&2\left(r^2{\bar\omega}e^{-2\nu}\right)^{-\half(l-2)}
\frac{e^{-(\nu+\lambda)}}{r^2(\ep+p)}
\partial_r\left[\left(r^2{\bar\omega}e^{-2\nu}\right)^{\frac{l}{2}}
(\ep+p)e^{(\nu+\lambda)}rW_l\right]
\eea
with solution,
\be
W_l= K \left(\bom e^{-2\nu}\right)^{-\frac{l}{2}}
\frac{e^{-(\nu+\lambda)}}{(\ep+p)}r^{-(l+1)}
\ee
(for some constant $K$) which is singular as $r\rightarrow 0$.

When $m=0$ this argument fails to establish that $l$ cannot be equal to 
1, because Eq. (\ref{polar_proof}) is trivially satisfied for $l=1$ as a 
result of the overall $l-1$ factor.  Instead, the argument proves that 
$l$ cannot be greater than 1 in this case and therefore that $l=1$.



\chapter{Algebraic Equations: Relativistic}

In this appendix, we make use of the following definitions.
\begin{eqnarray}
a_l &\equiv& (l+1)Q_l^2 - l Q_{l+1}^2 \\
b_l &\equiv& m^2 - l(l+1)\left(1- Q_l^2 - Q_{l+1}^2\right)
\end{eqnarray}
and we repeat the definitions
\bea
\kappa  &\equiv& \frac{(\sigma + m\Omega)}{\Omega} \\
Q_l     &\equiv& \left[ \frac{(l+m)(l-m)}{(2l-1)(2l+1)} 
\right]^{\half}.
\end{eqnarray}
We will also make use of the definition
\be
\Theta(k) \equiv \cases{0 &$k<0$\cr 1 &$k\geq 0$\cr}.
\ee

\noindent
The regular power series expansions of the perturbation 
variables about the center of the star, $r=0$, are 
\begin{eqnarray}
h_l(r) &=&
\sum^\infty_{i=0} h_{l,i} \left(\frac{r}{R}\right)^{l+1+2i},  
\label{h_i}
\\
U_l(r) &=&
\sum^\infty_{i=0} u_{l,i} \left(\frac{r}{R}\right)^{l+1+2i},  
\label{U_i}
\\
W_l(r) &=&
\sum^\infty_{i=0} w_{l,i} \left(\frac{r}{R}\right)^{l+2i},  
\label{W_i}
\\
V_l(r) &=&
\sum^\infty_{i=0} v_{l,i} \left(\frac{r}{R}\right)^{l+2i},  
\label{V_i}
\end{eqnarray}
while about the surface of the star, $r=R$, they are
\begin{eqnarray}
h_l(r) &=&
\sum^\infty_{k=0} \tilde h_{l,k} \left(1-\frac{r}{R}\right)^k,  
\label{h_k}
\\
U_l(r) &=&
\sum^\infty_{k=0} \tilde u_{l,k} \left(1-\frac{r}{R}\right)^k,  
\label{U_k}
\\
W_l(r) &=&
\sum^\infty_{k=1} \tilde w_{l,k} \left(1-\frac{r}{R}\right)^k,  
\label{W_k}
\\
V_l(r) &=&
\sum^\infty_{k=0} \tilde v_{l,k} \left(1-\frac{r}{R}\right)^k.  
\label{V_k}
\end{eqnarray}
The boundary condition (\ref{bc_on_W}) is automatically 
satisfied by the form of the surface expansion of $W_l(r)$, Eq. 
(\ref{W_k}).  On the other hand, The matching condition
(\ref{match_cond}) is not automatically satisfied by the form of 
the surface expansion of $h_l(r)$, Eq. (\ref{h_k}).  Instead, 
condition (\ref{match_cond}) places the following non-trivial 
restriction on the series expansion (\ref{h_k}),
\be
0 = \tilde h_{l,0} \left[ \sum_{s=0}^\infty 
(l+s) {\hat h}_{l,s} \right]
- \tilde h_{l,1} \left[ \sum_{s=0}^\infty 
{\hat h}_{l,s} \right],
\label{match_h_R}
\ee
where the constants $\hat h_{l,s}$ are given by the 
recursion relation (\ref{h_ext_soln}) up to normalization.
The normalization factor $\hat h_{l,0}$ is then fixed by the 
matching condition (\ref{cont_cond}) once the interior solution 
is known. 

The series expansions about $r=0$ must agree with those about 
$r=R$ everywhere in the interior of the star.  To ensure this 
agreement we impose the matching condition that the series 
(\ref{h_i})-(\ref{V_i}) truncated at $i_{\mbox{\tiny max}}$
be equal at the point $r=r_0$ to the corresponding series 
(\ref{h_k})-(\ref{V_k}) truncated at $k_{\mbox{\tiny max}}$.  
That is,
\be
0 = \sum^{i_{\mbox{\tiny max}}}_{i=0} h_{l,i} 
\left(\frac{r_0}{R}\right)^{l+1+2i}
- \sum^{k_{\mbox{\tiny max}}}_{k=0} \tilde h_{l,k} 
\left(1-\frac{r_0}{R}\right)^k,
\label{match_h1}
\ee

\be
0 = \sum^{i_{\mbox{\tiny max}}}_{i=0} u_{l,i} 
\left(\frac{r_0}{R}\right)^{l+1+2i}
- \sum^{k_{\mbox{\tiny max}}}_{k=0} \tilde u_{l,k} 
\left(1-\frac{r_0}{R}\right)^k,
\label{match_u}
\ee

\be
0 = \sum^{i_{\mbox{\tiny max}}}_{i=0} w_{l,i} 
\left(\frac{r_0}{R}\right)^{l+2i}
- \sum^{k_{\mbox{\tiny max}}}_{k=1} \tilde w_{l,k} 
\left(1-\frac{r_0}{R}\right)^k,
\label{match_w}
\ee

\be
0 = \sum^{i_{\mbox{\tiny max}}}_{i=0} v_{l,i} 
\left(\frac{r_0}{R}\right)^{l+2i}
- \sum^{k_{\mbox{\tiny max}}}_{k=0} \tilde v_{l,k} 
\left(1-\frac{r_0}{R}\right)^k.
\label{match_v}
\ee
Furthermore, since the function $h_l(r)$ obeys a second
order differential equation, we must also impose a
matching condition on its derivative $h_l'(r)$; namely,
\be
0 = \sum^{i_{\mbox{\tiny max}}}_{i=0} (l+1+2i) h_{l,i} 
\left(\frac{r_0}{R}\right)^{l+2i}
+ \sum^{k_{\mbox{\tiny max}}}_{k=1} k \tilde h_{l,k} 
\left(1-\frac{r_0}{R}\right)^{k-1}.
\label{match_h2}
\ee

We now consider the perturbation equations 
(\ref{h_0''})-(\ref{r_th}).  We substitute for the 
equilibrium variables in these equations their power
series expansions (\ref{pi_i})-(\ref{mu_k}).  We substitute 
for the perturbation variables in these equations their power
series expansions (\ref{h_i})-(\ref{V_k}).  Then, by applying
straightforward rules for the multiplication of power series,
we extract the series expansions of the perturbation equations,
themselves.  The requirement that the coefficients of these 
expansions vanish independently then gives us the algebraic
equations for the unknown constants $h_{l,i}$, $u_{l,i}$, 
$w_{l,i}$, $v_{l,i}$, $\tilde h_{l,k}$, $\tilde u_{l,k}$, 
$\tilde w_{l,k}$ and $\tilde v_{l,k}$, (for all allowed values 
of $l$, $i$ and $k$) discussed in Sect. 4.1.2.  

When we substitute the series (\ref{pi_i})-(\ref{mu_i}) and
(\ref{h_i})-(\ref{V_i}) into Eq. (\ref{h_0''}) the resulting
series expansion about $r=0$ is
\be
0 = \sum_{i=1}^\infty \Biggl\{ \ba[t]{l}
[(l+2i) (l+2i+1)-l(l+1)]\, h_{l,i} \\
 \\
- \ds{\sum_{j=0}^{i-1}} \left[ (\nu_{j+1}+\lambda_{j+1})(l+2i-2j+1) +
(l^2+l-2)E_{j+1}\right] \, h_{l,i-j-1} \\
 \\
- \ds{\sum_{j=0}^{i-1}} 4(\nu_{j+1}+\lambda_{j+1}) \,u_{l,i-j-1}
\Biggr\} \ds{\left(\rx\right)^{l+1+2i}}.
\ea
\label{alg1}
\ee

When we substitute the series (\ref{pi_i})-(\ref{mu_i}) and
(\ref{h_i})-(\ref{V_i}) into Eq. (\ref{Vsub}) the resulting
series expansion about $r=0$ is
\be
0 = \sum_{i=0}^\infty \Biggl\{ (l+2i+1)\,w_{l,i} + \sum_{j=0}^{i-1}
(\nu_{j+1}+\lambda_{j+1}+\pi_{j+1}) \,w_{l,i-j-1}
- l(l+1)\,v_{l,i} \Biggr\} \ds{\left(\rx\right)^{l+2i}}.
\label{alg2}
\ee

When we substitute the series (\ref{pi_i})-(\ref{mu_i}) and
(\ref{h_i})-(\ref{V_i}) into Eq. (\ref{th_ph}) the resulting
series expansion about $r=0$ is
\be
0 = \sum_{i=0}^\infty \Biggl\{ \ba[t]{l}
l(l+1)\kappa \, h_{l,i}+[l(l+1)\kappa -2m\omega_0]\,u_{l,i} 
-2m \ds{\sum_{j=0}^{i-1}} \omega_{j+1}\,u_{l,i-j-1} \\
\\
+(l+1)Q_l \biggl[ \ba[t]{l}
	2\omega_0\,w_{l-1,i+1} +
	\ds{\sum_{j=0}^{i-1}} (2\omega_{j+1}+\mu_{j+1})\,w_{l-1,i-j} \\
	\\
	+\mu_{i+1}\,w_{l-1,0} - 2(l-1) \ds{\sum_{j=0}^i}\omega_j
	\,v_{l-1,i-j+1}\biggr]
	\ea  \\
\\
-lQ_{l+1} \biggl[ \ba[t]{l}
	2\omega_0\,w_{l+1,i} +\ds{\sum_{j=0}^{i-1}} (2\omega_{j+1}
	+\mu_{j+1})\,w_{l+1,i-j-1} \\
	\\
	+ 2(l+2) \ds{\sum_{j=0}^i} \omega_j\,v_{l+1,i-j}\biggr]
	\Biggr\} \ds{\left(\rx\right)^{l+1+2i}}.
	\ea 
\ea
\label{alg3}
\ee

When we substitute the series (\ref{pi_i})-(\ref{mu_i}) and
(\ref{h_i})-(\ref{V_i}) into Eq. (\ref{r_th}) the resulting
series expansion about $r=0$ is
\bea
0 &=& \sum_{i=0}^\infty \Biggl\{ 
\Bigl\{ \ba[t]{c}
	-4(i+1)\omega_0 \, u_{l-2,i+1} 
	+\ds{\sum_{j=0}^{i-1}} [-4(i-j)\omega_{j+1} 
	+ (l-3)\mu_{j+1}] \,u_{l-2,i-j} \\
	\\
	+ (l-3)\mu_{i+1}\, u_{l-2,0}
	\Bigr\} (l-2)l(l+1)Q_{l-1}Q_l 
	\ea 
\label{alg4} \\
&&\nn \\
&&+\Bigl\{ \ba[t]{l}
	[(l-1)l(l+2i+1)\kappa  -
	4ml(i+1)\omega_0]\,v_{l-1, i+1} \\
	\\
	+ \ds{\sum_{j=0}^{i-1}}
	[m(l-3)l\mu_{j+1}-2(l-1)l\kappa \nu_{j+1}
	-4ml(i-j)\omega_{j+1}] \,v_{l-1,i-j} \\
	\\
	+ [  \ba[t]{l}
		m(l-3)l\mu_{i+1}-2(l-1)l\kappa \nu_{i+1} \\
		\\
		- (l-1)^2l\kappa E_{i+1}+4(l-1)\kappa (\nu_{i+1}
		+\lambda_{i+1}] \,v_{l-1,0} 
		\ea \\
	\\
	- (l-1)l\kappa  \,w_{l-1,i+1} \\
	\\
	+ \ds{\sum_{j=0}^{i-1}}[4\kappa (\nu_{j+1}
	+\lambda_{j+1})-(l-1)l\kappa E_{j+1}]\,w_{l-1,i-j}
	\Bigr\} (l+1)Q_l 
	\ea\nn  \\
&&\nn \\
&&+\Bigl\{ \ba[t]{l}
	m\kappa (l+1+2i)\, h_{l,i} - 2m\kappa 
	\ds{\sum_{j=0}^{i-1}} \nu_{j+1}\, h_{l,i-j-1} \\
	\\
	+ \Bigl[ m\kappa (l+1+2i)
	+2\omega_0[(l+1+2i)a_l+b_l]\Bigr] \,u_{l,i} \\
	\\
	+ \ds{\sum_{j=0}^{i-1}} \Bigl[ \ba[t]{l}
		-2m\kappa \nu_{j+1}
		+2\omega_{j+1}[(l+2i-2j-1)a_l+b_l] \\
		\\
		+\mu_{j+1}(2a_l+b_l)\Bigr] \,u_{l,i-j-1}
	\Bigr\} l(l+1)
	\ea
\ea\nn   \\
&&\nn \\
&&-\Bigl\{ \ba[t]{l}
	(l+1)[(l+2)(l+2i+1)\kappa +2m(2l+2i+3)\omega_0] \,v_{l+1,i} \\
	\\
	+ \ds{\sum_{j=0}^{i-1}} (l+1)[ \ba[t]{l}
		m(l+4)\mu_{j+1}-2(l+2)\kappa \nu_{j+1} \\
		\\
		+ 2m(2l+2i-2j+1)\omega_{j+1}] \,v_{l+1, i-j-1}
		\ea \\
	\\
	-(l+1)(l+2)\kappa \,w_{l+1,i} \\
	\\
	+ \ds{\sum_{j=0}^{i-1}} [4\kappa (\nu_{j+1}+\lambda_{j+1})
	-(l+1)(l+2)\kappa E_{j+1}] \,w_{l+1,i-j-1}
	\Bigr\} lQ_{l+1}
	\ea\nn  \\
&&\nn \\
&&+\Theta(i-1)  \Bigl\{ \ba[t]{l}
	2(2l+2i+3)\omega_0\,u_{l+2,i-1} \\
	\\
	+ \ds{\sum_{j=0}^{i-2}} [ \ba[t]{l}
		2(2l+2i-2j+1)\omega_{j+1} \\
		\\
		+(l+4)\mu_{j+1}] \,u_{l+2,i-j-2}
		\Bigr\} \\
		\\
		\times l(l+1)(l+3)Q_{l+1} Q_{l+2}
\Biggr\} \ds{\left(\rx\right)^{l+1+2i}}.
\ea
\ea \nn
\eea


When we substitute the series (\ref{pi_k})-(\ref{mu_k}) and
(\ref{h_k})-(\ref{V_k}) into Eq. (\ref{h_0''}) the resulting
series expansion about $r=R$ is
\be
0 = \sum_{q=0}^\infty \Biggl\{ \ba[t]{l}
(q+1)(q+2)\,\tilde h_{l,q+2} +
(q+1)(\tilde\nu_0+\tilde\lambda_0-2q)\,\tilde h_{l,q+1} \\
\\
+ [ \ba[t]{l}
	(q+1)(q-2)-(l^2+l-2)\tilde E_0 \\
	\\
	-(q+2)(\tilde\nu_0+\tilde\lambda_0) 
	+ q(\tilde\nu_1+\tilde\lambda_1)]\,\tilde h_{l,q} 
	\ea \\
\\
-\ds{\sum_{j=0}^{q-2}} [ \ba[t]{l}
		(l^2+l-2)\tilde E_{j+1}
		+(q-j+1)(\tilde\nu_{j+1}+\tilde\lambda_{j+1}) \\
		\\
		-(q-j-1)(\tilde\nu_{j+2}
		+\tilde\lambda_{j+2})]\,\tilde h_{l,q-j-1}
 		\ea \\
\\
-\Theta (q-1) [(l^2+l-2)\tilde E_q
+2(\tilde\nu_q+\tilde\lambda_q)]\,\tilde h_{l,0} \\
\\
-\ds{\sum_{j=0}^q} 4(\tilde\nu_j+\tilde\lambda_j)\tilde
u_{l,q-j}\Biggr\} \ds{\left(\ry\right)^q}.
\ea
\label{alg5} 
\ee

When we substitute the series (\ref{pi_k})-(\ref{mu_k}) and
(\ref{h_k})-(\ref{V_k}) into Eq. (\ref{Vsub}) the resulting
series expansion about $r=R$ is
\be
0 = \sum_{q=0}^\infty \Biggl\{ \ba[t]{l}
(\tilde\pi_{-1}-q-1)\,\tilde w_{l,q+1} +
\Theta (q-1) (\tilde\pi_0+\tilde\nu_0
+\tilde\lambda_0+q+1)\,\tilde w_{l,q} \\
\\
+ \ds{\sum_{j=0}^{q-2}} 
(\tilde\pi_{j+1}+\tilde\nu_{j+1}+\tilde\lambda_{j+1})\tilde
w_{l,q-j-1}-l(l+1)\,\tilde v_{l,q}\Biggr\} 
\ds{\left(\ry\right)^q}.
\ea
\label{alg6} 
\ee

When we substitute the series (\ref{pi_k})-(\ref{mu_k}) and
(\ref{h_k})-(\ref{V_k}) into Eq. (\ref{th_ph}) the resulting
series expansion about $r=R$ is
\be
0 = \sum_{q=0}^\infty \Biggl\{ \ba[t]{l}
l(l+1)\kappa \,\tilde h_{l,q} +
[l(l+1)\kappa -2m\tilde\omega_0] \,\tilde u_{l,q} -2m\ds{\sum_{j=0}^{q-1}}
\tilde\omega_{j+1}\,\tilde u_{l,q-j-1} \\
\\
+(l+1)Q_l \left[ \ds{\sum_{j=0}^{q-1}}
(2\tilde\omega_j+\tilde\mu_j)\,\tilde w_{l-1,q-j} -2(l-1)\ds{\sum_{j=0}^q}
\tilde\omega_j\,\tilde v_{l-1,q-j}\right] \\
\\
-lQ_{l+1} \left[\ds{\sum_{j=0}^{q-1}}
(2\tilde\omega_j+\tilde\mu_j)\,\tilde w_{l+1,q-j}+2(l+2) \ds{\sum_{j=0}^q}
\tilde\omega_j \,\tilde v_{l+1,q-j}\right]\Biggr\} 
\ds{\left(\ry\right)^q}.
\ea
\label{alg7} 
\ee

When we substitute the series (\ref{pi_k})-(\ref{mu_k}) and
(\ref{h_k})-(\ref{V_k}) into Eq. (\ref{r_th}) the resulting
series expansion about $r=R$ is
\bea
0 &=& \sum_{i=0}^\infty \Biggl\{ 
\Bigl\{ \ba[t]{l}
	2(q+1)\tilde\omega_0\,\tilde u_{l-2,q+1} \\
	\\
	+ \ds{\sum_{j=0}^{q-1}}
	[2(q-j)\tilde\omega_{j+1}+2(l-q+j-1)\tilde\omega_j +
	(l-3)\tilde\mu_j]\,\tilde u_{l-2, q-j} \\
	\\
	+ [2(l-1)\tilde\omega_q + (l-3)\tilde\mu_q]\,\tilde u_{l-2,0}
	\Bigr\} (l-2)l(l+1)Q_{l-1}Q_l 
	\ea
\label{alg8} \\
&&\nn \\
&&+\Bigl\{ \ba[t]{l}
	-(q+1)l[(l-1)\kappa -2m\tilde\omega_0]\,\tilde v_{l-1,q+1} \\
	\\
	+ l[ \ba[t]{l}
		(l-1)q\kappa +2m(l-q-1)\tilde\omega_0 
		+2mq\tilde\omega_1 \\
		\\
		+m(l-3)\tilde\mu_0
		-2(l-1)\kappa \tilde\nu_0]\,\tilde v_{l-1,q} 
		\ea \\
	\\
	+ \ds{\sum_{j=0}^{q-2}} l[ \ba[t]{l}
		2m(l-q+j)\tilde\omega_{j+1}+2m(q-j-1)\tilde\omega_{j+2} \\
		\\
		+m(l-3)\tilde\mu_{j+1} - 2(l-1)\kappa \tilde\nu_{j+1}]
		\,\tilde v_{l-1,q-j-1} 
		\ea \\
	\\
	+ \Theta(q-1)l[2m(l-1)\tilde\omega_q+m(l-3)\tilde\mu_q
	-2(l-1)\kappa \tilde\nu_q]\,\tilde v_{l-1,0} \\
	\\
	+ \ds{\sum_{j=0}^{q-1}}
	[4\kappa (\tilde\nu_j+\tilde\lambda_j)
	-(l-1)l\kappa \tilde E_j] \,\tilde w_{l-1,q-j}
	\Bigr\} (l+1)Q_l 
	\ea\nn  \\
&&\nn \\
&&+\Bigl\{ \ba[t]{l}
	-m\kappa (q+1)\,\tilde h_{l,q+1}
	+m\kappa (q-2\tilde\nu_0)\,\tilde h_{l,q}\\
	\\
	 - \ds{\sum_{j=0}^{q-1}} 
	2m\kappa \tilde\nu_{j+1}\,\tilde h_{l,q-j-1}
	- (q+1)(m\kappa +2a_l\tilde\omega_0)\,\tilde u_{l,q+1} \\
	\\
	+[mq\kappa +2(qa_l+b_l)\tilde\omega_0-2qa_l\tilde\omega_1
	+(2a_l+b_l)\tilde\mu_0-2m\kappa \tilde\nu_0] \,\tilde u_{l,q} \\
	\\
	+ \ds{\sum_{j=0}^{q-2}} \Bigl[ \ba[t]{l}
		2[(q-j-1)a_l+b_l]\tilde\omega_{j+1} -
		2(q-j-1)a_l\tilde\omega_{j+2} \\
		\\
		+ (2a_l+b_l)\tilde\mu_{j+1}
		-2m\kappa \tilde\nu_{j+1}\Bigr]\,\tilde u_{l,q-j-1} 
		\ea \\
	\\
	+ \Theta(q-1)[2b_l\tilde\omega_q
	+(2a_l+b_l)\tilde\mu_q-2m\kappa \tilde\nu_q] \,\tilde u_{l,0}
	\Bigr\} l(l+1)
	\ea\nn \\
&&\nn \\
&&-\Bigl\{ \ba[t]{l}
	-(l+1)(q+1)[(l+2)\kappa 
	+2m\tilde\omega_0] \,\tilde v_{l+1,q+1} \\
	\\
	+ (l+1)[ \ba[t]{l}
		(l+2)q\kappa +2m(l+q+2)\tilde\omega_0 \\
		\\
		-2mq\tilde\omega_1+m(l+4)\tilde\mu_0
		-2(l+2)\kappa \tilde\nu_0] \,\tilde v_{l+1,q}
		\ea \\
	\\
	+ \ds{\sum_{j=0}^{q-2}} (l+1) [ \ba[t]{l}
		2m(l+q-j+1)\tilde\omega_{j+1}
		-2m(q-j-1)\tilde\omega_{j+2} \\
		\\
		+ m(l+4)\tilde\mu_{j+1}
		-2(l+2)\kappa \tilde\nu_{j+1}]\,\tilde v_{l+1,q-j-1}
		\ea \\
	\\
	+\Theta(q-1)(l+1)[2m(l+2)\tilde\omega_q
	+m(l+4)\tilde\mu_q-2(l+2)\kappa \tilde\nu_q]\,\tilde v_{l+1,0} \\
	\\
	+ \ds{\sum_{j=0}^{q-1}}
	[4\kappa (\tilde\nu_j+\tilde\lambda_j)
	-(l+1)(l+2)\kappa \tilde E_j] \,\tilde w_{l+1,q-j}
	\Bigr\} lQ_{l+1}
	\ea\nn \\
&&\nn \\
&&+ \Bigl\{ \ba[t]{l}
	-2(q+1)\tilde\omega_0\,\tilde u_{l+2,q+1} \\
	\\
	+ \ds{\sum_{j=0}^{q-1}} [2(l+q-j+2)\tilde\omega_j
	-2(q-j)\tilde\omega_{j+1}+(l+4)\tilde\mu_j]\,\tilde u_{l+2,q-j}\\
	\\
	+ [2(l+2)\tilde\omega_q + (l+4)\tilde\mu_q] \,\tilde u_{l+2,0}
	\Bigr\} l(l+1)(l+3)Q_{l+1} Q_{l+2}
\Biggr\} \ds{\left(\ry\right)^{l+1+2i}}.
\ea \nn
\eea

Eqs. (\ref{match_h_R})-(\ref{alg8}) make up the 
algebraic system (\ref{GR:linalg}) determining the eigenvalues 
of the axial- and polar-led hybrid modes.  As in the 
Newtonian case, one truncates the angular and radial series 
expansions at indices $l_{\mbox{\tiny max}}$, $i_{\mbox{\tiny max}}$ 
and $k_{\mbox{\tiny max}}$ and constructs the matrix $A$ by keeping 
the appropriate number of equations for the number of unknown 
coefficients $h_{l,i}$, $u_{l,i}$, $w_{l,i}$, $v_{l,i}$, 
$\tilde h_{l,k}$, $\tilde u_{l,k}$, $\tilde w_{l,k}$ and 
$\tilde v_{l,k}$.  Just as in the Newtonian case, however, one must
be aware of a certain linear dependence in the expansions about
$r=0$.  With the set of relativistic equations we have chosen to 
work with, this linear dependence arises only for the axial-led
hybrids and may be seen as follows.

For a given choice of $q\in [0,1,2,\ldots]$ the set of equations,
\[
\begin{array}{ll}
\mbox{(\ref{alg2})} & \mbox{for all} \ \ l=m+1, m+3,\ldots,m+2q+1 
\ \ \mbox{with} \ \ i=0; \\
\mbox{(\ref{alg3})} & \mbox{for all} \ \ i=0,1,\ldots,q \ \ \mbox{with} 
\ \ l=m+2q-2i; \ \  \mbox{and} \\
\mbox{(\ref{alg4})} & \mbox{for all} \ \ i=0,1,\ldots,q \ \ \mbox{with} 
\ \ l=m+2q-2i
\end{array}
\]
can be shown to be linearly dependent for arbitrary $\kappa$ and 
for any equilibrium stellar model.  For example, taking the 
simplest case of $q=0$, one finds that Eq. (\ref{alg2}) with 
$l=m+1$ and $i=0$ becomes,
\be
0 = w_{m+1,0} - (m+1) v_{m+1,0},
\label{lindep1}
\ee
Eq. (\ref{alg3}) with $l=m$ and $i=0$ becomes,
\be
0 = \ba[t]{l}
m(m+1)\kappa h_{m,0} 
+ m\left[(m+1)\kappa-2\bom_0\right]u_{m,0} \\
\\
- 2 m \bom_0 Q_{m+1} \left[w_{m+1,0}+(m+2)v_{m+1,0}\right],
\ea
\label{lindep2}
\ee
Eq. (\ref{alg4}) with $l=m$ and $i=0$ becomes
\be
0 = \ba[t]{l}
m(m+1)\biggl\{m(m+1)\kappa h_{m,0} 
+ m\left[(m+1)\kappa-2\bom_0\right]u_{m,0}\biggr\} \\
\\
- m Q_{m+1} \biggl\{ \ba[t]{l}
	(m+1)\left[(m+1)(m+2)\kappa
	+2m(2m+3)\bom_0 \right]v_{m+1,0} \\
	\\
	-(m+1)(m+2)\kappa w_{m+1,0}
	\biggr\},
\ea
\ea
\label{lindep3}
\ee
and it is not difficult to show that
\be
0 = \mbox{(\ref{lindep1})} 
- m(m+1) \Bigl\{\mbox{(\ref{lindep2})}
+ Q_{m+1} \left[(m+2)\kappa+2m\bom_0\right] 
\mbox{(\ref{lindep3})} \Bigr\},
\ee
which is the claimed linear dependence.

As in the Newtonian case, this problem can be taken care of 
by eliminating one of the linearly dependent equations for 
each $q$.  To properly construct the algebraic system 
(\ref{GR:linalg}) for the axial-led hybrid modes we use 
Eqs. (\ref{match_h_R})-(\ref{alg8}) for all $i$ except 
Eq. (\ref{alg4}) with $i=0$, for all allowed $l$.

%
\clearpage
\addcontentsline{toc}{chapter}{Bibliography}

\begin{thebibliography}{99}

\bibitem{lapack}
	Anderson, E., Bai, Z., Bischof, C., Demmel, 
	J., Dongarra, J., Du Croz, J., Greenbaum, A., 
	Hammarling, S., McKenney, A., Ostrouchov, S. 
	and Sorensen D., 1994, ``LAPACK User's Guide 
	- Release 2.0''. \ This guide and publicly 
	available source code can be found on the web 
	at http://www.netlib.org/lapack/lug/lapack\_lug.html
\bibitem{a97}
	Andersson, N., 1998, \apj, 502, 708
\bibitem{aks98}
	Andersson, N., Kokkotas, K. and Schutz, B. F., 1999, 
	\apj, 510, 846
\bibitem{aea99}
	Andersson, N., Kokkotas, K., Friedman, J. L., Lockitch, 
	K. H., Schutz, B. F., Stergioulas, N., 1999, in preparation.
\bibitem{akst98}
	Andersson, N., Kokkotas, K. and Stergioulas, N., 1999, 
	\apj, 516, 307
\bibitem{alf99}
	Andersson, N., Lockitch, K. H. and Friedman, J. L., 1999, 
	paper in preparation
\bibitem{bk99}
	Beyer, H. R. and Kokkotas, K. D., 1999, ``On the r-mode 
	spectrum of relativistic stars'', [gr-qc/9903019]
\bibitem{b98}
	Bildsten, L., 1998, \apj, 501, L89
\bibitem{bc98}
	Brady, P. R. and Creighton, T., 1998, ``Searching for periodic
	sources with LIGO. II: Heirarchical searches'', [gr-qc/9812014]
\bibitem{b1889}
	Bryan, G. H., 1889, \ptrsl, A180, 187
\bibitem{ct70}
	Campolattaro, A. and Thorne, K. S., 1970, \apj, 159, 847
\bibitem{ch70}
	Chandrasekhar, S. 1970, \prl, 24, 611
\bibitem{cf91a}
	Chandrasekhar, S. and Ferrari, V., 1991, \prsl, 433, 423
\bibitem{cf91b}
	Chandrasekhar, S. and Ferrari, V., 1991, \prsl, 434, 449
\bibitem{cm74}
	Chandrasekhar, S. and Miller, J. C., 1974, \mnras, 167, 63
\bibitem{c41}
	Cowling, T. G., 1941, \mnras, 101, 367
\bibitem{cl87}
	Cutler, L. and Lindblom, L., 1987, \apj, 314, 234
\bibitem{di73}
	Detweiler, S. L. and Ipser J. R., 1973, \apj, 185, 685
\bibitem{dl77}
	Detweiler, S. L. and Lindblom, L., 1977, \apj, 213, 193
\bibitem{fms98}
	Ferrari, V., Mataresse, S. and Schneider, R., 1999, \mnras, 303, 258
\bibitem{f78} 
	Friedman, J. L., 1978, \cmp, 62, 247
\bibitem{f96}
	Friedman, J. L., 1996, {\it J. Astrophys. Astr.}, 17, 199
\bibitem{f98}
	Friedman, J. L., 1998, in Wald, R. M., ed., {\it Black Holes 
	and Relativistic Stars}, (Chicago: University of Chicago Press)
\bibitem{fi92}
	Friedman, J. L. and Ipser, J. R., 1992, \ptrsl, A340, 391
\bibitem{jfs97} 
	Friedman, J. L. and Morsink, S. M., 1998, \apj, 502, 714
\bibitem{fs75}
	Friedman, J. L. and Schutz, B. F., 1975, \apj, 200, 204
\bibitem{fs78a}
	Friedman, J. L. and Schutz, B. F., 1978, \apj, 221, 937
\bibitem{fs78b}
	Friedman, J. L. and Schutz, B. F., 1978, \apj, 222, 281
\bibitem{h67}
	Hartle, J. B., 1967, \apj, 150, 1005
\bibitem{ht68}
	Hartle, J. B. and Thorne K. S., 1968, \apj, 153, 807
\bibitem{h98}
	Hiscock, W. A., 1998, ``Gravitational waves from rapidly
	rotating white dwarves'', [gr-qc/9807036] 
\bibitem{il90}
	Ipser, J. R. and Lindblom, L., 1990, \apj, 355, 226
\bibitem{il91}
	Ipser, J. R. and Lindblom, L., 1991, \apj, 373, 213
\bibitem{im85}
	Ipser, J. R. and Managan, R. A., 1985, \apj, 292, 517
\bibitem{it73}
	Ipser, J. R. and Thorne, K. S., 1973, \apj, 181, 181
\bibitem{j75}
	Jackson, J. D., 1975, {\it Classical Electrodynamics},
	2nd edition, (Wiley).
\bibitem{k88}
	Kojima, Y., 1988, Prog. Theor. Phys., 79, 665
\bibitem{k98}
	Kojima, Y., 1998, \mnras, 293, 49
\bibitem{kh99}
	Kojima, Y. and Hosonuma, M., 1999, \apj, 520, 788, 
\bibitem{k96}
	Kokkotas, K., 1996, in Marck J. A., and Lasota, J. P., eds.,
    	{\it Relativistic Gravitation and Gravitational Radiation}, 
	(Cambridge University Press, Cambridge, 1996). Also available
	as preprint [gr-qc/9603024].  
\bibitem{ks99}
	Kokkotas, K., and Schmidt, B., 1999,  ``Quasi-normal Modes of 
	Black Holes and Stars'', review article in preparation for 
	\lr, [http://www.livingreviews.org]. 
\bibitem{ks86}
	Kokkotas, K. and Schutz, B. F., 1986, \grg, 18, 913
\bibitem{ks98}
	Kokkotas, K. and Stergioulas, N., 1999, \aap, 341, 110
\bibitem{ks80}
	K\"{u}nzle, H. P. and Savage, J. R., 1980, \grg, 12, 155
\bibitem{l98}
	Levin, Y., 1999, \apj, 517, 328
\bibitem{l92}
	Lindblom, L., 1992, \apj, 398, 569
\bibitem{l95}
	Lindblom, L., 1995, \apj, 438, 265
\bibitem{l99}
	Lindblom, L., 1999, ``Stability of the r-modes in white dwarf
	stars'', [gr-qc/9903042]
\bibitem{lh83}
	Lindblom, L. and Hiscock, W. A., 1983, \apj, 267, 384
\bibitem{li98}
	Lindblom, L. and Ipser, J. R., 1998, \prd, 59, 044009
\bibitem{lm95}
	Lindblom, L. and Mendell, G., 1995, \apj, 444, 804
\bibitem{lm99}
	Lindblom, L. and Mendell, G., 1999, in preparation.
\bibitem{lmo99}
	Lindblom, L., Mendell, G. and Owen, B. J., 1999, 
	``Second-order rotational effects on the r-modes of 
	neutron stars'' [gr-qc/9902052]
\bibitem{lom98}
	Lindblom, L., Owen, B. J. and Morsink, S. M., 1998, 
	\prl, 80, 4843
\bibitem{lf98}
	Lockitch, K. L. and Friedman, J. L., 1998, \apj \ in press, 
	scheduled to appear in vol. 521, no.2. Reproduced here as Ch. 2.
\bibitem{mad98}
	Madsen, J., 1998, \prl, 81, 3311
\bibitem{m98}
	Marshall, F. E., Gotthelf, E. V., Zhang, W., Middleditch, J. 
	and Wang, Q. D., 1998, \apj, 499, L179
\bibitem{mtw}
	Misner, C. W., Thorne, K. S. and Wheeler, J. A., 1973,
	{\it Gravitation} (San Francisco: Freeman)	
\bibitem{o98}
	Owen, B. J., Lindblom, L., Cutler, C., Schutz, B. F., Vecchio, 
	A. and Andersson, N., 1998, \prd, 58, 084020
\bibitem{pp78}
	Papalouizou, J. and Pringle, J. E., 1978, \mnras, 182, 423
\bibitem{pp78b}
	Papalouizou, J. and Pringle, J. E., 1978, \mnras, 184, 501
\bibitem{nr}
	Press, W. H., Flannery, B. P., Teukolsky, S. A. and 
	Vetterling, W. T., 1992, {\it Numerical Recipes in C}, 2nd ed. 
	(Cambridge: Cambridge University Press)
\bibitem{pt69}
	Price, R. and Thorne, K. S., 1969, \apj, 155, 163
\bibitem{pea81}
	Provost, J., Berthomieu, G. and Rocca, A., 1981, \aap, 94, 126
\bibitem{rw57}
	Regge, T. and Wheeler, J. A., 1957, \pr, 108, 1063
\bibitem{rls99}
	Rezzolla, L., Lamb, F. K. and Shapiro, S. L., 1999, private
	communication.
\bibitem{rea99}
	Rezzolla, L., Shibata, M., Asada, H., Baumgarte, T. W. and 
	Shapiro, S. L., 1999, ``Constructing a mass-current
	radiation-reaction force for numerical simulations'',
	[gr-qc/9905027]
\bibitem{s89}
	Sawyer, R. F., 1989, \prd, 39, 3804
\bibitem{s82}
	Saio, H., 1982, \apj, 256, 717
\bibitem{sfm99}
	Schneider, R., Ferrari, V. and Mataresse, S., 1999, 
	``Stochastic backgrounds of gravitational waves from cosmological
	populations of astrophysical sources'', [astro-ph/9903470]
\bibitem{st83}
	Schumaker, B. L. and Thorne, K. S., 1983, \mnras 203, 457
\bibitem{st83b}
	Shapiro, S. L. and Teukolsky, S., {\it Black Holes, White Dwarfs
	and Neutron Stars} (Wiley)
\bibitem{sm83}
	Smeyers, P. and Martens, L., 1983, \aap, 125, 193
\bibitem{s99}
	Spruit, H. C., 1999, \aap, 341, L1
\bibitem{s98}
	Stergioulas, N., 1998, \lr, available online at 
	[http://www.livingreviews.org]
\bibitem{t69a}
	Thorne, K. S., 1969, \apj, 158, 1
\bibitem{t69b}
	Thorne, K. S., 1969, \apj, 158, 997
\bibitem{th80}
	Thorne, K. S., 1980, \rmp, 52, 299
\bibitem{tc67}
	Thorne, K. S. and Campolattaro, A., 1967, \apj, 149, 591
\bibitem{vh80}
	Van Horn, H. M., 1980, \apj, 236, 899
\bibitem{w84}
	Wagoner, R. V., 1984, \apj, 278, 345
\bibitem{wald}
	Wald, R. M., 1984, {\it General Relativity}, (Chicago:
	University of Chicago Press)
\bibitem{yl99}
	Yoshida, S. and Lee, U., 1999, in preparation.
\end{thebibliography}

\clearpage
%
\birthplacedate{...}
\begin{startvita}
\end{startvita}
\begin{publications}

1. Lockitch, K. H. and Friedman, J. L., 1998 ``Where are the 
r-modes of Isentropic Stars?'', \apj \ in press, scheduled to 
appear in vol. 521.

2. Andersson, N., Friedman, J. L. and Lockitch, K. H., 1999 
``Rotational Modes of Slowly Rotating Relativistic Stars'', 
(\textit{in preparation}).

3. Friedman, J. L. and Lockitch, K. H., 1999, ``Vacuum Handles 
and the Cosmic Censorship Conjecture'', (\textit{in preparation}).

4. Friedman, J. L., Laguna, P. and Lockitch, K. H., 1999, 
``Stability of Scalar Fields in Rotating Black Hole Spacetimes'', 
(\textit{in preparation}).
\end{publications}
\begin{presentations}

1. July 1999, ``Gravitational waves from unstable neutron stars'', 
University of Wisconsin - Milwaukee, colloquium.

2. June 1999, ``Where are the r-modes of relativistic stars?'',
ITP Conference on Strong Gravitational Fields, Santa Barbara, CA,
contributed talk.

3. April 1999, ``Gravitational waves from hot, rapidly rotating 
neutron stars'',
University of British Columbia, seminar.

4. March 1999, ``Where have all the r-modes gone?'',
Bull. Am. Phys. Soc., \textbf{44}, 996.
American Physical Society Centennial Meeting, Atlanta, GA, 
contributed talk.

5. Nov. 1996, ``Vacuum Handles and the Cosmic Censorship Conjecture'',
Sixth Midwest Relativity Meeting, Bowling Green State University, Ohio, 
contributed talk.
\end{presentations}
\finishvita
%
\end{document}